\documentclass[12pt]{article}
\usepackage{dsfont}

\usepackage{amsmath,amssymb,graphicx,mathabx}
\usepackage{diagbox}
\usepackage{hyperref}
\usepackage[nosort]{cite}
\usepackage{tikz}
\usetikzlibrary{calc}
\usetikzlibrary{patterns}
\usetikzlibrary{decorations.pathreplacing}
\usetikzlibrary{decorations.markings}
\usetikzlibrary{decorations.pathmorphing}
\usetikzlibrary{decorations.text}
\usetikzlibrary{positioning}
\usetikzlibrary{arrows.meta}

\usepackage{color,xcolor}
\definecolor{darkred}{rgb}{0.65,0.15,0}
\definecolor{newgreen}{rgb}{0.2,0.62,0.14}
\hypersetup{pdfborder={0 0 0},colorlinks=true,urlcolor=blue,citecolor=blue,linkcolor=darkred,linktocpage=true}

\numberwithin{equation}{section}

\def\nn{\nonumber}

\def\spa#1.#2{\left\langle#1\,#2\right\rangle}
\def\spb#1.#2{\left[#1\,#2\right]}

\def\ep{\epsilon}

\def\TS{\rm TS}

\newcommand{\esvtau}[1]{{\cal E}^{\rm sv}\! \left[\begin{smallmatrix}#1\end{smallmatrix};\tau\right]}

\newcommand{\eeetau}[1]{{\cal E} \! \left[\begin{smallmatrix}#1\end{smallmatrix};\tau\right]}

\newcommand{\bsv}[1]{\beta^{\rm sv}\! \left[\begin{smallmatrix}#1\end{smallmatrix}\right]}
\newcommand{\bsvtau}[1]{\beta^{\rm sv}\! \left[\begin{smallmatrix}#1\end{smallmatrix};\tau\right]}

\newcommand{\EBR}[2]{{\cal E}\! \left[\begin{smallmatrix}#1\end{smallmatrix};#2\right]}

\newcommand{\EsvBR}[2]{{\cal E}^{\rm sv}\!  \left[\begin{smallmatrix}#1\end{smallmatrix};#2\right]}

\newcommand{\alphaBR}[2]{\alpha\! \left[\begin{smallmatrix}#1\end{smallmatrix};#2\right]}

\newcommand{\SM}[3]{
\begingroup
\setlength\arraycolsep{2pt}
\left(\begin{matrix}#1\\#2\end{matrix}\,;#3\right)
\endgroup
}

\newcommand{\ccb}{\left(\begin{array}{cc}}
\newcommand{\cce}{\end{array}\right)}
\newcommand{\cccb}{\left(\begin{array}{ccc}}
\newcommand{\ccce}{\end{array}\right)}
\newcommand{\ccccb}{\left(\begin{array}{cccc}}
\newcommand{\cccce}{\end{array}\right)}
\font\tenshuffle=shuffle10 \font\sevenshuffle=shuffle7 \font\fiveshuffle=shuffle7 at 5pt
\def\shuffle{{%
\def\Dshuffle{\mathbin{\hbox{\tenshuffle\char'001}}}%
\def\Sshuffle{\mathbin{\hbox{\sevenshuffle\char'001}}}%
\def\SSshuffle{\mathbin{\hbox{\fiveshuffle\char'001}}}%
\mathchoice{\Dshuffle}{\Dshuffle}{\Sshuffle}{\SSshuffle}}}

\def\beq{\begin{equation}}
\def\eeq{\end{equation}}
\let\Re\relax
\let\Im\relax
\DeclareMathOperator{\Re}{Re}
\DeclareMathOperator{\Im}{Im}

\def\modMZV{{\rm mod} \ {\rm MZVs}}

\newcommand{\eq}{\begin{equation}}
\newcommand{\eqe}{\end{equation}}
\newcommand{\eqa}{\begin{eqnarray}}
\newcommand{\eqae}{\end{eqnarray}}

\newcommand{\p}{\partial}

\newcommand{\bea}{\begin{eqnarray}}
\newcommand{\eea}{\end{eqnarray}}
\newcommand{\dd}{\mathrm{d}}

\newcommand{\emptyslot}{~}

\newcommand{\NN}{\mathbb N}
\newcommand{\ZZ}{\mathbb Z}
\newcommand{\QQ}{\mathbb Q}

\newcommand{\dplus}[1]{{\cal D}^+ \! \left[\begin{smallmatrix}#1\end{smallmatrix}\right]}
\newcommand{\cplus}[1]{{\cal C}^+ \! \left[\begin{smallmatrix}#1\end{smallmatrix}\right]}
\newcommand{\chatplus}[1]{\widehat{ {\cal C}}^+ \! \left[\begin{smallmatrix}#1\end{smallmatrix}\right]}
\newcommand{\aplus}[1]{{\cal A}^+ \! \left[\begin{smallmatrix}#1\end{smallmatrix}\right]}
\newcommand{\cminus}[1]{{\cal C}^- \! \left[\begin{smallmatrix}#1\end{smallmatrix}\right]}

\def\b{\beta}
\def\tet{\vartheta}

\def\no{\nonumber}
\def\nabchi{\nabla}

\newbox\charbox
\newbox\slabox
\def\s#1{{      % Feynman slash
        \setbox\charbox=\hbox{$#1$}
        \setbox\slabox=\hbox{$/$}
        \dimen\charbox=\ht\slabox
        \advance\dimen\charbox by -\dp\slabox
        \advance\dimen\charbox by -\ht\charbox
        \advance\dimen\charbox by \dp\charbox
        \divide\dimen\charbox by 2
        \raise-\dimen\charbox\hbox to \wd\charbox{\hss/\hss}
        \llap{$#1$}
}}

\newcommand{\ad}{\text{ad}}

\usepackage[shadow,textwidth=2.7cm]{todonotes}
\usepackage{ifthen}
\reversemarginpar
\newcounter{todocounter}

\colorlet{oscolor}{blue!20!white}

\newcommand{\osinline}[2][]{
  \ifthenelse { \equal {#1} {} }
    { \def\temp {#2} }  % if #1 == blank
    { \def\temp {#1} }   % else (not blank)
  \refstepcounter{todocounter}\todo[color=oscolor,inline,caption={\textbf{\thetodocounter. OS} \temp}]{\textbf{\thetodocounter. OS:} #2}{}}

\colorlet{bvcolor}{violet!50!white}

\newcommand{\bvinline}[2][]{
  \ifthenelse { \equal {#1} {} }
    { \def\temp {#2} }  % if #1 == blank
    { \def\temp {#1} }   % else (not blank)
  \refstepcounter{todocounter}\todo[color=bvcolor,inline,caption={\textbf{\thetodocounter. BV} \temp}]{\textbf{\thetodocounter. BV:} #2}{}}

\colorlet{mhcolor}{red!20!white}

\newcommand{\mhinline}[2][]{
  \ifthenelse { \equal {#1} {} }
    { \def\temp {#2} }  % if #1 == blank
    { \def\temp {#1} }   % else (not blank)
  \refstepcounter{todocounter}\todo[color=mhcolor,inline,caption={\textbf{\thetodocounter. MH} \temp}]{\textbf{\thetodocounter. MH:} #2}{}}

\begin{document}

{\flushright UUITP--34/22\\[5mm]}

\begin{center}

{\LARGE \bf Elliptic modular graph forms II \\[3mm] {\Large Iterated integrals}}\\[5mm]

\vspace{6mm}
\normalsize
{\large  \bf Martijn Hidding, Oliver Schlotterer and Bram Verbeek}

\vspace{8mm}
{\it  Department of Physics and Astronomy\\
Uppsala University, 75108 Uppsala, Sweden}

\vspace{8mm}

{\tt \small martijn.hidding@physics.uu.se} \\
{\tt \small oliver.schlotterer@physics.uu.se}\\
{\tt \small bram.verbeek@physics.uu.se}

\vspace{8mm}

{\bf Abstract}

\vspace{2mm}

\begin{quote}
Elliptic modular graph forms (eMGFs) are non-holomorphic modular forms depending on a modular parameter $\tau$ of a torus and marked points $z$ thereon. Traditionally, eMGFs are constructed from nested lattice sums over the discrete momenta on the worldsheet torus in closed-string genus-one amplitudes. In this work, we develop methods to translate the lattice-sum realization of eMGFs into iterated integrals over modular parameters $\tau$ of the torus with particular focus on cases with one marked point. Such iterated-integral representations manifest algebraic and differential relations among eMGFs and their degeneration limit $\tau \rightarrow i\infty$. From a mathematical point of view, our results yield concrete realizations of single-valued elliptic polylogarithms at arbitrary depth in terms of meromorphic iterated integrals over modular forms and their complex conjugates. The basis dimensions of eMGFs at fixed modular and transcendental weights are derived from a simple counting of iterated integrals and a generalization of Tsunogai's derivation algebra.
\end{quote}

\vspace{6mm}

\end{center}

\thispagestyle{empty}

\newpage
\setcounter{page}{1}

\setcounter{tocdepth}{2}
\tableofcontents

\newpage

%%%%%%%%%%%%%%%%%%%%%%%%%%%%%%%%%%%%%%%%%%%%%%%%%%%%%%%%%%%
%%%%%%%%%%%%%%%%%%%%%%%%%%%%%%%%%%%%%%%%%%%%%%%%%%%%%%%%%%%
\section{Introduction}
\label{sec:1}
%%%%%%%%%%%%%%%%%%%%%%%%%%%%%%%%%%%%%%%%%%%%%%%%%%%%%%%%%%%
%%%%%%%%%%%%%%%%%%%%%%%%%%%%%%%%%%%%%%%%%%%%%%%%%%%%%%%%%%%

String perturbation theory is a steady source of rich mathematical structures.
In particular, the low-energy expansion of open- and closed-string scattering amplitudes
has proven to be a valuable laboratory to study periods of configuration spaces
of punctured Riemann surfaces: As part of the traditional prescription for closed-string
amplitudes, marked points for the external states are integrated over Riemann surfaces
whose genus matches the order in perturbation theory. The expansion coefficients
of these configuration-space integrals in the inverse string tension $\alpha'$ 
span function spaces of joint interest to number theorists, algebraic geometers, string
theorists and particle physicists, such as e.g.\ various flavours of polylogarithms.

Starting from genus one, integration over the closed-string insertion points on
a fixed surface introduces infinite families of non-holomorphic modular functions 
and forms depending on the complex-structure moduli. By the diagrammatic 
organization of the $\alpha'$-expansion at genus one \cite{Green:1999pv, Green:2008uj, DHoker:2015gmr}, these modular 
objects became known as {\it modular graph forms} \cite{DHoker:2015wxz, DHoker:2016mwo} 
and were studied from a variety of perspectives in the physics literature \cite{Green:2013bza, DHoker:2015sve, Basu:2015ayg, Basu:2016xrt, Basu:2016kli, Basu:2016mmk, DHoker:2016quv, Kleinschmidt:2017ege, Basu:2017nhs, Broedel:2018izr, Ahlen:2018wng,Gerken:2018zcy, Gerken:2018jrq, DHoker:2019txf, Dorigoni:2019yoq, DHoker:2019xef, DHoker:2019mib, DHoker:2019blr, Basu:2019idd,  Gerken:2019cxz, Hohenegger:2019tii, Gerken:2020yii, Basu:2020kka, Vanhove:2020qtt, Basu:2020pey, Basu:2020iok, Gerken:2020xfv, Hohenegger:2020slq, Dorigoni:2021jfr, Dorigoni:2021ngn, DHoker:2021ous}
and mathematics literature \cite{Brown:mmv, Zerbini:2015rss, Brown:2017qwo, Brown:2017qwo2, DHoker:2017zhq,Zerbini:2018sox, Zerbini:2018hgs, Zagier:2019eus, Berg:2019jhh, Drewitt:2021}. The reader is referred to \cite{Gerken:review} for an overview
of modular graph forms as of fall 2020, to \cite{Berkovits:2022ivl, Dorigoni:2022iem, DHoker:2022dxx} for a discussion in a broader context, to \cite{Gerken:2020aju} 
for a {\sc Mathematica} package and to \cite{DHoker:2013fcx, DHoker:2014oxd, Pioline:2015qha, DHoker:2017pvk, DHoker:2018mys, Basu:2018bde, DHoker:2020tcq, DHoker:2020uid, Basu:2020goe, Basu:2021xdt} for generalizations beyond genus one.

\begin{figure}[h!]
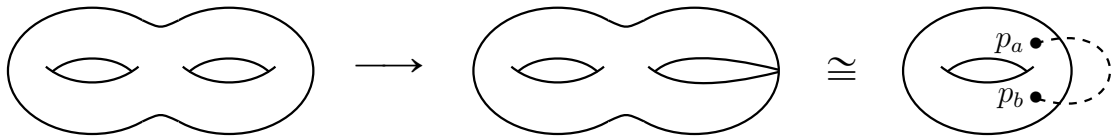

\begin{center}
\tikzpicture[scale=0.28,line width=0.30mm]
\scope[xshift=13.5cm]
\draw(0,0) ellipse  (4cm and 3cm);
\draw(-2.2,0.2) .. controls (-1,-0.8) and (1,-0.8) .. (2.2,0.2);
\draw(-1.9,-0.05) .. controls (-1,0.8) and (1,0.8) .. (1.9,-0.05);
\endscope
%%%
\scope[xshift=20cm]
\draw(0,0) ellipse  (4cm and 3cm);
\draw(-2.2,0.2) .. controls (-1,-0.8) and (1,-0.8) .. (2.2,0.2);
\draw(-1.9,-0.05) .. controls (-1,0.8) and (1,0.8) .. (1.9,-0.05);
\endscope
\draw[white,fill=white] (15.95,2.8)rectangle( 17.55,-2.8);
\draw(15.9,2.4) .. controls (16.75,2) .. (17.6,2.4);
\draw(15.9,-2.4) .. controls (16.75,-2) .. (17.6,-2.4);
\draw (27.6,0.1)node{\Large $ \longrightarrow$};
\scope[xshift=22.1cm]
\scope[xshift=13.5cm]
\draw(0,0) ellipse  (4cm and 3cm);
\draw(-2.2,0.2) .. controls (-1,-0.8) and (1,-0.8) .. (2.2,0.2);
\draw(-1.9,-0.05) .. controls (-1,0.8) and (1,0.8) .. (1.9,-0.05);
\endscope
\scope[xshift=20cm]
\draw(0,0) ellipse  (4cm and 3cm); %%%
\draw(-2.2,0.2) .. controls (-1,-0.8) and (1,-0.8) .. (4,-0.05);
\draw(-1.9,-0.05) .. controls (-1,0.8) and (1,0.8) .. (4,0.05);
\endscope
%
%%%
\draw[white,fill=white] (15.95,2.8)rectangle( 17.55,-2.8);
\draw(15.9,2.4) .. controls (16.75,2) .. (17.6,2.4);
\draw(15.9,-2.4) .. controls (16.75,-2) .. (17.6,-2.4);
\endscope
%%%% NOW THE NODAL STUFF
%
\draw (49.1,0.1)node{\Large $\cong$};
\scope[xshift=56cm]
\draw(0,0) ellipse  (4cm and 3cm);
\draw(-2.2,0.2) .. controls (-1,-0.8) and (1,-0.8) .. (2.2,0.2);
\draw(-1.9,-0.05) .. controls (-1,0.8) and (1,0.8) .. (1.9,-0.05);
\draw(2.3,1.3)node{$\bullet$}node[left]{ $p_a$};
\draw(2.3,-1.3)node{$\bullet$}node[left]{ $p_b$};
\draw[dashed](2.3,1.3) .. controls (7,2.8) and (7,-2.8) .. (2.3,-1.3);
\endscope
\endtikzpicture
\end{center}
\caption{The non-separating degeneration of a genus-two surface gives rise to
a genus-one surface with two nodal points $p_a$ and $p_b$ identified.}
	\label{gen2fig}
\end{figure}

Already at genus one, the expansion of modular graph forms around the cusp 
$\tau \rightarrow i \infty$ features (conjecturally single-valued) multiple zeta values
(MZVs) known from closed-string tree-level amplitudes \cite{Green:2008uj, DHoker:2015gmr, 
Zerbini:2015rss, DHoker:2015wxz, DHoker:2016quv, DHoker:2017zhq, DHoker:2019xef, 
Zagier:2019eus, Vanhove:2020qtt}. 
At higher genus, non-separating degenerations of modular
graph forms introduce a variant of their lower-genus counterparts where a pair 
of marked points is left unintegrated \cite{DHoker:2017pvk}. In particular, the
degeneration of the genus-two surface in figure~\ref{gen2fig} necessitates a 
generalization of modular graph forms at genus one that depends 
non-holomorphically on the difference $z= p_a{-}p_b$ of the nodal points on top
of the modular parameter $\tau$ \cite{DHoker:2018mys}. These generalizations
at genus one dubbed {\it elliptic modular graph forms} (eMGFs) offer valuable 
guidance in the simplification of the genus-two modular graph forms 
\cite{DHoker:2018mys, DHoker:2020tcq} and thereby streamline
the $\alpha'$-expansion of closed-string amplitudes.

Already conventional (i.e.\ $z$-independent) modular graph forms
at genus one obey an intriguing web of algebraic and differential relations
which often intertwine them with MZVs
\cite{DHoker:2015gmr, DHoker:2015sve, DHoker:2016mwo, 
Basu:2016kli, DHoker:2016quv, Gerken:2018zcy}. Many of these relations are obscured in the
original construction of modular graph forms as lattice sums over the discrete
momenta on a worldsheet torus. However, the systematics of their relations
can be manifested by rewriting modular graph forms as iterated integrals over 
holomorphic Eisenstein series and their complex conjugates 
\cite{Broedel:2018izr, Gerken:2020yii, Gerken:2020xfv, Dorigoni:2021jfr}.
The conversion of nested lattice sums over torus momenta to iterated integrals over 
modular parameters $\tau$ yields canonical representations of modular graph forms
in view of the linear-independence results of \cite{Nilsnewarticle}.  

The additional $z$-dependence of eMGFs leads to an even richer network of algebraic and 
differential relations involving echoes of both MZVs and conventional
modular graph forms \cite{DHoker:2020tcq, Basu:2020pey, Basu:2020iok, Dhoker:2020gdz}.
{\it The main result of this work is to expose these relations through an iterated-integral
description of eMGFs in one variable $z$}. In this way, we generalize the canonical 
representations of modular graph forms in terms of iterated Eisenstein integrals. Our key 
method is to solve the differential equations of eMGFs and their generating series 
in \cite{Dhoker:2020gdz} via more general classes of meromorphic iterated integrals over modular 
parameters and their complex conjugates: The integration kernels $G_{k}(\tau)$ of 
iterated Eisenstein integrals are augmented by the Kronecker-Eisenstein
coefficients~$f^{(k)}(u\tau{+}v|\tau)$. The latter are integrated at fixed comoving coordinates 
$u,v \in [0,1]$ of $z=u \tau {+} v$ and reduce to
Eisenstein series of congruence subgroups of $SL(2,\mathbb Z)$ if $u,v \in \mathbb Q$.

%%%%%%%%%%%%%%%%%%%%%%%%%%%%%%%%%%%%%%%%%%%%%%%%%%%%%%%%%%%
\subsection{Mathematical lines of motivation}
\label{sec:1.1}
%%%%%%%%%%%%%%%%%%%%%%%%%%%%%%%%%%%%%%%%%%%%%%%%%%%%%%%%%%%

Apart from the practical motivation of simplifying multiloop string amplitudes, the
results of this work reveal organizing principles and new examples of single-valued 
period functions \cite{brown2015notes, Brown:2018omk}. The simplest ``depth-one'' examples 
of eMGFs are Zagier's single-valued elliptic polylogarithms \cite{Ramakrish}.
The latter are known to be expressible as infinite sums over single-valued
polylogarithms at genus zero that essentially average over arguments $z {+} n \tau$
with $n \in \ZZ$. More general eMGFs are expected to cover
single-valued elliptic polylogarithms at arbitrary depth \cite{DHoker:2015wxz, Dhoker:2020gdz}.
Indeed, we pinpoint single-valued genus-zero polylogarithms of higher depth 
in the expansion of eMGFs around the cusp.

More importantly, our iterated-integral representations of eMGFs
amount to decomposing single-valued elliptic polylogarithms into meromorphic building
blocks and their complex conjugates. This was already the organizing principle of
Brown's single-valued genus-zero polylogarithms in one variable \cite{svpolylog} with
multivariable generalizations in \cite{Broedel:2016kls, DelDuca:2016lad}. 
From the viewpoint of closed-string amplitudes, the organization of eMGFs in this
work combines meromorphic and antimeromorphic open-string constituents, 
following the spirit of the celebrated Kawai-Lewellen-Tye (KLT) relations among
tree amplitudes \cite{Kawai:1985xq}.
More recently, it was investigated from a multitude of perspectives \cite{Schlotterer:2012ny,
Stieberger:2013wea, Stieberger:2014hba, Schlotterer:2018abc, Brown:2018omk, 
Vanhove:2018elu, Brown:2019wna} that the KLT formula for closed-string tree-level 
amplitudes is equivalent to the single-valued map of MZVs 
\cite{Schnetz:2013hqa, Brown:2013gia} in open-string $\alpha'$-expansions.

Even though a KLT-formula for closed-string amplitudes beyond genus zero is currently
unknown, iterated-integral representations of modular graph forms reveal striking
connections between open- and closed-string integrals at genus one \cite{Broedel:2018izr, 
Panzertalk, Zagier:2019eus, Gerken:2020xfv}. More specifically, the decomposition into 
meromorphic building blocks and their complex conjugates 
reduces modular graph forms to real-analytic combinations
of Enriquez' elliptic multiple zeta values (eMZVs) \cite{Enriquez:Emzv} or equivalently
meromorphic iterated Eisenstein integrals \cite{Broedel:2015hia}.
The reduction of modular graph forms to
eMZVs in the above references is a concrete showcase of 
single-valued integration at genus one \cite{Panzertalk},
and the precise connection of modular graph forms with Brown's single-valued
iterated Eisenstein integrals \cite{Brown:mmv, Brown:2017qwo, Brown:2017qwo2} will
be discussed in \cite{MGFinprogress}.

The iterated-integral representations of eMGFs developed in this work
uplift the relations between modular graph forms and eMZVs to functions
of both $z=u\tau{+}v$ and $\tau$. By the expansion methods for open-string
genus-one amplitudes in \cite{Broedel:2014vla, Broedel:2017jdo}, the Brown-Levin 
elliptic polylogarithms \cite{BrownLev} are the natural open-string counterparts to the eMGFs 
in closed-string integrals. At depth one, Zagier's single-valued elliptic polylogarithms have already 
been expressed \cite{Broedel:2019tlz} in terms of real-analytic combinations
of elliptic polylogarithms and eMZVs, and a path to higher-depth 
generalizations is discussed in \cite{Panzertalk}. 
The iterated $\tau$-integrals of $G_k(\tau)$ and $f^{(k)}(u\tau{+}v|\tau)$ to be employed as
meromorphic building blocks for eMGFs furnish an alternative organization
of Brown-Levin elliptic polylogarithms: Instead of their traditional definition via
iterated $z$-integrals, we encounter elliptic polylogarithms in a form that
results from solving their differential equations in $\tau$ \cite{Broedel:2018iwv}.

Similar to the case of modular graph forms \cite{Gerken:2020yii, Gerken:2020xfv, Dorigoni:2021jfr},
we find dropouts of certain iterated $\tau$-integrals from the space of eMGFs. These dropouts
are traced back to commutation relations in a Lie algebra of operators $b_k,\epsilon_k$ 
that generalizes Tsunogai's derivation algebra of $\epsilon_k^{\rm TS}$ ``dual'' 
to $G_k(\tau)$ \cite{Tsunogai, Pollack}. In the first place, we only have access to
matrix representations of $b_k$ and $\epsilon_k$ from the appearance of
$f^{(k)}(u\tau{+}v|\tau)$ and $G_k(\tau)$ in the KZB-type differential equations of 
generating series of eMGFs \cite{Dhoker:2020gdz}. As we will see, simple observations on 
the $z\rightarrow 0$ behaviour of eMGFs uplift the information from specific matrix
representations to determine large sets of commutation relations among $b_k$ 
and $\epsilon_k$. These commutators will be exploited to determine the counting
of independent eMGFs for a variety of transcendental and modular weights and
to present explicit bases.

The algebra of $b_k,\epsilon_k$ and its multi-variable generalization has 
been pioneered via open-string methods in \cite{Broedel:2020tmd, Kaderli:2022qeu}
and should also govern the counting of independent eMGFs in several variables $z_i$.
Moreover, polylogarithms and modular graph forms at higher genus are expected to 
involve even richer Lie-algebra structures, for example through the differential equations 
of suitably chosen generating series. We hope that the organization of eMGFs 
in \cite{Dhoker:2020gdz} and this work via generating series and meromorphic iterated
integrals will serve as a useful stepping stone
% prototype 
for future studies of configuration-space integrals on higher-genus surfaces
and their implications for string amplitudes.

%%%%%%%%%%%%%%%%%%%%%%%%%%%%%%%%%%%%%%%%%%%%%%%%%%%%%%%%%%%
\subsection{Summary of results}
\label{sec:1.2}
%%%%%%%%%%%%%%%%%%%%%%%%%%%%%%%%%%%%%%%%%%%%%%%%%%%%%%%%%%%

A central guiding principle in the writeup of this work was to illustrate the general structures
and the onset of new levels of complexity via pedagogical examples. The resulting 
length of this paper calls for a summary of its main results in this section.

%%%%%%%%%%%%%%%%%%%%%%%%%%%%%%%%%%%%%%%%%%%%%%%%%%%%%%%%%%%
\subsubsection{Brief recap of dihedral eMGFs}
\label{sec:1.2.1}
%%%%%%%%%%%%%%%%%%%%%%%%%%%%%%%%%%%%%%%%%%%%%%%%%%%%%%%%%%%

The simplest non-trivial examples of eMGFs are  
Zagier's single-valued elliptic polylogarithms \cite{Ramakrish}
\beq
\dplus{a  \\ b }\! (z|\tau)
= { (\Im \tau)^{a} \over \pi^{b }}  \sum _{p \in \Lambda'}   \,
 {  e^{2\pi i (nu - mv) }\over p^{a} \bar p^{b}} \, , \ \ \ \ a,b \in \mathbb Z 
 \label{summ.01}
\eeq
with modular parameter $\tau$ of the torus in the upper half plane
${\cal H}= \{\tau \in \mathbb C\, : \ \Im \tau > 0\}$.
Here and below, we refer to the marked point $z$ on a torus and the discretized lattice momentum
$p$ through its comoving coordinates $u,v \in [0,1]$
and the pair of integers $m,n \in \ZZ$ with $(m,n) \neq (0,0)$,
\beq
z= u \tau {+} v \in \mathbb C/(\mathbb Z \tau{+}\mathbb Z)
\, , \ \ \ \ \ \
p = m\tau{+}n 
%\, , \ \ \ \
\in \Lambda'  \, ,\ \ \ \ \ \
 \Lambda'  = (\mathbb Z \tau{+}\mathbb Z) \setminus\{0\}
 \label{summ.02}
\eeq
Generic eMGFs generalize (\ref{summ.01}) to multiple, nested
sums over lattice momenta $p_1,p_2,\ldots$, and their representatives associated with
dihedral graphs are given by \cite{Dhoker:2020gdz}
\beq
\cplus{a_1 &a_2 &\ldots &a_R \\ b_1 &b_2 &\ldots &b_R \\ z_1 &z_2 &\ldots &z_R}\! (\tau)
= { (\Im \tau)^{a_1+a_2+\ldots+a_R} \over \pi^{ b_1+b_2+\ldots+b_R }} \! \! \sum _{p_1,\ldots ,p_R \in \Lambda'} \! \! \delta \bigg(\sum_{s=1}^R p_s \bigg) \,
\prod _{r=1}^R { e^{2\pi i (n_r u_r - m_r v_r) }
 \over p_r ^{a_r} \bar p_r^{b_r}}
  \label{summ.03}
\eeq
See the reference for comments on trihedral and more general graph topologies.
While the exponents in (\ref{summ.03}) may in principle take arbitrary
integer values $a_j ,b_j \in \mathbb Z$, we shall mostly be interested in cases
with $a_j,b_j \geq 0$ and absolutely convergent sums with $a_r{+}a_{r'}{+}b_r{+}b_{r'}>2$
for each pair $1\leq r<r' \leq R$. In the normalization conventions of (\ref{summ.03}),
eMGFs are modular forms of weight $(0,\sum_{r=1}^R b_r{-}a_r)$ under 
$(z,\tau) \rightarrow (\frac{z}{\gamma \tau + \delta} , \frac{ \alpha \tau + \beta}{\gamma \tau + \delta})$ with $( \begin{smallmatrix}  \alpha & \beta \\ \gamma & \delta \end{smallmatrix}) \in SL(2,\ZZ)$, 
i.e.\ they transform with a purely antimeromorphic factor of $\prod_{r=1}^R (\gamma 
\bar \tau {+} \delta)^{b_r-a_r}$.

Modular graph forms are recovered from the
$z_r \rightarrow 0$ limit of (\ref{summ.03}) where the characters
$e^{2\pi i (n_r u_r - m_r v_r) }$ are set to unity. At vanishing $z_r$, modular graph forms 
are associated with one-particle irreducible Feynman graphs on the torus where 
each pair $(a_j,b_j) \in\mathbb Z^2$ specifies a decorated edge, and each vertex 
represents an integral of $z$ over the torus, $u,v \in [0,1]$. In case of
eMGFs at $z_r \neq 0$, these graphs are opened 
(see e.g.\ \cite{DHoker:2015wxz} for the one-variable case), and each 
distinct $z_r$ corresponds to an unintegrated vertex.

There is a huge variety of relations among eMGFs (\ref{summ.03}) with different 
sets of $a_j,b_j$ which appear mysterious
from the nested lattice sums. The main achievement of this work is to pass to
iterated-integral representations where the entirety of these relations is manifest. 
This will be done on the basis of the differential equations of (\ref{summ.03}) 
w.r.t.\ both $z$ and $\tau$, where repeated derivatives allow to peel off the
simpler lattice sums \cite{Dhoker:2020gdz}
\begin{align}
G_k(\tau) &= \sum _{p \in \Lambda'} \frac{1}{p^k} \, ,  \ \ \ \ k\geq 4   \label{summ.04} \\
f^{(k)}(u\tau{+}v|\tau) &= -  \sum _{p \in \Lambda'} \frac{ e^{2\pi i (nu - mv) } }{p^k}  \, ,  \ \ \ \ k\geq 2
\notag \end{align} 
i.e.\ holomorphic Eisenstein series $G_k$ and coefficients $f^{(k)}$ of the
doubly-periodic Kronecker-Eisenstein series. Both of them are modular forms of 
purely holomorphic weight $(k,0)$.

%%%%%%%%%%%%%%%%%%%%%%%%%%%%%%%%%%%%%%%%%%%%%%%%%%%%%%%%%%%
\subsubsection{Iterated integrals -- meromorphic and real-analytic}
\label{sec:1.2.2}
%%%%%%%%%%%%%%%%%%%%%%%%%%%%%%%%%%%%%%%%%%%%%%%%%%%%%%%%%%%

At fixed values of the comoving coordinates $u,v \in [0,1]$, both of
$G_k(\tau)$ and $f^{(k)}(u\tau{+}v|\tau)$ in (\ref{summ.04}) are meromorphic in $\tau$.
Hence, they give rise to homotopy-invariant iterated integrals 
\beq
 \int^\tau_{i\infty} \dd \tau_1 \, \tau_1^{j_1} f^{(k_1)}(u_1 \tau_1{+}v_1 | \tau_1)
  \int^{\tau_1}_{i\infty} \dd \tau_2 \, \tau_2^{j_2} f^{(k_2)}(u_2 \tau_2{+}v_2 | \tau_2)
   \int^{\tau_2}_{i\infty} \dd \tau_3  \ldots 
    \label{summ.05} 
   \eeq
with tangential-base-point regularization of the endpoint divergences 
at $\tau_j \rightarrow i\infty$ \cite{Brown:mmv}. By the differential equations
of eMGFs in \cite{Dhoker:2020gdz}, the exponents $j_i$
of the integrands $\tau_i$ only need to be considered in the range $0\leq j_i \leq k_{i}{-}2$,
in line with the selection rules on Brown's iterated Eisenstein integrals \cite{Brown:mmv}.

Given that $u= \frac{ \Im z}{\Im \tau}$, holomorphicity of the integrals (\ref{summ.05}) 
in $\tau$ is tied to keeping $u,v$ rather than $z,\bar z$ fixed. The iterated integrals (\ref{summ.05}) 
and their variants with some of the $ f^{(k_i)}(u_i \tau_i{+}v_i | \tau_i)$
replaced by $G_{k_i}(\tau_i)$ furnish the advertised meromorphic
building blocks of eMGFs. The coefficients of the iterated
integrals (\ref{summ.05}) in eMGFs will be found to comprise MZVs, rational
functions of $\tau$ and $\bar \tau$ as well as Bernoulli polynomials in $u$.

By the equal footing of holomorphic and antiholomorphic exponents in (\ref{summ.03}),
eMGFs close under complex conjugation (up to convention-dependent
powers of $\Im \tau$ and $\pi$) and necessitate real-analytic versions of (\ref{summ.05}). 
Following the real-analytic iterated Eisenstein integrals
\begin{align}
\bsvtau{j\\k} &=  \frac{(2\pi i)^{-1}}{(4 \pi \Im \tau)^{k-2-j}} \bigg\{ \int_{\tau}^{i \infty} \dd \tau_1 (\tau{-}\tau_1)^{k-2-j} (\bar\tau{-}\tau_1)^{j} G_k(\tau_1)  \label{summ.06} \\
&\quad \quad\quad \quad\quad \quad \quad \quad
 - \int_{\bar\tau}^{-i\infty}  \dd\bar\tau_1 (\tau{-}\bar\tau_1)^{k-2-j} (\bar\tau{-}\bar\tau_1)^{j} \overline{G_k(\tau_1)} \bigg\} \notag
\end{align}
encountered in modular graph forms \cite{Gerken:2020yii, Dorigoni:2021jfr}, the 
augmentation of integration kernels $G_{k_i}(\tau_i) \rightarrow - f^{(k_i)}(u_i \tau_i{+}v_i | \tau_i)$ 
leads to the $u,v$-dependent building block
\begin{align}
\bsvtau{j\\k \\ z} &=  \frac{(2\pi i)^{-1}}{(4 \pi \Im \tau)^{k-2-j}} \bigg\{
(-1)^k \int^{\tau}_{i \infty} \dd \tau_1 (\tau{-}\tau_1)^{k-2-j} (\bar\tau{-}\tau_1)^{j} f^{(k)}(u \tau_1{+}v | \tau_1)  \notag\\
&\quad \quad\quad \quad\quad \quad \quad \quad
 - \int^{\bar\tau}_{-i\infty}  \dd\bar\tau_1 (\tau{-}\bar\tau_1)^{k-2-j} (\bar\tau{-}\bar\tau_1)^{j} \overline{ f^{(k)}(u \tau_1{+}v | \tau_1) } \bigg\} \label{summ.07}
 \end{align}
with $0\leq j \leq k{-}2$ in both cases, and where $k\geq 2$ in (\ref{summ.07})
may also take odd values.
Generalizations to higher depth $\ell \in \mathbb N$ will be denoted by 
$\bsvtau{j_1 &j_2 &\ldots &j_\ell \\k_1 &k_2 &\ldots &k_\ell \\ z_1 &z_2 &\ldots &z_\ell}$
and mix holomorphic with antiholomorphic integration kernels as known
from the constituents of modular graph forms. As detailed in (\ref{eq:ebsv2}), the
combinations of integration kernels in the $u_1,u_2,v_1,v_2$-dependent generalization 
of $\bsvtau{j_1 &j_2\\k_1 &k_2}$ are $f^{(k_2)}f^{(k_1)}$ as well as $
f^{(k_2)}   \overline{ f^{(k_1)}  }$ and $\overline{f^{(k_1)} }   \overline{ f^{(k_2)} }$.
Moreover, $\beta^{\rm sv}$ at depth $\geq 2$ feature certain admixtures of 
antiholomorphic terms $\overline{ \alpha[\ldots;\tau]}$ in $\tau$ (see section \ref{sec:9.7}) 
where each contribution comprises a zeta-factor as known from
their $u,v$-independent counterparts in the discussion of 
modular graph forms in \cite{Gerken:2020yii, Dorigoni:2021jfr}.

Replacing one of the kernels $ f^{(k_i)}(u_i \tau_i{+}v_i | \tau_i)$ with $k_i\geq 4$ 
by $- G_{k_i}(\tau_i)$ amounts to setting $z_i\rightarrow 0$ at the level of the integrand
w.r.t.\ $\tau_i$. However, this limit does not commute with iterated integration over $\tau_i$, and
already at depth one, $\lim_{z \rightarrow 0}\bsvtau{j   \\k  \\ z  }$
differs from $\bsvtau{j \\k }$ by odd zeta values. The
analogous deviations at higher depth will be deduced from single-valued 
polylogarithms at genus zero \cite{svpolylog} which evaluate to single-valued MZVs
\cite{Schnetz:2013hqa, Brown:2013gia}. We shall use an empty-slot notation 
$\bsvtau{j_1 &\ldots &j_i &\ldots &j_\ell \\k_1 &\ldots &k_i &\ldots &k_\ell \\ 
z_1 &\ldots &\emptyslot &\ldots &z_\ell}$ to refer to an integration kernel $-G_{k_i}(\tau_i)$ 
in the place of $ f^{(k_i)}(u_i \tau_i{+}v_i | \tau_i)$.

The main result of this work is that eMGFs depending on $\tau$
and one marked point $z$ can be expressed via real-analytic iterated integrals
\beq
\bigg\{ \bsvtau{j_1 &j_2 &\ldots &j_\ell \\k_1 &k_2 &\ldots &k_\ell \\ z_1 &z_2 &\ldots &z_\ell}  
\, , \ \ \ \ 0 \leq j_i \leq k_i{-}2 \, , \ \ \ \
k_i \geq 2
\bigg\}
\label{summ.08} 
\eeq
with rational combinations of MZVs, integer powers of $\Im \tau$
and polynomials in $u$ as coefficients. Hence, the number of
independent eMGFs at fixed modular and transcendental weight
can be bounded by enumerating the elements of (\ref{summ.08}),
where kernels $G_{k_i}(\tau_i)$ only occur at even $k_i \geq 4$. In fact, we will derive
a simple dictionary in section \ref{sec:9.1.6} between the sums of exponents
$\sum_{r=1}^R a_r,\sum_{r=1}^R b_r$ in (\ref{summ.03}) 
encoding modular \& transcendental weight and the admissible values of
of $\ell, \sum_{i=1}^\ell j_i$ and $\sum_{i=1}^\ell k_i$ in (\ref{summ.08}). Our counting
is based on the working assumption that iterated integrals involving 
different kernels $\tau^{j_i} G_{k_i}$ and $\tau^{j_i} f^{(k_i)}$ are linearly independent
(which is established for cases without the $f^{(k_i)}$ \cite{Nilsnewarticle} and
can otherwise be tested at the level of $q$-expansions).

%%%%%%%%%%%%%%%%%%%%%%%%%%%%%%%%%%%%%%%%%%%%%%%%%%%%%%%%%%%
\subsubsection{Dropouts from iterated integrals and derivations}
\label{sec:1.2.3}
%%%%%%%%%%%%%%%%%%%%%%%%%%%%%%%%%%%%%%%%%%%%%%%%%%%%%%%%%%%

Not all of the real-analytic iterated integrals (\ref{summ.08}) are independently realized as
eMGFs. We will pinpoint the dropouts by studying generating series
of $n$-point configuration-space integrals with Koba-Nielsen factors akin to closed-string genus-one 
amplitudes that comprise all convergent eMGFs \cite{Dhoker:2020gdz}. 
The differential equations of these generating series are solved by the group-like element
\small
\begin{align}
&\Phi(z|\tau)= 1+ \sum_{k_1=2}^\infty \sum_{j_1=0}^{k_1-2} \frac{(-1)^{k_1-j_1}(k_1{-}1)}{(k_1{-}j_1{-}2)!}
\bigg(   \bsvtau{j_1 \\ k_1 \\  \emptyslot}  {\rm ad}_{\epsilon_0}^{k_1-j_1-2}(\ep_{k_1})
+\bsvtau{j_1 \\ k_1 \\ z }  {\rm ad}_{\epsilon_0}^{k_1-j_1-2}(b_{k_1})  \bigg)  \notag \\
&\ + \sum_{k_1,k_2=2}^\infty \sum_{j_1=0}^{k_1-2} \sum_{j_2=0}^{k_2-2} 
\frac{(-1)^{k_1+k_2-j_1-j_2}(k_1{-}1)(k_2{-}1)}{(k_1{-}j_1{-}2)! (k_2{-}j_2{-}2)!} 
 \bigg(  \bsvtau{j_1 &j_2 \\ k_1 &k_2 \\  \emptyslot&\emptyslot } {\rm ad}_{\epsilon_0}^{k_2-j_2-2} (\ep_{k_2})  {\rm ad}_{\epsilon_0}^{k_1-j_1-2}(\ep_{k_1}) 
\notag \\
&\quad \quad 
+ \bsvtau{j_1 &j_2 \\ k_1 &k_2 \\ z &\emptyslot }  {\rm ad}_{\epsilon_0}^{k_2-j_2-2}( \ep_{k_2} ) {\rm ad}_{\epsilon_0}^{k_1-j_1-2}( b_{k_1}  )
+ \bsvtau{j_1 &j_2 \\ k_1 &k_2 \\ \emptyslot &z } {\rm ad}_{\epsilon_0}^{k_2-j_2-2}( b_{k_2} )
{\rm ad}_{\epsilon_0}^{k_1-j_1-2}( \ep_{k_1} ) 
\notag \\
&\quad \quad 
+ \bsvtau{j_1 &j_2 \\ k_1 &k_2 \\ z &z }  {\rm ad}_{\epsilon_0}^{k_2-j_2-2}(b_{k_2} ) {\rm ad}_{\epsilon_0}^{k_1-j_1-2} (b_{k_1} )
\bigg) + \ldots
 \label{summ.09}
\end{align} \normalsize
acting on a series of MZVs, non-positive powers of $\Im \tau$ and
polynomials in $u$ that captures the asymptotics as $\tau \rightarrow i \infty$.
For the $n$-point generating series, the operators $\epsilon_0,\epsilon_{k\geq 4}$ and 
$b_{k\geq 2}$ are explicitly known $n!\times n!$ matrices with a simple dependence 
on auxiliary variables $s_{ij} ,\eta_j,\bar\eta_j$ and $\partial_{\eta_j}$, 
see \cite{Broedel:2020tmd, Dhoker:2020gdz} 
for details and \cite{Gerken:2019cxz, Gerken:2020yii} for the analogous generating 
series of modular graph forms. The ellipsis in (\ref{summ.09}) refers to an infinite tower of 
higher-depth terms where the correlation between the entries of $\beta^{\rm sv}$ and 
the composition of operators follows the patterns at depth two.

For each relation among the operators $\epsilon_{k},b_{k}$ that holds universally for  
the generating series at all $n\geq 2$, there is one linear combination of
$\beta^{\rm sv}$ which cannot arise from eMGFs.
As will be detailed in section \ref{sec:5.2}, representative instances of such
universal relations read
\begin{align}
0 &= [b_2,\ep_4] + [b_2,b_4]  \, , \ \ \ \ 0 = [b_4,\ep_4] + [b_2,b_6]+ 2 [b_3,b_5] 
\notag \\
 0 &= [b_3,  {\rm ad}_{\epsilon_0} (b_4)]+[b_3,  {\rm ad}_{\epsilon_0} (\ep_4)]
 +2 [b_4, {\rm ad}_{\epsilon_0} (b_3) ]+2 [\ep_4, {\rm ad}_{\epsilon_0} (b_3)]
- 5 [b_2,[b_2,b_3]] 
 \label{summ.10}
 \\
 0 &= [b_4,   {\rm ad}_{\epsilon_0}^2( \ep_4)]
- [\ep_4,  {\rm ad}_{\epsilon_0}^2(b_4)]
- [  {\rm ad}_{\epsilon_0}  (b_4),  {\rm ad}_{\epsilon_0} (\ep_4)]
-   20 [ b_2, [b_3,  {\rm ad}_{\epsilon_0}  (b_3)] ]
 \notag 
\end{align}
and one can form further relations via left- or right-action of arbitrary $ \epsilon_k, b_k$.

A closely related setup determined the counting of modular graph forms in section~6.2
of \cite{Gerken:2020yii} where the operators $b_k$ were absent. In this reference, the precursors
of $ \epsilon_k, b_k$ formed conjectural matrix representations of Tsunogai's derivation
algebra \cite{Tsunogai, Pollack} $\{\ep_k^{\rm TS}, \ k \in 2\mathbb N_0\}$. 
Accordingly, the larger set of generators $\{\epsilon_0,\epsilon_{k\geq 4},b_{k\geq 2}\}$ 
in (\ref{summ.09}) associated with eMGFs is interpreted as an extension of said derivation 
algebra. The characteristic commutation relations among $\ep_k^{\rm TS}$ should
be preserved under $\ep_{k}^{\rm TS} \rightarrow \ep_k + b_k$ if $k>0$ and 
$\ep_{0}^{\rm TS} \rightarrow \ep_0$ according to the discussions 
of \cite{Broedel:2020tmd, Kaderli:2022qeu} in an open-string context.

Tsunogai's $\ep_{k}^{\rm TS}$ can be realized as derivations acting on a free Lie algebra
in two generators $x,y$ that for instance occur in the differential equations of the elliptic
KZB associator \cite{KZB, EnriquezEllAss, Hain}. We leave it to the future to find a similar 
realization for the extension via $\ep_k , b_k$ which might determine relations such 
as (\ref{summ.10}) by acting on the tentative generators $x,y$. 
In this work, we instead use the compatibility of (\ref{summ.09}) 
with the singularity structure of eMGFs
and the matrix representations of \cite{Broedel:2020tmd, Dhoker:2020gdz} to derive
all commutator relations among $\epsilon_{k_i},b_{k_i}$ up to $\sum_{i} k_i=11$ that hold
universally for the generating series at all $n\geq 2$.

%%%%%%%%%%%%%%%%%%%%%%%%%%%%%%%%%%%%%%%%%%%%%%%%%%%%%%%%%%%
\subsubsection{Bases of eMGFs up to weight five}
\label{sec:1.2.4}
%%%%%%%%%%%%%%%%%%%%%%%%%%%%%%%%%%%%%%%%%%%%%%%%%%%%%%%%%%%

A major achievement in this work is to present bases of eMGFs (\ref{summ.03}) for a 
wide range of modular and transcendental weights, namely
$\sum_{r=1}^R(a_r{+}b_r) \leq 10$. An upcoming {\sc Mathematica}
package \cite{Hidding:2022zzz} will make the basis decompositions of other eMGFs in
this range publicly available. Our first guiding principle is to represent as many 
basis elements as possible via products of simpler objects -- eMGFs
of lower weights, modular graph forms and/or MZVs. In other words, the
actual goal is to span the {\it indecomposable} eMGFs at given weights
that cannot be reduced to such products.

At the level of real-analytic iterated integrals $\beta^{\rm sv}$, the quest for 
indecomposable eMGFs amounts to modding out by
their shuffle relations such as
\beq
\bsvtau{j_1\\k_1 \\ z_1}\bsvtau{j_2\\k_2 \\ z_2} = \bsvtau{j_1 &j_2 \\k_1 &k_2 \\ z_1 &z_2 }
+ \bsvtau{j_2 &j_1 \\k_2 &k_1 \\ z_2 &z_1 }
 \label{summ.31}
\eeq
Upon discarding shuffles, some of the $\beta^{\rm sv}$ in the set (\ref{summ.08}) 
that delimits the counting of eMGFs can be eliminated, e.g.\
$\bsvtau{j_1 &j_2 \\k_1 &k_2 \\ z_1 &z_2 } = -  \bsvtau{j_2 &j_1 \\k_2 &k_1 \\ z_2 &z_1 }$
modulo products. The generating series (\ref{summ.09}) becomes Lie-algebra valued
when modding out by shuffles, i.e.\ the operators ${\rm ad}_{\epsilon_0}^{k-j-2}( \ep_{k} ) $
and ${\rm ad}_{\epsilon_0}^{k-j-2}( b_{k} )$ conspire to nested commutators.
This ties in with the form of all the indecomposable relations among $\ep_k$ and $b_k$
such as (\ref{summ.10}), and their corollaries relevant for indecomposable 
eMGFs are obtained from adjoint action of further $\ep_k$ or $b_k$.
Note that the nilpotency properties $ {\rm ad}_{\epsilon_0}^{k-1} (\ep_{k})
=  {\rm ad}_{\epsilon_0}^{k-1} (b_{k})=0 $ akin to $ {\rm ad}_{\epsilon^{\rm TS}_0}^{k-1} (\ep^{\rm TS}_{k})=0$
in Tsunogai's derivation algebra imply that
such derived relations can never leave the operators ${\rm ad}_{\epsilon_0}^{j} (\ep_k)$
and ${\rm ad}_{\epsilon_0}^{j} (b_k)$ with $0\leq j\leq k{-}2$.

To make our strategy towards bases of indecomposable eMGFs (\ref{summ.03}) 
more concrete, we shall preview the counting of modular invariant representatives 
of transcendental weight $w=\sum_{r=1}^R a_r $ up to and including five. 
Up to terms involving MZVs, the entries of the $\beta^{\rm sv}$ in their 
iterated-integral representations obey the selection rules
\beq
\left. \begin{array}{c}
{\rm modular}\ {\rm invariant} \\
{\rm eMGFs} \ @\ {\rm weight} \ w 
\end{array} \right\} \
\longleftrightarrow  \ \left\{
\begin{array}{l}
 \beta^{\rm sv} \ {\rm in} \ (\ref{summ.08}) \ {\rm with}
\ {\rm depth} \ 0\leq \ell \leq \lfloor \frac{w}{2} \rfloor \ {\rm and} \\\ 
\; \sum_{i=1}^\ell j_i = w{-}\ell \ {\rm as} \ {\rm well} \ {\rm as}
\, \sum_{i=1}^\ell k_i = 2 w
\end{array} \right.
 \label{summ.11}
\eeq
At weight $w=3$, for instance, the following shuffle-independent $\beta^{\rm sv}$
compatible (\ref{summ.11}) are identified in section \ref{sec:5.8.3}:
\beq
\bsvtau{2 \\ 6 \\ z}, \ \bsvtau{2 \\ 6 \\ \emptyslot}, \ \bsvtau{1 &0 \\ 3 &3 \\ z &z }, \
\bsvtau{0 &1 \\ 2 &4 \\ z &z } , \ \bsvtau{0 &1 \\ 2 &4  \\ z &\emptyslot }
 \label{summ.12}
\eeq
However, the latter two do not occur independently in (\ref{summ.09}) 
by the corollary $[b_2,{\rm ad}_{\epsilon_0}(\ep_4)] 
+ [b_2,{\rm ad}_{\epsilon_0}(b_4)]=0$ of the first line of (\ref{summ.10}). 
As a consequence, we have four instead of five indecomposable modular invariant
eMGFs at weight $w= 3$, and the guiding principles detailed in section \ref{bassec.2.2} lead us 
to chose a basis (see (\ref{summ.01}) for the definition of
$\dplus{a \\ b}\! (z|\tau)$)
\beq
\dplus{3  \\ 3 }\! (z|\tau) , \ 
\dplus{3  \\ 3 }\! (0|\tau) , \
\cplus{1 &1 &1 \\ 1 &1 &1 \\ z  &0 &0}\! (\tau) , \
\cplus{2 &0 &1 \\ 0 &2 &1 \\ z &0 &0}\! (\tau) - \cplus{0 &2 &1 \\ 2 &0 &1 \\ z &0 &0}\! (\tau)
 \label{summ.13}
\eeq
where $\dplus{3  \\ 3 }\! (0|\tau)= E_3(\tau)$ is a non-holomorphic Eisenstein series,
and the last element is taken to be a difference to form an imaginary quantity.

The same counting method yields the basis dimensions of indecomposable 
modular invariant eMGFs up to and including $w= 5$ noted in table \ref{countinvs}.
Similar to the basis (\ref{summ.13}) at weight three, dihedral eMGFs in (\ref{summ.03})
turn out to be sufficient to express eMGFs of arbitrary topologies at weight $w\leq 5$.
Moreover, eMGFs with non-trivial modular weight $\sum_{r=1}^R a_r \neq \sum_{r=1}^R b_r$
can be organized through the same strategy, and we will construct basis elements
from derivatives of modular invariant eMGFs w.r.t.\ $z$ and $\tau$. Note, however, that
our counting is restricted to eMGFs that depend on a single marked point $z$ on the torus --
multi-variable eMGFs depending on $z_1,z_2,\ldots$ will in general admit further indecomposable
representatives.

\begin{table}[h]
\begin{center}
\begin{tabular}{c||c|c|c|c|c}
%\diagbox[innerwidth=1.5em,innerheight=1.5em]{$n$}{$w$} &0&1&2&3&4&5 \\\hline\hline
weight $w$ &1&2&3
&4&5  \\\hline \hline 
\# \ $\beta^{\rm sv}$ mod $\shuffle$ &1 &2 &5 &15 &51\\\hline 
\# \ relations among $\epsilon_k,b_k$ &0 &0 &1 &5 &22\\\hline 
\# \ indecomposable eMGFs &1 &2 &4 &10 &29
\end{tabular}
\end{center}
\caption{\textit{The counting of modular invariant indecomposable eMGFs at
transcendental weights $w\leq 5$.}}
\label{countinvs}
\end{table}

\subsection*{Organization}

The remainder of this work is organized as follows: After reviewing selected aspects
of (elliptic) modular graph forms in section \ref{sec:2}, we discuss properties of iterated 
integrals over $G_{k}$ and $f^{(k)}$ and the construction of real-analytic combinations
$\beta^{\rm sv}$ in section \ref{sec:3}. 
The counting of independent eMGFs based on an extension $\{\epsilon_{0},\epsilon_{k\geq 4},
b_{k\geq 2}\}$ of Tsunogai's derivation algebra is performed in section \ref{sec:5}.
Section \ref{sec:9} then offers two approaches to systematically convert eMGFs into 
linear combinations of $\beta^{\rm sv}$. This will be used in section~\ref{bassec}
to describe a choice of dihedral bases for eMGFs for a certain range of modular
and transcendental weights. Several appendices complement the discussion
of the main text via further review material, more technical extra information
and additional examples. The ancillary files of the arXiv submission gather
many of our results in machine-readable form, and a follow-up paper
\cite{Hidding:2022zzz} will describe a {\sc Mathematica} package implementing
our methods.

\subsection*{Acknowledgments}

We are grateful to Marco David, Eric D'Hoker, Daniele Dorigoni, Jan Gerken, Axel Kleinschmidt,
Boris Pioline and Carlos Rodriguez for stimulating discussions and collaboration on related topics. Eric D'Hoker and Axel Kleinschmidt are thanked for collaboration in early stages of the project. Furthermore, we are grateful to Daniele Dorigoni and Axel Kleinschmidt for valuable comments on the manuscript. The research of MH, OS and BV is supported by the European Research Council under ERC-STG-804286 UNISCAMP, and BV is furthermore supported by the Knut and Alice Wallenberg Foundation under grant KAW2018.0162. This research was supported by the Munich Institute for Astro-, Particle and BioPhysics (MIAPbP) which is funded by the Deutsche Forschungsgemeinschaft (DFG, German Research Foundation) under Germany's Excellence Strategy -- EXC-2094 -- 390783311. The authors would like to thank the Isaac Newton Institute for Mathematical Sciences for support and hospitality during the programme ``New connections in number theory and physics'' where part of the work on this paper was undertaken. This work was supported by EPSRC grant number EP/R014604/1.

\newpage

%%%%%%%%%%%%%%%%%%%%%%%%%%%%%%%%%%%%%%%%%%%%%%%%%%%%%%%%%%%
%%%%%%%%%%%%%%%%%%%%%%%%%%%%%%%%%%%%%%%%%%%%%%%%%%%%%%%%%%%
\section{Review}
\label{sec:2}
%%%%%%%%%%%%%%%%%%%%%%%%%%%%%%%%%%%%%%%%%%%%%%%%%%%%%%%%%%%
%%%%%%%%%%%%%%%%%%%%%%%%%%%%%%%%%%%%%%%%%%%%%%%%%%%%%%%%%%%

In this section we review basics and selected aspects of elliptic modular graph forms (eMGFs),
following the general formalism \cite{Dhoker:2020gdz} and earlier results in \cite{DHoker:2018mys}.

%%%%%%%%%%%%%%%%%%%%%%%%%%%%%%%%%%%%%%%%%%%%%%%%%%%%%%%%%%%
\subsection{Basic definitions}
\label{sec:2.0}
%%%%%%%%%%%%%%%%%%%%%%%%%%%%%%%%%%%%%%%%%%%%%%%%%%%%%%%%%%%

We start by reviewing the definition of eMGFs as non-holomorphic functions of a
modular parameter $\tau$ of a torus that lives in the upper half plane $\tau \in {\cal H}$
and punctures $z_1,z_2,\ldots$ on that torus $\mathbb C/(\ZZ \tau {+} \ZZ)$.
The basic instances and building blocks of eMGFs are the closed-string
Green function $g(z|\tau)$ and the Kronecker-Eisenstein coefficients $f^{(k)}(z|\tau)$
with $k \in \mathbb N_0$ whose definitions in terms of theta functions are reviewed
in appendix \ref{app:theta}.
In view of their double periodicity under $z\rightarrow z{+}1$ and
$z\rightarrow z{+}\tau$, they both admit lattice-sum
representations\footnote{The sum in the first line of (\ref{elemlattice}) follows from the 
second Kronecker limit formula, and that in the second line is absolutely convergent for
any $z \in \mathbb C/(\ZZ \tau {+} \ZZ)$ if $k\geq 3$. Given that $f^{(1)}(z|\tau)$ and $f^{(2)}(z|\tau)$
arise from $z$- and $\tau$-derivatives (\ref{gfderiv}) of the Green function, their lattice-sum 
representations are formally consistent with that of $g(z|\tau)$.}
\begin{align}
g(z|\tau) &=  \frac{ \Im \tau }{ \pi  }  \sum _{ p \in \Lambda '} \frac{  \chi _p (z|\tau) }{  |p|^{2} }
\label{elemlattice} \\
f^{(k)}(z|\tau) &=  - \sum_{p \in \Lambda'} \frac{ \chi_p(z|\tau) }{p^k}  \, , \ \ \ \ k\geq 1
\notag
\end{align}
The $z$-dependence is carried by the character
\begin{align}
\chi_p(z|\tau) 
= e^{ 2\pi i (nu-mv)} 
=  \exp \left( \frac{2\pi i}{\tau{-}\bar\tau} (  \bar{p} z {-} p\bar{z}) \right)
\label{basic.3}
\end{align}
which is written in terms of comoving coordinates $u,v$
as well as lattice momenta $p$
\begin{align}
z&=u \tau {+}v \, , &u,v &\in [0,1] \notag \\
p&=m\tau{+}n \, , & m,n &\in \ZZ
\label{basic.0}
\end{align}
Moreover, we use the
following notation for the summation range in (\ref{elemlattice})
\bea
\Lambda = \ZZ \tau + \ZZ \, , \ \ \ \ \ \
\Lambda' = \Lambda \setminus \{0\}
\label{revsec.1}
\eea
such that $p \in \Lambda'$ amounts to $(m,n) \in \ZZ^2$ with $(m,n ) \neq (0,0)$.
The lattice sums (\ref{elemlattice}) manifest the following modular properties
under $( \begin{smallmatrix}  \alpha & \beta \\ \gamma & \delta \end{smallmatrix}) \in SL(2,\ZZ)$\footnote{We use
the modular group $SL(2,\ZZ)$ rather than $PSL(2,\ZZ)$ because
the comoving coordinates $u,v$ and momentum components $m,n$
transform non-trivially under the element $( \begin{smallmatrix}  -1 &0 \\ 0& -1 \end{smallmatrix})$.}
\beq
g \bigg( \frac{z}{\gamma \tau {+} \delta} \bigg| \frac{ \alpha \tau {+} \beta}{\gamma \tau {+} \delta} \bigg)
= g(z|\tau) \, , \ \ \ \ \ \
f^{(k)} \bigg( \frac{z}{\gamma \tau {+} \delta} \bigg| \frac{ \alpha \tau {+} \beta}{\gamma \tau {+} \delta} \bigg)
= (\gamma \tau {+} \delta)^k f^{(k)}(z|\tau)
\label{revsec.2}
\eeq
More generally, functions $F(z|\tau)$ with the transformation law
\beq
F \bigg( \frac{z}{\gamma \tau {+} \delta} \bigg| \frac{ \alpha \tau {+} \beta}{\gamma \tau {+} \delta} \bigg)
= (\gamma\tau{+}\delta)^a (\gamma\bar \tau {+} \delta)^b F(z|\tau)
\label{revsec.3}
\eeq
under $SL(2,\ZZ)$ will be said to have modular weight $(a,b)$ -- they are generically
non-holomorphic Jacobi forms of weight $(a,b)$ and vanishing index.

%%%%%%%%%%%%%%%%%%%%
%%%%%%%%%%%%%%%%%%%%

\subsubsection{Zagier's single-valued elliptic polylogarithms and Eisenstein series}
\label{sec:2.1.1}

The simplest eMGFs beyond (\ref{elemlattice}) are the single-valued elliptic polylogarithms
introduced by Zagier \cite{Ramakrish}\footnote{Our conventions are related to the normalization of single-valued elliptic polylogarithms in \cite{Ramakrish} via
\[
D_{a,b}(z|\tau) = \frac{ (\tau{-}\bar \tau)^{a+b-1} }{2\pi i}
\sum_{p \in \Lambda'} \frac{ \chi_p(z|\tau ) }{p^a \bar p^b}  = (2i)^{a+b-2}(\pi \Im \tau)^{b-1} \dplus{a \\ b}\!(z|\tau)
\]}
\beq
\dplus{a \\ b}\!(z|\tau)
=  \frac{ (\Im \tau)^a}{\pi^b} \sum_{p \in \Lambda'} \frac{ \chi_{p}(z|\tau ) }{p^a \bar p^b}
\, , \ \ \ \ a,b \in \mathbb Z
\label{basic.13}
\eeq
along with their special cases
\bea
g_k(z|\tau) =  { (\Im \tau)^k \over \pi^k  }  \sum _{ p \in \Lambda '} {  \chi _p (z|\tau) \over  |p|^{2k} }  = \dplus{k \\ k}\!(z|\tau)
\label{basic.13a}
\eea
We recover the Green function and the Kronecker-Eisenstein coeffcients
via $g(z|\tau)=g_1(z|\tau)=\dplus{1 \\ 1}\!(z|\tau)$ and
$ (\Im \tau)^{k}f^{(k)}(z|\tau)= - \dplus{k \\ 0}\!(z|\tau)$, respectively. The powers of $\Im\tau$
in (\ref{basic.13}) and (\ref{basic.13a}) are chosen to arrive at purely antiholomorphic
modular weight $(0,b{-}a)$ for $\dplus{a \\ b}\!(z|\tau)$ such that the $g_k(z|\tau)$
in (\ref{basic.13a}) are modular invariant.

Holomorphic and non-holomorphic Eisenstein series $G_k$ and $E_k$ arise from evaluating
the above lattice sums at $z \in \Lambda$, say
\begin{align}
G_k(\tau)&= \sum_{p \in \Lambda'} \frac{ 1 }{p^k} = - f^{(k)}(0|\tau) \,, &k \geq4
\label{revsec.5}\\
E_k(\tau) &= { (\Im \tau)^k \over \pi^k  }  \sum _{ p \in \Lambda '} { 1 \over  |p|^{2k} } = g_k(0|\tau)\,, &k \geq2
\notag
\end{align}
leading to modular weights $(k,0)$ of $G_k$ and modular invariant $E_k$.

\subsubsection{Dihedral eMGFs}
\label{sec:2.1.2}

Dihedral eMGFs generalize the $\dplus{a \\ b}\!(z|\tau)$ in (\ref{basic.13}) to
multiple lattice sums. In the shorthand notation
\beq
\begin{array}{l}
A=[a_1,a_2,\ldots,a_R] \, ,  \\
B=[b_1,\, \hspace{-0.2mm}b_2,\ldots,\, \hspace{-0.2mm}b_R] \, , \\
Z = [z_1, z_2, \cdots, z_R]\, ,
\end{array} \hskip 0.5in
|A| = \sum_{r=1}^R a_r\, ,
\hskip 0.5in
|B| = \sum_{r=1}^R b_r
\label{basic.15}
\eeq
with exponents $a_r,b_r \in \ZZ$ and elliptic variables $z_r \in \mathbb C/(\ZZ \tau {+} \ZZ)$,
we define dihedral eMGFs with $R$ columns by the following multiple Kronecker-Eisenstein
sums,
\beq
\cplus{A \\ B \\ Z}\! (\tau) = \frac{ (\Im \tau)^{|A|} }{ \pi^{ |B|}}  \sum _{p_1,\ldots ,p_R \in \Lambda '} \delta \bigg(\sum_{s=1}^R p_s \bigg) \,
\prod _{r=1}^R 
\frac{ \chi_{p_r}(z_r|\tau)  }{ p_r ^{a_r} \bar p_r^{b_r}}
\label{gen.66}
\eeq
The sums are absolutely convergent if $a_r+a_{r'}+b_r+b_{r'} >2$ for any pair $1\leq r,r' \leq R$
and can be related to polynomials in the Green function (\ref{elemlattice}) in case of
a pair with $a_r+a_{r'}+b_r+b_{r'}=2$ with $z_r {-} z_{r'} \notin \mathbb Z\tau{+}\mathbb Z$, 
see for instance section \ref{sec:edgecase}. The
powers of $\Im \tau$ ensure purely antiholomorphic modular weights $(0,|B|{-}|A|)$ in
${\cal C}^+$, see appendix \ref{app:A.1.3} for the alternative convention ${\cal C}^-$ to
attain purely holomorphic modular weights. The eMGF in (\ref{gen.66}) is easily
seen to be permutation invariant in its columns, i.e.\ under simultaneous swap
of any two triplets $a_r,b_r,z_r$ and $a_{r'},b_{r'},z_{r'}$.

The single-valued elliptic polylogarithms (\ref{basic.13}) and (\ref{basic.13a})
arise as the special cases of two-column eMGFs, i.e.\ (\ref{gen.66}) at $R=2$,
\begin{align}
\cplus{a_1 &a_2 \\ b_1 &b_2 \\ z_1 &z_2}\! (\tau)
&= (-1)^{a_2+b_2} \, \cplus{a_1+a_2 &0 \\ b_1+b_2 &0 \\ z_1-z_2 &0}\! (\tau)
\notag \\
\cplus{a &0\\b &0\\z &0}\!(\tau)&=\dplus{a \\ b}\!(z|\tau)
\label{revsec.6}
\end{align}
whereas one-column eMGFs vanish by momentum conservation, $\cplus{a\\b\\z}\!(\tau)=0$.
Cases with $R\geq 3$ columns may depend on multiple variables $z_1,z_2,\ldots$, but
their iterated-integral description is only developed for the one-variable
case $z_r \in \{0,z\}$ in this work.
A brief review of eMGF identities including translation invariance and momentum conservation
used in (\ref{revsec.6}) can be found in appendix \ref{app:A.1}.

As detailed in section 2.5.2 of \cite{Dhoker:2020gdz}, eMGFs are defined for arbitrary decorated
graphs. Apart from dihedral eMGFs in (\ref{gen.66}) and trihedral ones in appendix A
of \cite{Dhoker:2020gdz}, more complicated graph topologies including box-, kite- and
tetrahedral ones can be addressed by straightforwardly adapting the notation
of \cite{Gerken:2020aju} to the $z_r$-dependent case. One of our main results in this
work is that eMGFs in one variable up to and including $|A|{+}|B|=10$ can be reduced to dihedral ones, see section \ref{bassec.4}.

Note that modular graph forms (MGFs) \cite{DHoker:2015wxz, DHoker:2016mwo} are
recovered when all the $z_r$ of an eMGFs are set to zero\footnote{The
normalization conventions of (e)MGFs ${\cal C}^+[\ldots]$ in this work are
identical to those of \cite{Dhoker:2020gdz}. However, the MGFs $\cplus{A\\B}$
of \cite{DHoker:2016mwo, DHoker:2016quv, DHoker:2019txf, DHoker:2021ous} are normalized
with a factor of $\pi^{-\frac{|A|}{2}-\frac{|B|}{2}}$ in the
place of $\pi^{- |B|}$ in (\ref{gen.66}).}
\beq
\cplus{a_1 &a_2 &\ldots &a_R \\ b_1 &b_2 &\ldots &b_R }\! (\tau)
= \cplus{a_1 &a_2 &\ldots &a_R \\ b_1 &b_2 &\ldots &b_R \\ 0 &0 &\ldots &0}\! (\tau)
\label{MGFlimit}
\eeq
While MGFs with odd $|A|{+}|B|$ vanish by the reflection identity (\ref{compap.8}),
eMGFs with odd $|A|{+}|B|$ are generically non-trivial in presence of $z_r \neq 0$.

%%%%%%%%%%%%%%%%%%%%
%%%%%%%%%%%%%%%%%%%%

\subsubsection{Shorthands for real eMGFs}
\label{sec:2.1.3}

It will be convenient to introduce the following shorthand
\beq
C_{a_1,a_2,\ldots,a_s| a_{s+1},\ldots,a_R}(z|\tau)
= \cplus{a_1 &\ldots &a_s &a_{s+1} &\ldots &a_R \\
a_1 &\ldots &a_s &a_{s+1} &\ldots &a_R \\
z &\ldots &z &0 &\ldots &0}\! (\tau)
\label{cabc.1}
\eeq
for eMGFs in one variable with identical holomorphic and antiholomorphic
exponents $a_r=b_r$. The eMGFs (\ref{cabc.1}) enjoy a residual permutation symmetry
in $\{a_1,\ldots,a_s\}$ and $\{ a_{s+1},\ldots,a_R\}$, and one can swap
$C_{a_1,\ldots,a_s| a_{s+1},\ldots,a_R}=C_{ a_{s+1},\ldots,a_R| a_1,\ldots,a_s}$ as
a consequence of (\ref{compap.2}) and (\ref{compap.8}). Moreover, the
eMGFs (\ref{cabc.1}) are real and can be constructed from convolutions of the
Green function (\ref{elemlattice}). The MGFs resulting from the limit $z\rightarrow0$
of (\ref{cabc.1}) were initially introduced as modular graph {\it functions} \cite{DHoker:2015wxz}
and will be denoted by $C_{a_1,\ldots,a_R}$ (without vertical bar in the subscript)
following \cite{DHoker:2015gmr}.

The simplest cases of the real eMGFs in (\ref{cabc.1}) are $g_k
= C_{a|k-a}$ (which are in fact independent on $a$) and the three- or
four-column examples,
\begin{align}
C_{a|b,c}(z|\tau) &= \cplus{a &b &c \\ a &b &c \\ z &0 &0}\!(\tau) = \bigg( \frac{\Im \tau}{\pi} \bigg)^{a+b+c} \! \! \! \sum_{p_1,p_2,p_3 \in \Lambda'} \frac{
 \delta(p_1{+}p_2{+}p_3) \chi_{p_1}(z|\tau) }{ |p_1|^{2a}  |p_2|^{2b}  |p_3|^{2c}  } \label{cabc.2}
\\
C_{a|b,c,d}(z|\tau) &= \cplus{a &b &c &d \\ a &b &c &d \\ z &0 &0 &0}\!(\tau)
= \bigg( \frac{\Im \tau}{\pi} \bigg)^{a+b+c+d} \! \! \! \! \! \sum_{p_1,p_2,p_3,p_4 \in \Lambda'}  \! \!\frac{ \delta(p_1{+}p_2{+}p_3{+}p_4) \chi_{p_1}(z|\tau) }{ |p_1|^{2a}  |p_2|^{2b}  |p_3|^{2c} |p_4|^{2d}  } \notag
\\
C_{a,b|c,d}(z|\tau) &= \cplus{a &b &c &d \\ a &b &c &d \\ z &z &0 &0}\!(\tau)
= \bigg( \frac{\Im \tau}{\pi} \bigg)^{a+b+c+d} \! \! \! \! \! \sum_{p_1,p_2,p_3,p_4 \in \Lambda'}  \! \!\frac{ \delta(p_1{+}p_2{+}p_3{+}p_4) \chi_{p_1+p_2}(z|\tau) }{ |p_1|^{2a}  |p_2|^{2b}  |p_3|^{2c} |p_4|^{2d}  } \notag
\end{align}
that reduce to MGFs $C_{a,b,c} = \cplus{a &b &c \\ a &b &c }$
and $C_{a,b,c,d} = \cplus{a &b &c &d \\ a &b &c &d }$ as $z\rightarrow 0$.
The real eMGFs in (\ref{cabc.2}) will appear in the bases of eMGFs presented
in section \ref{bassec}.

%%%%%%%%%%%%%%%%%%%%%%%%%%%%%%%%%%%%%%%%%%%%%%%%%%%%%%%%%%%
\subsection{Differential equations of eMGFs}
\label{sec:2.1}
%%%%%%%%%%%%%%%%%%%%%%%%%%%%%%%%%%%%%%%%%%%%%%%%%%%%%%%%%%%

The lattice-sum representations (\ref{gen.66}) of eMGFs are convenient to
evaluate their derivatives in $z$ and $\tau$. Throughout this work, derivatives
w.r.t.\ $\tau$ are taken at constant comoving coordinates $u,v$ in $z=u\tau{+}v$
(rather than at fixed $z$) and fixed integers $m,n$ of the
lattice momenta $p=m\tau{+}n$,
\bea
\nabchi _\tau = 2 i (\Im \tau)^2 \p_{\tau}  \, , \ \ \ \ \ \ \ \ \hbox{ for } (u,v) \hbox{ and } (m,n) \hbox{ fixed}
\label{crdrv}
\eea
such that $\nabchi _\tau \chi_p (z|\tau) =  0 $. The Cauchy-Riemann derivative or Maass operator
$\nabchi _\tau$ maps the space of eMGFs of modular weights $(0,\mu)$ to those of
weights $(0,\mu{-}2)$. Its complex conjugate is given by $\overline{\nabchi}_{ \tau} = - 2 i (\Im \tau)^2 \p_{\bar \tau}$ and shifts modular weights $(\mu,0)$ to $(\mu{-}2,0)$.

We will furthermore employ the derivatives in $z$ and $\bar z$,
at fixed $\tau$ and $p$,
\begin{align}
\nabla_z & =   \partial_u - \bar \tau \partial_v =  2i (\Im \tau) \, \partial_z
\label{diff.1} \\
\overline{\nabla}_{ z} & = \partial _u - \tau \partial_v =  - 2 i (\Im \tau) \, \p_{\bar z}
\no
\end{align}
They act on characters (\ref{basic.3}) via $\nabla_z \chi_p(z|\tau) =2\pi i \bar p \chi_p(z|\tau)  $
and $\overline{\nabla}_z \chi_p(z|\tau) =2\pi i p \chi_p(z|\tau)  $ such that $\nabla_z$
maps eMGFs of weight $(0,\mu)$ to those of weight $(0,\mu{-}1)$.

Note that the associated Laplacian actions on modular invariant eMGFs are given by
\begin{align}
\Delta_z &= 4 \Im \tau \partial_z \partial_{\bar z} = \frac{(\partial_u {-} \bar \tau \partial_v)(\partial_u {-} \tau \partial_v) }{\Im \tau} = \frac{  \overline{\nabla}_z \nabla_z }{\Im \tau}
\no \\
\Delta_\tau &= 4 (\Im \tau)^2 \partial_\tau \partial_{\bar \tau} =  \overline{\nabla}_\tau (\Im \tau)^{-2} \nabla_\tau
\label{laplz.1}
\end{align}
and both preserve modular weights $(0,0)$.

%%%%%%%%%
%%%%%%%%%
%%%%%%%%%

\subsubsection{$\tau$-derivatives of eMGFs}
\label{sec:2.2.1}

With the notation $S_r=[\vec{0}^{r-1},1,\vec{0}^{R-r}]$ for the unit vector in the $r^{\rm th}$ direction,
the $\tau$-derivatives of dihedral eMGFs are compactly encoded in
\begin{align}
\pi \nabchi_\tau \, \cplus{A \\ B \\Z}\! (\tau)  & =  \sum_{r=1}^R a_r \, \cplus{A+S_r \\ B-S_r\\ Z}\! (\tau)
\label{gen.66Ctau}
\end{align}
They follow the analogous differential equations of MGFs \cite{DHoker:2016mwo}
and specialize as follows for the two-column eMGFs defined in (\ref{revsec.6}):
\begin{align}
(\pi \nabchi_\tau)^m \, \dplus{a  \\ b }\! (z|\tau) & =  {(a{+}m{-}1)! \over (a{-}1)!}   \dplus{a+m  \\ b-m  }\! (z|\tau)
\, , \ \ \ \ \ \ 0<m<b \notag\\
(\pi \nabchi_\tau)^b \, \dplus{a  \\ b  }\! (z|\tau)&=
- \frac{(a{+}b{-}1)!}{(a{-}1)!} (\Im \tau)^{a+b} f^{(a+b)}(z|\tau)
\label{revsec.8}
\end{align}
The simplest examples of the Laplacian $\Delta_\tau$ in (\ref{laplz.1}) are
the eigenvalue equations
\beq
\big( \Delta_\tau - k(k{-}1) \big) g_k(z|\tau) = 0
\label{laplz.0}
\eeq
also see section 3.5 of \cite{Dhoker:2020gdz} for inhomogeneous Laplace equations of the
three-column eMGFs $C_{a|b,c}(z|\tau)$ in (\ref{cabc.2}).
Note that (\ref{revsec.8}) and (\ref{laplz.0}) reduce to the well-known Cauchy-Riemann
and Laplace equations of non-holomorphic Eisenstein series (\ref{revsec.5}) as
$z\rightarrow 0$,
\begin{align}
(\pi \nabla_\tau)^k E_k(\tau) &= \frac{ (2k{-}1)! }{(k{-}1)!} (\Im \tau)^{2k} G_{2k}(\tau)
\, , \ \ \ \ \ \
\big( \Delta_\tau - k(k{-}1) \big) E_k(\tau) = 0
\label{laplz.0a}
\end{align}

%%%%%
%%%%%
%%%%%

\subsubsection{$z$-derivatives of eMGFs}
\label{sec:2.2.2}

Covariant $z$-derivatives of dihedral eMGFs take a similarly
compact form as their $\tau$-derivatives in (\ref{gen.66Ctau}),
\beq
 \nabla_z \, \cplus{A\\B\\Z}\!(\tau) = 2i \sum_{r=1}^R \frac{ \partial z_r}{\partial z}
\cplus{A\\B-S_r\\Z}\!(\tau)
\label{revsec.9}
\eeq
i.e.\ repeated occurrence of the differentiation variable $z$ (distinct from
$z_{s+1},\ldots,z_R$) leads~to
\beq
 \nabla_z \, \cplus{a_1 &\ldots &a_s &a_{s+1} &\ldots &a_R \\ b_1 &\ldots &b_s &b_{s+1} &\ldots &b_R \\ z &\ldots &z &z_{s+1} &\ldots &z_R}
=
2   i  \sum_{r=1}^s \cplus{a_1 &\ldots &a_r &\ldots &a_s &a_{s+1} &\ldots &a_R \\ b_1 &\ldots &b_r-1 &\ldots &b_s &b_{s+1} &\ldots &b_R \\ z &\ldots &z  &\ldots &z &z_{s+1} &\ldots &z_R}
\label{diff.2}
\eeq
in particular
\begin{align}
(  \nabla_z)^m \dplus{a  \\ b }\! (z|\tau)  & =  (2  i)^m \, \dplus{a  \\ b-m }\! (z|\tau) \, ,
\ \ \ \ \ \ 0<m<b
\no \\
(  \nabla_z)^b \dplus{a  \\ b }\! (z|\tau)  & =  - (2  i)^b  (\Im \tau)^a f^{(a)}(z|\tau)
\label{revsec.10}
\end{align}
The simplest examples of the Laplacian $\Delta_z$ in (\ref{laplz.1}) are
\begin{align}
\Delta_z\chi_p(z|\tau) &= - \frac{ 4\pi^2 }{\Im \tau} |p|^2 \chi_p(z|\tau)
\notag \\
\Delta_z g_k(z|\tau)  &= - 4\pi g_{k-1}(z|\tau)\, , \ \ \ \ k \geq 2
\label{laplz.2} \\
\Delta_z C_{a|b,c}(z|\tau)  &= \left\{ \begin{array}{cl} - 4\pi C_{a-1|b,c}(z|\tau) &: \ a\geq 2
\\
4\pi \big( E_{b+c}(\tau) - g_b(z|\tau)  g_c(z|\tau)  \big)&: \ a=1
\end{array} \right.
\notag
\end{align}
Similarly, we have:
\begin{align}
     \overline{\nabla}_z \, \cplus{A\\B\\Z}\!(\tau) = 2\pi i \, \Im\tau \sum_{r=1}^R \frac{ \partial \bar{z}_r}{\partial \bar{z}}
\cplus{A-S_r\\B\\Z}\!(\tau)
\label{revsec.11}
\end{align}

%%%
%%%
%%%

\subsubsection{Relations among eMGFs from $\nabla_\tau$-derivatives}
\label{sec:2.2.3}

Repeated $\nabla_\tau$-derivatives of arbitrary
eMGFs introduce factors of $f^{(a)}$ or $G_k$ as in (\ref{revsec.8}) and (\ref{laplz.0a}).
They can be traced back to vanishing antiholomorphic exponents in
the entries of $B{-}S_r$ on the right-hand side of (\ref{gen.66Ctau}).
The procedure to expose the resulting $f^{(a)}$ or $G_k$ is known
as {\it holomorphic subgraph reduction} and presented in detail
for dihedral and trihedral eMGFs in section 3.3 and
appendix A of \cite{Dhoker:2020gdz}.\footnote{The same techniques for
holomorphic subgraph reduction also expose factors of $f^{(a)}$ in
repeated $\nabla_z$-derivatives of eMGFs.}
Holomorphic subgraph reduction has been initially developed to obtain
$G_k$ in the Cauchy-Riemann derivatives of MGFs \cite{DHoker:2016mwo, Gerken:2018zcy}
and can also be derived from the Fay identities of the $f^{(a)}$ \cite{Gerken:2020aju}.

A main achievement of this work is to solve Cauchy-Riemann equations
such as (\ref{revsec.8}) via iterated integrals in $\tau$. In particular, we will focus
on situations with non-trivial eMGFs multiplying the $f^{(a)}$ or $G_k$ on the right-hand side.
The resulting iterated integrals are said to have higher depth, see
\cite{Broedel:2018izr, Gerken:2020yii} and section 3.4 of \cite{Dhoker:2020gdz}
for the assignment of depth to MGFs and eMGFs, respectively.

By expressing eMGFs in terms of iterated integrals in $\tau$ and supplementing
initial values as $\tau \rightarrow i \infty$, their intricate network of algebraic relations
becomes manifest. With the notation of section \ref{sec:2.1.3} for real MGFs
$C_{a,\ldots}$ and eMGFs $C_{a,\ldots| \ldots}$, the simplest examples
of non-trivial algebraic relations among MGFs are \cite{DHoker:2015gmr}
\begin{align}
C_{1,1,1}(\tau)&= E_3(\tau) + \zeta_3
\label{revsec.13}\\
C_{1,1,1,1}(\tau)&= 24 C_{2,1,1}(\tau) - 18 E_4(\tau) + 3 E_2(\tau)^2 \notag
\end{align}
which reduce the loop order of the defining graph.
The term $\zeta_3$ in the first line can be reconstructed from the
asymptotics as $\tau \rightarrow i\infty$, and similar relations at higher
weight have been proven through the sieve algorithm for MGFs
\cite{DHoker:2016mwo, DHoker:2016quv}. A database of such relations
for a large class of MGFs can be found in the {\sc Mathematica} package
\cite{Gerken:2020aju}.

The methods of this work will allow to generate the analogous relations
among eMGFs, and a public {\sc Mathematica} package implementing these methods 
will be released in a follow-up paper \cite{Hidding:2022zzz}. 
The iterated-integral representations of eMGFs in section \ref{bassec.3}
will expose the relation \cite{Basu:2020pey}
\begin{align}
\frac{1}{6} C_{1|1,1,1}(z|\tau) - \frac{1}{8} C_{1,1|1,1}(z|\tau)
&= C_{2|1,1}(z|\tau)  - g_4(z|\tau)   + \frac{1}{4} E_4(\tau) \label{teas.9} \\
&\quad - \frac{1}{4} g_2(z|\tau)^2
+ \frac{1}{2} E_2(\tau) g_2(z|\tau) - \frac{1}{8} E_2(\tau)^2
\notag
\end{align}
which reduces to the MGF identity in the second line of (\ref{revsec.13}) as $z \rightarrow 0$.
Note that $C_{1|1,1,1}, C_{1,1|1,1}$
and $C_{2|1,1}$ are also denoted by $D_4^{(1)},D_4^{(2)}$ and $D_4^{(1,1,2)}$
in the literature.

%%%%%%%%%%%%%%%%%%%%%%%%%%%%%%%%%%%%%%%%%%%%%%%%%%%%%%%%%%%
\subsection{Generating series $Y$ of eMGFs}
\label{sec:2.2}
%%%%%%%%%%%%%%%%%%%%%%%%%%%%%%%%%%%%%%%%%%%%%%%%%%%%%%%%%%%

Besides working out the Cauchy-Riemann equations for individual
eMGFs, we will also study their generating series \cite{Dhoker:2020gdz} that
resemble the Koba-Nielsen integrals in closed-string\footnote{The generating-series approach
to the low-energy expansion of configuration-space integrals in genus-one string
amplitudes was initiated in the open-string sector \cite{Mafra:2019ddf, Mafra:2019xms},
also in presence of one \cite{Broedel:2019gba, Broedel:2020tmd} or several
elliptic variables \cite{Kaderli:2022qeu}.} genus-one amplitudes.
As will be reviewed below, the Cauchy-Riemann derivatives of the generating series
are known with all factors of $f^{(k)}$ and $G_k$ manifest, without any need
to perform holomorphic subgraph reduction. The generating series below contain
any convergent eMGF in one variable that cannot be simplified using
holomorphic subgraph reduction, regardless of the topology of the defining graph.

More importantly, we will infer iterated-integral representations of eMGFs
in section \ref{sec:3} by solving the differential equations of their
generating series perturbatively, i.e.\ order by order in certain expansion
variables to be specified below. We will focus on the action of the holomorphic derivatives $\nabla_\tau$
while extracting the information of the antiholomorphic ones $\overline{\nabla}_\tau$
from the known complex-conjugation properties of eMGFs and their generating series.
This strategy was already crucial for MGFs whose iterated-integral
structure was unravelled in \cite{Gerken:2020yii} based on the differential equations
of their generating series in \cite{Gerken:2019cxz}.

\subsubsection{Two-point case}
\label{sec:2.3.1}

In the simplest generating series of eMGFs, we integrate a single variable
$z_2$ over the torus $\Sigma = \mathbb C / \Lambda$, similar to a two-point
closed-string genus-one amplitude:
\begin{align}
\label{eq:Y2pt}
Y_{ij} &(z_0,\eta,\bar\eta|\tau) = 2i\Im\tau \int_\Sigma \frac{\dd^2 z_2}{\Im\tau} e^{s_{02} g(z_{02}|\tau) + s_{12}g(z_{12}|\tau)}
\\
&\quad \times
\begin{pmatrix} \Omega(z_{12}, (\tau{-}\bar\tau)\eta|\tau) \overline{\Omega(z_{12},\eta|\tau)} & \Omega(z_{02}, (\tau{-}\bar\tau)\eta|\tau) \overline{\Omega(z_{12},\eta|\tau)}\\
\Omega(z_{12}, (\tau{-}\bar\tau)\eta|\tau) \overline{\Omega(z_{02},\eta|\tau)} & \Omega(z_{02}, (\tau{-}\bar\tau)\eta|\tau)\overline{\Omega(z_{02},\eta|\tau)}\end{pmatrix}_{ij} \nn
\end{align}
By translation invariance on the torus, the
integral can only depend on the differences $z_{ij} = z_i{-}z_j$ of the unintegrated
punctures $z_1,z_0$, so we fix $z_1=0$ throughout this section. The expansion
variables $s_{ij} \in \mathbb C$ multiplying the Green functions $g(z_{ij}|\tau)$
in the exponents are interpreted as products of external momenta in
a string-amplitude context. Finally, the $2\times 2$ matrix of non-holomorphic
Kronecker-Eisenstein series
\beq
\Omega(z,\eta|\tau) = \sum_{k=0}^\infty \eta^{k-1} f^{(k)}(z|\tau)
\label{revsec.21}
\eeq
depends on yet another expansion variable $\eta \in \mathbb C$ and its complex
conjugate.

As detailed in section 4 of \cite{Dhoker:2020gdz}, each entry of $Y_{ij} $
in (\ref{eq:Y2pt}) admits a Laurent expansion in $s_{02},s_{12},\eta,\bar \eta$
with dihedral eMGFs in its coefficients. The arguments $\bar \eta$ and $(\tau{-}\bar \tau)\eta$
of the Kronecker-Eisenstein series in (\ref{eq:Y2pt}) are engineered to obtain the
eMGFs in the ${\cal C}^+$ convention of (\ref{gen.66}), with vanishing holomorphic
modular weight and simple action of $\nabla_\tau$. Indeed, the $Y_{ij}$ obey the linear
and homogeneous differential equations in $\tau$ \cite{Dhoker:2020gdz}
\begin{align}
-4\pi \nabla_\tau Y_{ij}(z_0,\eta,\bar\eta|\tau) = \sum_{\ell=1}^2 \bigg[
{-} R_\eta(\ep_0) &+ \sum_{k=2}^\infty (k{-}1) (\tau{-}\bar \tau)^k f^{(k)}(z_{01}|\tau) R_\eta(b_k)
\label{revsec.22} \\
 &+
\sum_{k=4}^\infty (1{-}k) (\tau{-}\bar \tau)^k G_k(\tau) R_\eta(\ep_k)
\bigg]_{j \ell} Y_{i\ell}(z_0,\eta,\bar\eta|\tau) \notag
\end{align}
where the entries of the $2\times 2$ matrices $R_{\eta}(\cdot)$ are
linear in $\{s_{02},s_{12},\bar \eta\}$
and contain powers of and derivatives in $\eta$,
\begin{align}
R_\eta(\ep_0) &= \frac{1}{\eta^2} \ccb s_{12} &s_{02} \\ s_{12} &s_{02} \cce
 - \frac{1}{2}(s_{02}{+}s_{12}) \partial_\eta^2 - 2\pi i \bar \eta \partial_\eta
 \notag \\
R_\eta(\ep_k) &= \eta^{k-2} \ccb s_{12} &0 \\ 0 &s_{02} \cce \, , \ \ \ \ \ \ k\geq 4
 \label{revsec.23} \\
 %%%
 R_\eta(b_2) &= \ccb -s_{02} &s_{02} \\ s_{12} &-s_{12} \cce \notag \\
R_\eta(b_k) &= \eta^{k-2} \ccb 0 &(-1)^{k}s_{02} \\ s_{12} &0 \cce \, , \ \ \ \ \ \ k\geq 3  \notag
\end{align}

\subsubsection{Higher-point case}
\label{sec:2.3.2}

In order to cover one-variable eMGFs associated with graphs of {\it arbitrary} topologies,
we generalize the two-point generating series (\ref{eq:Y2pt}) to the $n!\times n!$ matrix
\begin{align}
\label{high.4}
Y(\begin{smallmatrix} M \\ N \end{smallmatrix} |  \begin{smallmatrix} K \\ L \end{smallmatrix} )
&= (2i\Im\tau)^{n-1} \int\limits_{\Sigma^{n-1}} \bigg( \prod_{j=2}^n \frac{\dd^2 z_j}{\Im\tau}  \bigg)
\prod_{0\leq i<j \atop{(i,j) \neq (0,1)}}^n e^{s_{ij} g(z_{ij}|\tau) }
\\
&\hspace{10mm}\times \varphi_{(\tau-\bar \tau) \vec{\eta}}(1,K)\varphi_{(\tau-\bar \tau) \vec{\eta}}(0,L)
\overline{\varphi_{ \vec{\eta}}(1,M)\varphi_{\vec{\eta}}(0,N)}
\nonumber
\end{align}
built from torus integrals over $z_2,z_3,\ldots ,z_n$ akin to those in 
$n$-point closed-string genus-one amplitudes. Instead of a single Kronecker-Eisenstein 
series (\ref{revsec.21}), the integrand of each entry involves products of the form
\beq
\varphi_{\vec{\eta}}(a_1,a_2,\ldots,a_r|\tau) = \Omega(z_{a_1 a_2},\eta_{a_2 a_3\ldots a_r}|\tau)
\Omega(z_{a_2 a_3},\eta_{a_3 a_4\ldots a_r}|\tau) \ldots
\Omega(z_{a_{r-1} a_r},\eta_{a_r}|\tau)
\label{high.1}
\eeq
with $\eta_{ab \ldots c}=\eta_a{+}\eta_b{+}\ldots {+}\eta_c$,
for a total of $n{-}1$ factors of $\Omega$ and $\bar \Omega$ each. For ordered
sets $M,N,K,L$ of length zero and one in (\ref{high.4}), the products (\ref{high.1}) specialize
to $\varphi_{\emptyset}(a|\tau) =1$ and $\varphi_{\eta_b}(a,b|\tau) = \Omega(z_{ab},\eta_b|\tau)$,
respectively. In this way, one recovers the $2\times2$ matrix (\ref{eq:Y2pt})
by choosing both $(K,L)$ and $(M,N)$ to be one of $(2,\emptyset),(\emptyset,2)$.

The $n! \times n!$ matrix structure of (\ref{high.4}) stems from the
$n!$ possibilities to distribute the elements $\{2,3,\ldots,n\}$ into the
two ordered sets $K,L$ which jointly form the column indices. Similarly,
the rows are indexed by the $n!$ arrangements of $\{2,3,\ldots,n\}$ into
ordered sets $M,N$.\footnote{At $n{=}3$ points, for instance, the both indices
$\begin{smallmatrix} K \\ L \end{smallmatrix}$ \& $\begin{smallmatrix} M \\ N \end{smallmatrix}$ range over the six choices $
 \{   \begin{smallmatrix} 23 \\ \emptyset \end{smallmatrix},
 \begin{smallmatrix} 32 \\ \emptyset \end{smallmatrix},
  \begin{smallmatrix} 2 \\ 3 \end{smallmatrix},
   \begin{smallmatrix} 3 \\ 2 \end{smallmatrix},
 \begin{smallmatrix} \emptyset \\ 32 \end{smallmatrix},
  \begin{smallmatrix} \emptyset \\ 23 \end{smallmatrix} \}$.} Finally,
we suppress the dependence of $Y(\begin{smallmatrix} M \\ N \end{smallmatrix} |  \begin{smallmatrix} K \\ L \end{smallmatrix} )$ on $z_0 \in \Sigma, \,\tau \in {\cal H}$
and $s_{ij},\eta_j \in \mathbb C$ to avoid cluttering.

Similar to the two-point case, the $n$-point generating series (\ref{high.4}) admits
Laurent expansions in (sums of) the $\frac{1}{2}(n {+} 2) (n {-} 1)$ variables $s_{ij} \in \mathbb C$
and $(n{-}1)$ variables $\eta_j \in \mathbb C$. The coefficients of this expansion
are eMGFs in one variable $z_0$ whose topologies vary with the number of integrations: The
double integrals over $z_2,z_3$ at multiplicity $n=3$ generically yield trihedral eMGFs
(see section 5.1 of \cite{Dhoker:2020gdz} for details), while $n=4$ additionally allows
for box-, kite- and tetrahedral topologies.

In fact, none of the eMGFs in the $s_{ij}$- and $\eta_j$-expansion of the
generating series (\ref{high.4}) are amenable to holomorphic subgraph
reduction.\footnote{This can be seen from the absence of cycles
in the first arguments of the $\Omega(z_{ij},\ldots)$ and separately
of the $\overline{\Omega(z_{ij},\ldots)}$ in the building blocks (\ref{high.1}) of the 
integrands. Integrating cycles such as $\Omega(z_{ij},\ldots)\Omega(z_{jk},\ldots)
\Omega(z_{ki},\ldots)$ over $z_i,z_j,z_k$ would generate eMGFs that can
be simplified via holomorphic subgraph reduction \cite{Dhoker:2020gdz}.} 
For instance, all the dihedral $ \cplus{A \\ B \\Z}$ of (\ref{gen.66}) 
in this expansion have at most one entry $a_i=0$ and $b_i=0$ each,
with all other exponents $a_j,b_j$ positive. These are the types of eMGFs that will 
appear in the bases of indecomposable representatives in later sections.

The $n$-point generalization of the Cauchy-Riemann equation (\ref{revsec.22}) still
involves $f^{(k)}$ with $k\neq 1$ and $G_k$ with $k\geq 4$, but the accompanying
$R_{\vec{\eta}}(\cdot)$ are now $n! \times n!$ matrices:
\begin{align}
-4\pi \nabchi_\tau Y( \begin{smallmatrix} M \\ N \end{smallmatrix} |  \begin{smallmatrix} K \\ L \end{smallmatrix}) = \sum_{P,Q}  \bigg( {-}R_{\vec{\eta}}(\epsilon_0)
&+ \sum_{k=2}^{\infty}(k{-}1) (\tau{-}\bar \tau)^k f^{(k)}(z_{01}|\tau) R_{\vec{\eta}}(b_k) \label{high.12} \\
&+ \sum_{k=4}^{\infty}(1{-}k) (\tau{-}\bar \tau)^k G_k(\tau) R_{\vec{\eta}}(\epsilon_k)
  \bigg)_{ \begin{smallmatrix} K \\ L \end{smallmatrix} \big|  \begin{smallmatrix} P \\ Q \end{smallmatrix}}
Y( \begin{smallmatrix} M \\ N \end{smallmatrix} |  \begin{smallmatrix} P \\ Q \end{smallmatrix}) \notag
\end{align}
The summation range for $P,Q$ here and below refers to the $n!$ 
distributions of $\{2,3,\ldots,n\}$ into two ordered sets as explained
above, and the entries of $R_{\vec{\eta}}(\cdot)$ are linear in $\{s_{ij},\bar \eta_j\}$ and
differential operators in $\eta_j$. The detailed form of these entries is known at all multiplicities
and almost identical to similar $n!\times n!$ matrices $r_{\vec{\eta}}(\cdot)$ in the open-string
computations of \cite{Broedel:2020tmd}, see section 5.2 of \cite{Dhoker:2020gdz} for the dictionary.
The $6\times 6$ matrices $R_{\eta_2,\eta_3}(\epsilon_k),R_{\eta_2,\eta_3}(b_k)$ will play
an important role for the arguments in section \ref{sec:5.2.2} and can therefore be found
in the ancillary file of the arXiv submission of this work.

Upon expansion in $s_{ij}, \eta_j,\bar\eta_j$, (\ref{high.12}) generates the Cauchy-Riemann
equations of large classes of dihedral eMGFs and exposes the $f^{(k)},G_k$
that would also follow from holomorphic subgraph reduction in an order-by-order
computation of the $\tau$-derivatives via (\ref{gen.66Ctau}). A differential equation for $\nabla_z Y( \begin{smallmatrix} M \\ N \end{smallmatrix} |  \begin{smallmatrix} K \\ L \end{smallmatrix})$ similar to (\ref{high.12}) can be found in
\cite{Dhoker:2020gdz} which generates expressions for $\nabla_z$-derivatives (\ref{revsec.9})
of eMGFs. Since the main focus of this work is on iterated-$\tau$-integral representations
of eMGFs, we shall only discuss solutions of (\ref{high.12}) and relegate the analogous
integrations of the $z$-derivatives to future work.

%%%%%%%%%%
%%%%%%%%%%
\subsubsection{Relations among the matrices $R_{\vec{\eta}}(\cdot)$}
\label{sec:2.3.3}

The matrices $R_{\vec{\eta}}(\cdot)$ on the right-hand side of (\ref{high.12})
obey a variety of relations, and we shall here focus on those that are believed
to hold at all multiplicity. First of all, commutativity of the derivatives of $\partial_{z_0}$
and $\partial_{\tau}$ implies that
\beq
R_{\vec{\eta}}([b_{w},\ep_k])  = - \sum_{a=0}^{w-2} {w{-}2 \choose a}
R_{\vec{\eta}}([b_{a+2},b_{k+w-a-2}])\, , \ \ \ \
w\geq 2 \, , \ \ \ \ k\geq 4
 \label{revsec.25}
\eeq
with the shorthands $R_{\vec{\eta}}(ab) = R_{\vec{\eta}}(a) R_{\vec{\eta}}(b) $
and $R_{\vec{\eta}}([a,b]) = R_{\vec{\eta}}(a) R_{\vec{\eta}}(b)
-R_{\vec{\eta}}(b) R_{\vec{\eta}}(a) $, see \cite{Broedel:2020tmd} and
\cite{Dhoker:2020gdz} for discussions in an open- and closed-string context.

Moreover, by analogy with the corresponding generating series of MGFs
\cite{Gerken:2019cxz}, the $R_{\vec{\eta}}(\cdot)$ are expected to
enjoy the nilpotency property
\begin{align}
R_{\vec{\eta}}\big(\ad_{\epsilon_0}^{k-1} (\epsilon_k)\big) &= 0  \, , \ \ \ \ \ \ k\geq 4  \notag
\\
R_{\vec{\eta}}\big(\ad_{\epsilon_0}^{k-1} (b_k)\big) &= 0  \, , \ \ \ \ \ \ k\geq 2
\label{nilpot}
\end{align}
w.r.t.\ repeated adjoint action of $R_{\vec{\eta}}(\ep_0)$,
\beq
R_{\vec{\eta}}({\rm ad}_{x}^N y) =  {\rm ad}_{R_{\vec{\eta}}(x)}^N R_{\vec{\eta}}( y)
\label{defnilpot}
\eeq
The second line of (\ref{nilpot}) can be proven by the same arguments that apply
in an open-string context \cite{Broedel:2020tmd}. The first line of (\ref{nilpot}) in turn
is in general conjectural and has been tested through the explicit form of the
$n$-point matrices $R_{\vec{\eta}}(\epsilon_k) $ for a wide range of $k$ \& $n$.

Similar to the generators $\{ \epsilon^{\TS}_{k}, \ k \in 2\NN_0\}$ in Tsunogai's
derivation algebra \cite{Tsunogai, Pollack}, the $R_{\vec{\eta}}(\cdot)$
turn out to obey a variety of further relations besides (\ref{revsec.25})
and (\ref{nilpot}) that preserve the total weight $w= \sum_i k_i$ of $R_{\vec{\eta}}(b_{k_i})$
and $R_{\vec{\eta}}(\ep_{k_i})$. A key achievement in the later section~\ref{sec:5.2}
is to assemble the complete set of relations at fixed total weight $w \leq 11$,
based on the assumption (\ref{nilpot}). These relations in turn will be the key to anticipate
independent eMGFs in one variable.

%%%%%%%%%%%%%%%%%%%%%%%%%%%%%%%%%%%%%%%%%%%%%%%%%%%%%%%%%%%
\subsection{Expanding eMGFs around the cusp}
\label{sec:2.4}
%%%%%%%%%%%%%%%%%%%%%%%%%%%%%%%%%%%%%%%%%%%%%%%%%%%%%%%%%%%

In order to determine eMGFs from their differential equations in $\tau$,
one needs to supplement their behaviour at the cusp $\tau \rightarrow i \infty$
or $q\rightarrow 0$ for $q = e^{2\pi i \tau}$.
In the same way as the derivative $\nabla_\tau$ is taken at fixed comoving coordinates
$u,v$ of $z = u \tau{+}v$, we also hold $u,v$ fixed as $\tau \rightarrow i \infty$.
Moreover, we shall perform this limit at non-zero $0<u<1$ which suppresses
terms of the order $q^u$ due to powers of $e^{2\pi i z} = q^u e^{2\pi i v}$. 

Similar to MGFs, a variety of eMGFs were shown to yield Laurent polynomials in
\beq
y = \pi \Im \tau
\label{defyvar}
\eeq
upon expansion around the cusp. The coefficients in these Laurent polynomials of MGFs 
are MZVs \cite{Panzertalk} and conjecturally belong to the single-valued
subclass of MZVs \cite{Zerbini:2015rss, DHoker:2015wxz}.
For eMGFs, the coefficients in the Laurent polynomial at zeroth order in $q$
additionally involve Bernoulli polynomials $B_k(u)$ in $u$ such as $B_0(u)=1, \
B_1(u) = u-\frac{1}{2}$ or $B_2(u)=u^2-u+\frac{1}{6}$, see appendix \ref{app:Bern}
for a brief recap. 
However, setting $u \rightarrow 0$ in the Laurent polynomial of an eMGF
does {\it not} recover the Laurent polynomial of the MGF in its $z\rightarrow 0$ limit (\ref{MGFlimit}).  
The deviations stem from terms $q^{u},q^{2u},\ldots$ in the expansion of eMGFs 
that will be discussed in more detail in section \ref{sec:3.3}.

%%%%%%%%%%
%%%%%%%%%%
\subsubsection{Examples of eMGFs at the cusp}
\label{sec:2.4.1}

The explicit form of the Laurent polynomials capturing the behaviour of eMGFs
at the cusp is known for several examples \cite{DHoker:2018mys}.
The single-valued elliptic polylogarithms $\dplus{a \\ b}$ in (\ref{basic.13})
are well-known to degenerate to
\begin{align}
\dplus{a \\ b}\!(z|\tau) &= - \frac{(2i)^{b-a} B_{a+b}(u) (-4y)^a}{(a{+}b)!}+ {\cal O}(q^u,\bar q^u)
\notag \\
g_k(z|\tau) &= - \frac{ B_{2k}(u) (-4y)^k }{(2k)!} +{\cal O}(q^u,\bar q^u)
 \label{revsec.29}
\end{align}
where ${\cal O}(q^u,\bar q^u)$ refers to contributions with
at least one power of $q^u$ or $\bar q^u$, i.e.\ with a falloff
$\sim e^{-2yu} =  e^{-2\pi u \Im \tau}$ at the cusp. For the
multi-column eMGFs of section \ref{sec:2.1.3}, examples
of Laurent polynomials known from \cite{DHoker:2018mys} include
\begin{align}
C_{1 | 1,1}(z|\tau) &=
y^3 \bigg(  {-} \frac{8 B_{6}(u)}{15} - \frac{4 B_{4}(u)}{9} \bigg) + 2 B_{2}(u) \zeta_{3} + \frac{ \zeta_{5}}{4 y^2} +{\cal O}(q^u,\bar q^u)
\notag\\
C_{1|1,1,1}(z|\tau) &=  - y^4 \bigg( \frac{ 4}{7} B_8(u) + \frac{16}{15} B_6(u) + \frac{2}{9} B_4(u) \bigg)  + 2 \zeta_3 B_2(u) y
\notag \\
& \quad + \frac{ 3 \zeta_5}{2y} \bigg( B_2(u) + \frac{1}{6} \bigg) - \frac{ 3 \zeta_3^2 }{4y^2} + \frac{ 9 \zeta_7}{8y^3}+ {\cal O}(q^u,\bar q^u) \label{teas.11}\\
%%%
%%%
C_{1,1|1,1}(z|\tau) &= {-}y^4 \bigg( \frac{ 8}{35} B_8(u) + \frac{32}{45} B_6(u) + \frac{ 8}{27} B_4(u) - \frac{1}{2025} \bigg)
\notag \\
&\quad
+ \zeta_3 y \bigg( 8 B_4(u) + \frac{ 8}{3} B_2(u) +  \frac{2}{45} \bigg) + \frac{ \zeta_5}{y} \bigg( 6 B_2(u) + \frac{1}{3} \bigg) + \frac{ 3 \zeta_7 }{4y^3} +{\cal O}(q^u,\bar q^u)
\notag
\end{align}
We have made minor corrections to the expressions in \cite{DHoker:2018mys}
by adding a minus sign to the $y^4$-order of $C_{1|1,1,1}(z|\tau)$ and dividing
the contributions $\sim   \zeta_3 y$ to $C_{1,1|1,1}(z|\tau) $ by two.

Following our earlier comment, the limit $u \rightarrow 0$ of these expressions
(reducing Bernoulli polynomials to Bernoulli numbers $B_k=B_k(0)$) does {\it not}
reproduce the Laurent polynomial of the associated MGFs (recall that $E_k(\tau)=g_k(0|\tau)$),
\begin{align}
E_k(\tau) &= - \frac{ B_{2k} (-4y)^k }{(2k)!} + \frac{ 4 (2k{-}3)! \zeta_{2k-1} }{(k{-}2)! (k{-}1)! (4y)^{k-1}}
+{\cal O}(q,\bar q)
\notag \\
C_{1,1,1}(\tau) &=  \frac{2y^3}{945} + \zeta_3 + \frac{ 3 \zeta_5 }{4y^2} +{\cal O}(q,\bar q)
 \label{revsec.31} \\
C_{1,1,1,1}(\tau) &=  \frac{y^4}{945} +\frac{2 \zeta_3  y}{3} + \frac{ 10 \zeta_5 }{y}
- \frac{ 3 \zeta_3^2 }{y^2} + \frac{ 9 \zeta_7}{4y^3}+{\cal O}(q,\bar q)
\notag
\end{align}
where ${\cal O}(q,\bar q)$ refers to terms with at least one power of $q$ or $\bar q$,
i.e.\ with a falloff $\sim e^{-2y} = e^{-2\pi \Im \tau}$ at the cusp.

%%%%%%%%%%
%%%%%%%%%%
\subsubsection{Higher orders in expanding eMGFs around the cusp}
\label{sec:2.4.2}

The full expansion of MGFs around the cusp features Laurent polynomials
$ \sum_{\ell=\ell_{\rm min}}^{\ell_{\rm max}}
c^+_{\ell,m,n} y^\ell$ in $\Im \tau$ at each order in $q^m\bar q^n$
with $m,n\in \NN_0$,
\beq
\cplus{A \\ B}\!(\tau) = \sum_{m,n=0}^\infty \sum_{\ell=\ell_{\rm min}}^{\ell_{\rm max}}
c^+_{\ell,m,n}\big[ \begin{smallmatrix} A \\ B \end{smallmatrix} \big]
y^\ell q^m \bar q^n
 \label{revsec.32}
\eeq
The coefficients $c^+_{\ell,m,n}$ are again (conjecturally single-valued) MZVs,
and their explicit form can be conveniently obtained from the
iterated-integral representations to be reviewed in section \ref{sec:2.5}.
The expansion (\ref{revsec.32}) is invaluable for numerical
evaluations of MGFs since the sums over $m,n$ converge rapidly for a wide region
of ${\cal H} \ni \tau$ centered around $i$.

As will become apparent in section \ref{sec:3}, the expansion of eMGFs around
the cusp generalizes (\ref{revsec.32}) in two ways:
\begin{itemize}
\item the coefficients $c^+_{\ell,m,n}$ in general become polynomials in $u$
with $\mathbb Q$-linear combinations of MZVs as coefficients -- see
(\ref{revsec.29}) and (\ref{teas.11}) for an organization in
terms of Bernoulli polynomials
\item apart from integer powers $m,n \geq 0$ of $q$ and $\bar q$, one additionally
encounters series in $q^u$ and $q,q^{1\pm u}$ as well as complex conjugates
\end{itemize}
Based on the iterated-integral representations of eMGFs in later sections, the
coefficients in this expansion are accessible in all detail. In this way, one can
perform numerical evaluations at high precision to crosscheck the modular
properties or determine certain integration constants.

%%%%%%%%%%%%%%%%%%%%%%%%%%%%%%%%%%%%%%%%%%%%%%%%%%%%%%%%%%%
\subsection{MGFs from iterated Eisenstein integrals}
\label{sec:2.5}
%%%%%%%%%%%%%%%%%%%%%%%%%%%%%%%%%%%%%%%%%%%%%%%%%%%%%%%%%%%

We shall now review the iterated-Eisenstein-integral representations of MGFs that
expose their algebraic relations over $\QQ$-linear combinations of MZVs
and their expansion (\ref{revsec.32}) around the cusp. The central building
blocks $\bsvtau{j_1 &j_2 &\ldots &j_\ell \\ k_1 &k_2 &\ldots &k_\ell }$ introduced
in \cite{Gerken:2020yii} combine iterated integrals over holomorphic Eisenstein
series $G_{k_i}$ with their complex conjugates, for instance
\begin{align}
\bsvtau{j \\ k} &=  \frac{(2\pi i)^{-1}}{(4y)^{k-2-j}} \bigg\{ \int_{\tau}^{i \infty} \dd \tau_1 (\tau{-}\tau_1)^{k-2-j} (\bar\tau{-}\tau_1)^{j} G_k(\tau_1) \notag \\
&\hspace{2cm} - \! \int_{\bar\tau}^{-i\infty} \! \dd\bar\tau_1 (\tau{-}\bar\tau_1)^{k-2-j} (\bar\tau{-}\bar\tau_1)^{j} \overline{G_k(\tau_1)} \bigg\}
\label{eq:bsv1}\\
\bsvtau{j_1 &j_2\\ k_1 &k_2}&=
\sum_{p_1=0}^{k_1{-}2{-}j_1} \sum_{p_2=0}^{k_2{-}2{-}j_2} \frac{\binom{k_1{-}2{-}j_1}{p_1}\binom{k_2{-}2{-}j_2}{p_2}}{(4y)^{p_1+p_2}} \overline{\alphaBR{j_1 +p_1&j_2+p_2\\k_1 &k_2}{\tau}}
 + \frac{(2\pi i)^{-2}}{(4y)^{k_1+k_2-j_1-j_2-4}} \notag \\
&\! \! \!  \! \!  \! \!  \!  \! \! \! \times \bigg\{
\int^{i\infty}_\tau \dd\tau_2  (\tau{-}\tau_2)^{k_2-j_2-2} (\bar\tau{-}\tau_2)^{j_2} G_{k_2}(\tau_2) \int^{i\infty}_{\tau_2} \dd\tau_1 (\tau{-}\tau_1)^{k_1-j_1-2}(\bar\tau{-}\tau_1)^{j_1}  G_{k_1}(\tau_1) \nn\\
&\! \! \! \! \! \! \!  \!  \! \!  \! \quad  -  \int^{i\infty}_\tau \dd\tau_2(\tau{-}\tau_2)^{k_2-j_2-2} (\bar\tau{-}\tau_2)^{j_2}  G_{k_2}(\tau_2) \! \int^{-i\infty}_{\bar\tau} \! \dd\bar\tau_1  (\tau{-}\bar\tau_1)^{k_1-j_1-2}(\bar\tau{-}\bar\tau_1)^{j_1}\overline{G_{k_1}(\tau_1)}\nn\\
&\! \! \! \! \! \! \!  \!  \! \! \! \quad  + \! \int^{-i\infty}_{\bar\tau} \! \dd\bar\tau_1(\tau{-}\bar\tau_1)^{k_1-j_1-2}(\bar\tau{-}\bar\tau_1)^{j_1} \overline{G_{k_1}(\tau_1)} \! \int^{-i\infty}_{\bar\tau_1} \! \dd\bar\tau_2  (\tau{-}\bar\tau_2)^{k_2-j_2-2} (\bar\tau{-}\bar\tau_2)^{j_2} \overline{G_{k_2}(\tau_2)}\bigg\}\nn
\end{align}
where the endpoint divergences as $\tau_j \rightarrow i\infty$ are regularized via
tangential base points \cite{DeligneTBP, Brown:mmv}\footnote{Tangential-base-point regularization
is required for the zero-mode contributions $G_{k}(\tau)=2\zeta_k+{\cal O}(q)$ of
the holomorphic Eisenstein series in the integrand of (\ref{eq:bsv1}) and boils
down to discarding the endpoint singularity of $\int_\tau^{i\infty} \tau_1^j \dd \tau_1
= - \frac{\tau^{j+1}}{j{+}1}$ with $j\geq 0$.}. The integer entries of the
$\beta^{\rm sv}[\ldots]$ are in the range $k_i \geq 4$ and $0\leq j_i \leq k_i{-}2$, and the
number of columns $\begin{smallmatrix} j_i \\ k_i\end{smallmatrix}$ is referred to as
the {\it depth}. Starting from the depth-two example in (\ref{eq:bsv1}), one
encounters antiholomorphic contributions $\overline{\alphaBR{j_1 &j_2 &\ldots &j_\ell
\\ k_1 &k_2 &\ldots &k_\ell}{\tau}}$ dubbed integration constants that vanish
at the cusp. They always carry three or more units of transcendental
weight via MZVs and can be viewed as genus-one analogues of the
contributions $\mathcal{Z}^{\rm sv}$ to single-valued multiple polylogarithms reviewed 
in appendix~\ref{app:svpoly}.

The $\beta^{\rm sv}[\ldots]$ are invariant under the modular $T$ transformation
$\tau \rightarrow \tau{+}1$ but transform inhomogeneously under the $S$ transformation
$\tau \rightarrow -\frac{1}{\tau}$ \cite{Gerken:2020yii, Dorigoni:2021ngn}.
As will be exemplified below in (\ref{fkreps.19}),
the integration constants $\overline{\alpha[\ldots]}$ depend on $\bar \tau$ via
antiholomorphic combinations of iterated Eisenstein integrals that preserve
the $T$ invariance of $\beta^{\rm sv}[\ldots]$.

Any convergent MGF can be expressed in terms of $\beta^{\rm sv}$ with non-positive
powers of $y$ and $\mathbb Q$-linear combinations of (conjecturally single-valued)
MZVs in their coefficients \cite{Gerken:2020yii}.
This has been shown in the reference by embedding MGFs of arbitrary graph topology into
generating series similar to (\ref{eq:Y2pt}) and (\ref{high.4}) such that their holomorphic
$\tau$-derivative takes the form of (\ref{revsec.22}) with $z_{01}=0$. The $\beta^{\rm sv}$
parametrize a perturbative solution of (\ref{revsec.22}) at $z_{01}=0$ and thereby
obey the following holomorphic differential equation
\begin{align}
-4\pi \nabla_\tau \bsvtau{j_1 &j_2 &\ldots &j_\ell \\ k_1 &k_2 &\ldots &k_\ell  }
&= \sum_{i=1}^\ell (k_i{-}j_i{-}2) \bsvtau{j_1  &\ldots &j_{i}+1 &\ldots &j_\ell \\ k_1 &\ldots &k_i &\ldots &k_\ell  } \label{nabbsv.1}  \\
&\ \ \ \
- \delta_{j_\ell,k_\ell-2} (\tau{-}\bar \tau)^{k_\ell} G_{k_\ell}(\tau)
\bsvtau{j_1 &j_2 &\ldots &j_{\ell-1} \\ k_1 &k_2 &\ldots &k_{\ell-1}} \notag
\end{align}
which is readily confirmed for the examples (\ref{eq:bsv1}) at depth $\ell \leq 2$.
We will introduce a generalization of the $\beta^{\rm sv}$ in section \ref{sec:3.2}
with kernels $f^{(k)}$ in the place of the holomorphic Eisenstein series $G_{k}$
in (\ref{eq:bsv1}) and (\ref{nabbsv.1}). In this way, we will construct a perturbative
solution of (\ref{high.12}) at generic $z_{01} \in \Sigma$ and arrive at a canonical
representation of eMGFs.

%%%%%%%%%%
%%%%%%%%%%
\subsubsection{$\beta^{\rm sv}$ versus holomorphic iterated Eisenstein integrals}
\label{sec:2.5.1}

As can be anticipated from the examples at depth $\ell \leq 2$ in (\ref{eq:bsv1}),
the building blocks $\beta^{\rm sv}$ are expressible via real-analytic combinations
of Brown's holomorphic iterated Eisenstein integrals \cite{Brown:mmv}
\beq
\eeetau{j_1 &j_2 &\ldots &j_\ell \\ k_1 &k_2 &\ldots &k_\ell  }
=  (2\pi i )^{1+j_\ell -k_\ell} \int_\tau^{i\infty} \dd \tau_\ell \, \tau_\ell^{j_\ell} G_{k_\ell}( \tau_\ell)
\EBR{j_1 &j_2 &\ldots &j_{\ell-1} \\ k_1 &k_2 &\ldots &k_{\ell-1} }{\tau_\ell} \label{gen.36a}
\eeq
with $\eeetau{ \emptyset \\ \emptyset  } = 1 $, entries $k_i \geq 4 $ and $ 0 \leq j_i \leq k_i{-}2$
as well as tangential-base-point regularization of the endpoint divergences.
The recursive definition manifests the differential equation
\begin{align}
2\pi i \partial_\tau
\eeetau{j_1 &j_2 &\ldots &j_\ell \\ k_1 &k_2 &\ldots &k_\ell  }
&= - (2\pi i)^{2-k_\ell + j_\ell} \tau^{j_\ell} G_{k_\ell}( \tau)
\eeetau{j_1 &j_2 &\ldots &j_{\ell-1} \\ k_1 &k_2 &\ldots &k_{\ell-1} }
\label{gen.35x}
\end{align}
which may be viewed as a holomorphic antecedent of (\ref{nabbsv.1}).
Given that holomorphic iterated Eisenstein integrals (\ref{gen.36a}) are linearly
independent for different choices of $j_i,k_i$ \cite{Nilsnewarticle}, the same is true for
their real-analytic combination $\beta^{\rm sv}[\ldots]$. Similarly, the shuffle property
\beq
{\cal E}[A;\tau] {\cal E}[B;\tau] = \sum_{C \in A\shuffle B} {\cal E}[C;\tau]
\label{EEshuffle}
\eeq
for words $A,B$ in the combined letters $\begin{smallmatrix} j_i \\ k_i \end{smallmatrix}$
(see the explanation below (\ref{Gshuffle}) for the shuffle product $A \shuffle B$)  
propagate to the $\beta^{\rm sv}$,
\beq
\beta^{\rm sv}[A;\tau] \beta^{\rm sv}[B;\tau] = \sum_{C \in A\shuffle B} \beta^{\rm sv}[C;\tau]
\label{bsvshuffle}
\eeq
assuming that the explicit form of the integration constants $\overline{ \alpha[\ldots]}$
\cite{Gerken:2020yii, Dorigoni:2021jfr} does not introduce any obstructions.
By the absence of $\overline{ \alpha[\begin{smallmatrix}
j \\ k \end{smallmatrix};\tau]}$ at depth one, shuffle relations of
$\bsvtau{j_1 &j_2  \\ k_1 &k_2  }$ are equivalent to the
antisymmetry $\overline{\alphaBR{j_1 &j_2 \\ k_1 &k_2}{\tau}}
=-\overline{\alphaBR{j_2 &j_1  \\ k_2 &k_1}{\tau}}$ at depth two.

It is straightforward to reduce the $\beta^{\rm sv}[\ldots]$ to ${\cal E}[\ldots]$
and their complex conjugates by binomial expansion of all the $(\tau{-}\tau_i)^{k_i-j_i-2}$
and $(\bar \tau{-}\tau_i)^{j_i}$ in (\ref{eq:bsv1}). In earlier references
\cite{Gerken:2020yii, Gerken:2020xfv}, these expansions have been performed
via intermediate objects ${\cal E}^{\rm sv}[\ldots]$
that are reviewed in appendix \ref{sec:appesv}.

%%%%%%%%%%
%%%%%%%%%%
\subsubsection{$q$-series from holomorphic iterated Eisenstein integrals}
\label{sec:2.5.2}

Instead of considering powers of $\tau$ among the integration kernels of (\ref{gen.36a}),
one can admit one additional kernel $G_0 = -1$ and define
\begin{align}
    \mathcal{E}(k_1 , k_2 , \ldots , k_r ;\tau) &=   (2 \pi i)^{1-k_r} \int^{i \infty}_{\tau}   \mathrm{d} \tau_{r}   \, G_{k_r}(\tau_{r} ) \mathcal{E}(k_1,k_2, \ldots ,k_{r-1};\tau_{r}) \label{fkreps.5a}
    \end{align}
with $ \mathcal{E}( \emptyset ;\tau) =1$ and $k_i\geq 0$. Depth-$\ell$ integrals
over kernels $\tau^j G_{k\geq 4}$ in (\ref{gen.36a}) can be straightforwardly expanded
in terms of the modified integrals (\ref{fkreps.5a}) with $\ell$ non-zero entries $k_i$, e.g.\
\cite{Broedel:2018izr}
\beq
\eeetau{j_1  \\ k_1  } = j_1! {\cal E}(\vec{0}^{j_1},k_1;\tau) \, , \ \ \ \
\eeetau{j_1 &j_2  \\ k_1 &k_2  }  = j_2! \sum_{a=0}^{j_2}\frac{ (j_1{+}a)! }{a!}
{\cal E}(\vec{0}^{j_1+a},k_1, \vec{0}^{j_2-a},k_2;\tau)
 \label{revsec.41}
\eeq
In order to facilitate $q$-expansions and identify $T$-invariant building blocks, one
subtracts the zero mode of
\beq
G_k(\tau)
= 2 \zeta_{k}+ \frac{2 (2 \pi i )^k }{(k{-}1)!}  \sum_{n=1}^\infty \frac{ n^{k-1} q^n}{1-q^n}
= \frac{ (2 \pi i )^k }{(k{-}1)!} \bigg[ {-} \frac{ B_k }{k} + 2\sum_{n=1}^\infty \frac{ n^{k-1} q^n}{1-q^n} \bigg]
\label{fkreps.4}
\eeq
to form modified kernels $G_k^0(\tau)$ with associated iterated integrals ${\cal E}_0(\ldots)$
\cite{Broedel:2015hia},
\begin{align}
G_k^0(\tau) &= G_k(\tau) - 2 \zeta_k =  \frac{2 (2 \pi i )^k }{(k{-}1)!}  \sum_{n=1}^\infty \frac{ n^{k-1} q^n}{1-q^n} \, , \ \ \ \ k \geq 4
 \label{fkreps.5b}  \\
\mathcal{E}_0(k_1 , k_2 , \ldots , k_r ;\tau) &=   (2 \pi i)^{1-k_r} \int^{i \infty}_{\tau}   \mathrm{d} \tau_{r}   \, G^0_{k_r}(\tau_{r} ) \mathcal{E}_0(k_1,k_2, \ldots ,k_{r-1};\tau_{r})  \notag
\end{align}
while preserving $\mathcal{E}_0(\emptyset ;\tau)=1$ and the kernel $G_0 = G_0^0 = -1$. 
Both of ${\cal E}(\ldots)$ and ${\cal E}_0(\ldots)$ obey shuffle relations, and the regularized value
\beq
\mathcal{E}(0;\tau)=\mathcal{E}_0(0;\tau)=2\pi i \tau = \log(q)
\label{regval}
\eeq
can be used to enforce a non-zero first entry, for instance
\begin{align}
&\mathcal{E}_0( \vec{0}^{p_1-1} ,  k_1 ;\tau) = \sum_{n = 0}^{p_1-1}\frac{(-1)^{n} }{ \left(p_1{-}n{-}1\right)!}  \mathcal{E}_0(k_1 , \vec{0}^{n};\tau) \log(q)^{p_1-n-1}
    \label{fkreps.80}
\end{align}
Any $\mathcal{E}_0(k_1,\ldots;\tau)$ with $k_1 \neq 0$ is defined in terms of convergent
integrals (\ref{fkreps.5b}), and its $T$-invariance is manifest in the simple $q$-expansion \cite{Broedel:2015hia}
\begin{align}
&{\cal E}_0(k_1,\vec{0}^{p_1-1},k_2,\vec{0}^{p_2-1},\ldots,k_r,\vec{0}^{p_r-1};\tau) =(-2)^r
  \bigg(  \prod_{j=1}^{r} \frac{ 1 }{(k_j{-}1)!} \bigg)
\label{qgamma1}\\
& \ \ \ \ \ \ \ \times \sum_{m_i,n_i=1}^{\infty} \frac{m_1^{k_1-1} m_2^{k_2-1} \ldots m_r^{k_r-1}  q^{m_1n_1+m_2n_2+\ldots +m_rn_r}}{(m_1 n_1)^{p_1} (m_1n_1+m_2n_2)^{p_2} \ldots (m_1n_1+m_2n_2+\ldots +m_rn_r)^{p_r}}   \notag
\end{align}
The convergent $\mathcal{E}_0(k_1 , \vec{0}^{n};\tau),\ k_1\geq 4$ of depth one on the 
right-hand side of (\ref{fkreps.80}) can in fact be expanded in terms of the polylogarithms 
${\rm Li}_p(q^n)$ in (\ref{fkreps.11}),
\beq
{\cal E}_0(k_1,\vec{0}^{p_1-1};\tau) = -\frac{2}{(k_1{-}1)!} \sum_{n=1}^\infty
n^{k_1-p_1-1} {\rm Li}_p(q^n)
 \label{revsec.42}
\eeq
The conversions of Brown's iterated Eisenstein integrals ${\cal E}[\ldots]$ at higher depth
to convergent ${\cal E}_0(\ldots)$ and hence the $q$-series in (\ref{qgamma1}) are
discussed in section 3.3 and appendix D of \cite{Broedel:2018izr} as well as
appendix G of \cite{Gerken:2020yii}.

%%%%%%%%%%
%%%%%%%%%%
\subsubsection{$q$-series for $\beta^{\rm sv}$}
\label{sec:2.5.3}

In order to illustrate the relevance of the $q$-series ${\cal E}_0(\ldots)$
for MGFs, we recall the closed formula at depth one \cite{Dorigoni:2021jfr}
\begin{align}
\bsvtau{j \\k} &\nn=  \frac{B_{k}  j! (k{-}2{-}j)! (-4y)^{j+1} }{k! \, (k{-}1)!}  +\sum_{a=0}^{j} (k{-}j{-}2{+}a)! \binom{j}{a} (4y)^{2+2j-k-a} {\cal E}_0(k,0^{k-j-2+a})\\
&\quad\label{eq:betasvE0}+\sum_{b=0}^{k-j-2} (j{+}b)! \binom{k{-}2{-}j}{b} (4y)^{-b}\overline{ {\cal E}_0(k,0^{j+b})}
\end{align}
Moreover, the integration constants $\overline{\alphaBR{j_1 &j_2 &\ldots &j_\ell
\\ k_1 &k_2 &\ldots &k_\ell}{\tau}}$ at depth $\ell\geq 2$ are expressible in terms
of ${\cal E}_0(k_1,\ldots)$ at $k_1\neq 0$, and typical depth-two results are
\begin{align}
 \overline{\alphaBR{2&0 \\ 4 &4}{\tau}} &= \frac{2 \zeta_3}{3} \Big( \overline{ \eeetau{0 \\ 4  }  } - \frac{ i \pi \bar\tau}{360}  \Big) =  \frac{2 \zeta_3}{3}  \overline{ {\cal E}_0(4;\tau)  }\notag \\
  \overline{\alphaBR{2&2 \\ 6 &4}{\tau}} &= -\frac{ \zeta_3}{315} \overline{{\cal E}_0(4,0,0;\tau)}
   - \frac{4 \zeta_3}{3} \overline{ {\cal E}_0(6,0,0;\tau)}
  \label{fkreps.19} \\
  \overline{\alphaBR{4&2 \\ 6 &4}{\tau}} &=
- 16 \zeta_3 \overline{ {\cal E}_0(6,0,0,0,0;\tau)}
+ \frac{ 4 \zeta_5}{5} \overline{ {\cal E}_0(4,0,0;\tau)  }
 \notag
\end{align}
Following earlier examples in \cite{Gerken:2020yii, Gerken:2020xfv}, the expressions for
all $\overline{\alphaBR{j_1&j_2 \\ k_1 &k_2}{\tau}}$ with $k_1{+}k_2\leq 28$
in terms of $\zeta_{2a+1} {\cal E}_0(2b, \vec{0}^p;\tau)$ can be
found in the ancillary file of \cite{Dorigoni:2021jfr}. Closed formulae
at depth two and three are under investigation.

\newpage

%%%%%%%%%%%%%%%%%%%%%%%%%%%%%%%%%%%%%%%%%%%%%%%%%%%%%%%%%%%
%%%%%%%%%%%%%%%%%%%%%%%%%%%%%%%%%%%%%%%%%%%%%%%%%%%%%%%%%%%
\section{Iterated integrals for eMGFs}
\label{sec:3}
%%%%%%%%%%%%%%%%%%%%%%%%%%%%%%%%%%%%%%%%%%%%%%%%%%%%%%%%%%%
%%%%%%%%%%%%%%%%%%%%%%%%%%%%%%%%%%%%%%%%%%%%%%%%%%%%%%%%%%%

In this section, we construct $z$-dependent generalizations
of the $\beta^{\rm sv}$ in section \ref{sec:2.5} to describe eMGFs in one variable. The differential 
equations of eMGFs and their
generating series in $\tau$ together with their reality properties and asymptotics at the cusp
induce unique iterated-integral representations with kernels $\tau^j G_k(\tau)$
and $\tau^j f^{(k)}(u\tau{+}v,\tau)$ at fixed $u,v$. The goals of this section are to
present several viewpoints on such $u,v$-dependent iterated integrals, to illustrate
the subtle role of contributions $q^u$ in their asymptotics at the cusp and to obtain
the $q$-expansions of the $\beta^{\rm sv}$.

%%%%%%%%%%%%%%%%%%%%%%%%%%%%%%%%%%%%%%%%%%%%%%%%%%%%%%%%%%%
\subsection{Generating series of eMGFs in terms of generalized $\beta^{\rm sv}$}
\label{sec:3.1}
%%%%%%%%%%%%%%%%%%%%%%%%%%%%%%%%%%%%%%%%%%%%%%%%%%%%%%%%%%%

The differential equation (\ref{high.12}) of the generating series $Y$ of eMGFs in (\ref{high.4}) can
be streamlined to remove the term $\sim R_{\vec{\eta}}(\epsilon_0)$ without any
accompanying $\tau$-dependent kernel. Following the analogous construction for generating series of MGFs in section 3.1 of \cite{Gerken:2020yii}, we redefine the $n!\times n!$ matrix
$Y( \begin{smallmatrix} M \\ N \end{smallmatrix} |  \begin{smallmatrix} K \\ L \end{smallmatrix})$
\beq
\widehat Y( \begin{smallmatrix} M \\ N \end{smallmatrix} |  \begin{smallmatrix} K \\ L \end{smallmatrix}) = \sum_{P,Q} \exp\bigg( \frac{ R_{\vec{\eta}}(\epsilon_0) }{4y} \bigg)_{ \begin{smallmatrix} K \\ L \end{smallmatrix} \big|  \begin{smallmatrix} P \\ Q \end{smallmatrix}} Y( \begin{smallmatrix} M \\ N \end{smallmatrix} |  \begin{smallmatrix} P \\ Q \end{smallmatrix})
\label{gen.31}
\eeq
such that $\widehat Y$ obeys the following modified version of (\ref{high.12})
(recall that $y=\pi \Im\tau$ and  $R_{\vec{\eta}}(ab)= R_{\vec{\eta}}(a) R_{\vec{\eta}}(b)$, and
see the discussion below (\ref{high.1}) for the range of $P,Q$)
\begin{align}
2\pi i \partial_\tau \widehat Y( \begin{smallmatrix} M \\ N \end{smallmatrix} |  \begin{smallmatrix} K \\ L \end{smallmatrix})= \sum_{P,Q} \Big[
 &\sum_{k=2}^\infty (k{-}1) (\tau{-}\bar \tau)^{k-2} f^{(k)}(z_{01}|\tau) R_{\vec{\eta}}\big( e^{\frac{  \epsilon_0 }{4y}}  b_k e^{-\frac{  \epsilon_0 }{4y}}  \big) \label{gen.32} \\
 +&\sum_{k=4}^\infty (1{-}k) (\tau{-}\bar \tau)^{k-2} G_k(\tau) R_{\vec{\eta}}\big( e^{\frac{  \epsilon_0 }{4y}}  \epsilon_k e^{-\frac{  \epsilon_0 }{4y}}  \big)
\Big]_{ \begin{smallmatrix} K \\ L \end{smallmatrix} \big|  \begin{smallmatrix} P \\ Q \end{smallmatrix}} \widehat Y( \begin{smallmatrix} M \\ N \end{smallmatrix} |  \begin{smallmatrix} P \\ Q \end{smallmatrix})
\notag
\end{align}
By the conjectural nilpotency properties (\ref{nilpot}) of $R_{\vec{\eta}}( \epsilon_k)$
\& $R_{\vec{\eta}}( b_k)$, the exponentials
\beq
 R_{\vec{\eta}}( e^{\frac{\epsilon_0}{4y}}  x e^{-\frac{  \epsilon_0 }{4y}} ) = \sum_{j=0}^\infty \frac{1}{j!} \Big(
\frac{1}{4y} \Big)^j  R_{\vec{\eta}}\big({\rm ad}^j_{\epsilon_0}(x)\big)
\label{gen.33}
\eeq
truncate to a finite number of terms with $j=0,1,2,\ldots,k{-}2$:
\begin{align}
2\pi i \partial_\tau \widehat Y( \begin{smallmatrix} M \\ N \end{smallmatrix} |  \begin{smallmatrix} K \\ L \end{smallmatrix}) &= \sum_{P,Q} \Big[
 \sum_{k=2}^\infty (k{-}1)  f^{(k)}(z_{01}|\tau) \sum_{j=0}^{k-2} \frac{ (-1)^j}{j! (2\pi i )^j} (\tau{-}\bar \tau)^{k-2-j} R_{\vec{\eta}}\big(  {\rm ad}^j_{\epsilon_0}( b_k)   \big) \label{gen.34} \\
 &\ \ \ \
+\sum_{k=4}^\infty (1{-}k)  G_k(\tau) \sum_{j=0}^{k-2}  \frac{ (-1)^j}{j! (2\pi i )^j}  (\tau{-}\bar \tau)^{k-2-j} R_{\vec{\eta}}\big(
{\rm ad}^j_{\epsilon_0}( \epsilon_k)    \big)
\Big]_{ \begin{smallmatrix} K \\ L \end{smallmatrix} \big|  \begin{smallmatrix} P \\ Q \end{smallmatrix}} \widehat Y( \begin{smallmatrix} M \\ N \end{smallmatrix} |  \begin{smallmatrix} P \\ Q \end{smallmatrix})
\notag
\end{align}
%

%%%%%%%%%%%%%%%%%%%%%%%%%%%%%%%%%%%%%%%%%%%%%%%%%%%%%%%%%%%
\subsubsection{Perturbative solution for the redefined series $\widehat Y$}
\label{sec:3.1.1}
%%%%%%%%%%%%%%%%%%%%%%%%%%%%%%%%%%%%%%%%%%%%%%%%%%%%%%%%%%%

One can construct a formal solution to the differential equation (\ref{gen.34})
in terms of iterated integrals ${\cal E}^{\rm sv}[\ldots]$ that generalize the ones
over holomorphic Eisenstein series in appendix \ref{sec:appesv}. By extending
the set of kernels $\tau^j G_{k}$ with $k\geq 4$ and $0\leq j\leq k{-}2$
to $\tau^j f^{(k)}$ with $k\geq 2$ and $0\leq j\leq k{-}2$, we reduce (\ref{gen.34})
to the simpler problem of solving
\begin{align}
2\pi i \partial_\tau
\esvtau{j_1 &j_2 &\ldots &j_\ell \\ k_1 &k_2 &\ldots &k_\ell \\ z_1 &z_2 &\ldots &z_\ell}
&=
(2\pi i)^{2-k_\ell + j_\ell} (\tau {-} \bar \tau)^{j_\ell} f^{(k_\ell)}(z_\ell | \tau)
\esvtau{j_1 &j_2 &\ldots &j_{\ell-1} \\ k_1 &k_2 &\ldots &k_{\ell-1} \\ z_1 &z_2 &\ldots &z_{\ell-1}}
\label{gen.35z}
\end{align}
with $\esvtau{ \emptyset \\ \emptyset  \\ \emptyset} = 1 $ and $z_i \in \{0,z\}$.
Just as in the differential equation (\ref{high.12}) of the generating series $Y$,
the $\tau$-derivative in (\ref{gen.35z}) is performed at fixed co-moving coordinates
$u,v$ of $z=u\tau{+}v$. Given that $G_k(\tau)=
- f^{(k)}(0|\tau)$, the special case $z_\ell=0$ of (\ref{gen.35z}) lines up with
(\ref{gen.35}) for building blocks of MGFs,
\begin{align}
2\pi i \partial_\tau
\esvtau{j_1 &j_2 &\ldots &j_{\ell-1} &j_\ell \\ k_1 &k_2 &\ldots &k_{\ell-1} &k_\ell \\ z_1 &z_2 &\ldots
&z_{\ell-1} &\emptyslot }
&=
 - (2\pi i)^{2-k_\ell + j_\ell} (\tau {-} \bar \tau)^{j_\ell} G_{k_\ell}( \tau)
\esvtau{j_1 &j_2 &\ldots &j_{\ell-1} \\ k_1 &k_2 &\ldots &k_{\ell-1} \\ z_1 &z_2 &\ldots &z_{\ell-1}}
\label{gen.35za}
\end{align}
where the vanishing $z_\ell$ has been omitted in the notation on the left-hand side.
As we will discuss in detail in section~\ref{sec:new.1.disc},
the limit $\lim_{z \to 0}f^{(k)}(z|\tau)=-G_k(\tau)$ does not commute with $\tau$-integration.
Hence, we shall in general replace $z_r$ in (\ref{gen.35z}) or (\ref{gen.35za}) by an
empty slot when referring to an integration kernel
${-}(\tau {-} \bar \tau)^{j_r} G_{k_r}( \tau)$ instead of
$(\tau {-} \bar \tau)^{j_r} f^{(k_r)}(z_r | \tau)$ with $z_r\neq 0$.
This notation unifying iterated $\tau$-integrals over $G_k$ and $f^{(k)}$
relates to the $z$-independent case via
\beq
\esvtau{j_1 &j_2 &\ldots &j_\ell \\ k_1 &k_2 &\ldots &k_\ell \\ \emptyslot & \emptyslot &\ldots & \emptyslot } = \esvtau{j_1 &j_2 &\ldots &j_\ell \\ k_1 &k_2 &\ldots &k_\ell  } \, , %\ \ \ \ \ \
\label{eq:emptyslot}
\eeq
and leads to an unambiguous bookkeeping for the solutions of~\eqref{gen.35za}.
The number $\ell$ of columns will be referred to as the {\it depth} of ${\cal E}^{\rm sv}$
in (\ref{gen.35z}) and (\ref{gen.35za}).

Based on (\ref{gen.35z}) and
\begin{align}
c_{j,k} = \frac{(-1)^{k-j} (k{-}1) }{(k{-}j{-}2)!} \, , \ \ \ \ \ \ \EsvBR{\ldots &j &\ldots \\ \ldots &k &\ldots \\ \ldots &\emptyslot  &\ldots }{\tau} = 0 \ \ {\rm if} \ k=2 \ {\rm or} \ k \ {\rm odd}
\label{gen.34b}
\end{align}
it is easy to check that (\ref{gen.34}) is solved by the series \small
\begin{align}
\widehat{Y}&( \begin{smallmatrix} M \\ N \end{smallmatrix} |  \begin{smallmatrix} K \\ L \end{smallmatrix} | \tau)  = \sum_{P,Q} \bigg\{ 1 +  \sum_{k_1=2}^\infty \sum_{j_1=0}^{k_1-2} c_{j_1,k_1}
\bigg(   \EsvBR{j_1 \\ k_1 \\  \emptyslot }{\tau} R_{\vec{\eta}}\big( {\rm ad}_{\ep_0}^{k_1-j_1-2}(\ep_{k_1}) \big)\notag \\
&\ \ \ \ \ \ \ \ \ \ \ \ \ \ \ \ \ \
+ \EsvBR{j_1 \\ k_1 \\ z }{\tau} R_{\vec{\eta}}\big( {\rm ad}_{\ep_0}^{k_1-j_1-2}(b_{k_1}) \big)\bigg)  \notag \\
&+ \sum_{k_1,k_2=2}^\infty \sum_{j_1=0}^{k_1-2} \sum_{j_2=0}^{k_2-2} c_{j_1,k_1}c_{j_2,k_2}
\bigg(  \EsvBR{j_1 &j_2 \\ k_1 &k_2 \\  \emptyslot & \emptyslot }{\tau} R_{\vec{\eta}}\big( {\rm ad}_{\ep_0}^{k_2-j_2-2}(\ep_{k_2}) {\rm ad}_{\ep_0}^{k_1-j_1-2}(\ep_{k_1}) \big) \notag \\
&\ \ \ \ \ \ \ \ \ \ \ \ \ \ \ \ \ \
+ \EsvBR{j_1 &j_2 \\ k_1 &k_2 \\ z & \emptyslot }{\tau} R_{\vec{\eta}}\big( {\rm ad}_{\ep_0}^{k_2-j_2-2}(\ep_{k_2}) {\rm ad}_{\ep_0}^{k_1-j_1-2}(b_{k_1}) \big) \label{gen.34a} \\
&\ \ \ \ \ \ \ \ \ \ \ \ \ \ \ \ \ \
+ \EsvBR{j_1 &j_2 \\ k_1 &k_2 \\ \emptyslot &z }{\tau} R_{\vec{\eta}}\big(
 {\rm ad}_{\ep_0}^{k_2-j_2-2}(b_{k_2}) {\rm ad}_{\ep_0}^{k_1-j_1-2}(\ep_{k_1}) \big) \notag \\
 &\ \ \ \ \ \ \ \ \ \ \ \ \ \ \ \ \ \
+ \EsvBR{j_1 &j_2 \\ k_1 &k_2 \\ z &z }{\tau} R_{\vec{\eta}}\big(  {\rm ad}_{\ep_0}^{k_2-j_2-2}(b_{k_2}) {\rm ad}_{\ep_0}^{k_1-j_1-2}(b_{k_1}) \big)
\bigg) + \ldots
\bigg\}_{ \begin{smallmatrix} K \\ L \end{smallmatrix} \big|  \begin{smallmatrix} P \\ Q \end{smallmatrix}} \, \widehat{Y}( \begin{smallmatrix} M \\ N \end{smallmatrix} |  \begin{smallmatrix} P \\ Q \end{smallmatrix} | i \infty) \notag
\end{align} \normalsize
The ellipsis comprises solutions ${\cal E}^{\rm sv}[\ldots]$ to (\ref{gen.35z}) and (\ref{gen.35za})
of depth $\ell \geq 3$, and specifying their interplay
with the $n!\times n!$ matrices
$R_{\vec{\eta}}(\ep_k) $ and $R_{\vec{\eta}}(b_k) $ is mostly a
notational exercise: The contribution at depth $\ell$ is a multiple
sum over $k_1,k_2,\ldots,k_\ell \geq 2$ with associated $0\leq j_i \leq k_i{-}2$, and the summand
gathers all the $2^\ell$ possibilities to insert empty slots in the place of the $z$'s in $\EsvBR{j_1  &\ldots &j_\ell \\ k_1 &\ldots &k_\ell \\ z &\ldots &z }{\tau}$
while replacing the accompanying $b_{k_i}$ by $\ep_{k_i}$.
The dependence of both $\widehat{Y}( \begin{smallmatrix} M \\ N \end{smallmatrix} |  \begin{smallmatrix} K \\ L \end{smallmatrix} | \tau)$ and $\widehat{Y}( \begin{smallmatrix} M \\ N \end{smallmatrix} |  \begin{smallmatrix} P \\ Q \end{smallmatrix} | i \infty)$ on $u,v$ is suppressed to avoid cluttering.

The leftover steps in making the solution (\ref{gen.34a}) explicit are
\begin{itemize}
\item to construct $z$-dependent real-analytic ${\cal E}^{\rm sv}[\ldots]$ subject
to (\ref{gen.35z}) with the appropriate reality properties of eMGFs
\item to find a suitable initial value $\widehat Y(i\infty)$ which is independent of $\tau$ but may depend on $(u,v)$ to reproduce the behaviour of eMGFs at the cusp as reviewed in section~\ref{sec:2.4} (see section~\ref{sec:9.1} for two-point examples of $\widehat Y(i\infty)$).
\end{itemize}

%%%%%%%%%%%%%%%%%%%%%%%%%%%%%%%%%%%%%%%%%%%%%%%%%%%%%%%%%%%
\subsubsection{Perturbative solution for the original series $Y$}
\label{sec:3.1.2}
%%%%%%%%%%%%%%%%%%%%%%%%%%%%%%%%%%%%%%%%%%%%%%%%%%%%%%%%%%%

As in \cite{Gerken:2020yii}, the redefinition (\ref{gen.31}) can be undone by forming the linear combinations
\begin{align}
\bsvtau{j_1 &j_2 &\ldots &j_\ell \\ k_1 &k_2 &\ldots &k_\ell \\ z_1&z_2 &\ldots &z_\ell}
&= \sum_{p_1=0}^{k_1-j_1-2}  \sum_{p_2=0}^{k_2-j_2-2} \ldots  \sum_{p_\ell=0}^{k_\ell-j_\ell-2}
{ k_1{-}j_1{-}2 \choose p_1}  { k_2{-}j_2{-}2 \choose p_2} \ldots { k_\ell{-}j_\ell{-}2 \choose p_\ell} \notag \\
&\ \ \ \ \times \Big( \frac{1}{4y} \Big)^{p_1+p_2+\ldots +p_\ell}
\esvtau{j_1+p_1 &j_2+p_2 &\ldots &j_\ell + p_\ell \\ k_1 &k_2 &\ldots &k_\ell \\ z_1&z_2 &\ldots &z_\ell}
\label{gen.38}
\end{align}
of the expansion coefficients ${\cal E}^{\rm sv}[\ldots]$ in (\ref{gen.34a})
which also applies to empty slots in the place of $z_r$ as explained
below (\ref{gen.35za}). The number $\ell$ of columns is again referred to as the
{\it depth} of the $\beta^{\rm sv}$ in (\ref{gen.38}). The expansion of the original series $Y$
instead of $\widehat Y$ is then given by  \small
\begin{align}
Y(\begin{smallmatrix} M \\ N \end{smallmatrix} |  \begin{smallmatrix} K \\ L \end{smallmatrix} |\tau) &= 
\sum_{P,Q,R,S}\bigg\{ 1+ \sum_{k_1=2}^\infty \sum_{j_1=0}^{k_1-2} c_{j_1,k_1}
\bigg(   \bsvtau{j_1 \\ k_1 \\  \emptyslot} R_{\vec{\eta}}\big( {\rm ad}_{\ep_0}^{k_1-j_1-2}(\ep_{k_1}) \big)
\notag \\
&\ \ \ \ \ \quad\quad\quad\quad\quad+\bsvtau{j_1 \\ k_1 \\ z }  R_{\vec{\eta}}\big( {\rm ad}_{\ep_0}^{k_1-j_1-2}(b_{k_1}) \big)\bigg)  \notag \\
&\quad + \sum_{k_1,k_2=2}^\infty \sum_{j_1=0}^{k_1-2} \sum_{j_2=0}^{k_2-2} c_{j_1,k_1}c_{j_2,k_2}
\bigg(  \bsvtau{j_1 &j_2 \\ k_1 &k_2 \\  \emptyslot&\emptyslot } R_{\vec{\eta}}\big( {\rm ad}_{\ep_0}^{k_2-j_2-2}(\ep_{k_2}) {\rm ad}_{\ep_0}^{k_1-j_1-2}(\ep_{k_1}) \big) \notag \\
&\ \ \ \ \
\quad\quad\quad\quad\quad+ \bsvtau{j_1 &j_2 \\ k_1 &k_2 \\ z & \emptyslot } R_{\vec{\eta}}\big( {\rm ad}_{\ep_0}^{k_2-j_2-2}(\ep_{k_2}) {\rm ad}_{\ep_0}^{k_1-j_1-2}(b_{k_1}) \big) \label{gen.34c} \\
&\ \ \ \ \
\quad\quad\quad\quad\quad+ \bsvtau{j_1 &j_2 \\ k_1 &k_2 \\ \emptyslot &z } R_{\vec{\eta}}\big(
 {\rm ad}_{\ep_0}^{k_2-j_2-2}(b_{k_2}) {\rm ad}_{\ep_0}^{k_1-j_1-2}(\ep_{k_1}) \big) \notag \\
 &\ \ \ \ \
\quad\quad\quad\quad\quad+ \bsvtau{j_1 &j_2 \\ k_1 &k_2 \\ z &z } R_{\vec{\eta}}\big(  {\rm ad}_{\ep_0}^{k_2-j_2-2}(b_{k_2}) {\rm ad}_{\ep_0}^{k_1-j_1-2}(b_{k_1}) \big)
\bigg) + \ldots
\bigg\}_{\begin{smallmatrix} K \\ L \end{smallmatrix} \big|  \begin{smallmatrix} P \\ Q \end{smallmatrix}} \notag\\
&\quad \times \exp\bigg( {-}\frac{ R_{\vec{\eta}}(\ep_0)}{4y} \bigg)_{\begin{smallmatrix} P \\ Q \end{smallmatrix} \big|  \begin{smallmatrix} R \\ S \end{smallmatrix}}
 \widehat Y(
\begin{smallmatrix} M \\ N \end{smallmatrix} |  \begin{smallmatrix} R \\ S \end{smallmatrix}|i\infty) \notag
\end{align} \normalsize
The ellipsis at the end of the curly bracket again refers to $\beta^{\rm sv}$ of
depth $\ell \geq 3$ (with the $2^\ell$ possibilities to replace subsets of the 
entries $z$ in the third line of $\beta^{\rm sv}[\ldots]$
by empty slots while adjusting the associated $b_{k_i} \rightarrow \ep_{k_i}$). The infinite series
enclosed by $\{\ldots\}$ in (\ref{gen.34c}) will also be referred to as the {\it path-ordered
exponential}, and the explicit form of its eight types of depth-three terms
can be found in appendix \ref{POEd3}.

As a direct verification of the differential equation (\ref{high.12}) of the
path-ordered exponential (\ref{gen.34c}), one can exploit the following corollary of
(\ref{gen.35z}) and (\ref{gen.38})
\begin{align}
-4\pi \nabla_\tau \bsvtau{j_1 &j_2 &\ldots &j_\ell \\ k_1 &k_2 &\ldots &k_\ell \\ z_1 &z_2 &\ldots &z_\ell}
&= \sum_{i=1}^\ell (k_i{-}j_i{-}2) \bsvtau{j_1  &\ldots &j_{i}+1 &\ldots &j_\ell \\ k_1 &\ldots &k_i &\ldots &k_\ell \\ z_1  &\ldots &z_i &\ldots &z_\ell } \label{poesec.1} \\
&\ \ \ \
+ \delta_{j_\ell,k_\ell-2} (\tau{-}\bar \tau)^{k_\ell} f^{(k_\ell)}(z_\ell|\tau)
\bsvtau{j_1 &j_2 &\ldots &j_{\ell-1} \\ k_1 &k_2 &\ldots &k_{\ell-1} \\ z_1 &z_2 &\ldots &z_{\ell-1}}
\notag
\end{align}
where $\bsvtau{ \emptyset \\ \emptyset  \\ \emptyset} = 1 $, and the Cauchy-Riemann 
derivative (\ref{crdrv}) is again performed
at fixed $u,v$ in $z=u\tau{+}v$. Similar to (\ref{gen.35za}), the limit $z_r\rightarrow 0$
of the $f^{(k_r)}(z_r|\tau)$ does not commute with $\tau$-integration and we
again employ empty slots when referring to integration kernels
$-G_{k_r}(\tau)$ instead of $f^{(k_r)}(z_r|\tau)$, e.g.
\begin{align}
-4\pi \nabla_\tau \bsvtau{j_1 &j_2 &\ldots &j_\ell \\ k_1 &k_2 &\ldots &k_\ell \\ z_1 &z_2 &\ldots &\emptyslot}
&= \sum_{i=1}^\ell (k_i{-}j_i{-}2) \bsvtau{j_1  &\ldots &j_{i}+1 &\ldots &j_\ell \\ k_1 &\ldots &k_i &\ldots &k_\ell \\ z_1  &\ldots &z_i &\ldots &\emptyslot } \label{poesec.1a} \\
&\ \ \ \
- \delta_{j_\ell,k_\ell-2} (\tau{-}\bar \tau)^{k_\ell} G_{k_\ell}(\tau)
\bsvtau{j_1 &j_2 &\ldots &j_{\ell-1} \\ k_1 &k_2 &\ldots &k_{\ell-1} \\ z_1 &z_2 &\ldots &z_{\ell-1}}
\notag
\end{align}
Hence, (\ref{poesec.1}) and (\ref{poesec.1a}) (possibly with empty slots among
$z_1,z_2,\ldots,z_{\ell-1}$) generalize the differential equation (\ref{nabbsv.1}) of
the $\beta^{\rm sv}[\ldots]$ with two rows of entries that govern iterated-integral representations
of MGFs. The depth-one and depth-two examples of (\ref{poesec.1}) are given by
\begin{align}
-4\pi \nabla_\tau \bsvtau{j \\ k \\ z}
&=   (k{-}j{-}2) \bsvtau{j{+}1 \\ k\\ z}
+ \delta_{j,k-2} (\tau{-}\bar \tau)^{k} f^{(k)}(z|\tau)
 \label{nabbsv.2} \\
 -4\pi \nabla_\tau \bsvtau{j_1  &j_2 \\ k_1 &k_2 \\ z_1 &z_2}
&=   (k_1{-}j_1{-}2)\bsvtau{j_1+1  &j_2 \\ k_1 &k_2 \\ z_1 &z_2}
+  (k_2{-}j_2{-}2)\bsvtau{j_1  &j_2+1 \\ k_1 &k_2 \\ z_1 &z_2} \notag \\
&\ \ \ \ \ \
+ \delta_{j_2,k_2-2} (\tau{-}\bar \tau)^{k_2} f^{(k_2)}(z_2|\tau) \bsvtau{j_1 \\ k_1 \\ z_1 }
\notag
\end{align}
and the expression for  $-4\pi \nabla_\tau \bsvtau{j_1  &j_2 \\ k_1 &k_2 \\ z_1 &\emptyslot}$
would feature $-G_{k_2}(\tau)$ instead of $f^{(k_2)}(z_2|\tau)$ in the last line.

%%%%%%%%%%%%%%%%%%%%%%%%%%%%%%%%%%%%%%%%%%%%%%%%%%%%%%%%%%%
\subsubsection{Implications for the counting of eMGFs}
\label{sec:3.1.3}
%%%%%%%%%%%%%%%%%%%%%%%%%%%%%%%%%%%%%%%%%%%%%%%%%%%%%%%%%%%

As will be discussed in detail in section \ref{sec:5}, the expansion (\ref{gen.34c}) of the
generating series of eMGFs in one variable has important implications for the counting
of independent representatives. The path-ordered exponential in (\ref{gen.34c})
applies to $n$-point generating series which comprise all convergent eMGFs in
one variable of arbitrary topology that do not admit simplifications
via holomorphic subgraph reduction. The counting of $\beta^{\rm sv}[\ldots]$ with
$0\leq j_i \leq k_i{-}2$ and given values of $\sum_{i=1}^\ell k_i$ sets an upper bound
on the basis dimensions (over $\mathbb Q$-linear combinations of MZVs) of
one-variable eMGFs $\cplus{A \\ B \\ Z}$ at fixed $|A|+|B|= \sum_{i=1}^\ell k_i$.

However, the upper bound on the independent eMGFs will be lowered by the
commutation relations among the matrices $R_{\vec{\eta}}(\ep_k)$ and $R_{\vec{\eta}}(b_k)$
such as (\ref{revsec.25}) and generalizations to be discussed in section \ref{sec:5.2}.
Each relation of this type implies a dropout among the linear combinations of
$\beta^{\rm sv}[\ldots]$ that are realized via eMGFs. The analogous
counting of MGFs is for instance discussed in section 5.2 of \cite{Gerken:2020yii}.

%%%%%%%%%%%%%%%%%%%%%%%%%%%%%%%%%%%%%%%%%%%%%%%%%%%%%%%%%%%
\subsection{Solving the $\nabla_{\tau} \beta^{\rm sv}$ equation via iterated $\tau$-integrals}
\label{sec:3.2}
%%%%%%%%%%%%%%%%%%%%%%%%%%%%%%%%%%%%%%%%%%%%%%%%%%%%%%%%%%%

In this section, we describe meromorphic and real-analytic building blocks
for the expansion coefficients ${\cal E}^{\rm sv}[\ldots]$ and $\beta^{\rm sv}[\ldots]$
in the generating series (\ref{gen.34a}) and (\ref{gen.34c}) of eMGFs. We will encounter
combinations of their real and imaginary parts that generalize the real-analytic
iterated Eisenstein integrals $\beta^{\rm sv}$ underlying MGFs \cite{Gerken:2020yii}.
Similar to the $\beta^{\rm sv}$ representations of MGFs, this approach reduces
eMGFs to iterated integrals over $\tau$ and manifests their algebraic relations as
well as their expansion around the cusp.

%%%%%%%%%%%%%%%%%%%%%%%%%%%%%%%%%%%%%%%%%%%%%%%%%%%%%%%%%%%
%%%%%%%%%%%%%%%%%%%%%%%%%%%%%%%%%%%%%%%%%%%%%%%%%%%%%%%%%%%
\subsubsection{Meromorphic building blocks of Brown-type}
\label{sec:new.1.1}
%%%%%%%%%%%%%%%%%%%%%%%%%%%%%%%%%%%%%%%%%%%%%%%%%%%%%%%%%%%
%%%%%%%%%%%%%%%%%%%%%%%%%%%%%%%%%%%%%%%%%%%%%%%%%%%%%%%%%%%

The first step is to generalize the meromorphic iterated Eisenstein integrals (\ref{gen.36a}) of Brown
to kernels $\tau^j G_k(\tau) \rightarrow - \tau^j f^{(k)}(z_i|\tau)$ that additionally
depend on elliptic variables $z_i=u_i\tau{+}v_i$. For this purpose, we introduce {\it iterated Kronecker-Eisenstein (KE) integrals} of depth $\ell$ by
\beq
\eeetau{j_1 &j_2 &\ldots &j_\ell \\ k_1 &k_2 &\ldots &k_\ell \\ z_1 &z_2 &\ldots &z_\ell}
= (2\pi i )^{1+j_\ell -k_\ell} \int^\tau_{i\infty} \dd \tau_\ell \, \tau_\ell^{j_\ell} f^{(k_\ell)}(u_\ell \tau_\ell{+}v_\ell | \tau_\ell)
\EBR{j_1 &j_2 &\ldots &j_{\ell-1} \\ k_1 &k_2 &\ldots &k_{\ell-1} \\ z_1 &z_2 &\ldots &z_{\ell-1}}{\tau_\ell}
\label{gen.36}
\eeq
with $\eeetau{ \emptyset \\ \emptyset  \\ \emptyset} = 1 $ as well as $k_i \geq 2$ and $0 \leq j_i \leq k_i{-}2$, where the $\tau_\ell$-integration is performed at fixed comoving coordinates
$(u_\ell,v_\ell)$ of $z_\ell = u_\ell \tau_\ell {+} v_\ell$. Since the $f^{(k)}(u\tau{+}v|\tau)$ at
fixed $u,v$ are meromorphic in $\tau$, see for instance (\ref{elemlattice}),
the iterated integrals resulting from (\ref{gen.36}) are homotopy invariant
and yield meromorphic functions of $\tau$.

The recursive integral definition (\ref{gen.36}) manifests the differential equations
\begin{align}
2\pi i \partial_\tau
\eeetau{j_1 &j_2 &\ldots &j_\ell \\ k_1 &k_2 &\ldots &k_\ell  \\ z_1 &z_2 &\ldots &z_{\ell} }
&=  (2\pi i)^{2-k_\ell + j_\ell} \tau^{j_\ell} f^{(k_\ell)}(z_\ell| \tau)
\eeetau{j_1 &j_2 &\ldots &j_{\ell-1} \\ k_1 &k_2 &\ldots &k_{\ell-1}  \\ z_1 &z_2 &\ldots &z_{\ell-1}}
\label{gen.35y}
\end{align}
generalizing (\ref{gen.35x}) for iterated Eisenstein integrals with kernels $G_{k}(\tau)$.
On the one hand, the kernels in (\ref{gen.36a}) and (\ref{gen.36}) at $k\geq 4$ are related by
$\lim_{z \rightarrow 0} f^{(k)}(z|\tau) = - G_k(\tau)$. On the other hand, we will see below that the
iterated integrals $\eeetau{j_1 &j_2 &\ldots &j_\ell \\ k_1 &k_2 &\ldots &k_\ell  } $ are not the
direct outcome of setting $z_i \rightarrow 0$ in
$\eeetau{j_1 &j_2 &\ldots &j_\ell \\ k_1 &k_2 &\ldots &k_\ell \\ z_1 &z_2 &\ldots &z_\ell} $.
This behaviour of iterated KE integrals (\ref{gen.36}) as $z\rightarrow 0$ can be
traced back to the second term in\footnote{The expressions in (\ref{fkreps.1}) can be assembled
from the Fourier expansion of the meromorphic Kronecker-Eisenstein coefficients $g^{(k)}(z|\tau)$ in section 3.3.3 of \cite{Broedel:2014vla}.}
\begin{align}
f^{(k)}(z| \tau) &=  \frac{(2 \pi i)^{k}}{(k{-}1)!}\bigg[
\frac{B_k(u)}{k} +\frac{u^{k-1}}{1-e^{-2 i \pi z}}
+ \sum_{n=1}^{\infty} q^n\bigg(
\frac{(u+n)^{k-1}}{q^n - e^{-2 \pi i z}}
- \frac{(u-n)^{k-1}}{q^n - e^{2 \pi i z}}
\bigg)
\bigg]
\label{fkreps.1}
\end{align}
which results from the simple pole in $z$ of the Kronecker-Eisenstein series (\ref{Omega}).
The Bernoulli polynomials $B_k(u)$ in the first term are reviewed in
appendix \ref{app:Bern} and reduce to the Bernoulli numbers $B_k$ in the Fourier expansion
(\ref{fkreps.4}) of holomorphic Eisenstein series at $k\geq 4$ and $u\rightarrow0$.

Similar to the zero modes $2\zeta_k$ of $G_k(\tau)$, the
contributions $\sim B_k(u)$ to the integration kernels $f^{(k)}(u\tau{+}v|\tau)$ introduce endpoint
divergences into the integrals (\ref{gen.36}) as $\tau_\ell \rightarrow~i\infty$.
Just as for iterated Eisenstein integrals (\ref{gen.36a}),
we regularize these divergences via tangential base points \cite{Brown:mmv} which effectively assigns the value $\frac{ \tau^{j+1}}{j{+}1}$ to
the integrals $\int^\tau_{i\infty} \dd \tau_\ell \, \tau_\ell^{j}$ with $j\geq 0$.

As will be detailed in section \ref{sec:qeMGFs}, the following rewriting of (\ref{fkreps.1}) 
will be instrumental to obtain the $q$-expansion of iterated KE integrals (\ref{gen.36}):
\begin{align}
f^{(k)}(z| \tau) &= \frac{(2 \pi i)^{k}}{ (k{-}1)!}\bigg[ \frac{B_{k}(u)}{k}-  u^{k-1}\sum_{m=1}^\infty e^{2\pi i m z}   \label{fkreps.1a} \\
 &\quad\quad+   \sum_{m,n = 1}^\infty q^{m n}    \left((u{-}n)^{k-1}e^{-2\pi i m z}-(u{+}n)^{k-1}e^{2\pi i m z}\right)  \bigg] \nonumber
\end{align}
As one can see from $e^{2\pi i m z}= q^{m u} e^{2\pi i m v}$,
the sums over $m,n$ converge as long as $0<u<1$. There is
no loss of generality in imposing this range of $u$ since
\begin{itemize}
\item $f^{(k)}(z| \tau)$ are doubly periodic as $z\cong z{+}1 \cong z{+}\tau$
\item the case of $u=0$ encountered
for $k\geq3$ is covered by the form (\ref{fkreps.4}) of the holomorphic
Eisenstein series which vanish for odd $k$
\end{itemize}
We will see below that iterated integration of (\ref{fkreps.1a}) leads to
$q$-expansions of eMGFs that generalize the expansion (\ref{revsec.32}) of MGFs
and combine the series in $e^{2\pi i mz}$ in the first line to multiple polylogarithms.

%%%%%%%%%%%%%%%%%%%%%%%%%%%%%%%%%%%%%%%%%%%%%%%%%%%%%%%%%%%
%%%%%%%%%%%%%%%%%%%%%%%%%%%%%%%%%%%%%%%%%%%%%%%%%%%%%%%%%%%
\subsubsection{Alternative meromorphic building blocks}
\label{sec:new.1.2}
%%%%%%%%%%%%%%%%%%%%%%%%%%%%%%%%%%%%%%%%%%%%%%%%%%%%%%%%%%%
%%%%%%%%%%%%%%%%%%%%%%%%%%%%%%%%%%%%%%%%%%%%%%%%%%%%%%%%%%%

Similar to the alternative description of iterated Eisenstein integrals in (\ref{fkreps.5a}),
one can bypass the powers of $\tau$ in the kernels $\tau^j f^{(k)}(z_i|\tau)$ of
iterated KE integrals (\ref{gen.36}) by admitting the kernel $f^{(0)}=-G_0= 1$ in
\begin{align}
\mathcal{E}\!\SM{k_1 & k_2 & \ldots & k_r}{z_1 & z_2 & \ldots & z_r }{\tau} &=  2 \pi i \int_{i \infty}^{\tau} \frac{  \mathrm{d} \tau_{r} }{(2 \pi i)^{k_{r}}}\,  f^{(k_r)}\left(u_{r} \tau_{r}{+}v_{r}|\tau_{r}\right) \mathcal{E}\!\SM{k_1 & k_2 & \ldots & k_{r-1}}{z_1 & z_2 & \ldots & z_{r-1} }{\tau_{r}} \label{fkreps.5}
    \end{align}
with $ \mathcal{E}( \begin{smallmatrix}  \emptyset \\ \emptyset \end{smallmatrix};\tau) =1$
and $k_i\geq 0$. Iterated KE integrals (\ref{gen.36}) can be readily expanded in
terms of the modified ones in (\ref{fkreps.5}) by the same techniques that apply
to the $G_k$-kernels \cite{Broedel:2018izr}, e.g.
\beq
\eeetau{j_1  \\ k_1\\ z_1  } = j_1! \, {\cal E}\!\SM{\vec{0}^{j_1} &k_1}{ \vec{0}^{j_1} &z_1}{ \tau} \, , \ \ \ \
\eeetau{j_1 &j_2  \\ k_1 &k_2 \\ z_1 &z_2 }  = j_2! \sum_{a=0}^{j_2}\frac{ (j_1{+}a)! }{a!}
{\cal E}\!\SM{\vec{0}^{j_1+a}&k_1&\vec{0}^{j_2-a}&k_2}{\vec{0}^{j_1+a}&z_1&\vec{0}^{j_2-a}&z_2}{\tau} \label{revsec.41a}
\eeq
This straightforwardly generalizes the relations (\ref{revsec.41}) between
different formulations of iterated Eisenstein integrals, and the depth $\ell$
in (\ref{gen.36}) again translates into the number of non-zero $k_i$ in (\ref{fkreps.5}).

Given that both variants (\ref{gen.36}) and (\ref{fkreps.5}) of iterated KE
integrals obey shuffle relations, one can enforce $k_1 \neq 0$ in the first entry.
The regularized value $\mathcal{E}(\begin{smallmatrix} \vec{0}^p \\
\vec{0}^p \end{smallmatrix};\tau) =\frac{1}{p!} \log(q)^p$ for instance yields
\begin{align}
&\mathcal{E}\!\SM{ \vec{0}^{p_1-1} &  k_1 }{ \vec{0}^{p_1-1} &  z_1 }{\tau} = \sum_{n = 0}^{p_1-1}\frac{(-1)^{n} }{ \left(p_1{-}n{-}1\right)!}  \mathcal{E}\!\SM{k_1 & \vec{0}^{n}}{z_1 & \vec{0}^{n}}{\tau} \log(q)^{p_1-n-1}
    \label{fkreps.8}
\end{align}
see appendix \ref{app:exps1} for analogous relations at depth two and three.
Based on the expansion (\ref{fkreps.1}) of the $f^{(k)}$ kernels, the depth-one integrals on
the right-hand side can be performed in terms of the polylogarithms $\text{Li}_p$
in (\ref{fkreps.11}),
\begin{align}
    &\mathcal{E}\!\SM{k_1 &\vec{0}^{p_1-1}}{z &\vec{0}^{p_1-1}}{\tau} = \frac{1}{(k_1{-}1)!}\bigg\{\frac{B_{k_1}(u) \log(q)^{p_1}}{k_1 p_1!}-u^{k_1-p_1-1} \text{Li}_{p_1}(e^{2 i \pi  z})  \label{fkreps.6} \\
    &\quad  +\sum_{n=1}^\infty \Big[ (-1)^{k_1+1} (n{-}u)^{k_1-p_1-1} \text{Li}_{p_1}(q^n e^{-2 i \pi  z})
    -(n{+}u)^{k_1-p_1-1} \text{Li}_{p_1} (q^n e^{2 i \pi  z}) \Big]\bigg\}  \nonumber
\end{align}
One can alternatively start from the representation (\ref{fkreps.1a}) of the $f^{(k)}$
and identify the $\text{Li}_{p}$ through their series representation
\beq
\text{Li}_p(z) = \sum_{m=1}^\infty \frac{ z^m }{m^p}
\eeq
which implies (\ref{fkreps.6}) to comprise series in $q^u$, $q^{n-u}$ and $q^{n+u}$
with $n\geq 1$. At higher depth, we have not been able to express the
complete $q$-expansion of (\ref{fkreps.5}) in terms of multiple polylogarithms.
However, section \ref{sec:3.3} is dedicated to a subsector of the $q$-expansion
which boils down to multiple polylogarithms at arbitrary depth.
Note that the $p_1=1$ instances $\mathcal{E}(\begin{smallmatrix} k_1 \\
z \end{smallmatrix};\tau)$ of (\ref{fkreps.6}) are closely
related to the functions $\Omega^{(k_1-1)}$ of $z$ and $\tau$ that govern 
the construction of the symbol prime of elliptic polylogarithms in \cite{Wilhelm:2022wow}.

%%%%%%%%%%%%%%%%%%%%%%%%%%%%%%%%%%%%%%%%%%%%%%%%%%%%%%%%%%%
%%%%%%%%%%%%%%%%%%%%%%%%%%%%%%%%%%%%%%%%%%%%%%%%%%%%%%%%%%%
\subsubsection{Non-commutativity of $z_i\rightarrow 0$ with integration and hybrid integrals}
\label{sec:new.1.disc}
%%%%%%%%%%%%%%%%%%%%%%%%%%%%%%%%%%%%%%%%%%%%%%%%%%%%%%%%%%%
%%%%%%%%%%%%%%%%%%%%%%%%%%%%%%%%%%%%%%%%%%%%%%%%%%%%%%%%%%%

We note that the requirement $0\leq j_i\leq k_i{-}2$ in (\ref{gen.36}) leads to the bound
$p_1 \leq k_1{-}1$ on the number of zero entries $\vec{0}^{p_1-1}$ in (\ref{fkreps.6}).
Hence, the exponent in the second term $\sim u^{k_1-p_1-1}$
in (\ref{fkreps.6}) is non-negative and the limit $z\rightarrow 0$ or $u,v \rightarrow 0$ is
non-singular. The case of $p_1=k_1{-}1$ is particularly interesting since the
$z\rightarrow 0$ limit of $\text{Li}_{p_1}(e^{2 i \pi  z})$ then introduces the
zeta value $\zeta_{k_1-1}$, see (\ref{fkreps.12})
\begin{align}
\lim_{z_1\rightarrow 0}
\mathcal{E}\!\SM{k_1 &\vec{0}^{p_1-1}}{z_1 &\vec{0}^{p_1-1}}{\tau}
= \left\{
\begin{array}{cl}
0 &: \ k_1 \geq 3 \ {\rm odd} , \ p_1<k_1{-}1 \\
-\frac{ \zeta_{k_1-1} }{(k_1{-}1)!} &: \ k_1 \geq 3 \ {\rm odd} , \ p_1=k_1{-}1 \\
{\cal E}(k_1,\vec{0}^{p_1-1};\tau)  &: \ k_1 \geq 4 \ {\rm even} , \ p_1<k_1{-}1 \\
{\cal E}(k_1,\vec{0}^{p_1-1};\tau) - \frac{ \zeta_{k_1-1} }{(k_1{-}1)!} &: \ k_1  \geq 4 \ {\rm even} , \ p_1=k_1{-}1
\end{array} \right.
\label{discont.1}
\end{align}
where we have used the vanishing of odd Bernoulli numbers $B_3,B_5,\ldots$ and
the cancellations between the $q^n$-corrections in the second line of (\ref{fkreps.6})
for odd $k_1$. Since the integrals (\ref{fkreps.6}) at $k_1=2$ necessarily
involve $p_1=1$ and thereby a contribution of $\text{Li}_{1}(e^{2 i \pi  z})$,
we find a logarithmic divergence in the second term of $\mathcal{E}(\begin{smallmatrix} 2 \\
z \end{smallmatrix};\tau)$ as $z \rightarrow 0$:
\beq
\mathcal{E}\!\SM{2}{z }{\tau} = \frac{1}{2} B_2(u) \log(q) + \log(1{-}e^{2\pi i z})
+ \sum_{n=1}^{\infty} \big[
\log(1{-}q^n e^{2\pi i z}) + \log(1{-}q^n e^{-2\pi i z})
\big]
\label{logdiv}
\eeq
An important corollary of (\ref{discont.1}) is that the iterated KE integrals
in (\ref{fkreps.5}) generically do not reduce to the iterated Eisenstein integrals
${\cal E}(k_1,\ldots;\tau)$ of section \ref{sec:2.5.2} as $z_i \rightarrow0$ even
though the kernels obey $\lim_{z_i \rightarrow 0}f^{(k_i)}(z_i|\tau) = - G_{k_i}(\tau)$
at $k_i\geq 3$. Hence, the limit $z_i \rightarrow0$ of the KE kernels
$f^{(k_i)}(z_i|\tau)$ does not commute with iterated integration over $\tau$ at fixed $u_i,v_i$.
Accordingly, the iterated KE integrals (\ref{gen.36}) do not reproduce the iterated Eisenstein
integrals $\eeetau{j_1 &j_2 &\ldots &j_\ell \\ k_1 &k_2 &\ldots &k_\ell  } $ as $z_i \rightarrow 0$.

Just as in section \ref{sec:3.1.1}, we employ empty slots
$\begin{smallmatrix} \ldots &j &\ldots \\
 \ldots &k &\ldots \\
  \ldots &\emptyslot &\ldots  \end{smallmatrix}$ instead of
  $\begin{smallmatrix} \ldots &j &\ldots \\
 \ldots &k &\ldots \\
  \ldots &0 &\ldots  \end{smallmatrix}$ when the integration kernel
  is $-G_k$ instead of $f^{(k)}$. The differential equations (\ref{gen.36})
  and (\ref{fkreps.5}) generalize to {\it hybrid integrals} with a mix of kernels
  $-G_k$ instead of $f^{(k)}$,
\begin{align}
\eeetau{j_1 &j_2 &\ldots &j_\ell \\ k_1 &k_2 &\ldots &k_\ell \\ z_1 &z_2 &\ldots &\emptyslot}
&= - (2\pi i )^{1+j_\ell -k_\ell} \int^\tau_{i\infty} \dd \tau_\ell \, \tau_\ell^{j_\ell} G_{k_\ell}(\tau_\ell)
\EBR{j_1 &j_2 &\ldots &j_{\ell-1} \\ k_1 &k_2 &\ldots &k_{\ell-1} \\ z_1 &z_2 &\ldots &z_{\ell-1}}{\tau_\ell}
\label{altgen.36}
\\
\mathcal{E}\!\SM{k_1 & k_2 & \ldots & k_r}{z_1 & z_2 & \ldots &\emptyslot }{\tau} &= -  2 \pi i \int_{i \infty}^{\tau} \frac{  \mathrm{d} \tau_{r} }{(2 \pi i)^{k_{r}}}\,  G_{k_r}(\tau_{r}) \mathcal{E}\!\SM{k_1 & k_2 & \ldots & k_{r-1}}{z_1 & z_2 & \ldots & z_{r-1} }{\tau_{r}} \label{altfkreps.5}
\end{align}
where we furthermore allow empty slots in the place of $z_1,z_2,\ldots $ on both sides.
In this way, we attain a unified notation for $z$-dependent iterated KE integrals and
the iterated Eisenstein integrals of section \ref{sec:2.5}. In particular,
iterated Eisenstein integrals are recovered as in \eqref{eq:emptyslot}
\begin{align}
\eeetau{j_1 &j_2 &\ldots &j_\ell \\ k_1 &k_2 &\ldots &k_\ell \\ \emptyslot & \emptyslot &\ldots & \emptyslot } &= \eeetau{j_1 &j_2 &\ldots &j_\ell \\ k_1 &k_2 &\ldots &k_\ell  }
\label{discont.2}\\
\mathcal{E}\!\SM{k_1 & k_2 & \ldots & k_r}{\emptyslot & \emptyslot & \ldots & \emptyslot }{\tau}  &= \mathcal{E}(k_1, k_2, \ldots, k_r;\tau)
\notag
\end{align}
By the non-commutativity of limits $z_i\rightarrow 0$ with
integration exemplified in (\ref{discont.1}) we will in general
have to distinguish
\begin{align}
\lim_{z \rightarrow 0} \eeetau{\ldots &j_i &\ldots \\
\ldots &k_i &\ldots\\ \ldots &z &\ldots } =  \eeetau{\ldots &j_i &\ldots \\
\ldots &k_i &\ldots\\ \ldots &0 &\ldots } &\neq \eeetau{\ldots &j_i &\ldots \\
\ldots &k_i &\ldots\\ \ldots &\emptyslot &\ldots }   \label{discont.3} \\
\lim_{z \rightarrow 0}\mathcal{E}\!\SM{ \ldots &k_i &\ldots }{ \ldots &z &\ldots }{\tau}=\mathcal{E}\!\SM{ \ldots &k_i &\ldots }{ \ldots &0 &\ldots }{\tau}  &\neq
\mathcal{E}\!\SM{ \ldots &k_i &\ldots }{ \ldots & \emptyslot &\ldots }{\tau}
\notag
\end{align}
At depth two, the constraints $j_1\leq k_1{-}2$ and $j_2\leq k_2{-}2$ in (\ref{gen.36})
do not guarantee the absence of negative powers of $u$ from the integrals
${\cal E}(\begin{smallmatrix} k_1 &\ldots \\ z_1 &\ldots \end{smallmatrix};\tau)$
on the right-hand side of (\ref{revsec.41a}). As we will see,
the differential equations of eMGFs can only realize combinations of iterated KE
integrals where all negative powers of $u$ cancel. This will be exploited
in section~\ref{sec:5.2.1} where we infer relations among the
generators $b_k,\ep_k$ in the path-ordered exponential (\ref{gen.34c}) by requiring
cancellation of poles as $u\rightarrow 0$.

%%%%%%%%%%%%%%%%%%%%%%%%%%%%%%%%%%%%%%%%%%%%%%%%%%%%%%%%%%%
%%%%%%%%%%%%%%%%%%%%%%%%%%%%%%%%%%%%%%%%%%%%%%%%%%%%%%%%%%%
\subsubsection{Real-analytic iterated KE integrals ${\cal E}^{\rm sv}$}
\label{sec:new.1.3}
%%%%%%%%%%%%%%%%%%%%%%%%%%%%%%%%%%%%%%%%%%%%%%%%%%%%%%%%%%%
%%%%%%%%%%%%%%%%%%%%%%%%%%%%%%%%%%%%%%%%%%%%%%%%%%%%%%%%%%%

While the path-ordered exponential in (\ref{gen.34a}) is engineered to attain a holomorphic
derivative $\partial_\tau Y(z|\tau)$ as required by (\ref{high.12}), we have not yet used
any information on the antiholomorphic $\tau$-derivatives. Following
the strategy in \cite{Gerken:2020yii}, one can impose the reality properties of eMGFs and
the generating series instead of prescribing antiholomorphic differential equations
to determine antiholomorphic terms in the solutions ${\cal E}^{\rm sv}[\ldots]$
to the holomorphic differential equations (\ref{gen.35z}). As will be confirmed
on a case-by-case basis in section \ref{sec:9.1}, the combinations of iterated Eisenstein
integrals ${\cal E}[\ldots]$ and $\overline{{\cal E}[\ldots]}$ in the $z$-independent
case (\ref{fkreps.17}) can be uplifted to the $z$-dependent iterated KE integrals (\ref{gen.36})
\begin{align}
\EsvBR{j_1 \\ k_1 \\ z_1}{\tau} &= \sum_{r_1=0}^{j_1} (-2\pi i \bar\tau)^{r_1} \binom{j_1}{r_1}
\bigg\{ \EBR{j_1-r_1 \\ k_1 \\ z_1}{\tau}
+(-1)^{j_1-r_1} \overline{  \EBR{j_1-r_1 \\ k_1 \\ z_1}{\tau} } \bigg\}
\notag \\
\EsvBR{j_1 &j_2 \\ k_1 &k_2 \\ z_1 &z_2}{\tau} &= \overline{\alphaBR{j_1 &j_2 \\ k_1 &k_2 \\ z_1 &z_2}{\tau}}+ \sum_{r_1=0}^{j_1}\sum_{r_2=0}^{j_2} (-2\pi i \bar\tau)^{r_1+r_2}
\binom{j_1}{r_1}\binom{j_2}{r_2}
\bigg\{ \EBR{j_1 -r_1&j_2-r_2 \\ k_1 &k_2 \\ z_1 &z_2}{\tau}
\label{fkreps.18}  \\
&\hspace{10mm} +(-1)^{j_1-r_1} \overline{ \EBR{j_1-r_1 \\ k_1 \\ z_1}{\tau} }  \EBR{j_2-r_2 \\ k_2 \\ z_2}{\tau}
+(-1)^{j_1+j_2-r_1-r_2} \overline{ \EBR{j_2-r_2 &j_1-r_1 \\ k_2 &k_1 \\ z_2 &z_1}{\tau} }
\bigg\}   \notag
\end{align}
also see appendix \ref{app:exps3} for the analogous depth-three expressions. The interplay
between meromorphic and antimeromorphic constituents in (\ref{fkreps.18}) and
(\ref{fkreps.18c}) renders the $z$-dependent
${\cal E}^{\rm sv}$ invariant under the modular $T$ transformation $\tau\rightarrow \tau{+}1$.

Similar to the situation for MGFs, (\ref{fkreps.18}) leaves certain depth-two objects
$\overline{ \alpha[\ldots;\tau]}$
undetermined which are antiholomorphic in $\tau$ at fixed $u_i,v_i$ but will be nevertheless
referred to as {\it integration constants}. They are furthermore required to be $T$-invariant
and to vanish at the cusp since the ${\cal E}[\ldots]$ and $\overline{{\cal E}[\ldots]}$
in (\ref{fkreps.18}) and (\ref{fkreps.18c}) already have the required monodromies
and asymptotics at $\tau \rightarrow i\infty$.
Just like the analogous integration constants $ \overline{\alphaBR{j_1 &j_2 \\ k_1 &k_2}{\tau}} $
in MGFs exemplified in (\ref{fkreps.19}), their $z_i$-dependent
analogues in (\ref{fkreps.18}) carry at least three units of transcendental weights via
MZVs, and their dependence on $\bar \tau,u,v$ as well as simple examples will
be discussed in section \ref{sec:9.1.ex}.

However, since the $\overline{ \alpha[\ldots;\tau]}$ will be inferred from properties
of eMGFs, we will only determine their linear combinations that are realized
as within eMGFs, i.e.\ that enter the path-ordered exponential (\ref{gen.34a}).
The dropouts of individual ${\cal E}^{\rm sv}[\ldots]$ from (\ref{gen.34a}) through the relations of the
$R_{\vec{\eta}}(\epsilon_k)$ and $R_{\vec{\eta}}(b_k)$ lead to dropouts
among the $\overline{ \alpha[\ldots;\tau]}$ that are individually accessible
to eMGF methods.

In fact, the analogous limitation for the $z$-independent
$\overline{\alphaBR{j_1 &j_2 \\ k_1 &k_2}{\tau}}$ encountered in MGFs at
depth two was overcome in \cite{Dorigoni:2021jfr, Dorigoni:2021ngn}:
By extending the space of depth-two MGFs via Poincar\'e series with
iterated-integral descriptions, the references gave rise to expressions for
all the individual $\overline{\alphaBR{j_1 &j_2 \\ k_1 &k_2}{\tau}}$ with $k_1{+}k_2\leq 28$
and $0\leq j_i \leq k_i{-}2$. It would be interesting if a Poincar\'e-series
approach to eMGFs can also fix the expressions for individual $z$-dependent
$\overline{ \alpha[\ldots;\tau]}$.

%%%%%%%%%%%%%%%%%%%%%%%%%%%%%%%%%%%%%%%%%%%%%%%%%%%%%%%%%%%
%%%%%%%%%%%%%%%%%%%%%%%%%%%%%%%%%%%%%%%%%%%%%%%%%%%%%%%%%%%
\subsubsection{Real-analytic iterated KE integrals $\beta^{\rm sv}$}
\label{sec:new.1.4}
%%%%%%%%%%%%%%%%%%%%%%%%%%%%%%%%%%%%%%%%%%%%%%%%%%%%%%%%%%%
%%%%%%%%%%%%%%%%%%%%%%%%%%%%%%%%%%%%%%%%%%%%%%%%%%%%%%%%%%%

Since we are ultimately interested in the original generating series $Y$ rather
than the redefined one $\widehat Y$, the real-analytic coefficients $\beta^{\rm sv}$ in the
path-ordered (\ref{gen.34c}) will be discussed in more detail. In the first place, the
all-depth relation (\ref{gen.38}) determines $\beta^{\rm sv}$ in terms of ${\cal E}^{\rm sv}$
which in turn boil down to iterated KE integrals and their complex conjugates
by expressions like (\ref{fkreps.18}). The assembly of all the ${\cal E}[\ldots]$ and
$\overline{{\cal E}[\ldots]}$ at depth one and two turns out to be equivalent to
\begin{align}
\bsvtau{j_1 \\ k_1 \\ z_1} &= - \frac{(2\pi i)^{-1}}{(4y)^{k_1-2-j_1}} \bigg\{(-1)^{k_1} \int_{\tau}^{i \infty} \dd \tau_1 (\tau{-}\tau_1)^{k_1-2-j_1} (\bar\tau{-}\tau_1)^{j_1} f^{(k_1)} (u_1\tau_1{+}v_1|\tau_1) \notag \\
&\hspace{2.5cm} - \! \int_{\bar\tau}^{-i\infty} \! \dd\bar\tau_1 (\tau{-}\bar\tau_1)^{k_1-2-j_1} (\bar\tau{-}\bar\tau_1)^{j_1} \overline{f^{(k_1)}(u_1 \tau_1{+}v_1|\tau_1)} \bigg\}
\label{eq:ebsv1}
\end{align}
as well as\small
\begin{align}
\bsvtau{j_1 &j_2\\ k_1 &k_2 \\ z_1 &z_2}&=
\sum_{p_1=0}^{k_1{-}2{-}j_1} \sum_{p_2=0}^{k_2{-}2{-}j_2} \frac{\binom{k_1{-}2{-}j_1}{p_1}\binom{k_2{-}2{-}j_2}{p_2}}{(4y)^{p_1+p_2}} \overline{\alphaBR{j_1 +p_1&j_2+p_2\\k_1 &k_2 \\ z_1 &z_2}{\tau}}
+ \frac{(2\pi i)^{-2}}{(4y)^{k_1+k_2-j_1-j_2-4}} \label{eq:ebsv2} \\
&\! \! \!  \! \!  \! \!\! \! \!  \! \!  \! \!\! \! \!  \! \!  \! \! \!\! \! \!  \! \!  \! \!
\!\! \! \!  \! \!  \! \!  \times \bigg\{ 
\int\limits^{i\infty}_\tau \dd\tau_2  (\tau_2{-}\tau)^{k_2-j_2-2} (\tau_2{-}\bar\tau)^{j_2}
f^{(k_2)} (u_2\tau_2{+}v_2|\tau_2)   \int\limits^{i\infty}_{\tau_2} \dd\tau_1 (\tau_1{-}\tau)^{k_1-j_1-2}(\tau_1{-}\bar\tau)^{j_1}
f^{(k_1)} (u_1\tau_1{+}v_1|\tau_1)  \nn\\
&\! \! \! \! \! \! \! \! \! \!  \! \!  \! \!\! \! \!  \! \!  \! \! \!\! \! \!  \! \!  \! \!
\!\! \! \!  \! \!  \! \!  \quad \! \! \! \!  -    \int\limits^{i\infty}_\tau \dd\tau_2(\tau_2{-}\tau)^{k_2-j_2-2} (\tau_2{-}\bar\tau)^{j_2} f^{(k_2)} (u_2\tau_2{+}v_2|\tau_2)  \! \int\limits^{-i\infty}_{\bar\tau} \! \dd\bar\tau_1  (\tau{-}\bar\tau_1)^{k_1-j_1-2}(\bar\tau{-}\bar\tau_1)^{j_1}\overline{ f^{(k_1)} (u_1\tau_1{+}v_1|\tau_1)  }\nn\\
&\! \! \! \! \! \! \! \! \! \!  \! \!  \! \!\! \! \!  \! \!  \! \!  \!\! \! \!  \! \!  \! \!
\!\! \! \!  \! \!  \! \!  \quad \! \! \! \! + \! \int\limits^{-i\infty}_{\bar\tau} \! \dd\bar\tau_1(\tau{-}\bar\tau_1)^{k_1-j_1-2}(\bar\tau{-}\bar\tau_1)^{j_1} \overline{f^{(k_1)} (u_1\tau_1{+}v_1|\tau_1) } \! \int\limits^{-i\infty}_{\bar\tau_1} \! \dd\bar\tau_2  (\tau{-}\bar\tau_2)^{k_2-j_2-2} (\bar\tau{-}\bar\tau_2)^{j_2} \overline{ f^{(k_2)} (u_2\tau_2{+}v_2|\tau_2)  }\bigg\}\nn
\end{align}  \normalsize
The kernels $ (\tau{-}\tau_1)^{k_1-2-j_1} (\bar\tau{-}\tau_1)^{j_1} f^{(k_1)} (u_1\tau_1{+}v_1|\tau_1)$ 
are meromorphic in the integration variable $\tau_1$ at fixed $u_1,v_1$,
so the appearance of $\bar \tau$ does not conflict with homotopy invariance.

Both of ${\cal E}^{\rm sv}$ and $\beta^{\rm sv}$ are believed to inherit shuffle relations
from their meromorphic building blocks
\beq
{\cal E}^{\rm sv}[A;\tau] {\cal E}^{\rm sv}[B;\tau] = \sum_{C \in A\shuffle B} {\cal E}^{\rm sv}[C;\tau]
\, , \ \ \ \
\beta^{\rm sv}[A;\tau] \beta^{\rm sv}[B;\tau] = \sum_{C \in A\shuffle B} \beta^{\rm sv}[C;\tau]
\label{besv.shuff}
\eeq
with words $A,B$ in the combined letters $\begin{smallmatrix}j_i \\ k_i \\ z_i \end{smallmatrix}$.
This is manifest for the ${\cal E}[\ldots]$ and $\overline{{\cal E}[\ldots]}$-contributions to
${\cal E}^{\rm sv}$ and $\beta^{\rm sv}$ in (\ref{fkreps.18}) and (\ref{eq:ebsv1}), (\ref{eq:ebsv2}),
and the integration constants are conjectured to preserve the (\ref{besv.shuff}) via
identities such as
\beq
\overline{\alphaBR{j_1 &j_2 \\ k_1 &k_2 \\ z_1 &z_2}{\tau}} = -\overline{\alphaBR{j_2 &j_1 \\ k_2 &k_1 \\ z_2 &z_1}{\tau}} \, , \ \ \ \
\overline{\alphaBR{j_1 &j_2 &j_3 \\ k_1 &k_2 &k_3 \\ z_1 &z_2 &z_3}{\tau}}
+\overline{\alphaBR{j_2 &j_1 &j_3 \\ k_2 &k_1 &k_3 \\ z_2 &z_1 &z_3}{\tau}}
+\overline{\alphaBR{j_2 &j_3 &j_1 \\ k_2 &k_3 &k_1 \\ z_2 &z_3 &z_1}{\tau}}=0
\label{alpshffl}
\eeq
Similar to the ${\cal E}[\ldots]$, the real-analytic combinations ${\cal E}^{\rm sv}[\ldots]$
and $\beta^{\rm sv}[\ldots]$ do not reduce to the
building blocks of MGFs at $z_i= 0$ as was made evident by the notation in \eqref{eq:emptyslot},
\begin{align}
%{\rm in} \ {\rm general:} \ \ \ \
&\lim_{z \rightarrow 0} \esvtau{\ldots &j_i &\ldots \\
\ldots &k_i &\ldots\\ \ldots &z &\ldots } = \esvtau{\ldots &j_i &\ldots \\
\ldots &k_i &\ldots\\ \ldots &0 &\ldots } \neq \esvtau{\ldots &j_i &\ldots \\
\ldots &k_i &\ldots\\ \ldots &\emptyslot &\ldots }
 \notag \\
&\lim_{z \rightarrow 0} \bsvtau{\ldots &j_i &\ldots \\
\ldots &k_i &\ldots\\ \ldots &z &\ldots } = \bsvtau{\ldots &j_i &\ldots \\
\ldots &k_i &\ldots\\ \ldots &0 &\ldots } \neq \bsvtau{\ldots &j_i &\ldots \\
\ldots &k_i &\ldots\\ \ldots &\emptyslot &\ldots }
\label{discont.5}
\end{align}
As a depth-one example of such inequalities
we shall give the real-analytic analogue of (\ref{discont.1}) in the next section.
Note that the same non-commutativity of limits $z_i\rightarrow 0$ with iterated
$\tau$ integration applies to the integration constants
$\overline{\alpha[\ldots]}$.

%%%%%%%%%%%%%%%%%%%%%%%%%%%%%%%%%%%%%%%%%%%%%%%%%%%%%%%%%%%
\subsection{$\beta^{\rm sv}$ at depth one as single-valued elliptic polylogarithms}
\label{sec:3.4}
%%%%%%%%%%%%%%%%%%%%%%%%%%%%%%%%%%%%%%%%%%%%%%%%%%%%%%%%%%%

In this section, the depth-one instances of the $z$-dependent $\beta^{\rm sv}$ in (\ref{eq:ebsv1})
will be shown to reproduce Zagier's single-valued elliptic polylogarithms
in (\ref{basic.13}).

%%%%%%%%%%%%%%%%%%%%%%%%%%%%%%%%%%%%%%%%%%%%%%%%%%%%%%%%%%%
%%%%%%%%%%%%%%%%%%%%%%%%%%%%%%%%%%%%%%%%%%%%%%%%%%%%%%%%%%%
\subsubsection{Rewriting the $\beta^{\rm sv}$}
\label{sec:3.4.1}
%%%%%%%%%%%%%%%%%%%%%%%%%%%%%%%%%%%%%%%%%%%%%%%%%%%%%%%%%%%
%%%%%%%%%%%%%%%%%%%%%%%%%%%%%%%%%%%%%%%%%%%%%%%%%%%%%%%%%%%

The expression (\ref{eq:ebsv1}) for $\beta^{\rm sv}$ at depth one
can be straightforwardly rewritten in terms of the meromorphic iterated KE
integrals ${\cal E}[\ldots]$ in (\ref{gen.36}) and their complex conjugates
via binomial expansion of factors like $(\bar \tau{-}\tau_1)^{j_1}$ in the integrand.
In fact, the alternative meromorphic building blocks ${\cal E}(\ldots)$ in (\ref{fkreps.5}) yield
even more beneficial representations of the $\beta^{\rm sv}$, namely\footnote{One
can readily derive (\ref{discont.6}) from (\ref{eq:ebsv1}) by employing the 
following corollary of (\ref{revsec.41a}) and (\ref{fkreps.8}):\[
\mathcal{E}\!\SM{k &\vec{0}^{n}}{z &\vec{0}^{n}}{\tau}
= \frac{(2\pi i)^{n-k+1}}{n!} \int^\tau_{i\infty} \dd \tau_1 \, (\tau{-}\tau_1)^n \, f^{(k)}(u \tau_1{+}v|\tau_1)
\]}
\begin{align}
\bsvtau{j \\ k \\ z}  &= (-1)^k \sum_{a=0}^j (k{-}j{-}2{+}a)! { j \choose a} (4y)^{2+2j-k-a}
\mathcal{E}\!\SM{k &\vec{0}^{k-j-2+a}}{z &\vec{0}^{k-j-2+a}}{\tau} \notag \\
&\quad + \sum_{b=0}^{k-j-2} (j{+}b)! { k{-}2{-}j \choose b} (4y)^{-b}
\overline{ \mathcal{E}\!\SM{k &\vec{0}^{j+b}}{z &\vec{0}^{j+b}}{\tau} }
\label{discont.6}
\end{align}
Upon expanding the ${\cal E}(\begin{smallmatrix} k &\vec{0}^j \\ z &\vec{0}^j
\end{smallmatrix};\tau)$ around the cusp via (\ref{fkreps.6}), the combinations
in (\ref{discont.6}) reorganize the ${\rm Li}_p$ functions and their complex
conjugates into single-valued polylogarithms (\ref{svpolycl}) at depth one
(with ${\cal Z}^{\rm sv}=0$):
\begin{align}
\bsvtau{j \\ k  \\ z }  &= \frac{j!(k{-}j{-}2)!}{(k{-}1)!}\bigg(\frac{B_k(u)}{k!}(-4y)^{j+1} + (-4y)^{2+j-k}
G^{\rm sv}(\vec{0}^{k-j-2},1,\vec{0}^{j};e^{2\pi i z}) \label{nwbsv.6}\\
&\! \! \! \! \! +  (-4y)^{2+j-k} \sum_{n=1}^{\infty} \big[
G^{\rm sv}(\vec{0}^{k-j-2},1,\vec{0}^{j};e^{2\pi i z}q^n)
+(-1)^{k} G^{\rm sv}(\vec{0}^{k-j-2},1,\vec{0}^{j};e^{-2\pi i z}q^n) \big]\bigg)
\nonumber
\end{align}
The notation $G^{\rm sv}$ refers to Brown's single-valued polylogarithms at
genus zero \cite{svpolylog}, see appendix \ref{app:svpoly} for a general review
and (\ref{comrel.27e}) for their relation to ${\rm Li}_p$ functions. The expression
(\ref{nwbsv.6}) is invariant under the modular $T$ transformation since
all of $z$, $y= \pi \Im \tau$ and integer powers of $q,\bar q$ are. Intermediate
steps towards (\ref{nwbsv.6}) make use of the identities
\beq
G^{\rm sv}(0;e^{2\pi i z}) = -4uy
\, , \ \ \ \ \ \
G^{\rm sv}(0;e^{\pm 2\pi i z}q^n) = -4(n{\pm}u)y
\label{intsteps}
\eeq
as well as the shuffle property (\ref{Gsvshuffle}) of single-valued genus-zero polylogarithms.

%%%%%%%%%%%%%%%%%%%%%%%%%%%%%%%%%%%%%%%%%%%%%%%%%%%%%%%%%%%
%%%%%%%%%%%%%%%%%%%%%%%%%%%%%%%%%%%%%%%%%%%%%%%%%%%%%%%%%%%
\subsubsection{Matching with single-valued elliptic polylogarithms}
\label{sec:3.4.2}
%%%%%%%%%%%%%%%%%%%%%%%%%%%%%%%%%%%%%%%%%%%%%%%%%%%%%%%%%%%
%%%%%%%%%%%%%%%%%%%%%%%%%%%%%%%%%%%%%%%%%%%%%%%%%%%%%%%%%%%

Zagier's single-valued elliptic polylogarithms
in (\ref{basic.13}) are also available as infinite sums of
single-valued genus-zero polylogarithms \cite{Ramakrish}.
By comparing expansions in the reference with (\ref{nwbsv.6}), we conclude
\begin{align}
\bsvtau{j \\ k  \\ z } &= - (2i)^{2j-k+2} \frac{j!(k{-}2{-}j)!}{(k{-}1)!} \dplus{j+1 \\ k-1-j}\!(z|\tau)
\label{nwbsv.7}
\end{align}
or conversely
\beq
\dplus{a \\ b}\!(z|\tau) = -\frac{ (2i)^{b-a} (a{+}b{-}1)! }{(a{-}1)!(b{-}1)!}
\bsvtau{a-1 \\ a+b  \\ z } \label{invbsv.7}
\eeq
where the simplest examples at $k \leq 4$ are
\begin{align}
\bsvtau{0 \\ 2  \\ z } &= -  \dplus{1 \\ 1}\!(z|\tau)\, ,
&\bsvtau{0 \\ 4  \\ z } & = \frac{1}{12}  \dplus{1 \\ 3}\!(z|\tau)
\notag \\
\bsvtau{0 \\ 3  \\ z } & = \frac{i}{4} \dplus{1 \\ 2}\!(z|\tau)\, ,
&\bsvtau{1 \\ 4  \\ z } &= - \frac{1}{6} \dplus{2 \\ 2}\!(z|\tau)
  \label{simpins} \\
\bsvtau{1 \\ 3  \\ z } & = -i  \dplus{2 \\ 1}\!(z|\tau)\, , \ \ \
&\bsvtau{2 \\ 4  \\ z } &= \frac{4}{3} \dplus{3 \\ 1}\!(z|\tau)
\notag
\end{align}
In this case, we relied on direct computations and results of \cite{Ramakrish}
to match the lattice sum (\ref{basic.13}) defining
$\dplus{a \\ b}\!(z|\tau)$ with real-analytic combination of iterated integrals.
However, the main theme of the paper will be to express eMGFs ${\cal C}^+$
in terms of $\beta^{\rm sv}$ via differential equations and the perturbative
solution (\ref{gen.34c}). We will revisit the depth-one computations in
section \ref{sec:9.1} while justifying the antiholomorphic terms in (\ref{fkreps.18}),
and later sections open up a variety of perspectives on the higher-depth translation
between ${\cal C}^+$ and $\beta^{\rm sv}$.

Given that the depth-one instances of $\beta^{\rm sv}$ reproduce
Zagier's single-valued elliptic polylogarithms, we interpret
combinations of $\bsvtau{j_1 &j_2 &\ldots &j_\ell \\ k_1 &k_2 &\ldots &k_\ell \\ z_1&z_2 &\ldots &z_\ell}$
at $\ell \geq 2$ that form eMGFs as higher-depth generalizations of single-valued
elliptic polylogarithms. A variety of showcases can be found in later sections; in particular,
the expansion of $\beta^{\rm sv}$ around the cusp turns out to involve single-valued
genus-zero polylogarithms at arbitrary depth, see section \ref{sec:3.3}.

In this work, single-valued elliptic polylogarithms at higher depth
are encountered as iterated integrals in $\tau$, but the differential equations of eMGFs
reviewed in section \ref{sec:2.1} imply that one can alternatively employ iterated integrals
in $z$. At depth one, Zagier's single-valued elliptic polylogarithms have
been expressed in terms of iterated $z$-integrals of meromorphic Kronecker-Eisenstein
coefficients \cite{Broedel:2019tlz}, a variant of the elliptic polylogarithms of Brown
and Levin \cite{BrownLev}. It would be interesting to determine the analogous
$z$-integral representations of eMGFs at higher depth.

%%%%%%%%%%%%%%%%%%%%%%%%%%%%%%%%%%%%%%%%%%%%%%%%%%%%%%%%%%%
%%%%%%%%%%%%%%%%%%%%%%%%%%%%%%%%%%%%%%%%%%%%%%%%%%%%%%%%%%%
\subsubsection{Comparison with depth-one MGFs}
\label{sec:new.1.5a}
%%%%%%%%%%%%%%%%%%%%%%%%%%%%%%%%%%%%%%%%%%%%%%%%%%%%%%%%%%%
%%%%%%%%%%%%%%%%%%%%%%%%%%%%%%%%%%%%%%%%%%%%%%%%%%%%%%%%%%%

By equating the expressions (\ref{nwbsv.6}) and (\ref{nwbsv.7}) for $\bsvtau{j \\ k  \\ z }$,
we can make an important observation on their limit $z\rightarrow 0$: On the one hand,
the eMGFs $ \dplus{a \\ b}\!(z|\tau)$ reduce to non-holomorphic Eisenstein series
(\ref{revsec.5}) and their Cauchy-Riemann derivatives \cite{DHoker:2015wxz}
whose $\bsvtau{j \\ k }$ representations are known from \cite{Gerken:2020yii},
\begin{align}
(\pi \nabla_\tau)^m E_k(\tau) &= \Big( {-}\frac{1}{4} \Big)^{m} \frac{ (2k{-}1)! }{(k{-}1)! (k{-}1{-}m)!}
\bigg\{
{-} \bsvtau{ k-1+m\\ 2k} + \frac{ 2 \zeta_{2k-1} }{(2k{-}1) (4y)^{k-1-m} }
\bigg\}\notag
\\
\frac{(\pi \overline{\nabla}_\tau)^m E_k(\tau)}{y^{2m}} &=  \frac{({-}4)^{m} (2k{-}1)! }{(k{-}1)! (k{-}1{-}m)!}
\bigg\{
{-} \bsvtau{ k-1-m\\ 2k} + \frac{ 2 \zeta_{2k-1} }{(2k{-}1) (4y)^{k-1+m} }
\bigg\}
\label{nwbsv.11}
\end{align}
On the other hand, the odd zeta values in (\ref{nwbsv.11}) are recovered from the
single-valued polylogarithms $G^{\rm sv}$ at argument $e^{2 \pi i z}$ in the
first line of (\ref{nwbsv.6}) via
\beq
G^{\rm sv}(\vec{0}^{k-j-2},1,\vec{0}^j;1)
= (-1)^{j+1} {k{-}2 \choose j}
\!\times \! \left\{ \begin{array}{cl}  \!2 \zeta_{k-1} \! &: \, k \  \textrm{even} \\ 0 &: \ k \,  \textrm{odd} \end{array} \right.
\label{nwbsv.12}
\eeq
As a consequence, the limit $z \rightarrow 0$ relates $\bsvtau{j \\ k  \\ z }$ and $\bsvtau{j \\ k \\ \emptyslot}$ via
\beq
\lim_{z \rightarrow0}\bsvtau{j \\ k  \\ z } =
\left\{ \begin{array}{cl} \displaystyle
\bsvtau{j \\ k\\ \emptyslot } - \frac{ 2 \zeta_{k-1} }{(k{-}1) (4y)^{k-j-2} } &: \ k\geq 4 \  \textrm{even} \\ 0 &: \ k \geq 3 \  \textrm{odd} \end{array} \right.
\label{nwbsv.13}
\eeq
and furnishes a depth-one example of the non-commutativity (\ref{discont.5}) of the limit
$z\rightarrow 0$ with $\tau$-integration (see
(\ref{discont.1}) for the meromorphic analogue of this non-commutativity). By
combining (\ref{nwbsv.13}) with (\ref{nwbsv.6}),
we infer a new representation for the building blocks of MGFs at depth one:
\beq
\bsvtau{j \\ k \\ \emptyslot } = \frac{ j! (k{-}j{-}2)! }{(k{-}1)!} \bigg\{ \frac{ B_k }{k!} (-4y)^{j+1} + 2 (-4y)^{2+j-k} \sum_{n=1}^\infty G^{\rm sv}(\vec{0}^{k-j-2},1,\vec{0}^{j};q^n) \bigg\}
\label{nwbsv.14}
\eeq
The case of $k=2$ (which only leaves $j=0$) has been excluded from (\ref{nwbsv.13})
since the term $G^{\rm sv}(\vec{0}^{k-j-2},1,\vec{0}^j;e^{2\pi i z}) $ in the first line
of (\ref{nwbsv.6}) then specializes to $- 2 \Re {\rm Li}_1(e^{2\pi i z}) = \log|1- e^{2\pi i z}|^2$
and diverges as $z \rightarrow 0$. This is expected since
\begin{align}
\bsvtau{0 \\ 2  \\ z } &= - 2 y B_2(u) - 2 \Re \bigg\{
{\rm Li}_1(e^{2\pi i z}) +\sum_{n= 1}^{\infty} \big[  {\rm Li}_1(e^{2\pi i z}q^n)
+ {\rm Li}_1(e^{-2\pi i z}q^n) \big] \bigg\} \notag \\
&= - g(z|\tau)
\label{nwbsv.15}
\end{align}
can be identified as minus the closed-string Green function $g(z|\tau)$ \cite{DHoker:2015wxz}
which has a logarithmic singularity at the origin by (\ref{app:theta.2}). This is consistent with the
differential equations $\pi \nabla_\tau g(z|\tau)= - (\Im \tau)^2 f^{(2)}(z|\tau)$ and $- 4\pi \nabla_\tau \bsvtau{0 \\ 2  \\ z }=(\tau {-}\bar \tau)^2  f^{(2)}(z|\tau)$ following
from (\ref{revsec.8}) and (\ref{nabbsv.2}), respectively.

%%%%%%%%%%%%%%%%%%%%%%%%%%%%%%%%%%%%%%%%%%%%%%%%%%%%%%%%%%%
%%%%%%%%%%%%%%%%%%%%%%%%%%%%%%%%%%%%%%%%%%%%%%%%%%%%%%%%%%%

\subsubsection{Preview of higher-depth examples}
\label{sec:3.5}
%%%%%%%%%%%%%%%%%%%%%%%%%%%%%%%%%%%%%%%%%%%%%%%%%%%%%%%%%%%

The representation (\ref{invbsv.7}) of Zagier's single-valued elliptic
polylogarithms as a single $\beta^{\rm sv}$ translates two-column eMGFs into
real-analytic iterated KE integrals. Generic eMGFs with three or more columns
will translate into $\beta^{\rm sv}$ of higher depth, usually
accompanied by lower-depth terms like the zeta values in (\ref{nwbsv.11}).
Later sections will introduce systematic methods to make the dictionary explicit,
and we shall already give some preview examples here.

For the modular invariant three-column eMGFs (\ref{cabc.2}) constructed from Green functions,
the simplest examples of their $\beta^{\rm sv}$ representations are
\begin{align}
C_{1|1,1}(z|\tau)  &= 8 \bsvtau{1& 0\\3& 3\\z& z}
- 10 \bsvtau{2\\6\\ \emptyslot} -
 20 \bsvtau{2\\6\\z}
+ 2  \zeta_{3} B_{2}(u)+\frac{ \zeta_{5}}{4 y^2}
\notag \\
C_{1 | 2,1}(z|\tau) &=
{-} 18 \bsvtau{2& 0\\4& 4\\z& z} +
 24 \bsvtau{1& 1\\3& 5\\z& z} +
 24 \bsvtau{2& 0\\5& 3\\z& z}
\notag \\
 & \quad - 70 \bsvtau{3\\8\\ \emptyslot} - 56 \bsvtau{3\\8\\z}
 +\frac{3  \zeta_{5}}{2 y} B_{2}(u) + \frac{5 \zeta_{7}}{16 y^3}  \label{prevex.1}\\
%%%
C_{2 | 1,1}(z|\tau) &= {-} 18 \bsvtau{2& 0\\4& 4\\ \emptyslot& z} -
 18 \bsvtau{2& 0\\4& 4\\z& \emptyslot} +
 18 \bsvtau{2& 0\\4& 4\\z& z}
 \notag \\
 & \quad + 14 \bsvtau{3\\8\\ \emptyslot} -
 140 \bsvtau{3\\8\\z} +
 12 \zeta_{3} \bsvtau{0\\4\\z}
 - \frac{ \zeta_{5}}{2 y} B_{2}(u)  - \frac{ \zeta_{7}}{16 y^3}  \notag
\end{align}
The analogous expressions for their Cauchy-Riemann derivatives
can be obtained from the differential equations (\ref{poesec.1})
and (\ref{poesec.1a}) of the $\beta^{\rm sv}$. Note that the $\beta^{\rm sv}$ representations
of the associated MGFs $C_{a,b,c}(\tau) = C_{a|b,c}(z{=}0|\tau)$ are \cite{Gerken:2020yii}
\begin{align}
C_{1,1,1}(\tau) &= - 30  \bsvtau{2\\6\\ \emptyslot}  + \zeta_3 + \frac{3 \zeta_5}{4y^2}
 \label{prevex.3} \\
 C_{2,1,1}(\tau) &= -18  \bsvtau{2&0\\4&4\\ \emptyslot}  - 126 \bsvtau{3\\8\\ \emptyslot}
 + 12 \zeta_3  \bsvtau{0\\4\\ \emptyslot}
  +  \frac{ 5 \zeta_5 }{12y} - \frac{ \zeta_3^2}{4y^2}
  + \frac{ 9 \zeta_7 }{16y^3}
 \notag
\end{align}
Representative depth-two examples involving odd $k_1{+}k_2$ arise
from $z$-derivatives
\begin{align}
\nabla_z C_{1|1,1}(z|\tau) &= 4 \bsvtau{1 &0 \\ 3 &2 \\ z &z} - 8 \bsvtau{2 \\ 5 \\ z}  + 4 \zeta_3 B_1(u) \label{prevex.2}  \\
\nabla_z C_{2|1,1}(z|\tau) &=
12 \bsvtau{2 &0 \\ 4 &3 \\ z &z}
-12 \bsvtau{2	&0 \\ 4 &3 \\ \emptyslot &z}
 -60 \bsvtau{3 \\ 7 \\ z}
+ 8 \zeta_3 \bsvtau{ 0 \\3 \\ z} - \frac{ \zeta_5}{y} B_1(u)
\notag\\
\nabla_z C_{1|2,1}(z|\tau) &=   12 \bsvtau{1 &1 \\ 3 &4 \\ z &z}
-6  \bsvtau{2 &0 \\ 4 &3 \\ z &z} +12  \bsvtau{2 &0 \\ 5 &2 \\ z &z}
- 24 \bsvtau{3 \\ 7 \\ z}
+ \frac{3  \zeta_5}{y} B_1(u) \notag
\end{align}
where the $\nabla_z$-action on $\beta^{\rm sv}$ does not
follow a comparably simple formula as (\ref{poesec.1}) for the
$\nabla_\tau$-derivatives.

%%%%%%%%%%%%%%%%%%%%%%%%%%%%%%%%%%%%%%%%%%%%%%%%%%%%%%%%%%%
\subsection{Iterated KE integrals of higher depth at the cusp}
\label{sec:3.3}
%%%%%%%%%%%%%%%%%%%%%%%%%%%%%%%%%%%%%%%%%%%%%%%%%%%%%%%%%%%

For meromorphic and real-analytic iterated KE integrals at depth one,
we have given simple closed formulae (\ref{fkreps.6}) and (\ref{nwbsv.6}) for all
orders in the expansion around the cusp. At higher depth, there is no bottleneck
in obtaining arbitrary orders of the $q$-expansion from direct integration of (\ref{fkreps.1a})
and performing numerical evaluations. However, it is not clear to us whether the
entire $q$-expansions can be lined up with infinite series of (single-valued) multiple polylogarithms
at higher depth.

We shall now identify a non-trivial subsector of the $q$-expansion of iterated KE
integrals that is accessible in terms of multiple polylogarithms at arbitrary depth.
These are the terms of order $q^0, q^u, q^{2u},\ldots$ as opposed to $q^n, q^{n\pm u},
q^{2n},q^{2n\pm u},q^{2n\pm 2u},\ldots$ with $n \geq 1$ (and their $q\leftrightarrow \bar q$ analogues)
which contribute to the leading Laurent polynomials of MGFs as $z\rightarrow 0$.
The entirety of the orders $q^0, q^u, q^{2u},\ldots$ and $\bar q^0, \bar q^u, \bar q^{2u},\ldots$
in the $q$-expansions of eMGFs or
iterated KE integral will be referred to as their {\it leading terms}.

In the limit $z \rightarrow 0$ where eMGFs reduce to MGFs, the polylogarithms
capturing the series in $q^u$ and $\bar q^u$ in the leading terms of eMGFs
will be shown to contribute single-valued MZVs to their Laurent polynomials.
Accordingly, the polylogarithmic contributions to the leading terms in this section
play a central role to deduce the Laurent polynomials of eMGFs from those of
MGFs, see section \ref{sec:9.2}.

The results of this section also exemplify that leading terms of iterated KE integrals (\ref{gen.36})
or $\beta^{\rm sv}$ of depth $\ell \geq 2$ may exhibit poles in $u$. These poles will play
a crucial role in section \ref{sec:5.2.1} to infer all-multiplicity relations among the
matrices $R_{\vec{\eta}}(\ep_k),R_{\vec{\eta}}(b_k)$ in (\ref{high.12}) or (\ref{gen.34c}).

%%%%%%%%%%%%%%%%%%%%%%%%%%%%%%%%%%%%%%%%%%%%%%%%%%%%%%%%%%%
\subsubsection{Leading terms of meromorphic iterated KE integrals}
\label{sec:3.3.1}
%%%%%%%%%%%%%%%%%%%%%%%%%%%%%%%%%%%%%%%%%%%%%%%%%%%%%%%%%%%

The leading terms of meromorphic iterated KE integrals are most conveniently
presented at the level of the ${\cal E}(\ldots)$ in (\ref{fkreps.5}) rather than the
${\cal E}[\ldots]$ in (\ref{gen.36}). At depth one, we extract the leading terms
\begin{align}
    \mathcal{E}\!\SM{k_1 &\vec{0}^{p_1-1}}{z &\vec{0}^{p_1-1}}{\tau} &= \frac{1}{(k_1{-}1)!}\bigg\{\frac{B_{k_1}(u) \log(q)^{p_1}}{k_1 p_1!}-u^{k_1-p_1-1} \text{Li}_{p_1}(e^{2 i \pi  z} ) + {\cal O}(q^{1-u})\bigg\}
    \label{fkreps.9.0}
    \end{align}
from the all-order expansion (\ref{fkreps.6}) by integrating the first two
terms in (\ref{fkreps.1}),\footnote{We have used that the
additional terms $\sum_{n=1}^{\infty}
\frac{(u+n)^{k-1} q^n}{q^n - e^{-2 \pi i z}}
$ and $\sum_{n=1}^{\infty} \frac{(u-n)^{k-1} q^n}{q^n - e^{2 \pi i z}}$
in (\ref{fkreps.1}) are suppressed with $q^{1+u}$ and $q^{1-u}$, respectively, as
$\tau \rightarrow i \infty$ at fixed $u,v$.}
\begin{align}
f^{(k)}(z| \tau) &=  \frac{(2 \pi i)^{k}}{(k{-}1)!}\bigg\{
\frac{B_k(u)}{k} +\frac{u^{k-1}}{1-e^{-2 i \pi z}}
+  {\cal O}(q^{1-u})
\bigg\}
    \label{ext.9.0}
    \end{align}
By restricting depth-two integrals to the same contributions of the integration kernels, we
arrive at the leading terms
\begin{align}
    \mathcal{E}\!\SM{k_1 & \vec{0}^{p_1-1} & k_2 &\vec{0}^{p_2-1}}{z &\vec{0}^{p_1-1} & z &\vec{0}^{p_2-1}}{\tau} &= \frac{1}{\left(k_{1}{-}1\right) !\left(k_{2}{-}1\right) !} \bigg\{
    \frac{B_{k_{1}}(u) B_{k_{2}}(u)}{k_{1} k_{2}} \frac{\log (q)^{p_{1}+p_2}}{(p_1{+}p_2)!}   \notag \\
    &\quad -\frac{B_{k_{2}}(u)}{k_{2}} u^{k_{1}-1-p_{1}-p_2} \operatorname{Li}_{p_{1}+p_2}(e^{2 \pi i z})    \label{fkreps.9} \\
&\quad - \frac{u^{k_{2}-p_2-1}B_{k_{1}}(u)}{ k_{1} \left(p_2{-}1\right)!} \sum_{n=0}^{p_1}
\frac{\left(n{+}p_2{-}1\right)!}{n! \left(p_1{-}n\right)!} \frac{ \log(q)^{p_1-n} }{(-u)^{n} } \text{Li}_{n+p_2}(e^{2 i \pi  z}) \notag \\
&\quad +u^{k_{1}+k_{2}-p_{1}-p_{2}-2} G(\vec{0}^{p_{2}-1}, 1, \vec{0}^{p_{1}-1}, 1 ; e^{2 \pi i z})
  + {\cal O}(q^{1-u})  \bigg\} \notag
\end{align}
The last line features multiple polylogarithms $G(\ldots;e^{2\pi i z})$ of depth
two following the general definition (\ref{fkreps.10}). The analogous depth-three expression
is given in appendix \ref{app:exps2} in terms of multiple polylogarithms
up to and including depth three. Moreover, the derivation of (\ref{fkreps.9.0}), (\ref{fkreps.9})
and (\ref{appd3}) as well as the all-depth systematics can be
found in appendix \ref{sec:laurentpolynomialsarbitrarydepth}.
By the contribution $\sim u^{k-1}(1{-}e^{-2\pi i z})^{-1}$ to $f^{(k)}(z|\tau)$ in (\ref{fkreps.1}),
the leading terms of iterated KE integrals at depth $\ell$ involve multiple polylogarithms
up to and including depth~$\ell$.

A similar logic applies to hybrid iterated KE integrals involving
both $G_k$ and $f^{(k)}$ kernels. In the same way as the leading terms
in the $z$-independent depth-one case
\beq
\mathcal{E}\!\SM{k_1 & \vec{0}^{p_1-1} }{\emptyslot &\vec{0}^{p_1-1} }{\tau}
= {\cal E}(k_1, \vec{0}^{p_1-1};\tau ) =\frac{ B_{k_1} \log(q)^{p_1}}{k_1! p_1!} + {\cal O}(q)
\label{onlygk.1}
\eeq
are lacking the $\text{Li}_{p_1}(e^{2 i \pi  z} ) $ contribution to (\ref{fkreps.9.0}),
leading terms at higher depth become shorter for each $G_k$ kernel in the
place of $f^{(k)}$, for instance
\begin{align}
   \mathcal{E}\!\SM{k_1 & \vec{0}^{p_1-1} & k_2 &\vec{0}^{p_2-1}}{z &\vec{0}^{p_1-1} & \emptyslot &\vec{0}^{p_2-1}}{\tau} &= \frac{ B_{k_{2}}}{\left(k_{1}{-}1\right) !\left(k_{2}{-}1\right) !} \bigg\{
    \frac{B_{k_{1}}(u)}{k_{1} k_{2}} \frac{\log (q)^{p_{1}+p_2}}{(p_1{+}p_2)!}   \notag \\
    &\quad-\frac{1}{k_{2}} u^{k_{1}-1-p_{1}-p_2} \operatorname{Li}_{p_{1}+p_2} (e^{2 \pi i z})
  + {\cal O}(q^{1-u})  \bigg\} \label{hybrid.5} \\
 %%%
 %%%
   \mathcal{E}\!\SM{k_1 & \vec{0}^{p_1-1} & k_2 &\vec{0}^{p_2-1}}{\emptyslot &\vec{0}^{p_1-1} & z &\vec{0}^{p_2-1}}{\tau} &= \frac{B_{k_{1}}}{\left(k_{1}{-}1\right) !\left(k_{2}{-}1\right) !} \bigg\{
    \frac{ B_{k_{2}}(u)}{k_{1} k_{2}} \frac{\log (q)^{p_{1}+p_2}}{(p_1{+}p_2)!}   \notag \\
    &\hspace{-3cm}
- \frac{u^{k_{2}-p_2-1}}{ k_{1} \left(p_2{-}1\right)!} \sum_{n=0}^{p_1}  \frac{\left(n{+}p_2{-}1\right)!}{n! \left(p_1{-}n\right)!} \frac{ \log(q)^{p_1-n} }{(-u)^{n}} \text{Li}_{n+p_2} (e^{2 i \pi  z})  + {\cal O}(q^{1-u})  \bigg\}   \notag
\end{align}
Upon comparison with (\ref{fkreps.9}), the leading terms at depth two with
$f^{(k_i)} \rightarrow - G_{k_i}$ are obtained from $B_{k_i}$ times the
coefficient of $B_{k_i}(u)$ in $\mathcal{E}(\begin{smallmatrix}
k_1 & \vec{0}^{p_1-1} & k_2 &\vec{0}^{p_2-1}  &\ldots \\
z &\vec{0}^{p_1-1} & z &\vec{0}^{p_2-1} &\ldots \end{smallmatrix}; {\tau})$.
By the same rule, one can extract hybrid integrals at depth three involving
one or two $G_{k_i}$ kernels from the coefficients of the respective
$B_{k_i}(u)$ in (\ref{appd3}). With only $G_{k_i}$ kernels, the higher-depth
generalization of (\ref{onlygk.1}) reads
\beq
{\cal E}(k_1,\vec{0}^{p_1-1},k_2,\vec{0}^{p_2-1},\ldots,k_r,\vec{0}^{p_r-1};\tau)
= \frac{1}{(p_1{+}p_2{+}\ldots{+}p_r)!} \bigg( \prod_{j=1}^r \frac{ B_{k_j} \log(q)^{p_j} }{k_j!} \bigg)
+ {\cal O}(q)
\label{onlygk.2}
\eeq
Note that the iterated KE integrals
$\mathcal{E}(\begin{smallmatrix}
k_1 & \vec{0}^{p_1-1} & k_2 &\vec{0}^{p_2-1} \\
z_1 &\vec{0}^{p_1-1} &z_2 &\vec{0}^{p_2-1}\end{smallmatrix}; {\tau})$
(or their analogues with $z_1$ or $z_2$ replaced by an empty slot)
may feature negative powers of $u$ via contributions such as
$u^{k_{1}-1-p_{1}-p_2} \operatorname{Li}_{p_{1}+p_2} (e^{2 \pi i z}) $
in (\ref{fkreps.9}) and (\ref{hybrid.5}).
These poles still occur for entries subject to $p_1{+}p_2\leq k_1{+}k_2{-}2$
which are compatible with the constraint $j\leq k{-}2$ on the kernels $\tau^j G_k$
and $\tau^j f^{(k)}$ according to (\ref{revsec.41a}).

%%%%%%%%%%%%%%%%%%%%%%%%%%%%%%%%%%%%%%%%%%%%%%%%%%%%%%%%%%%
\subsubsection{Leading terms of real-analytic iterated KE integrals}
\label{sec:3.3.2}
%%%%%%%%%%%%%%%%%%%%%%%%%%%%%%%%%%%%%%%%%%%%%%%%%%%%%%%%%%%

Based on the leading terms of the meromorphic ${\cal E}(\ldots)$, one can infer those
of the real-analytic $\beta^{\rm sv}$. At depth one for instance, the
all-order expansion (\ref{nwbsv.6}) obtained from (\ref{discont.6}) identifies leading terms
\begin{align}
\bsvtau{j \\ k  \\ z }  &= \frac{j!(k{-}j{-}2)!}{(k{-}1)!}\bigg\{\frac{B_k(u)}{k!}(-4y)^{j+1} \label{dpt1LT} \\
&\quad + (-4y)^{2+j-k}
G^{\rm sv}(\vec{0}^{k-j-2},1,\vec{0}^{j};e^{2\pi i z}) +{\cal O}(q^{1-u},\bar q^{1-u}) \bigg\}
\notag
\end{align}
where the notation ${\cal O}(q^{1-u},\bar q^{1-u})$ refers to
terms $ {\cal O}(e^{-2y(1-u)}) $, see section \ref{sec:2.4.1}.
Starting from depth two, it remains to combine the meromorphic leading terms of
section \ref{sec:3.3.1} with the higher-depth
generalization of (\ref{discont.6}), for instance
\begin{align}
\bsvtau{j_1 &j_2\\ k_1 &k_2 \\ z_1 &z_2}&= \sum_{p_1=0}^{k_1{-}2{-}j_1} \sum_{p_2=0}^{k_2{-}2{-}j_2} \frac{\binom{k_1{-}2{-}j_1}{p_1}\binom{k_2{-}2{-}j_2}{p_2}}{(4y)^{p_1+p_2}} \overline{\alphaBR{j_1 +p_1&j_2+p_2\\k_1 &k_2 \\ z_1 &z_2}{\tau}} \notag \\
%%%%%
&+(-1)^{k_1+k_2} \sum_{c=0}^{k_1-j_1-2} {k_1{-}j_1{-}2 \choose c} \sum_{d=0}^{j_1}
{j_1 \choose d} \sum_{a=0}^{j_2+d} {j_2{+}d \choose a}
(k_2{-}j_2{-}2{+}a{+}c)!  \notag \\
&\ \ \times  (k_1{-}2{-}c{-}d)! (4y)^{4+d-a +j_1+2j_2-k_1-k_2}
\mathcal{E}\!\SM{k_1 &\vec{0}^{k_1{-}2{-}c{-}d} &k_2 &\vec{0}^{k_2{-}j_2{-}2{+}a{+}c}  }{
z_1 &\vec{0}^{k_1{-}2{-}c{-}d} &z_2 &\vec{0}^{k_2{-}j_2{-}2{+}a{+}c} }{\tau}
\notag \\
%%%%%
&+(-1)^{k_2} \sum_{a=0}^{j_2}{j_2\choose a}  \sum_{b=0}^{k_1-2-j_1} {k_1{-}2{-}j_1 \choose b}
(j_1{+}b)! (k_2{-}j_2{-}2{+}a)! \label{d2bsv}\\
&\ \ \times (4y)^{2-a-b+2j_2-k_2}
\mathcal{E}\!\SM{k_2 &\vec{0}^{k_2-2-j_2+a}  }{
z_2 &\vec{0}^{k_2-2-j_2+a}  }{\tau}
\overline{\mathcal{E}\!\SM{k_1 & \vec{0}^{j_1+b} }{
z_1 & \vec{0}^{j_1+b}}{\tau} }
\notag \\
%%%%%
&+\sum_{c=0}^{k_2-j_2-2} {k_2{-}j_2{-}2 \choose c} \sum_{d=0}^{j_2} {j_2 \choose d}
\sum_{b=0}^{k_1-j_1-2+c} { k_1{-}j_1{-}2{+}c \choose b} \notag \\
&\ \ \times (4y)^{2+c-b+j_2-k_2}
\overline{\mathcal{E}\!\SM{k_2 &\vec{0}^{k_2-2-c-d} &k_1 & \vec{0}^{j_1+b+d} }{
z_2 &\vec{0}^{k_2-2-c-d} &z_1 & \vec{0}^{j_1+b+d}}{\tau} }
\notag
\end{align}
and a similar formula at depth three in appendix \ref{app:exps7}.

The multiple polylogarithms and their complex conjugates in the leading terms
of $\beta^{\rm sv}$ turn out to conspire to their single-valued versions $G^{\rm sv}$
reviewed in appendix \ref{app:svpoly}. This can be straightforwardly tested on a
case-by-case basis from the composition rules (\ref{svpolycl}) of $G^{\rm sv}$ in terms
of meromorphic polylogarithms and their complex conjugates. As a shortcut towards the leading
terms of $\beta^{\rm sv}$ in terms of single-valued polylogarithms, one can
effectively set $\log(q) \rightarrow \log|q|^2 = -4y$ and $G(\ldots;e^{2\pi i z}) \rightarrow
G^{\rm sv}(\ldots;e^{2\pi i z})$ in expressions like (\ref{fkreps.9.0}) and
(\ref{fkreps.9}), e.g.
\small
\begin{align}
    \mathcal{E}\!\SM{k_1 &\vec{0}^{p_1-1}}{z_1 &\vec{0}^{p_1-1}}{\tau} &\rightarrow \frac{1}{(k_1{-}1)!}\bigg\{\frac{B_{k_1}(u) (-4y)^{p_1}}{k_1 p_1!} \notag \\
    &\quad +u^{k_1-p_1-1} G^{\rm sv}(\vec{0}^{p_1-1},1  ;e^{2 i \pi  z} ) + {\cal O}(q^{1-u},\bar q^{1-u})  \bigg\} \notag \\
    \mathcal{E}\!\SM{k_1 & \vec{0}^{p_1-1} & k_2 &\vec{0}^{p_2-1}}{z &\vec{0}^{p_1-1} & z &\vec{0}^{p_2-1}}{\tau} &\rightarrow \frac{1}{\left(k_{1}{-}1\right) !\left(k_{2}{-}1\right) !} \bigg\{
    \frac{B_{k_{1}}(u) B_{k_{2}}(u)}{k_{1} k_{2}} \frac{(-4y)^{p_{1}+p_2}}{(p_1{+}p_2)!}    \label{effrule.1a} \\
    &\quad +\frac{B_{k_{2}}(u)}{k_{2}} u^{k_{1}-1-p_{1}-p_2} G^{\rm sv}(\vec{0}^{p_{1}+p_2-1},1;e^{2 \pi i z})   \notag \\
&\quad +\frac{u^{k_{2}-p_2-1}B_{k_{1}}(u)}{ k_{1} \left(p_2{-}1\right)!} \sum_{n=0}^{p_1}
\frac{\left(n{+}p_2{-}1\right)!}{n! \left(p_1{-}n\right)!} \frac{(-4y)^{p_1-n} }{(-u)^{n} }
G^{\rm sv} (\vec{0}^{n+p_2-1},1;e^{2 i \pi  z}) \notag \\
&\quad +u^{k_{1}+k_{2}-p_{1}-p_{2}-2} G^{\rm sv}(\vec{0}^{p_{2}-1}, 1, \vec{0}^{p_{1}-1}, 1 ; e^{2 \pi i z}) + {\cal O}(q^{1-u},\bar q^{1-u})   \bigg\} \notag
\end{align} \normalsize
while discarding the integration constants and all the antimeromorphic
iterated KE integrals in (\ref{d2bsv}),
\beq
\overline{\alphaBR{j_1  &j_2 &\ldots &j_\ell \\k_1 &k_2 &\ldots &k_\ell \\ z_1 &z_2 &\ldots &z_{\ell} }{\tau}}\rightarrow 0
\, , \ \ \ \
\overline{\mathcal{E}\!\SM{k_1 &k_2 &\ldots &k_r}{z_1 &z_2 &\ldots &z_r}{\tau} }\rightarrow 0
\label{effrule.2}
\eeq
The analogous shortcut for hybrid integrals (\ref{hybrid.5}) yields
\begin{align}
    \mathcal{E}\!\SM{k_1 & \vec{0}^{p_1-1} & k_2 &\vec{0}^{p_2-1}}{z &\vec{0}^{p_1-1} & \emptyslot &\vec{0}^{p_2-1}}{\tau} &\rightarrow \frac{B_{k_{2}}}{\left(k_{1}{-}1\right) !\left(k_{2}{-}1\right) !} \bigg\{
    \frac{B_{k_{1}}(u) }{k_{1} k_{2}} \frac{(-4y)^{p_{1}+p_2}}{(p_1{+}p_2)!}  \notag \\
    &\quad+\frac{1}{k_{2}} u^{k_{1}-1-p_{1}-p_2}
G^{\rm sv}(\vec{0}^{p_1+p_2-1},1;e^{2 \pi i z})+ {\cal O}(q^{1-u},\bar q^{1-u})   \bigg\}  \notag \\
 %%%
 %%%
   \mathcal{E}\!\SM{k_1 & \vec{0}^{p_1-1} & k_2 &\vec{0}^{p_2-1}}{\emptyslot &\vec{0}^{p_1-1} & z &\vec{0}^{p_2-1}}{\tau} &\rightarrow \frac{B_{k_{1}}}{\left(k_{1}{-}1\right) !\left(k_{2}{-}1\right) !} \bigg\{
    \frac{B_{k_{2}}(u)}{k_{1} k_{2}} \frac{(-4y)^{p_{1}+p_2}}{(p_1{+}p_2)!}   \label{effrule.1b} \\
    &\quad \hspace{-2.7cm} + \frac{u^{k_{2}-p_2-1}}{ k_{1} \left(p_2{-}1\right)!} \sum_{n=0}^{p_1}  \frac{\left(n{+}p_2{-}1\right)!}{n! \left(p_1{-}n\right)!} \frac{ (-4y)^{p_1-n} }{(-u)^{n}}
G^{\rm sv}(\vec{0}^{n+p_2-1},1; e^{2 i \pi  z}) + {\cal O}(q^{1-u},\bar q^{1-u})  \bigg\}   \notag
\end{align} 
We have tested for all cases up to and including $k_1{+}k_2=11$
that applying the effective rules (\ref{effrule.1a}) and (\ref{effrule.1b}) 
to the expansions (\ref{discont.6}) and (\ref{d2bsv})
of $\beta^{\rm sv}$ in terms of ${\cal E}$ and $\overline{{\cal E}}$ is
equivalent to assembling the meromorphic leading terms of section \ref{sec:3.3.1}
and their complex conjugates.
Such effective rules are plausible for arbitrary $k_1,k_2,\ldots,k_\ell$
and depth $\ell$: Any tentative single-valued map at genus one which encapsulates the
transition from ${\cal E}$ to ${\cal E}^{\rm sv}$ (or their linear combinations
 (\ref{gen.38}) defining $\beta^{\rm sv}$) is expected to be compatible with the single-valued
 map at genus zero upon expansion around the cusp. It will be an interesting
 question for the future if the full $\tau$-dependence of the real-analytic
 iterated KE integrals $\beta^{\rm sv}$ can be understood as the result of
 a such a single-valued map at genus one, in the same way as it was proposed
 for the iterated Eisenstein integrals entering MGFs \cite{Gerken:2020xfv}.

As we will see below, the effective rules (\ref{effrule.1a}) and (\ref{effrule.1b}) already anticipate
the leading terms of the integration constants
$\overline{\alpha[\ldots]}$: These antimeromorphic leading terms
reproduce the MZVs entering
single-valued polylogarithms $G^{\rm sv}$ at depth $\geq 2$
through the terms ${\cal Z}^{\rm sv}$ in (\ref{svpolycl}).

%%%%%%%%%%%%%%%%%%%%%%%%%%%%%%%%%%%%%%%%%%%%%%%%%%%%%%%%%%%
\subsubsection{Examples of real-analytic leading terms and poles in $u$}
\label{sec:3.3.2ex}
%%%%%%%%%%%%%%%%%%%%%%%%%%%%%%%%%%%%%%%%%%%%%%%%%%%%%%%%%%%

The simplest shuffle-independent leading terms at depth two
constructed from (\ref{eq:ebsv2}) and (\ref{fkreps.9}) are given by
\begin{align}
\bsvtau{0 &0 \\ 2 &3  \\ z &z } &= \frac{2}{9} y^2 B_2(u) B_3(u) +
  \frac{1}{4} B_2(u) G^{\rm sv}(0, 1;e^{2\pi i z}) + \bigg( \frac{u}{12 y} - \frac{1}{16 y}
   \bigg) G^{\rm sv}(0, 0,  1;e^{2\pi i z}) \notag \\
   &\quad  -  \frac{ G^{\rm sv}(0, 1, 1;e^{2\pi i z})}{8 y} + {\cal O}(q^{1-u},\bar q^{1-u})  \notag \\
\bsvtau{0 &1 \\ 2 &3  \\ z &z } &= -\frac{16}{9} y^3 B_2(u) B_3(u) + 4 u y^2 B_2(u) G^{\rm sv}(1;e^{2\pi i z}) -
2 u y G^{\rm sv}(1,1;e^{2\pi i z}) \notag \\
 &\quad +
 \frac{ u}{3} (4 u-3 ) y G^{\rm sv}(0, 1;e^{2\pi i z}) + \bigg( \frac{ u }{3} - \frac{ 1}{4} \bigg) G^{\rm sv}(0, 0, 1;e^{2\pi i z})   \label{comrel.32} \\
 &\quad - \frac{1}{2} G^{\rm sv}(0, 1, 1;e^{2\pi i z})+ {\cal O}(q^{1-u},\bar q^{1-u})
\notag \\
 \bsvtau{1 &0 \\ 3 &3  \\ z &z } &=
- \frac{2}{27}y^3 B_3(u)^2 -  \frac{y}{6} B_3(u)   G^{\rm sv}(0,1;e^{2\pi i z}) +
\frac{  G^{\rm sv}(0,1;e^{2\pi i z})^2 }{32 y}  \notag \\
&\quad+  \bigg({-} \frac{ 1}{24}  +
 \frac{ u}{8}    - \frac{ u^2}{12} \bigg)  G^{\rm sv}(0,0,1;e^{2\pi i z})+
\frac{ u}{4}  G^{\rm sv}(0,1,1;e^{2\pi i z}) + {\cal O}(q^{1-u},\bar q^{1-u})
\notag
   \end{align}
Here and below, we use shuffle relations of the single-valued polylogarithms
to attain the form of $G^{\rm sv}(\ldots, 1;e^{2\pi i z})$ while
replacing $G^{\rm sv}(0;e^{2\pi i z}) = - 4 u y$ in this process.
We have assumed the integration constants $\overline{\alpha[\ldots]}$
to only contribute to terms ${\cal O}(q^{1-u},\bar q^{1-u}) $
in (\ref{comrel.32}) since the $G^{\rm sv}$
do not involve any MZVs here. However, the next example
\begin{align}
\bsvtau{0 &0 \\ 2 &4  \\ z &z } &=
 \frac{y^2}{36}  B_2(u) B_4(u) -  \frac{ B_2(u) G^{\rm sv}(0, 0, 1;e^{2\pi i z})}{24 y}
 - \frac{ B_2(u) G^{\rm sv}(0, 0, 0, 1;e^{2\pi i z})}{32 u y^2}  \notag \\
 &\quad +
\frac{ B_4(u) G^{\rm sv}(0, 0, 0, 1;e^{2\pi i z})}{192 u^3 y^2}+
\frac{ G^{\rm sv}(0, 0, 1, 1;e^{2\pi i z})}{48 y^2}+ {\cal O}(q^{1-u},\bar q^{1-u})
\label{comrel.34}
 \end{align}
necessitates contributions of the $\overline{ \alpha[\ldots] }$ to its leading terms
in order to reconstruct the first term $\sim \zeta_3$ in the following single-valued polylogarithm
of depth two (see (\ref{comrel.27}))
\begin{align}
&G^{\rm sv}(0, 0, 1, 1;e^{2\pi i z}) = 2 \zeta_3 \overline{G(1;e^{2\pi i z})}+ G(0,0,1,1;e^{2\pi i z})
+G(0,0,1;e^{2\pi i z}) \overline{G(1;e^{2\pi i z})}
\notag \\
&\ \ +G(0,0;e^{2\pi i z}) \overline{G(1,1;e^{2\pi i z})}
+G(0;e^{2\pi i z}) \overline{G(1,1,0;e^{2\pi i z})}
+\overline{G(1,1,0,0;e^{2\pi i z})}
\label{comrel.34sv}
\end{align}
With the analogous leading terms of $\bsvtau{0 &1 \\ 2 &4  \\ z &z }$ and
$\bsvtau{0 &2 \\ 2 &4  \\ z &z }$ given in appendix \ref{app.lead.1} and the
$\zeta_3$-terms in their single-valued polylogarithms, one can predict
\begin{align}
 \overline{\alphaBR{0 &0 \\ 2 &4 \\ z &z}{\tau}} &={\cal O}(\bar q^{1-u})  \, , \ \ \ \
  \overline{\alphaBR{0 &1\\ 2 &4 \\ z &z}{\tau}} = {\cal O}(\bar q^{1-u})   \notag \\
 \overline{\alphaBR{0 &2 \\ 2 &4 \\ z &z}{\tau}} &= \frac{2}{3}  \zeta_3 \overline{ G(1; e^{2\pi i z}) } + {\cal O}(\bar q^{1-u})
  \label{baralp.1}
\end{align}
which will be confirmed by the independent computations in section \ref{sec:9.1.ex}.

For the leading terms (\ref{hybrid.5}) of the meromorphic hybrid integrals with one kernel
$G_{k}$ and $f^{(k)}$ each, the simplest real-analytic counterparts are given by
\begin{align}
 \bsvtau{0 &0 \\ 2 &4  \\ z & \emptyslot } &=
\frac{y^2}{36}  B_4 B_2(u) + \frac{B_4 G^{\rm sv}(0, 0, 0, 1;e^{2\pi i z})}{ 192 u^3 y^2} + {\cal O}(q^{1-u},\bar q^{1-u})
 \notag \\
 %%%
  \bsvtau{0 &1 \\ 2 &4  \\ z & \emptyslot } &=
-\frac{y^3}{9}  B_4 B_2(u) + \frac{ B_4 G^{\rm sv}(0, 0, 1;e^{2\pi i z})}{24 u^2} +
\frac{ B_4 G^{\rm sv}(0, 0, 0, 1;e^{2\pi i z})}{48 u^3 y}+ {\cal O}(q^{1-u},\bar q^{1-u})
\notag \\
%%%%
 \bsvtau{0 &2 \\ 2 &4  \\ z & \emptyslot } &=
\frac{4}{3} y^4 B_4 B_2(u) + \frac{2 y^2 B_4 G^{\rm sv}(0, 1;e^{2\pi i z})}{3 u}+
\frac{ y B_4 G^{\rm sv}(0, 0, 1;e^{2\pi i z})}{3 u^2} \label{comrel.38} \\
&\quad +
\frac{ B_4 G^{\rm sv}(0, 0, 0, 1;e^{2\pi i z})}{12 u^3}+ {\cal O}(q^{1-u},\bar q^{1-u})
\notag
 \end{align}
Since the single-valued polylogarithms are of depth one, the
associated $ \overline{\alphaBR{0 &j \\ 2 &4 \\ z & \emptyslot}{\tau}}$ are not
expected to contribute any leading terms.

The expressions (\ref{comrel.34}) and (\ref{comrel.38}) for
$ \bsvtau{0 &0 \\ 2 &4  \\ z &z } $ and $ \bsvtau{0 &0 \\ 2 &4  \\ z &\emptyslot } $
individually exhibit poles in $u$ that cannot be removed
via shuffle-relations of $G^{\rm sv}(\ldots)$. The residue of these poles
within the path-ordered exponential (\ref{gen.34c}) is proportional
to $R_{\vec{\eta}}( {\rm ad}^2_{\ep_0}[b_2,\ep_4+b_4])$ which vanishes
by the integrability condition (\ref{revsec.25}). Equivalently, the combination
\begin{align}
&\bsvtau{0 &0 \\ 2 &4  \\ z &z } -  \bsvtau{0 &0 \\ 2 &4  \\ z & \emptyslot } =
 \frac{y^2}{36}  \big( B_4-B_4(u) \big) B_2(u)
+ \frac{ B_2(u) G^{\rm sv}(0, 0, 1;e^{2\pi i z})}{24 y}   \label{comrel.36} \\
&\ \ \ \ \ \
+ \bigg(  \frac{5 u}{192 y^2} -\frac{1}{48 y^2} \bigg)  G^{\rm sv}(0, 0, 0, 1;e^{2\pi i z})
  -  \frac{ G^{\rm sv}(0, 0, 1, 1;e^{2\pi i z})}{48 y^2}+ {\cal O}(q^{1-u},\bar q^{1-u})
\notag
  \end{align}
left by inserting the relation $R_{\vec{\eta}}( {\rm ad}^2_{\ep_0}[b_2,\ep_4+b_4])=0$
into (\ref{gen.34c}) is non-singular as $z\rightarrow 0$.
Hence, the poles in $u$ in some of the earlier leading terms do not appear in
eMGFs. The analogous non-singular expressions for
$\bsvtau{0 &j \\ 2 &4  \\ z &z } -  \bsvtau{0 &j \\ 2 &4  \\ z &\emptyslot }$
at $j=1,2$ can be found in appendix \ref{app.lead.2}.

%%%%%%%%%%%%%%%%%%%%%%%%%%%%%%%%%%%%%%%%%%%%%%%%%%%%%%%%%%%
\subsubsection{Contributions to the Laurent polynomial}
\label{sec:3.3.3}
%%%%%%%%%%%%%%%%%%%%%%%%%%%%%%%%%%%%%%%%%%%%%%%%%%%%%%%%%%%

While the leading terms in the above discussion retain contributions $\sim q^u, q^{2u},\ldots$
and $\sim \bar q^u, \bar q^{2u},\ldots$ in an expansion around the cusp,
the Laurent polynomials of section \ref{sec:2.4.1} are defined as the coefficients
of $q^0 \bar q^0$. The contributions of meromorphic and real-analytic iterated
KE integrals to the Laurent polynomials solely arise from setting
$f^{(k)}(z|\tau) \rightarrow \frac{(2 \pi i)^{k}}{ k! }  B_{k}(u) $
and $G_k(\tau)\rightarrow - \frac{(2 \pi i)^{k}}{ k! }  B_{k} $. In
the meromorphic case, these terms integrate to
\begin{align}
\eeetau{j_1 \\ k_1   \\ z_1  } &= \frac{ B_{k_1}(u_1) \log(q)^{j_1+1} }{k_1!(j_1{+}1)}
+ {\cal O}(q^u)  \label{lpmero.1} \\
\eeetau{j_1 &j_2  \\ k_1 &k_2   \\ z_1 &z_2  }  &=
 \frac{ B_{k_1}(u_1)B_{k_2}(u_2)  \log(q)^{j_1+j_2+2} }{k_1! k_2!(j_1{+}1)(j_1{+}j_2{+}2)}
+ {\cal O}(q^u) \notag
\end{align}
at depth $\ell=1,2$ and follow a simple general formula equivalent to (\ref{onlygk.2}) at higher depth
\beq
\eeetau{j_1 &j_2 &\ldots &j_\ell \\ k_1 &k_2 &\ldots &k_\ell  \\ z_1 &z_2 &\ldots &z_{\ell} }  = \prod_{i=1}^\ell  \frac{ B_{k_i}(u_i) \log(q)^{j_i+1} }{k_i!(i + \sum_{m=1}^i j_m)} + {\cal O}(q^u)
 \label{lpmero.2}
\eeq
In the real-analytic case, the contributions of $\beta^{\rm sv}$ to Laurent polynomials
are most conveniently assembled from the ${\cal E}^{\rm sv}$ via (\ref{gen.38}).
As a consequence of (\ref{fkreps.18}) we can obtain
\begin{align}
\esvtau{j_1 \\ k_1   \\ z_1  } &= \frac{ B_{k_1}(u_1) (-4y)^{j_1+1} }{k_1!(j_1{+}1)}
+ {\cal O}(q^u,\bar q^u)  \label{lpmero.3} \\
\esvtau{j_1 &j_2  \\ k_1 &k_2   \\ z_1 &z_2  }  &=
 \frac{ B_{k_1}(u_1)B_{k_2}(u_2) (-4y)^{j_1+j_2+2} }{k_1! k_2!(j_1{+}1)(j_1{+}j_2{+}2)}
+ {\cal O}(q^u,\bar q^u) \notag
\end{align}
from replacing $\log(q) \rightarrow \log|q|^2 = -4y$ in (\ref{lpmero.1}),
where the notation $ {\cal O}(q^u,\bar q^u)$ refers to terms ${\cal O}(e^{-2yu})$.
The same substitution rule should apply at general depth,
\beq
\esvtau{j_1 &j_2 &\ldots &j_\ell \\ k_1 &k_2 &\ldots &k_\ell  \\ z_1 &z_2 &\ldots &z_{\ell} }  = \prod_{i=1}^\ell  \frac{ B_{k_i}(u_i) (-4y)^{j_i+1} }{k_i!(i + \sum_{m=1}^i j_m)} + {\cal O}(q^u,\bar q^u)
 \label{lpmero.4}
\eeq
Even though the iterated Eisenstein integrals ${\cal E}\! \left[\begin{smallmatrix}
\ldots &j &\ldots \\ \ldots &k &\ldots \\ \ldots & &\ldots
\end{smallmatrix} \right]$ and ${\cal E}^{\rm sv}\! \left[\begin{smallmatrix}
\ldots &j &\ldots \\ \ldots &k &\ldots\\ \ldots & &\ldots
\end{smallmatrix} \right]$
are in general different from the $z \rightarrow 0$ limits of ${\cal E}\! \left[\begin{smallmatrix}
\ldots &j &\ldots \\ \ldots &k &\ldots\\ \ldots &z &\ldots
\end{smallmatrix} \right]$ and ${\cal E}^{\rm sv}\! \left[\begin{smallmatrix}
\ldots &j &\ldots \\ \ldots &k &\ldots \\ \ldots &z &\ldots
\end{smallmatrix} \right]$, the Laurent polynomials of hybrid integrals with some of $z_i$
replaced by empty slots are obtained by truncating Bernoulli polynomials to the
associated Bernoulli numbers, $B_{k_i}(u_i) \rightarrow B_{k_i}$, e.g.
\begin{align}
\eeetau{j_1 &j_2  \\ k_1 &k_2   \\ \emptyslot &z_2  }  &=
 \frac{ B_{k_1} B_{k_2}(u_2) \log(q)^{j_1+j_2+2} }{k_1! k_2!(j_1{+}1)(j_1{+}j_2{+}2)}
+ {\cal O}(q^u)
 \label{lpmero.5} \\
\esvtau{j_1 &j_2  \\ k_1 &k_2   \\ \emptyslot &z_2  }  &=
 \frac{ B_{k_1} B_{k_2}(u_2) (-4y)^{j_1+j_2+2} }{k_1! k_2!(j_1{+}1)(j_1{+}j_2{+}2)}
+  {\cal O}(q^u,\bar q^u) \notag
\end{align}
At depth one, the Laurent polynomials of the ${\cal E}^{\rm sv}$ in (\ref{lpmero.3}) can be
conveniently summed to obtain
\beq
\bsvtau{j \\ k  \\ z }  = \frac{j!(k{-}j{-}2)!}{(k{-}1)! k!} B_k(u) (-4y)^{j+1} +  {\cal O}(q^u,\bar q^u)
 \label{lpmero.6}
\eeq
as in (\ref{nwbsv.6}). For $\beta^{\rm sv}$ at depth two in turn, their definition
(\ref{gen.38}) via double sums introduces hypergeometric $_3F_2$ functions that evaluate
to rational numbers such as \cite{Gerken:2020yii}
\begin{align}
\bsvtau{j_1 &j_2  \\ k_1 &k_2   \\ z_1 &z_2  } &= \frac{
(j_1{+}j_2{+}1)! (k_2{-}2{-}j_2)!
B_{k_1}(u_1)B_{k_2}(u_2)
(-4y)^{j_1+j_2+2} }{(j_1{+}1) k_1! k_2! (k_2{+}j_1)!} \notag \\
&\quad \times{} _3F_2\Big[
\begin{smallmatrix}
1 {+} j_1,\ 2 {+} j_1 {+} j_2, \  2 {+} j_1 {-} k_1 \\
2 {+} j_1,\ 1 {+} j_1 {+} k_2
\end{smallmatrix} ; 1
\Big]
+  {\cal O}(q^u,\bar q^u)
 \label{lpmero.81}
\end{align}

%%%%%%%%%%%%%%%%%%%%%%%%%%%%%%%%%%%%%%%%%%%%%%%%%%%%%%%%%%%
\subsubsection{Extended Laurent polynomial from $z\rightarrow 0$ limits}
\label{sec:3.ext}
%%%%%%%%%%%%%%%%%%%%%%%%%%%%%%%%%%%%%%%%%%%%%%%%%%%%%%%%%%%

We shall now illustrate the departure (\ref{discont.5}) of $\lim_{z \rightarrow 0} \bsvtau{\ldots &j_i &\ldots \\
\ldots &k_i &\ldots\\ \ldots &z &\ldots }$ from $\bsvtau{\ldots &j_i &\ldots \\
\ldots &k_i &\ldots\\ \ldots &\emptyslot &\ldots }$ at the level of
the Laurent polynomials. The polylogarithms in the leading terms
of $\beta^{\rm sv}$ reduce to single-valued MZVs
\beq
 \lim_{z \rightarrow 0}  G^{\rm sv}(\vec{0}^{p_r-1},1, \ldots,\vec{0}^{p_2-1},1,\vec{0}^{p_1-1},1;e^{2\pi i z}) = (-1)^r\zeta^{\rm sv}_{p_1,p_2,\ldots,p_r}
\label{svMZVlim}
\eeq
accompanied by powers of $y$ and thereby contribute to the $q^0 \bar q^0$ order.
In this way, the leading terms (\ref{dpt1LT}) at depth one yield the $z \rightarrow 0$ limit
\beq
\lim_{z \rightarrow 0 }\bsvtau{j \\ k  \\ z }  = \frac{j!(k{-}j{-}2)! B_k}{(k{-}1)! k! } (-4y)^{j+1}
- \frac{\zeta^{\rm sv}_{k-1} }{(k{-}1) (4y)^{k-j-2} }
+  {\cal O}(q,\bar q) \, , \ \   \ \ k\geq 3
 \label{lpmero.82}
 \eeq
with an extra zeta value in comparison to the Laurent polynomial (\ref{lpmero.6})
taken at finite~$z$. We have excluded $k=2$ in view of the logarithmic singularity
of (\ref{nwbsv.15}) as $z\rightarrow 0$, and the single-valued zeta values
vanish for odd $k$ by $\zeta_{2n}^{\rm sv}=0$ and $\zeta_{2n+1}^{\rm sv}=
2 \zeta_{2n+1}$ for $n\in \mathbb N$.

More generally, combinations of higher-depth $\beta^{\rm sv}$ with finite
limit $z \rightarrow 0$ lead to the same type of expansion (\ref{revsec.32})
around the cusp as MGFs: powers of $q^m\bar q^n$ accompanied by Laurent
polynomials in $y$ with MZVs in the coefficients. At the $q^0\bar q^0$ order of these limits, the
contributions of section \ref{sec:3.3.3} from the $q^0\bar q^0$ order of the parental
eMGF are augmented by the single-valued MZVs (\ref{lpmero.82}) due to the
polylogarithms in the leading terms. Hence, the $q^0\bar q^0$ terms of finite
$z\rightarrow0$ limits are referred to as {\it extended Laurent polynomials}.

At depth two for instance, the leading terms in (\ref{comrel.32}) are
individually finite and yield extended Laurent polynomials
\begin{align}
\lim_{z \rightarrow0}\bsvtau{0 &0 \\ 2 &3  \\ z &z } &=  -\frac{\zeta_3}{8y} +  {\cal O}(q,\bar q)  \, , \ \ \ \
\lim_{z \rightarrow0}\bsvtau{0 &1 \\ 2 &3  \\ z &z } =  -\frac{ \zeta_3}{2} +  {\cal O}(q,\bar q)
 \label{lpmero.83}
   \end{align}
as well as
\begin{align}
\lim_{z \rightarrow0} \bsvtau{1 &0 \\ 3 &3  \\ z &z } = \frac{ \zeta_3}{12}+  {\cal O}(q,\bar q)
 \label{lpmero.84}
   \end{align}
However, generic $z$-dependent $\beta^{\rm sv}$ feature poles in $u$ as discussed
in section \ref{sec:3.3.2ex} and one cannot extract an extended Laurent polynomial
term by term. Non-singular combinations such as (\ref{comrel.36}), (\ref{comrel.39})
and (\ref{comrel.42}) in turn yield
\begin{align}
\lim_{z \rightarrow0} \bigg\{
\bsvtau{0 &0 \\ 2 &4  \\ z &z } -  \bsvtau{0 &0 \\ 2 &4  \\ z & \emptyslot } \bigg\}
&= \frac{\zeta_3}{72y} +  {\cal O}(q,\bar q)  \notag\\
\lim_{z \rightarrow0} \bigg\{
\bsvtau{0 &1 \\ 2 &4  \\ z &z } -  \bsvtau{0 &1 \\ 2 &4  \\ z & \emptyslot } \bigg\}
&= \frac{\zeta_3}{36} +  {\cal O}(q,\bar q)  \label{lpmero.85} \\
\lim_{z \rightarrow0} \bigg\{
\bsvtau{0 &2 \\ 2 &4  \\ z &z } -  \bsvtau{0 &2 \\ 2 &4  \\ z & \emptyslot } \bigg\}
&= {\cal O}(q,\bar q)  \notag
   \end{align}
As will be detailed in the next section, the generating series of eMGFs
in (\ref{gen.34c}) does not feature any poles in $u$ and will
at worst exhibit the logarithmic singularity of $g(z|\tau) = - \bsvtau{0\\ 2  \\ z}$
as $z\rightarrow 0$. Hence, the combinations with finite $z\rightarrow 0$
limit as in (\ref{lpmero.85}) appear naturally in setting up bases of eMGFs.

%%%%%%%%%%%%%%%%%%%%%%%
%%%%%%%%%%%%%%%%%%%%%%%
\subsection{Towards $q$-expansions of eMGFs}
\label{sec:qeMGFs}

This section is dedicated to determining the $q$-expansion of
eMGFs beyond the leading terms. The key idea is to rely on decompositions
such as (\ref{discont.6}) and (\ref{d2bsv}) of real-analytic iterated KE integrals
$\beta^{\rm sv}$ into
meromorphic building blocks ${\cal E}$ and their complex conjugates.
By virtue of $z$-dependent analogues of the iterated Eisenstein integrals ${\cal E}_0$
in section \ref{sec:2.5.2}, the $q$-expansion of ${\cal E}$ and therefore
$\beta^{\rm sv}$ will be available to any desired order. With the methods
of later sections to express eMGFs in terms of $\beta^{\rm sv}$, the
subsequent $q$-expansions can be used for numerical checks of their 
modular properties.

%%%%%%%%%%%%%%%%%%
%%%%%%%%%%%%%%%%%%
\subsubsection{$z$-dependent variants of ${\cal E}_0$}
\label{qexpsec.A}

Similar to the discussion of $z$-independent
iterated integrals ${\cal E}_0$ in section \ref{sec:2.5.2}, we
isolate the $q$-series by subtracting
the constant term $\sim B_k(u)$ in the expansion (\ref{fkreps.1}) of the $\tau$ integrands
$f^{(k\neq 0)}(u\tau{+}v|\tau)$ around the cusp,
\begin{align}
f_0^{(k)}(z| \tau) &=  f^{(k)}(z| \tau)  - \frac{(2 \pi i)^{k}}{k!} B_k(u)
\label{ezero.01} \\
&=  \frac{(2 \pi i)^{k}}{(k{-}1)!}\bigg[
\frac{u^{k-1}}{1-e^{-2 i \pi z}}
+ \sum_{n=1}^{\infty} q^n\bigg(
\frac{(u+n)^{k-1}}{q^n - e^{-2 \pi i z}}
- \frac{(u-n)^{k-1}}{q^n - e^{2 \pi i z}}
\bigg)
\bigg]
\, , \ \ \ \  \ \  k\neq 0
\notag
\end{align}
while preserving the kernel $f^{(0)}= f_0^{(0)}=1$. We then adapt the recursive definition
(\ref{fkreps.5}) to the new kernels (\ref{ezero.01})
\begin{align}
\mathcal{E}_0\!\SM{k_1 & k_2 & \ldots & k_r}{z_1 & z_2 & \ldots & z_r }{\tau} &=  2 \pi i \int_{i \infty}^{\tau} \frac{  \mathrm{d} \tau_{r} }{(2 \pi i)^{k_{r}}}\,  f_0^{(k_r)}\left(u_{r} \tau_{r}{+}v_{r}|\tau_{r}\right) \mathcal{E}_0\!\SM{k_1 & k_2 & \ldots & k_{r-1}}{z_1 & z_2 & \ldots & z_{r-1} }{\tau_{r}}
\label{ezero.02}
    \end{align}
with $ \mathcal{E}_0( \begin{smallmatrix}  \emptyset \\ \emptyset \end{smallmatrix};\tau) =1$
and $k_i\geq 0$, which implies identical shuffle identities for ${\cal E}$ and ${\cal E}_0$.
For instance, by $\mathcal{E}(\begin{smallmatrix} \vec{0}^p \\
\vec{0}^p \end{smallmatrix};\tau) = \mathcal{E}_0(\begin{smallmatrix} \vec{0}^p \\
\vec{0}^p \end{smallmatrix};\tau) =\frac{1}{p!} \log(q)^p$,
the depth-one relation (\ref{fkreps.8}) readily carries over to ${\cal E} \rightarrow {\cal E}_0$.
The main difference between the two variants of the resulting integrals
$\mathcal{E}(\begin{smallmatrix} k &\vec{0}^{p-1} \\
z &\vec{0}^{p-1} \end{smallmatrix};\tau)$ and $\mathcal{E}_0(\begin{smallmatrix} k&\vec{0}^{p-1} \\
z&\vec{0}^{p-1} \end{smallmatrix};\tau) $ is the removal of
the contribution $\sim B_k(u) \log(q)^p$ to the depth-one ${\cal E}$ in (\ref{fkreps.6}):
\begin{align}
&\mathcal{E}_0\!\SM{k &\vec{0}^{p-1}}{z &\vec{0}^{p-1}}{\tau} =
\mathcal{E}\!\SM{k &\vec{0}^{p-1}}{z &\vec{0}^{p-1}}{\tau} - \frac{ B_{k}(u) \log(q)^{p} }{k ! p!} \notag \\
&\quad= \frac{1}{(p{-}1)!} \sum_{j=0}^{p-1} (-1)^j {p{-}1 \choose j}
\log(q)^{p-1-j} \eeetau{j  \\ k\\ z  } - \frac{ B_{k}(u) \log(q)^{p} }{k ! p!}
\label{inival.31}\\
&\quad = {-} \frac{1}{(k{-}1)!}\bigg\{u^{k-p-1} \text{Li}_{p}(e^{2 i \pi  z})   \notag \\
&\quad\quad\quad
+\sum_{n=1}^\infty \Big[ (-1)^{k} (n{-}u)^{k-p-1} \text{Li}_{p}(q^n e^{-2 i \pi  z})
    +(n{+}u)^{k-p-1} \text{Li}_{p} (q^n e^{2 i \pi  z}) \Big]\bigg\}
    \notag
\end{align}
More generally, one can rewrite any ${\cal E}_0$ in (\ref{ezero.02})
in terms of representatives with $k_1\neq 0$ in the first entry, possibly
accompanied by powers of $\log(q)$. The required shuffle-identities are
identical to those of ${\cal E}$. However, as an extra virtue of the ${\cal E}_0$,
their instances with $k_1\neq 0$ are invariant under the modular $T$
transformation $\tau \rightarrow \tau{+}1$ since ($k_r\neq 0$)
\begin{itemize}
\item[(i)] the constituents $z,u$ and $q$ of the integration kernels $f_0^{(k_r)}$ in (\ref{ezero.01})
are $T$-invariant
\item[(ii)] iterated integration against $\dd \tau_r \, f_0^{(k_r)}(u_r\tau_r{+}v_r|\tau_r)$
or $\dd \tau_r$ preserves $T$-invariance
\end{itemize}
As can be anticipated from the depth-one example (\ref{inival.31}),
the corresponding ${\cal E}$ with $k_1\neq 0$ are not $T$-invariant
since the $\tau$-integral of their constant term $B_k(u)$ leads to powers
of $\log(q) = 2\pi i \tau$, i.e.\ the above property (ii) is violated for the kernels $f^{(k_r)}$.

We emphasize that generic ${\cal E}_0$ with $k_1\neq 0$ still retain some of the
polylogarithmic contributions $G(\ldots,1;e^{2\pi i z})$ to the leading terms of ${\cal E}$
as in (\ref{fkreps.9.0}) and (\ref{fkreps.9}). For instance, the leading term of
$\mathcal{E}_0(\begin{smallmatrix} k&\vec{0}^{p-1} \\
z&\vec{0}^{p-1} \end{smallmatrix};\tau) $ at depth one in (\ref{inival.31})
still features a multiple of $u^{k-p-1} {\rm Li}_p(e^{2\pi i z})$, and the one-variable
case $z_j=z$ admits a simple closed formula for arbitrary depth:
\begin{align}
 &\mathcal{E}_0\!\SM{k_1 &\vec{0}^{p_1-1} &k_2 &\vec{0}^{p_2-1} &\ldots &k_r &\vec{0}^{p_r-1}}{z &\vec{0}^{p_1-1} &z &\vec{0}^{p_2-1} &\ldots &z &\vec{0}^{p_r-1}}{\tau} = \prod_{i=1}^r \frac{ u^{k_i-1-p_i} }{(k_i{-}1)!}
\label{closedE0} \\
&\quad \quad \times G(\vec{0}^{p_r-1},1, \ldots, \vec{0}^{p_2-1},1,\vec{0}^{p_1-1},1;e^{2\pi i z})
  + {\cal O}(q^{1-u})
  \notag
\end{align}
Accordingly, the $z\rightarrow 0$ limits of the $z$-dependent ${\cal E}_0$ in (\ref{ezero.02})
do not reproduce the iterated Eisenstein integrals ${\cal E}_0$ in section
\ref{sec:2.5.2}. At depth one, for instance, the deviations between
$\lim_{z\rightarrow 0}\mathcal{E}_0(\begin{smallmatrix} k&\vec{0}^{p-1} \\
z&\vec{0}^{p-1} \end{smallmatrix};\tau) $ and ${\cal E}_0(k,\vec{0}^{p-1};\tau)$
by odd zeta values are identical to those in (\ref{discont.1}) with ${\cal E}_0$
in the place of ${\cal E}$.

The expansions (\ref{discont.6}) and (\ref{d2bsv}) of $\beta^{\rm sv}$ into
${\cal E}$ and their complex conjugates already have $k_1\neq 0$ in their
first entry. We can convert each term into manifestly $T$-invariant ${\cal E}_0$
and $\bar{{\cal E}_0}$ using the straightforward uplifts
of the analogous formulae for iterated Eisenstein integrals in appendix D
of \cite{Broedel:2018izr} as well as
appendix G of \cite{Gerken:2020yii}. At depth two, for instance,
 \small
\begin{align}
&\mathcal{E}\!\SM{k_1 &\vec{0}^{p_1-1} &k_2 &\vec{0}^{p_2-1}}{z_1 &\vec{0}^{p_1-1} &z_2 &\vec{0}^{p_2-1}}{\tau}
= \mathcal{E}_0\!\SM{k_1 &\vec{0}^{p_1-1} &k_2 &\vec{0}^{p_2-1}}{z_1 &\vec{0}^{p_1-1} &z_2 &\vec{0}^{p_2-1}}{\tau}
+ \frac{ B_{k_2}(u_2)}{k_2!}  \mathcal{E}_0\!\SM{k_1 &\vec{0}^{p_1+p_2-1}  }{z_1 &\vec{0}^{p_1+p_2-1}  }{\tau}
\label{etoeod2} \\
&\
+ \frac{ B_{k_1}(u_1)}{k_1! (p_2{-}1)!}
  \sum_{n=0}^{p_1}  (-1)^n \,\frac{(n{+}p_2{-}1)!}{n! (p_1{-}n)!}
 \log(q)^{p_1-n}
\mathcal{E}_0\!\SM{k_2 &\vec{0}^{p_2+n-1}  }{z_2 &\vec{0}^{p_2+n-1}  }{\tau}
%%%%%%%
+ \frac{ B_{k_1}(u_1)B_{k_2}(u_2)}{k_1! k_2!} \frac{ \log(q)^{p_1+p_2} }{(p_1{+}p_2)!}
 \notag
\end{align} \normalsize
as can be cross-checked by comparison of the leading terms (\ref{closedE0})
and (\ref{fkreps.9}) in the one-variable case $z_1=z_2=z$. The generalization
of this conversion formula to depth three can be found in appendix \ref{app:etoe0}.

One can naturally define a hybrid version of the $z$-dependent ${\cal E}_0$ integral
by adapting (\ref{altfkreps.5}) to
\begin{align}
\mathcal{E}_0\!\SM{k_1 & k_2 & \ldots & k_r}{z_1 & z_2 & \ldots &\emptyslot }{\tau} &= -  2 \pi i \int_{i \infty}^{\tau} \frac{  \mathrm{d} \tau_{r} }{(2 \pi i)^{k_{r}}}\,  G^0_{k_r}(\tau_{r}) \mathcal{E}_0\!\SM{k_1 & k_2 & \ldots & k_{r-1}}{z_1 & z_2 & \ldots & z_{r-1} }{\tau_{r}}
\label{hybride0}
\end{align}
where one may replace any subset of $z_1,\ldots,z_{r-1}$ by empty slots to allow
for multiple instances of the kernels $G_{k_i}^0(\tau)$ in (\ref{fkreps.5b}). Decompositions
of ${\cal E}$ into ${\cal E}_0$ as in (\ref{etoeod2}) straightforwardly carry over to the
hybrid case by replacing $B_{k_i}(u_i) \rightarrow B_{k_i}$ for each $z_i$ that is converted
to an empty slot. Since there
is no analogue of $f_0^{(k)}(z| \tau) =   \frac{(2 \pi i)^{k}}{(k{-}1)!}
\frac{u^{k-1}}{1-e^{-2 i \pi z}}+{\cal O}(q^{1-u})$ in $G_{k_i}^0(\tau)$, the leading term
of $ \mathcal{E}_0(\begin{smallmatrix} k_1 &\vec{0}^{p_1-1} &k_2 &\vec{0}^{p_2-1} &\ldots &k_r &\vec{0}^{p_r-1} \\ z_1 &\vec{0}^{p_1-1} &z_2 &\vec{0}^{p_2-1} &\ldots &z_r &\vec{0}^{p_r-1} \end{smallmatrix};\tau)$ vanishes if one or more of the $z_i$ are replaced by empty slots.

%%%%%%%%%%%%%%%%%%
%%%%%%%%%%%%%%%%%%
\subsubsection{$q$-expansion of ${\cal E}_0$}
\label{qexpsec.B}

Since the kernels $G_{k}^0,f^{(k)}_0$ of the iterated integrals ${\cal E}_0$
no longer feature the constant terms $\sim B_k,B_k(u)$, their $q$-expansion
takes a simple form to all orders. For instance, setting $e^{2\pi i z} = q^u e^{2\pi i v}$
in the expansion of the kernels $f^{(k)}$ in (\ref{fkreps.1a}) without the term $B_k(u)$
and performing the straightforward $\tau$-integrals leads to
\begin{align}
&\mathcal{E}_0\!\SM{k_1 &\vec{0}^{p_1-1}}{z_1 &\vec{0}^{p_1-1}}{\tau} = - \frac{1}{(k_1{-}1)!}
\sum_{m_1=1}^\infty  \frac{1}{m_1^{p_1}} \bigg\{  u_1^{k_1-p_1-1}   q^{u_1 m_1} e^{2\pi i v_1 m_1}
\label{ezero.31} \\
 &\quad + \sum_{n_1=1}^\infty \bigg[
 (-1)^{k_1} (n_1{-}u_1)^{k_1-p_1-1} q^{(n_1-u_1) m_1} e^{-2\pi i v_1 m_1}
 + (n_1{+}u_1)^{k_1-p_1-1} q^{(n_1+u_1) m_1} e^{2\pi i v_1 m_1}
 \bigg]
 \bigg\} \notag
\end{align}
which can be alternatively obtained from (\ref{inival.31}) by expanding
${\rm Li}_p(z) = \sum_{m=1}^{\infty} \frac{ z^m }{m^p}$. The
extension to higher depth is mostly a notational challenge which we shall address
through the shorthand
\beq
\sum_{ \{\sigma_j,m_j,n_j\} } = \sum_{\sigma_j = \pm 1} \sum_{m_j=1}^\infty
\sum_{n_j=\frac{1}{2}(1-\sigma_j)}^\infty
\label{ezero.32}
\eeq
In this way, we can compactly write
\begin{align}
f_0^{(k_1)}(u_1\tau{+}v_1|\tau) &= -\frac{ (2\pi i)^{k_1} }{(k_1{-}1)!}
\sum_{ \{\sigma_1,m_1,n_1\} } \sigma_1^{k_1} (n_1 {+} \sigma_1 u_1)^{k_1-1}
q^{(n_1+ \sigma_1 u_1) m_1} e^{2\pi i \sigma_1 v_1 m_1}
\label{ezero.33} \\
\mathcal{E}_0\!\SM{k_1 &\vec{0}^{p_1-1}}{z_1 &\vec{0}^{p_1-1}}{\tau} &= - \frac{1}{(k_1{-}1)!}
\sum_{ \{\sigma_1,m_1,n_1\} }  \frac{ \sigma_1^{k_1} }{m_1^{p_1}} \,
  (n_1{+} \sigma_1 u_1)^{k_1-p_1-1} q^{(n_1+ \sigma_1 u_1) m_1} e^{2\pi i \sigma_1 v_1 m_1}
\notag
\end{align}
The two-term sum over $\sigma_j \in \{1,-1\}$ encodes the two contributions to
the summand w.r.t.\ $m,n$ in the second line of (\ref{fkreps.1a}) that are related
by $u \rightarrow -u$. Moreover, the lower bound $n_j\geq \frac{1}{2}(1{-}\sigma_j)$ of the last sum
in (\ref{ezero.32}) ensures that $\sigma_j=+1$ introduces an extra term with $n_j=0$.
This extra term takes into account that the contribution $ u^{k-1}\sum_{m=1}^\infty e^{2\pi i m z} $ 
to the first line of (\ref{fkreps.1a}) is an extension of the expression
$ \sum_{m = 1}^\infty q^{m n}  (u{+}n)^{k-1}e^{2\pi i m z}$ in the second line to $n=0$.

With this understanding of the summation prescription in (\ref{ezero.32}), we
can compactly present the results of the $\tau$-integration in ${\cal E}_0$ at
higher depth. If all of $k_1,k_2,\ldots,k_r$ are nonzero, we have
\begin{align}
    &\mathcal{E}_{0}\!\SM{k_1 & k_2 & \ldots & k_r}{z_1 & z_2 & \ldots & z_r }{\tau} =
     \frac{(-1)^r}{\prod_{j=1}^r (k_j{-}1) ! }
     \sum_{ \{ \sigma_1,m_1,n_1\} } \ldots \sum_{ \{ \sigma_r,m_r,n_r\} }
     \bigg( \prod_{j=1}^r \sigma_j^{k_j} e^{2\pi i \sigma_j m_j z_j}
\bigg) \label{ezero.35}\\
& \ \ \ \  \times \frac{(n_1{+}\sigma_1 u_1)^{k_1-1}(n_2{+}\sigma_2 u_2)^{k_2-1}\ldots (n_r{+}\sigma_r u_r)^{k_r-1}  q^{m_1n_1+m_2 n_2+ \ldots +m_r n_r}}{m_1(n_1{+} \sigma_1 u_1)\big[m_1(n_1{+} \sigma_1 u_1)+m_2(n_2{+} \sigma_2 u_2) \big] \ldots \big[ \sum_{i=1}^r m_i(n_i{+} \sigma_i u_i)\big]}\, ,
\ \ \ \ k_j \neq 0 \notag
\end{align}
In presence of kernels $f^{(0)}_0=1$, the denominator features
more general exponents $p_j\in \mathbb N$, \small
\begin{align}
&\mathcal{E}_{0} \! \SM{k_1 &\vec{0}^{p_1-1} & k_2 &\vec{0}^{p_2-1} & \ldots & k_r &\vec{0}^{p_r-1}}{z_1 &\vec{0}^{p_1-1} & z_2 &\vec{0}^{p_2-1} & \ldots & z_r&\vec{0}^{p_r-1} }{\tau}
    = \frac{(-1)^r}{\prod_{j=1}^r (k_j{-}1) ! }
 \! \sum_{ \{ \sigma_1,m_1,n_1\} } \! \! \!  \ldots \! \! \!  \sum_{ \{ \sigma_r,m_r,n_r\} } \! \!
\bigg( \prod_{j=1}^r  \sigma_j^{k_j} e^{2\pi i \sigma_j m_j z_j} \bigg) \notag  \\
& \ \ \ \ \ \ \ \times \frac{(n_1{+}\sigma_1 u_1)^{k_1-1}(n_2{+}\sigma_2 u_2)^{k_2-1}
\ldots (n_r{+}\sigma_r u_r)^{k_r-1}  q^{m_1n_1+m_2 n_2+ \ldots +m_r n_r}}{\big[m_1(n_1{+} \sigma_1 u_1)\big]^{p_1}\big[m_1(n_1{+} \sigma_1 u_1)+m_2(n_2{+} \sigma_2 u_2) \big]^{p_2} \ldots \big[ \sum_{i=1}^r m_i(n_i{+} \sigma_i u_i)\big]^{p_r}}
 \label{ezero.36}
\end{align} \normalsize
where we again demand $k_j\neq 0$. In the one-variable case $z_1=z_2=\ldots = z_r=z$,
the leading terms (\ref{closedE0}) can be identified as the contribution from
$n_1=n_2=\ldots = n_r=0$ and $\sigma_1=\sigma_2 = \ldots=\sigma_r=1$
to (\ref{ezero.36}).\footnote{
The leftover summations over $m_1,m_2,\ldots, m_r$ then yield the
series expansion of the multiple polylogarithm in (\ref{closedE0}),
\[
G(\vec{0}^{p_r-1},1, \ldots, \vec{0}^{p_2-1},1,\vec{0}^{p_1-1},1;e^{2\pi i z}) = (-1)^r
\sum_{m_1,m_2,\ldots,m_r=1}^{\infty} \frac{ e^{2\pi i ( m_1+m_2+\ldots +m_r) z} } {m_1^{p_1} (m_1{+}m_2)^{p_2}
\ldots (m_1{+}m_2{+}\ldots{+}m_r)^{p_r}}
\]}

The analogous expansions of hybrid integrals
such as (\ref{hybride0}) are once more distinct from a naive $z_j \rightarrow 0$ limit of
(\ref{ezero.36}). Instead, one has to drop the corresponding terms $n_j=0$ at
$\sigma_j=1$, i.e.\ modify the summation prescription (\ref{ezero.32})~to
\beq
\mathcal{E}_{0} \! \SM{k_1 &\ldots & k_j & \ldots & k_r }{z_1 &\ldots &\emptyslot &\ldots
&z_r }{\tau}\, : \ \ \ \
\sum_{ \{\sigma_j,m_j,n_j\} } \rightarrow 2  \sum_{m_j,n_j=1}^\infty
\label{ezero.37}
\eeq
while setting $u_j= z_j =0$ and $\sigma_j=1$ in the summand. This procedure
reproduces the $q$-expansions (\ref{qgamma1}) of iterated Eisenstein integrals
${\cal E}_0$, where all the $z_j$ of (\ref{ezero.36}) are replaced by empty slots.
Moreover, (\ref{ezero.37}) extracts the following hybrid integrals at depth two
from (\ref{ezero.36}):
\begin{align}
\mathcal{E}_{0} \! \SM{k_1 &\vec{0}^{p_1-1} & k_2 &\vec{0}^{p_2-1} }{z &\vec{0}^{p_1-1} &\emptyslot &\vec{0}^{p_2-1}  }{\tau} &= \frac{2}{(k_1{-}1)!(k_2{-}1)!} \sum_{ \{\sigma_1,m_1,n_1\} }
\frac{ \sigma_1^{k_1} }{m_1^{p_1}} \, (n_1{+}\sigma_1 u)^{k_1-p_1-1} e^{2\pi i \sigma_1 v m_1}\notag \\
&\quad \times
\sum_{m_2,n_2=1}^\infty \frac{ n_2^{k_2-1} q^{m_1(n_1+\sigma_1 u)+m_2 n_2}  }{ \big[
m_1(n_1{+}\sigma_1 u) + m_2 n_2
\big]^{p_2}}
\label{ezero.38}\\
%%%%
\mathcal{E}_{0} \! \SM{k_1 &\vec{0}^{p_1-1} & k_2 &\vec{0}^{p_2-1} }{\emptyslot &\vec{0}^{p_1-1} &z &\vec{0}^{p_2-1}  }{\tau} &= \frac{2}{(k_1{-}1)!(k_2{-}1)!} \sum_{m_1,n_1=1}^\infty
\frac{  n_1^{k_1-p_1-1} }{m_1^{p_1}}
\notag \\
&\quad \times
\sum_{ \{\sigma_2,m_2,n_2\} }
 \frac{  \sigma_2^{k_2} (n_2{+}\sigma_2 u)^{k_2-1} e^{2\pi i \sigma_2 v m_2}
 q^{m_1 n_1 + m_2(n_2+\sigma_2 u)}  }{ \big[
 m_1 n_1 + m_2(n_2{+}\sigma_2 u)
\big]^{p_2}}
\notag
\end{align}
In summary, the techniques of this section give rise to a fully explicit $q$-expansion
of real-analytic iterated KE integrals $\beta^{\rm sv}$ by combining the following
three steps:
\begin{itemize}
\item decompose $\beta^{\rm sv}$ into meromorphic ${\cal E}$ and their complex conjugates
via identities such as (\ref{discont.6}), (\ref{d2bsv}) and (\ref{d3bsvLT})
\item reduce ${\cal E}$ to $T$-invariant $\mathcal{E}_0(\begin{smallmatrix} k_1&\ldots \\
z_1&\ldots \end{smallmatrix};\tau)$ at $k_1\neq 0$ via
(\ref{inival.31}), (\ref{etoeod2}) and (\ref{etoeod3}) as well as generalizations to higher depth
\item import the $q$-expansion (\ref{ezero.36}) and its simplified
version (\ref{ezero.37}) for $G_k^0$ kernels in the place of $f_0^{(k)}$
\end{itemize}
By the decomposition of eMGFs into $\beta^{\rm sv}$ to be discussed in the next
sections, we get access to any order in the $q$-expansion of eMGFs. As long as $u \in [0,1]$
is not getting too close to~$1$, the $q$-series as in (\ref{ezero.36}) converge rapidly and
allow for efficient numerical evaluations of eMGFs. The leading term of (\ref{ezero.36}) is
known in terms of multiple polylogarithms (\ref{closedE0}) whose numerical evaluation
is for instance addressed in \cite{Bauer:2000cp, Vollinga:2004sn}.
The remaining contributions to (\ref{ezero.36}) are series in $q^{1-u}, q$ and $q^{1+u}$
where one can coordinate the respective cutoffs according to the numerical value of $u$.

\newpage

%%%%%%%%%%%%%%%%%%%%%%%%%%%%%%%%%%%%%%%%%%%%%%%%%%%%%%%%%%%
%%%%%%%%%%%%%%%%%%%%%%%%%%%%%%%%%%%%%%%%%%%%%%%%%%%%%%%%%%%
\section{Basis dimensions of eMGFs from $\{b_k,\ep_k\}$ algebra}
\label{sec:5}
%%%%%%%%%%%%%%%%%%%%%%%%%%%%%%%%%%%%%%%%%%%%%%%%%%%%%%%%%%%
%%%%%%%%%%%%%%%%%%%%%%%%%%%%%%%%%%%%%%%%%%%%%%%%%%%%%%%%%%%

In this section, we determine the counting of independent one-variable eMGFs $\cplus{A \\ B \\ Z}$
with fixed sums $|A|$, $|B|$ of exponents for the holomorphic and antiholomorphic
lattice momenta in (\ref{basic.15}). In order to avoid confusion with the modular weights
$(0,|B|{-}|A|)$ or the transcendental weight $|A|$ of an eMGF in the convention of
(\ref{gen.66}), we shall refer to $|A|+|B|$ as the {\it lattice weight}.

Our method relies on the fact that all convergent eMGFs of arbitrary topology
that cannot be simplified via holomorphic subgraph reduction
arise in the expansions of $n$-point generating series (\ref{high.4}) at sufficiently
large $n$ \cite{Dhoker:2020gdz}. These expansion coefficients are contained in the 
path-ordered exponential (\ref{gen.34c}) and therefore expressible in terms of the 
real-analytic iterated KE integrals $\beta^{\rm sv}$ along with
\begin{itemize}
\item non-positive powers of $y$ from the factor of
$\exp ( {-}\frac{ R_{\vec{\eta}}(\ep_0)}{4y} )$ and
\item MZVs as well as functions of $u,v$ from the
initial value $  \widehat Y(i\infty)$
\end{itemize}
However, not all the $\bsvtau{j_1 &j_2 &\ldots &j_\ell \\ k_1 &k_2 &\ldots &k_\ell \\ \ldots &\ldots &\ldots &\ldots}$ with different $k_i\geq 2$ and
$0\leq j_i\leq k_i{-}2$ as well as combinations of $z$ and 
empty slots in the last line are individually realized as eMGFs.
There are dropouts due to the relations among the $n! \times n!$
matrices $R_{\vec{\eta}}(\ep_k)$ and $R_{\vec{\eta}}(b_k)$ in the path-ordered exponential
some of which were reviewed in section \ref{sec:2.3.3}.

By combining the counting of $\beta^{\rm sv}$ in the path-ordered exponential
(\ref{gen.34c}) and their dropouts due to relations among the $R_{\vec{\eta}}(\ep_k)$
and $R_{\vec{\eta}}(b_k)$, we will find the basis dimensions of eMGFs
at given $|A|, |B|$ previewed in table \ref{allshuffir}. More specifically, the table counts the
numbers of linearly independent eMGFs including MGFs
w.r.t.\ $\mathbb Q$-linear combinations of MZVs
that cannot be reduced to (sums of) products of simpler objects such as
shuffle products (\ref{besv.shuff}) of $\beta^{\rm sv}$. We will refer to eMGFs
as {\it indecomposable} if they cannot be expressed in terms of
MZVs and eMGFs of lower lattice weight.

\begin{table}[h]
\begin{center}
\begin{tabular}{c||c|c|c|c|c|c|c|c|c}
%\diagbox[innerwidth=1.5em,innerheight=1.5em]{$n$}{$w$} &0&1&2&3&4&5 \\\hline\hline
\diagbox[]{$|A| \! \!$}{$\! \! |B|$} &1&2&3
&4&5&6 &7&8&9  \\\hline\hline
1 &1 &1 &2 &1 &2
&1 &2 &1 &2  \\\hline
2 &1 &2 &2 &3 &3 &4 &4 &6 &  \\\hline
3 &2 &2 &4 &5 &9 &9 &15 & &  \\\hline
4 &1 &3 &5 &10 &14 &24 & & &  \\\hline
5 &2 &3 &9 &14 &29 & & && \\\hline
6 &1 &4 &9 &24 & & &&& \\\hline
7 &2 &4 &15 & & &&& \\\hline
8 &1 &6 & & &&&& \\\hline
9 &2 & & &&&&&
\end{tabular}
\end{center}
\caption{\textit{The numbers of indecomposable eMGFs at given $|A|,|B|$
(including MGFs but excluding MZVs).}}
\label{allshuffir}
\end{table}

%%%%%%%%%%%%%%%%%%%%%%%%%%%%%%%%%%%%%%%%%%%%%%%%%%%%%%%%%%%
\subsection{Preliminaries}
\label{sec:5.1}
%%%%%%%%%%%%%%%%%%%%%%%%%%%%%%%%%%%%%%%%%%%%%%%%%%%%%%%%%%%

Before presenting our derivation of the counting in table \ref{allshuffir}, we clarify
the relation between lattice weights of eMGFs, the matrices $R_{\vec{\eta}}(\ep_k),
R_{\vec{\eta}}(b_k)$ in the path-ordered exponential and the commutator relations
of a tentative $\{b_k,\ep_k\}$-algebra.

%%%%%%%%%%%%%%%%%%%%%%%%%%%%%%%%%%%%%%%%%%%%%%%%%%%%%%%%%%%
\subsubsection{$\{b_k,\ep_k\}$-correspondents of lattice weight and transcendental weight}
\label{sec:5.1.1}
%%%%%%%%%%%%%%%%%%%%%%%%%%%%%%%%%%%%%%%%%%%%%%%%%%%%%%%%%%%

By their homogeneity degrees in $s_{ij},\eta_i,\bar \eta_i$,
relations among products of $R_{\vec{\eta}}(\ep_{k_i}),R_{\vec{\eta}}(b_{k_i})$
are graded by both the number of matrix factors and the sum of the $k_i$. Hence, the
dropouts of the accompanying
$\bsvtau{j_1 &j_2 &\ldots &j_\ell \\ k_1 &k_2 &\ldots &k_\ell \\ z_1&z_2 &\ldots &z_\ell}$
from the space of eMGFs can be similarly graded by $\sum_{i=1}^\ell k_i$ and
$\ell +\sum_{i=1}^\ell j_i$.

As we will see in section \ref{sec:9.1.6}, a simple counting of
homogeneity degrees in $s_{ij},\eta_i,\bar \eta_i$ implies the
correspondence
\beq
\cplus{A \\ B \\ Z} \longleftrightarrow
\bsvtau{j_1 &j_2 &\ldots &j_\ell \\ k_1 &k_2 &\ldots &k_\ell \\ z_1&z_2 &\ldots &z_\ell}
\ {\rm with} \
\left\{
\begin{array}{l}
|A|= \ell + \sum_{i=1}^\ell j_i
\\
|B|= {-}\ell+ \sum_{i=1}^\ell (k_i{-} j_i)
\end{array} \right\} \ \modMZV
\label{count.1}
\eeq
for the $\beta^{\rm sv}$ without any MZVs in their coefficients. Hence,
the quantities $\ell +\sum_{i=1}^\ell j_i$ and  $\sum_{i=1}^\ell k_i$
of the $\beta^{\rm sv}$ preserved by the relations among
$R_{\vec{\eta}}(\ep_k)$ and $R_{\vec{\eta}}(b_k)$ can be identified
as the transcendental weight $|A|$ and lattice-weight $|A|{+}|B|$
of the corresponding eMGF, respectively. At depth $\ell=1$, for instance,
the relation between $ \cplus{a &0 \\ b &0 \\ z &0}$
and $\bsvtau{j  \\ k  \\ z}$ in (\ref{invbsv.7}) identifies $a=j{+}1$
and $b=k{-}j{-}1$ as expected from (\ref{count.1}).

We will determine the complete set of relations among
nested commutators of $R_{\vec{\eta}}(\ep_{k_i})$ and
$R_{\vec{\eta}}(b_{k_i})$ up to and including $\sum_i k_i=11$
that are universal to all multiplicities $n$ of the generating series.
In this way, we infer the counting of indecomposable eMGFs at lattice weights
$|A|{+}|B|\leq 10$ previewed in table \ref{allshuffir}.\footnote{The analogous
counting at $|A|{+}|B| = 11$ is accessible as well from the relations 
among $R_{\vec{\eta}}(\ep_{k_i})$ and $R_{\vec{\eta}}(b_{k_i})$ at $\sum_i k_i=11$.
However, since the bases of eMGFs at $|A|{+}|B| = 11$ are most conveniently
constructed from $z$-derivatives of eMGFs at $|A|{+}|B| = 12$ beyond
the reach of our current analysis of $R_{\vec{\eta}}(\ep_{k_i})$ and $R_{\vec{\eta}}(b_{k_i})$,
the counting in table \ref{allshuffir} and some of the later tables is restricted to $|A|{+}|B|\leq 10$.}
An explicit construction of bases (over $\mathbb Q$-linear combinations of MZVs)
will be performed in section \ref{bassec}.
Since dihedral eMGFs (\ref{gen.66}) turn out to be sufficient to
span these bases, we conclude that all eMGFs of trihedral or
higher-point topology with lattice weights $|A|{+}|B|\leq 10$
can be reduced to dihedral ones. A similar result was found
for MGFs with lattice weights $|A|{+}|B|\leq 12$ \cite{Gerken:2020yii, Gerken:2020aju}.

In slight abuse of terminology, we will assign lattice weight $\sum_{i=1}^\ell k_i$
and transcendental weight $-\ell+\sum_{i=1}^\ell(k_i{-}j_i)$ to products of $\ell$
factors ${\rm ad}_{\ep_0}^{j_i}(\ep_{k_i})$ or ${\rm ad}_{\ep_0}^{j_i}(b_{k_i})$.
This is related to the transcendental weight $\ell +\sum_{i=1}^\ell j_i$ of $\beta^{\rm sv}$ via
$j_i \rightarrow k_i{-}j_i{-}2$ because of the pairing $\bsvtau{\ldots &j_i &\ldots \\
\ldots &k_i &\ldots \\ \ldots &z &\ldots} \leftrightarrow {\rm ad}_{\ep_0}^{k_i-j_i-2}(b_{k_i})$
and $\bsvtau{\ldots &j_i &\ldots \\
\ldots &k_i &\ldots \\ \ldots & \emptyslot &\ldots} \leftrightarrow {\rm ad}_{\ep_0}^{k_i-j_i-2}(\ep_{k_i})$
in the path-ordered exponential (\ref{gen.34c}).

%%%%%%%%%%%%%%%%%%%%%%%%%%%%%%%%%%%%%%%%%%%%%%%%%%%%%%%%%%%
\subsubsection{From matrix representations to an algebra}
\label{sec:5.1.2}
%%%%%%%%%%%%%%%%%%%%%%%%%%%%%%%%%%%%%%%%%%%%%%%%%%%%%%%%%%%

By analogy with the generating-series approach to
MGFs, the operators $R_{\vec{\eta}}(\ep_{k})$ and
$R_{\vec{\eta}}(b_{k})$ in the $n$-point differential equations (\ref{high.12}) of eMGFs
are interpreted as $n!\times n!$ matrix representations of an abstract algebra. For MGFs,
conjectural $(n{-}1)! \times (n{-}1)!$ matrix representations of Tsunogai's derivation
algebra $\{ \epsilon^{\TS}_{k}, \ k \in 2\NN_0\}$ \cite{Tsunogai, Pollack} were generated
from $n$-point genus-one integrals in \cite{Gerken:2019cxz}. These matrices were
checked to obey known relations \cite{LNT, Pollack, Broedel:2015hia}
among Tsunogai's derivations $\epsilon^{\TS}_{k}$ for a wide range of
$k$ and $n$.

In this work, we reverse the logic and infer unknown commutator relations in an extension
$\{ \epsilon_{k}, \ k \in 2\NN_0 \}\oplus \{b_{k}, \  k \in \NN{+}1  \}$
%\NN \ni k  \geq 2 \}$
of Tsunogai's algebra adapted to eMGFs in one
variable. Whenever a commutator relation holds among the $n!\times n!$ matrices
$R_{\vec{\eta}}(\ep_{k})$ and $R_{\vec{\eta}}(b_{k})$ for any multiplicity $n$,
we propose the analogous relation among the algebra generators $\ep_k$ and $b_k$.

A first collection (\ref{revsec.25}) of all-multiplicity relations among the $R_{\vec{\eta}}(\ep_{k})$ and
$R_{\vec{\eta}}(b_{k})$ follows from integrability
$\partial_{\tau} \partial_{z}Y =  \partial_{z} \partial_{\tau} Y$.
Hence, these relations are transferred to the algebra generators $\ep_k,b_k$
themselves, without any commitment to matrix representations $R_{\vec{\eta}}(\cdot)$,
\beq
[b_{w},\ep_k]  + \sum_{a=0}^{w-2} {w{-}2 \choose a}
[b_{a+2},b_{k+w-a-2}]=0 \, , \ \ \ \
w\geq 2 \, , \ \ \ \ k\geq 4
 \label{count.2}
\eeq
The examples of these relations at $w{+}k\leq 8$ are
\begin{align}
 [b_2,\ep_4] + [b_2,b_4] &= 0\, ,
& [b_3,\ep_4] + [b_2,b_5]+ [b_3,b_4] &= 0
\notag \\
 %%%
  [b_2,\ep_6] + [b_2,b_6]&= 0\, ,
&[b_4,\ep_4] + [b_2,b_6]+ 2 [b_3,b_5] &= 0
\label{count.3}
\end{align}
and contribute to the counting of eMGFs at lattice weight $|A|{+}|B|=6,7,8$.
The conjectural nilpotency conditions (\ref{nilpot}) among the
$R_{\vec{\eta}}(\ep_{k})$ and $R_{\vec{\eta}}(b_{k})$ in turn lead to the proposal
\begin{align}
 \ad_{\epsilon_0}^{k-1} (\epsilon_k) &= 0  \, , \ \ \ \ \ \ k\geq 4  \notag
\\
 \ad_{\epsilon_0}^{k-1} (b_k) &= 0  \, , \ \ \ \ \ \ k\geq 2
\label{count.4}
\end{align}
By the ubiquitous appearance of adjoint actions w.r.t.\ $\ep_0$, it will
be convenient to use the shorthand notation
\beq
\ep_k^{(j)} = {\rm ad}_{\ep_0}^j (\ep_k) \, , \ \ \ \ \ \
b_k^{(j)} = {\rm ad}_{\ep_0}^j (b_k)
\label{count.5}
\eeq
such that both of $\ep_k^{(j)}$ and $b_k^{(j)}$ at fixed $k$ form
a $(k{-}1)$-component vector $j=0,1,\ldots,k{-}2$. Note that by analogy with
Tsunogai's algebra one may also define an operator $\check{\epsilon}_{0}$
with shorthand $\widecheck{\rm ad}_{{\epsilon}_{0}}(\ast) = [\check{\epsilon}_{0},\ast]$
which acts inversely on the $\{\epsilon_k, b_k\}$ in the sense that
\begin{align}
\widecheck{\rm ad}_{{\epsilon}_{0}}(\epsilon_k^{(j)}) = j (k{-}j{-}1)\epsilon_k^{(j-1)} \, , \qquad k\geq4 \notag\\
\widecheck{\rm ad}_{{\epsilon}_{0}}(b_k^{(j)}) = j (k{-}j{-}1)b_k^{(j-1)}\, , \qquad k\geq2
\label{adje0rel}
\end{align}
The vectors $\ep_k^{(j)}$ and $b_k^{(j)}$ at fixed $k$ then form
$(k{-}1)$-dimensional representations of the $SL_2$ generated
by $\widecheck{\rm ad}_{{\epsilon}_{0}}$,  ${\rm ad}_{{\epsilon}_{0}}$
and the operator ${\rm ad}_{[ \check{\epsilon}_{0} , \epsilon_0] }
=\widecheck{\rm ad}_{{\epsilon}_{0}}  {\rm ad}_{{\epsilon}_{0}}-
{\rm ad}_{{\epsilon}_{0}} \widecheck{\rm ad}_{{\epsilon}_{0}}$ with
eigenvalues $k{-}2{-}2j$ on both $\epsilon_k^{(j)}$ and $b_k^{(j)}$.
Hence,  $\widecheck{\rm ad}_{{\epsilon}_{0}}$ and ${\rm ad}_{{\epsilon}_{0}}$ can be viewed as
raising and lowering operators for the eigenvalue of  
${\rm ad}_{[ \check{\epsilon}_{0} , \epsilon_0] }$, respectively.

%%%%%%%%%%%%%%%%%%%%%%%%%%%%%%%%%%%%%%%%%%%%%%%%%%%%%%%%%%%
\subsection{Assembling relations in the $\{b_k,\ep_k\}$ algebra}
\label{sec:5.2}
%%%%%%%%%%%%%%%%%%%%%%%%%%%%%%%%%%%%%%%%%%%%%%%%%%%%%%%%%%%

We shall now describe a procedure to determine the complete set of
relations among $\{b_k,\ep_k\}$ order by order in their lattice weight.
The key idea is to derive upper and lower bounds on the universal relations among
nested commutators of $R_{\vec{\eta}}(\ep_{k_i})$ and $R_{\vec{\eta}}(b_{k_i})$ at fixed
lattice weights $\sum_i k_i$. Since the lower and upper bounds fully agree 
at all $\sum_i k_i\leq 11$, we are led to a conclusive count of commutator relations 
and ultimately of indecomposable one-variable eMGFs in this range. We have
not yet tested whether the bounds still coincide at $\sum_i k_i\geq 12$ -- due to the limits of our
current {\sc Mathematica} implementation.

%%%%%%%%%%%%%%%%%%%%%%%%%%%%%%%%%%%%%%%%%%%%%%%%%%%%%%%%%%%
\subsubsection{Lower bounds from cancellations of poles in $u$}
\label{sec:5.2.1}
%%%%%%%%%%%%%%%%%%%%%%%%%%%%%%%%%%%%%%%%%%%%%%%%%%%%%%%%%%%

For the real-analytic iterated KE integrals $\beta^{\rm sv}$ as in (\ref{eq:ebsv1})
and (\ref{eq:ebsv2}), the leading
terms in their expansion around the cusp at fixed
$u,v$ have been discussed in section \ref{sec:3.3}. Starting from depth two,
expressions like (\ref{comrel.34}) or (\ref{comrel.38}) for their leading terms
exhibit simple or higher-order poles in $u$. However, eMGFs can only have the singularity
structure of $g(z|\tau) \rightarrow - \log|z|^2$ or $f^{(1)}(z|\tau) \rightarrow \frac{1}{z}$
at the origin and at any other lattice point $z \in \ZZ\tau{+} \ZZ$ 
and must be regular on the line segment where $v  \in (0,1)$ and
$u = \frac{z-\bar z}{\tau-\bar \tau} \rightarrow 0$. 
In fact, the eMGFs of interest to this work can only
have logarithmic singularities and do not feature any poles in $z$ or $\bar z$. 
(This applies to all the eMGFs in the expansion of the generating series (\ref{high.4}) 
which do not admit simplifications via holomorphic subgraph reduction.)

Hence, the residues of all poles in $u$ must vanish in the path-ordered exponential
(\ref{gen.34c}), in particular those in the leading terms that feature single-valued
polylogarithms $G^{\rm sv}(\ldots,1;e^{2\pi i z})$. We pass to an abstract version
\begin{align}
\Phi(z|\tau)&= 1+ \sum_{k_1=2}^\infty \sum_{j_1=0}^{k_1-2} c_{j_1,k_1}
\bigg(   \bsvtau{j_1 \\ k_1 \\  \emptyslot}   \ep^{(k_1-j_1-2)}_{k_1}
+\bsvtau{j_1 \\ k_1 \\ z }  b^{(k_1-j_1-2)}_{k_1}  \bigg)  \notag \\
&\quad + \sum_{k_1,k_2=2}^\infty \sum_{j_1=0}^{k_1-2} \sum_{j_2=0}^{k_2-2} c_{j_1,k_1}c_{j_2,k_2}
\bigg(  \bsvtau{j_1 &j_2 \\ k_1 &k_2 \\  \emptyslot&\emptyslot }  \ep^{(k_2-j_2-2)}_{k_2}  \ep^{(k_1-j_1-2)}_{k_1}  \notag \\
&\ \ \ \ \ \quad
+ \bsvtau{j_1 &j_2 \\ k_1 &k_2 \\ z &\emptyslot }   \ep^{(k_2-j_2-2)}_{k_2} b^{(k_1-j_1-2)}_{k_1} \label{count.6}\\
&\ \ \ \ \ \quad
+ \bsvtau{j_1 &j_2 \\ k_1 &k_2 \\ \emptyslot &z }  b^{(k_2-j_2-2)}_{k_2} \ep^{(k_1-j_1-2)}_{k_1}   \notag \\
 &\ \ \ \ \ \quad
+ \bsvtau{j_1 &j_2 \\ k_1 &k_2 \\ z &z } b^{(k_2-j_2-2)}_{k_2} b^{(k_1-j_1-2)}_{k_1}
\bigg) + \ldots
\notag
\end{align}
of the path-ordered exponential (see (\ref{POEd3.1})
for depth-three terms) since the analysis of singularities in $u$ implies relations
that must hold for {\it any} matrix representation $R_{\vec{\eta}}(\cdot)$.
By assembling the leading terms of $\bsvtau{0 &0 \\ 2 &4 \\ z &z}
- \bsvtau{0 &0 \\ 2 &4 \\ z &\emptyslot} $ from (\ref{comrel.34}) and (\ref{comrel.38}), for instance,
the coefficient of $G^{\rm sv}(0,0,0,1;e^{2\pi i z})/(y^2 u^3)$
is found to be proportional to $[b_2,b_4^{(2)}{+}\ep_4^{(2)}]$
which vanishes as a consequence of (\ref{count.3}).\footnote{In intermediate steps,
we have used regularity of $ \frac{ B_4(u)}{192 u^3}-\frac{ B_2(u)}{32 u}  -  \frac{B_4}{192 u^3}
= \frac{1}{48} - \frac{5 u}{192}$, so the singularity structure in $u$ usually hinges on
the explicit form of $B_k(u)$ and the values of $B_k$.}
Thus, inserting the leading terms of any $\beta^{\rm sv}$ up to some fixed lattice weight into the path-ordered exponential (\ref{count.6}) leads to large classes of relations among the 
operators $\epsilon_k$ and $b_k$: The $b_k^{(j)},\ep_k^{(j)}$-valued coefficients of $y^{a} u^{-b}
G^{\rm sv}(C,1;e^{2\pi i z})$ have to vanish separately for each $a\in \ZZ$, $b \in \NN$ and
word $C \in\{0,1\}^\times$.

Not all of the relations obtained in this way are independent: For instance, the
example $[b_2,b_4^{(2)}{+}\ep_4^{(2)}]=0$ due to (\ref{comrel.34}) and (\ref{comrel.38})
is a corollary of the simpler relation ${[b_2,b_4{+}\ep_4]=0}$ which follows both from
integrability and from poles in $u$ at lower lattice weight in (\ref{count.6}). One can form a variety
of derived relations in the algebra from (possibly repeated) adjoint action of arbitrary generators
$\ep_{0},\ep_{k\geq 4}, b_{k\geq 2}$. Relations which can not be formed from such
adjoint actions on simpler relations are referred to as \emph{indecomposable}.

In all cases up to and including $\sum_{i=1}^\ell k_i=9$, the coefficients of the
poles in $u$ exhaust both indecomposable and derived relations compatible with the
lattice weight. At $\sum_{i=1}^\ell k_i=10$, however, the derived relations
\beq
 [b_4, [b_2, b_4^{(2)} + \ep_4^{(2)}]] =0
\, , \ \ \ \ \ \
[b^{(2)}_4, [b_2, b_4 + \ep_4]] =0
\label{count.7}
\eeq
cannot be detected from poles in $u$ but must instead be tracked separately
as corollaries of imposing the sector with $\sum_{i=1}^\ell k_i=6$ to be
regular as $u \rightarrow 0$.

To mitigate the fact that not all derived relations can be immediately obtained from $u$-pole cancellation, it is advantageous to proceed iteratively in
lattice weight $\sum_{i=1}^\ell k_i$. Starting at low weight, one can obtain relations by plugging in the Laurent expansion of the relevant $\beta^{\rm sv}$ as before, and consequently construct all derived relations from this seed set up to the weights under consideration. Plugging this set of derived relations into the path-ordered exponential~(\ref{count.6}), one proceeds to the next combination of weights at which a pole in $u$ appears and repeats this process. In this way, we explicitly obtain only indecomposable relations from the path-ordered exponential, automatically completing to the full set of relations by acting with all derivations. The numbers of indecomposable relations among
$\epsilon_{k_i}^{(j_i)},b_{k_i}^{(j_i)}$ -- organized by both $\sum_ik_i,\sum_i j_i$
and total exponents $|A|,|B|$ of the lattice sums -- are given in table~\ref{indecrelcount}.

\begin{table}[h!]
\begin{center}
\begin{tabular}{c||c|c|c|c|c|c|c}
%\diagbox[innerwidth=1.5em,innerheight=1.5em]{$n$}{$w$} &0&1&2&3&4&5 \\\hline\hline
\diagbox[]{$J \! \!$}{$\! \! K$} &5&6&7
&8&9&10 &11  \\\hline\hline
0 &0 &1 &1 &2 &2
&3 &3    \\\hline
1 &0 &0 &1 &1 &2 &2 &3   \\\hline
2 &0 &0 &0 &1 &1 &2 &2    \\\hline
3 &0 &0 &0 &0 &0 &0 &1   \\\hline
4 &0 &0 &0 &0 &0 &0 &0
\end{tabular}
%%%
%%%
%%%
\hspace{1cm}
%%%
%%%
%%%
\begin{tabular}{c||c|c|c|c|c|c|c|c|c}
%\diagbox[innerwidth=1.5em,innerheight=1.5em]{$n$}{$w$} &0&1&2&3&4&5 \\\hline\hline
\diagbox[]{$|A| \! \!$}{$\! \! |B|$} &1&2&3
&4&5&6 &7&8&9  \\\hline\hline
1 &0 &0 &0 &0 &0
&0 &0 &0 &0  \\\hline
2 &0 &0 &0 &1 &1 &2 &2 &3 &3  \\\hline
3 &0 &0 &0 &1 &1 &2 &2 &3 &  \\\hline
4 &0 &0 &0 &1 &1 &2 &2 & &  \\\hline
5 &0 &0 &0 &0 &0 &1 & &&
\end{tabular}
\end{center}
\caption{\textit{The left panel displays
the number of indecomposable relations among
$\epsilon_{k_i}^{(j_i)},b_{k_i}^{(j_i)}$ at given
$K= \sum_ik_i$ and $J=\sum_i j_i$. In the right panel, the same counting
is reorganized according to the holomorphic and antiholomorphic exponents
$|A|,|B|$ of the ${\cal C}^+[\ldots]$ affected by these relations. Rows at higher $|A|$
which include nothing but zeroes were purposefully omitted from the range of this table.}}
\label{indecrelcount}
\end{table}

One should note that also the iterative scan of poles in $u$ in the path-ordered
exponential as described above is not guaranteed to yield all indecomposable
relations -- even in principle.
What is more, by the restriction to the leading terms in $\beta^{\rm sv}$,
it is conceivable that some of the poles in $u$ at higher order in $q$ yield
additional relations among the $\{b_{k_i},\ep_{k_i}\}$. In this sense, the analysis in this section
only gives a lower bound on the set of possible relations. However, as we will see next,
the lower bounds on the relations at $\sum_{i=1}^\ell k_i\leq 11$ obtained from $u$-pole cancellation saturate the upper bounds derived by other means in the next section \ref{sec:5.2.2}.

%%%%%%%%%%%%%%%%%%%%%%%%%%%%%%%%%%%%%%%%%%%%%%%%%%%%%%%%%%%
\subsubsection{Upper bounds from explicit representations $R_{\vec{\eta}}(\ep_{k})$ and $R_{\vec{\eta}}(b_{k})$ }
\label{sec:5.2.2}
%%%%%%%%%%%%%%%%%%%%%%%%%%%%%%%%%%%%%%%%%%%%%%%%%%%%%%%%%%%

As was stated in the previous section, regularity in the limit $u \to 0$ is not guaranteed 
to give a complete set of all-multiplicity relations among the $R_{\vec{\eta}}(\ep_{k})$ and 
$R_{\vec{\eta}}(b_{k})$. In order to obtain an upper bound of this set, we study the relations that 
apply to explicit representations at fixed multiplicity. In particular, we study relations of
the specific $6\times 6$ matrices
at $n=3$ points. These matrices have been determined in \cite{Broedel:2020tmd, Dhoker:2020gdz}
and are given in an ancillary file of
the arXiv submission of this work. Their relations are obtained by writing the most general
ansatz of nested commutators compatible with a given transcendental and lattice weight
(or homogeneity degree in $s_{ij},\eta_i,\bar \eta_i$) and setting
its action on the contributions (\ref{inival.07}) below to three-point initial values to zero.

The reason that we are not employing the $2\times 2$ matrices
(\ref{revsec.23}) at $n=2$ points is that they obey a variety of extra relations
that no longer hold at $n\geq 3$. For instance, given that all the $R_\eta(\ep_k)$
in (\ref{revsec.23}) with $k\geq 4$ are proportional to
$(\begin{smallmatrix} s_{12} & 0 \\ 0 &s_{02}\end{smallmatrix})$,
they commute with one another even though one has
$[R_{\vec{\eta}}(\ep_4), R_{\vec{\eta}}(\ep_6)]\neq0$ at $n=3$.
Hence, the representations at $n=3$ yield a much more stringent upper bound
on the $\{b_k,\ep_k\}$ relations at given lattice weight as compared to the ones at $n=2$.

It is conceivable that the $6\times 6$ matrices $R_{\eta_2,\eta_3}(\ep_k),
R_{\eta_2,\eta_3}(b_k)$ at $n=3$ still obey relations
at sufficiently high lattice weight that no longer hold at some $n\geq 4$.
However, we have checked up to and including $\sum_{i=1}^\ell k_i = 11$
that the span of relations of the $R_{\eta_2,\eta_3}(\cdot)$ exactly matches
those obtained from the cancellation of poles in $u$, when making sure to include all derived relations as detailed in the previous section.
Thus, we can present a reliable count of relations in table \ref{allrelcount} below.

\begin{table}[h!]
\begin{center}
\begin{tabular}{c||c|c|c|c|c|c|c|c|c}
\diagbox[]{$|A| \! \!$}{$\! \! |B|$} &1&2&3
&4&5&6 &7&8&9  \\\hline\hline
1 &0 &0 &0 &0 &0
&0 &0 &0 &0  \\\hline
2 &0 &0 &0 &1 &1 &2 &2 &3 &3  \\\hline
3 &0 &0 &1 &2 &4 &6 &10 &13 &  \\\hline
4 &0 &1 &2 &5 &9 &18 &28 & &  \\\hline
5 &0 &1 &4 &9 &22 &40 & && \\\hline
6 &0 &2 &6 &18 &40 & &&& \\\hline
7 &0 &2 &10 &28 & &&& \\\hline
8 &0 &3 &13 & &&&& \\\hline
9 &0 &3 & &&&&&
\end{tabular}
\end{center}
\caption{\textit{The full number of relations among $\epsilon_k,b_k$ relevant to
${\cal C}^+[\ldots]$ at given holomorphic and antiholomorphic contributions
$|A|,|B|$ to the lattice weight. The diagonals are lines of constant lattice
weight, with the first non-vanishing entries at $|A|{+}|B|=6$.}}
\label{allrelcount}
\end{table}

All the relations among products $\ep_{k_i}^{(j_i)}$, $b_{k_i}^{(j_i)}$ found
from the above procedure preserve the transcendental weight
$-\ell+\sum_{i=1}^\ell(k_i{-}j_i)$. Hence, we classify the counting
of $\{b_k,\ep_k\}$ relations at given lattice weight $\sum_{i=1}^\ell k_i$
by their transcendental weight in tables~\ref{indecrelcount} and~\ref{allrelcount} 
that ranges between $2$ and $-2+\sum_{i=1}^\ell k_i$. This amounts to organizing the relations
by the holomorphic and antiholomorphic contributions $|A|=-\ell+\sum_{i=1}^\ell(k_i{-}j_i)$ and
$|B|=\ell + \sum_{i=1}^\ell j_i$ to the lattice weight $|A|{+}|B|$, see the discussion
in section \ref{sec:5.1.1}. The absence of relations at transcendental weight $\leq 1$
crucially hinges on the nilpotency property (\ref{count.5}).

As already noted in \cite{Broedel:2020tmd, Dhoker:2020gdz}, the matrices
$R_{\vec{\eta}}(\ep_{k})$ in general do not obey the relations in Tsunogai's
derivation algebra $\{ \epsilon^{\TS}_{k}, \ k \in 2\NN_0 \}$. Instead, the references
give first evidence that the combinations
\beq
R_{\vec{\eta}}(\epsilon^{\TS}_{k}) = R_{\vec{\eta}}(\epsilon_{k}) + R_{\vec{\eta}}(b_{k})
\label{count.8}
\eeq
form matrix representations of Tsuongai's derivations, see \cite{Kaderli:2022qeu}
for generalizations to multiple unintegrated punctures. It is at this point
unclear whether all the relations among $\ep_k,b_k$ inherited from the $\epsilon^{\TS}_{k}$
can be detected from poles of (\ref{count.6}) in $u$ as detailed in section~\ref{sec:5.2.1}.
This will affect the analysis at lattice weight $\sum_{i=1}^\ell k_i\geq 14$, due to 
the first Tsunogai relation
$[\epsilon^{\TS}_{10},\epsilon^{\TS}_{4}]-3[\epsilon^{\TS}_{8},\epsilon^{\TS}_{6}]=0$
besides the all-weight families $[\epsilon^{\TS}_{2},\epsilon^{\TS}_{k}]=0$ and
${\rm ad}_{\epsilon^{\TS}_{0}}^{k-1}(\epsilon^{\TS}_{k})=0$.

%%%%%%%%%%%%%%%%%%%%%%%%%%%%%%%%%%%%%%%%%%%%%%%%%%%%%%%%%%%
\subsubsection{Examples of $\{b_k,\ep_k\}$ relations}
\label{sec:5.2.3}
%%%%%%%%%%%%%%%%%%%%%%%%%%%%%%%%%%%%%%%%%%%%%%%%%%%%%%%%%%%

The $\{b_k,\ep_k\}$ relations obtained from the above procedures certainly
reproduce those inferred from integrability, see (\ref{count.2}) for their general form
and (\ref{count.3}) for their examples at lattice weight $\leq 8$. Adjoint action
of $\ep_0$ readily yields derived relations at the same lattice
weight through the Leibniz rule
\beq
{\rm ad}_{\ep_0}^N\big( [ a,b] \big) = \sum_{j=0}^N {N \choose j} [ {\rm ad}_{\ep_0}^j(a),
{\rm ad}_{\ep_0}^{N-j}(b)  ]
\label{count.11}
\eeq
such as
\beq
{\rm lattice} \ {\rm weight} \ 6 \ \ \longleftrightarrow \ \ \left\{ \begin{array}{r}
 [b_2,\ep_4] + [b_2,b_4] = 0  \\
\phantom{x+ }  [b_2,\ep^{(1)}_4] + [b_2,b^{(1)}_4] = 0 \\
\phantom{x+ }   [b_2,\ep^{(2)}_4] + [b_2,b^{(2)}_4] = 0
\end{array} \right.
\label{count.12}
\eeq
By the nilpotency property $\ep^{(k-1)}_k=b^{(k-1)}_k=0$, integrability relations
involving $[\ep_{k_1},b_{k_2}]$ and $[b_{k_1},b_{k_2}]$ at lattice weight $k_{1}{+}k_2$
trivialize after $k_1{+}k_2{-}3$ actions of ${\rm ad}_{\ep_0}$. The order
of ${\rm ad}_{\ep_0}$ that annihilates a more elementary relation will be an
important guiding principle in the subsequent discussion.

Starting from lattice weight seven, there are additional $\{b_k,\ep_k\}$-relations besides
the corollaries of integrability and (\ref{count.11}). This can by anticipated by the
$1+2+2+1$ relations in table \ref{allrelcount} at $|A|{+}|B|=7$, more specifically by the
two relations at $(|A|,|B|)=(3,4)$ or $(4,3)$. They occur at the transcendental
weight of the first and second ${\rm ad}_{\ep_0}$-action on the integrability relation
$[b_3,\ep_4] {+} [b_2,b_5]{+} [b_3,b_4] = 0$. The additional independent relations 
can be taken to be
\beq
[b_3,b^{(1)}_4]+[b_3,\ep^{(1)}_4]+2 [b_4,b_3^{(1)}]+2 [\ep_4,b_3^{(1)}]
- 5 [b_2,[b_2,b_3]] = 0
\label{count.13}
\eeq
and its first ${\rm ad}_{\ep_0}$-action
\beq
[b_3,b^{(2)}_4]+[b_3,\ep^{(2)}_4]+ [b^{(1)}_4,b_3^{(1)}]+ [\ep^{(1)}_4,b_3^{(1)}]
- 5 [b_2,[b_2,b^{(1)}_3]] = 0
\label{count.14}
\eeq
while ${\rm ad}_{\ep_0}^2$ annihilates the left-hand side of (\ref{count.13})
solely as a consequence of $\ep^{(k-1)}_k=b^{(k-1)}_k=0$.
Together with ${\rm ad}^{\leq 3}_{\ep_0}$-action on
$[b_3,\ep_4] + [b_2,b_5]+ [b_3,b_4] =0$, (\ref{count.13}) and
(\ref{count.14}) exhaust the total of six relations at lattice weight seven
noted in table \ref{allrelcount}.

At lattice weight eight, the counting of $2+4+5+4+2$ relations
in table \ref{allrelcount} comes about from
\begin{itemize}
\item[(i)] two integrability relations $[b_2,\ep_6] + [b_2,b_6]= 0$
and $ [b_4,\ep_4] + [b_2,b_6]+ 2 [b_3,b_5] =0$
which are annihilated by ${\rm ad}_{\ep_0}^{5}$ 
along with their ${\rm ad}_{\ep_0}^{\geq 4}$ actions
\item[(ii)] two indecomposable relations involving terms with the weights of $[b_2,\ep^{(1)}_6]$
which are annihilated by ${\rm ad}_{\ep_0}^{3}$ along with their ${\rm ad}_{\ep_0}^{\geq 2}$
actions
\begin{align}
0 &= [b_2, [b_2, b_4{+}\ep_4]] \notag
\\
0 &=   2 [b_3,b_5^{(1)}] +
    6 [b_4,b_4^{(1)}] +
    3 [b_4, \ep_4^{(1)}]+
    6 [b_5, b_3^{(1)} ] +
    3 [\ep_4,b_4^{(1)}] \label{count.16}  \\
    & \quad -
    20 [b_3,[b_2,b_3]]  +
15 [b_2,[b_2,b_4]]
\notag
\end{align}
\item[(iii)] one indecomposable relation involving terms with the weights of $[b_2,\ep^{(2)}_6]$
which is annihilated by ${\rm ad}_{\ep_0}$
\begin{align}
0 &= [b_4, \ep_4^{(2)}]
- [\ep_4, b_4^{(2)}]
- [b_4^{(1)},\ep_4^{(1)}]
-   20 [ b_2, [b_3,b_3^{(1)}] ]
\label{count.17}
\end{align}
\end{itemize}
The first line of (\ref{count.16}) is of course a corollary of the relation
$0=[b_2,b_4{+}\ep_4]$ at lower lattice weight and could in principle be
added to the second relation in (\ref{count.16}) while keeping it in the
kernel of ${\rm ad}_{\ep_0}^{3}$.

%%%%%%%%%%%%%%%%%%%%%%%%%%%%%%%%%%%%%%%%%%%%%%%%%%%%%%%%%%%
\subsubsection{General structure of $\{b_k,\ep_k\}$ relations}
\label{sec:5.2.4}
%%%%%%%%%%%%%%%%%%%%%%%%%%%%%%%%%%%%%%%%%%%%%%%%%%%%%%%%%%%

To understand the structure of the relations we have uncovered, it is worthwhile to revisit 
aspects of Tsunogai's derivation algebra. The appearance of the nested commutator 
$[b_2,[b_2,b_3]]$ in (\ref{count.13}) is analogous to the last two terms (without $\epsilon_0^{\rm TS}$) in
\begin{align}
0&=80[\ep^{\TS}_{12},[\ep^{\TS}_4,\ep^{\TS}_{0}]]
+ 16 [\ep^{\TS}_4,[\ep^{\TS}_{12},\ep^{\TS}_0]]
- 250 [\ep^{\TS}_{10},[\ep^{\TS}_6,\ep^{\TS}_0]]\label{count.15}  \\
&\quad
- 125 [\ep^{\TS}_6,[\ep^{\TS}_{10},\ep^{\TS}_0]]
+ 280 [\ep^{\TS}_8,[\ep^{\TS}_8,\ep^{\TS}_0]]
- 462 [\ep^{\TS}_4,[\ep^{\TS}_4,\ep^{\TS}_8]]
- 1725 [\ep^{\TS}_6,[\ep^{\TS}_6,\ep^{\TS}_4]]
\notag
\end{align}
which is the simplest indecomposable relation among depth-three commutators in
Tsunogai's derivation algebra: Neither $[b_2,[b_2,b_3]]$ in (\ref{count.13}) nor
$[\ep^{\TS}_4,[\ep^{\TS}_4,\ep^{\TS}_8]] $ in (\ref{count.15})
can be removed by adding commutators of simpler relations.
The same applies to the nested commutators $[b_3,[b_2,b_3]],[b_2,[b_2,b_4]]$
and $[ b_2, [b_3,b_3^{(1)}] ]$ in (\ref{count.16}) and (\ref{count.17}).

In case of Tsunogai's derivations, the indecomposable relations with three and
more $\ep_k^{\TS}$ necessitate terms $[({\rm ad}_{\ep_0^{\TS}})^{j_1}\ep_{k_1}^{\TS},
({\rm ad}_{\ep_0^{\TS}})^{j_2}\ep_{k_2}^{\TS}]$ with $k_1,k_2\geq 4$
which will be referred to as depth two. By adding suitable ${\rm ad}_{\ep_0}$-actions
of simpler relations, the coefficients of the depth-two terms are determined by
period polynomials of holomorphic cusp forms \cite{Pollack}.

Once the indecomposable relations
among $\ep_k^{\TS}$ are organized according to the cusp-form analysis, they
form highest-weight vectors of the $SL_2$-algebra involving the
generator ${\rm ad}_{\ep_0^{\TS}}$. This simply means that the relations with depth-two
terms deduced from cusp forms are annihilated by a prescribed number of ${\rm ad}_{\ep_0^{\TS}}$:
In the case of (\ref{count.15}), the left-hand side vanishes under $({\rm ad}_{\ep_0^{\TS}})^{11}$
even though, term by term, only the vanishing under $({\rm ad}_{\ep_0^{\TS}})^{12}$ is manifest
(say for $[\ep^{\TS}_{12},[\ep^{\TS}_4,\ep^{\TS}_{0}]] $ by itself).
The dimensions of these $SL_2$ representations are in one-to-one
correspondence to the Laplace eigenvalues of the accompanying Poincar\'e
series ${\rm F}^{\pm(s)}_{k_1/2,k_2/2}$ in \cite{Dorigoni:2021jfr, Dorigoni:2021ngn}
whose iterated-integral representations involve holomorphic cusp forms.

Turning our attention to the $\{b_k,\ep_k\}$ relations relevant to eMGFs, it is tempting
to wonder if there is
any analogue to the period polynomials of holomorphic cusp forms which
might determine the coefficients of the terms $[b_{k_1}^{(j_1)},b_{k_2}^{(j_2)}]$ and
$[b_{k_1}^{(j_1)},\ep_{k_2}^{(j_2)}]$ in e.g.\ (\ref{count.13}) to (\ref{count.17}).
Regardless of the answer to this intriguing question, we follow the strategy
for $\ep^{\TS}_{k}$ and organize the indecomposable relations among $b_k$
and $\ep_k$ presented along with this paper into \emph{highest-weight vectors}. This means
that relations with nested commutators of three and more generators $\ep_{k\geq 4}$ and
$b_k$ are always combined with ${\rm ad}_{\ep_0}$-actions of suitable linear
combinations of simpler relations (for instance the ones from integrability): The 
coefficients in these combinations are chosen such that they are annihilated 
by the appropriate number of ${\rm ad}_{\ep_0}$-powers.

In the example (\ref{count.13}) at lattice weight seven,
we refrain from adding multiples of the corollary
${\rm ad}_{\ep_0}( [b_3,\ep_4] + [b_2,b_5]+ [b_3,b_4] )=0$ of integrability
to the left-hand side of (\ref{count.13}) to keep it in the kernel of ${\rm ad}_{\ep_0}^2$.
Similarly, the relation (\ref{count.17}) at lattice weight eight would no longer be annihilated
by ${\rm ad}_{\ep_0}$ if we added the ${\rm ad}_{\ep_0}$-image of (\ref{count.16})
or $({\rm ad}_{\ep_0})^2$ acting on the integrability relations in the second
line of (\ref{count.3}).

A more streamlined definition of highest-weight vectors can be based on the additional $SL_2$ generator $\widecheck{\rm ad}_{{\epsilon}_{0}}$ from (\ref{adje0rel}), which was identified as a raising operator below said equation,
\beq
\widecheck{\rm ad}_{{\epsilon}_{0}}({\cal R}) = 0 \ \ \Rightarrow \ \ {\cal R} \ {\rm is}
\ {\rm said} \ {\rm to}\ {\rm be} \ {\rm a} \ \textrm{highest-weight} \ {\rm vector}
\label{hwdef}
\eeq
In this way we bypass any reference to the size of the multiplet generated by repeated action of the lowering operator ${\rm ad}_{{\epsilon}_{0}}$
on the highest-weight vector.

The integrability relations (\ref{count.2}) are automatically highest-weight vectors, since its constituents make no reference to $\ep_0$ and thus are individually annihilated by $\widecheck{\rm ad}_{{\epsilon}_{0}}$.
Moreover, it is easy to check that the action of (\ref{adje0rel}) cancels between
the depth-two terms in the highest-weight vectors (\ref{count.13}),
(\ref{count.16}) and (\ref{count.17}).

Highest-weight vectors have the maximum eigenvalue under ${\rm ad}_{[\check{\ep}_0,\ep_0]}$ 
within multiplets of $SL_2$, and therefore cannot be obtained solely via action of the 
lowering operator ${\rm ad}_{\ep_0}$ on
other relations. However, decomposable relations obtained from action of
${\rm ad}_{\ep_k}$ at $k\geq 4$ or ${\rm ad}_{b_k}$ such as
$[b_2,[b_2,b_4{+}\ep_4]]=0$ are still counted as highest-weight vectors if they satisfy~\eqref{hwdef}.
Furthermore, one may construct highest-weight vectors by linear combinations of derived relations which descend
from different sequences of ${\rm ad}_{b_k}$, ${\rm ad}_{\ep_k}$ such as e.g.
\beq\label{eq:specHWV}
\big[ b_3, [ b_2, \ep_4^{(1)} + b_4^{(1)}] \big]
-2\, \big[ b_3^{(1)}, [ b_2, \ep_4 + b_4] \big] = 0
\eeq

In the case of eMGFs, the counting of independent
representatives for given powers of the {(anti-)holomorphic} lattice momenta
is clearly invariant under complex conjugation, i.e.\ $A\leftrightarrow B$. This has an echo at the level of
the $SL_2$-organisation of relations among $\{ \ep_k , b_k \}$: The collection of relations we have are invariant under
an operator $\mathcal{I}$ which maps $\mathcal{I}(\ep^{(j)}_k,b^{(j)}_k) = \frac{j!}{(k-j-2)!}(\ep_k^{(k-j-2)},b_k^{(k-j-2)})$ and reverses their noncommutative products. This operator composes with the established raising and lowering operators via $\widecheck{\rm ad}_{\ep_0}= \mathcal{I}^{-1}{\rm ad}_{\ep_0}\mathcal{I}$, such that it effectively exchanges the role of raising - and lowering operator. In this sense, it maps the highest-weight vector to its corresponding lowest-weight vector, annihilated by ${\rm ad}_{\ep_0}$.

Thus, the organisation of relations in terms of $SL_2$-multiplets is a natural presentation: 
The highest-weight-vector form of indecomposable relations such as (\ref{count.13}), (\ref{count.16})
or (\ref{count.17}) simplifies the counting of relations which may be generated via repeated ${\rm ad}_{\epsilon_0}$-action and thus manifests the invariance of table~\ref{allrelcount} under the action of complex conjugation.

As another advantage of the highest-weight vector form of the $\{b_k,\ep_k\}$
relations, it might facilitate the quest for the $u,v$-dependent analogues of
period polynomials of holomorphic cusp forms.

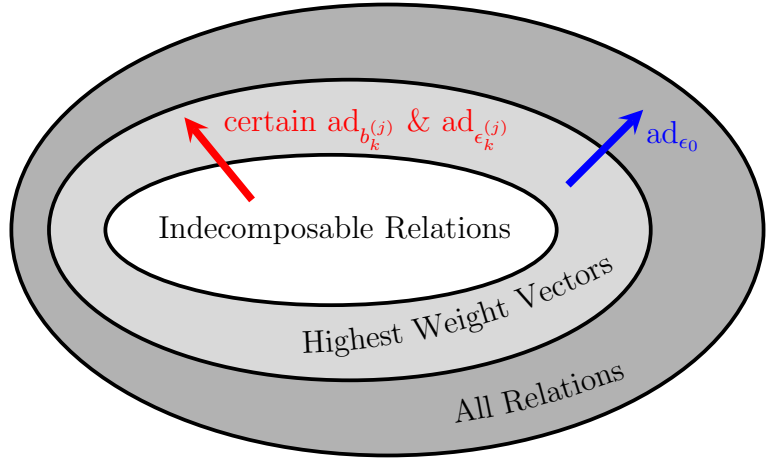
\begin{figure}[h!]
	\centering
\begin{tikzpicture}
    \draw[line width=0.5mm, fill=black!30]  (0,0) ellipse (5cm and 3cm);
    \draw[line width=0.5mm, fill=black!15] (-0.5,0) ellipse (4cm and 2cm);
    \draw[line width=0.5mm, fill=white] (-0.75,0) ellipse (3cm and 1cm);
    \draw[decoration={
            text along path,
            text={Highest Weight Vectors},
            text align={center}},decorate] (-3.1,-1.5) arc (260:310:9.5);
   \draw[decoration={
            text along path,
            text={All Relations},
            text align={center}},decorate] (-0.8,-2.7) arc (270:300:11);
    \draw (-0.7,0) node {Indecomposable Relations};
    \draw[-stealth,line width=1mm, red] (-1.8,0.4) -- (-2.7,1.5) node [pos=0.85,right=2mm] {certain 
 ${\rm ad}_{b^{(j)}_{k}}$ \& ${\rm ad}_{\epsilon^{(j)}_{k}}$};
   \draw[-stealth,line width=1mm, blue] (2.4,0.6) -- (3.4,1.6) node [pos=0.65,right=2mm] {${\rm ad}_{\epsilon_{0}}$};
\end{tikzpicture}
\caption{A schematic representation of the different sets of relations among the $\{\epsilon_k,b_k\}$ identified in this section. The indecomposable relations of table~\ref{indecrelcount} are those which cannot be generated from (linear combinations of) other relations using adjoint action by any element of the algebra. The highest-weight vectors are relations which cannot be generated from any other relation by adjoint action of $\epsilon_0$. The arrows in the diagram represent the operations by which the basic components of the respective larger sets may be generated, finally resulting in the total set of relations detailed in table~\ref{allrelcount}.}
	\label{relationSets}
\end{figure}

The classification of relations in the extension of Tsunogai's algebra is visualized 
in figure~\ref{relationSets}. In the ancillary files of this paper we present all 
indecomposable $\{b_k,\ep_k\}$-relations of lattice weight $\leq 11$ in the 
highest-weight-vector form.

%%%%%%%%%%%%%%%%%%%%%%%%%%%%%%%%%%%%%%%%%%%%%%%%%%%%%%%%%%%
\subsection{Deriving the counting of independent eMGFs}
\label{sec:5.8}
%%%%%%%%%%%%%%%%%%%%%%%%%%%%%%%%%%%%%%%%%%%%%%%%%%%%%%%%%%%

The purpose of this section is to deduce the counting of
one-variable eMGFs in table \ref{allshuffir} from the
results of the previous subsection.

%%%%%%%%%%%%%%%%%%%%%%%%%%%%%%%%%%%%%%%%%%%%%%%%%%%%%%%%%%%
\subsubsection{Counting $\beta^{\rm sv}$ with admissible $j_i,k_i$}
\label{sec:5.8.1}
%%%%%%%%%%%%%%%%%%%%%%%%%%%%%%%%%%%%%%%%%%%%%%%%%%%%%%%%%%%

The counting of indecomposable eMGFs hinges on the fact that 
all convergent one-variable eMGFs not amenable to holomorphic subgraph
reduction occur in the Laurent expansion of the $n$-point
generating series (\ref{high.4}) in $s_{ij},\eta_j,\bar \eta_j$. By the 
perturbative solution (\ref{gen.34c}) of the differential equations of
these generating series, eMGFs are expressible in terms of the real-analytic iterated
KE integrals $\beta^{\rm sv}$ discussed in sections \ref{sec:3.1.2} and \ref{sec:new.1.4}.
On the one hand, the coefficients of $\beta^{\rm sv}$ will involve MZVs and functions of $u,v$ due to
the initial value in (\ref{gen.34c}), see for instance the preview examples in section \ref{sec:3.5}.
On the other hand, the entries of the $\bsvtau{j_1 &j_2 &\ldots &j_\ell \\ k_1 &k_2 &\ldots &k_\ell \\ z_1&z_2 &\ldots &z_\ell}$ contributing to an eMGF $\cplus{A \\ B \\ Z} $
at given $|A|,|B|$ without any MZVs in their coefficients are 
constrained by (\ref{count.1}). Hence, a first step to count the indecomposable eMGFs at 
fixed $|A|,|B|$ is to enumerate all $\beta^{\rm sv}$ whose entries obey
\beq
|A|= \ell + \sum_{i=1}^\ell j_i
\, , \ \ \ \
|B|= {-}\ell+ \sum_{i=1}^\ell (k_i{-} j_i)
\label{againAB}
\eeq
In particular, the lattice weight is equal to $|A| {+} |B| = \sum_{i}k_i$,
and modular invariant eMGFs with $|A| = |B|$ have $j_i = \frac{1}{2}(k_i{-}2)$ on
average. In counting the solutions to (\ref{againAB}) at given $(|A|,|B|)$, one has to keep in mind
that columns $\bsvtau{\ldots &j &\ldots  \\ \ldots &k &\ldots  \\ \ldots &z &\ldots }$
admit $2\leq k \in \mathbb N$ and that columns
$\bsvtau{\ldots &j &\ldots  \\ \ldots &k &\ldots  \\ \ldots & &\ldots }$ have
even $k\geq 4$ (as usual with $0\leq j\leq k{-}2$ in both cases).

For the simplest non-trivial pairs of $(|A| , |B|)$ with $|A|,|B|\geq 1$, this leads 
to the counting in table \ref{exsmallAB}:

\begin{table}[h!]
\begin{center}

\begin{tabular}{c|c||c| c | c  }
$(|A|,|B|)$ &depth 1
&$(|A|,|B|)$ &depth 1&depth 2   \\\hline\hline
(1,1) &$\bigg. \bsvtau{0 \\ 2 \\ z } \bigg.$ &(1,3) &$\bigg. \bsvtau{0 \\ 4 \\ z },\bsvtau{0 \\ 4 \\ \emptyslot } \bigg.$
  \\\hline
(1,2)&$\bigg.\bsvtau{0 \\ 3 \\ z }\bigg.$ &(2,2) &$\bigg.\bsvtau{1 \\ 4 \\ z },\bsvtau{1 \\ 4 \\  \emptyslot }\bigg.$ &$\bigg.\bsvtau{0 &0 \\ 2& 2 \\ z&z }\bigg.$
\\\hline
(2,1) &$\bigg.\bsvtau{1 \\ 3 \\ z }\bigg.$ &(3,1) &$\bigg.\bsvtau{2 \\ 4 \\ z },\bsvtau{2 \\ 4 \\ \emptyslot }\bigg.$
\end{tabular}

\bigskip

\begin{tabular}{c|c|| c| c | c}
$(|A|,|B|)$  &depth 1
&$(|A|,|B|)$ &depth 1&depth 2   \\\hline\hline
(1,4) &$\bigg.\bsvtau{0 \\ 5 \\ z }\bigg.$
&(2,3) &$\bigg.\bsvtau{1 \\ 5 \\ z }\bigg.$ &$\bigg.\bsvtau{0 &0 \\ 2&3 \\ z &z }, \bsvtau{0 &0 \\ 3&2 \\ z &z }\bigg.$ \\\hline
(4,1) &$\bigg.\bsvtau{3 \\ 5 \\ z }\bigg.$
&(3,2) &$\bigg.\bsvtau{2 \\ 5 \\ z }\bigg.$  &$\bigg.\bsvtau{0 &1 \\ 2&3 \\ z &z }, \bsvtau{1 &0 \\ 3&2 \\ z &z }\bigg.$
\end{tabular}
\end{center}
\caption{\textit{Real-analytic KE integrals that govern the counting of eMGFs in one variable
at small lattice weights $|A|{+}|B|\leq 5$.}}
\label{exsmallAB}
\end{table}

At higher lattice weight, we encounter $\beta^{\rm sv}$ of depth up to and
including $\lfloor\frac{1}{2}|A| + \frac{1}{2} |B| \rfloor$. For even lattice
weight, this maximum depth is saturated by a power of the Green function
$\bsvtau{0 &0 &\ldots &0 \\ 2 &2 &\ldots &2 \\ z &z &\ldots &z }
\sim g(z|\tau)^{\frac{1}{2}|A| + \frac{1}{2} |B|}$. For odd lattice weights in
turn, there is a variety of maximum-depth terms such as
$\bsvtau{0 &0 &\ldots &0 &1 \\ 2 &2 &\ldots &2 &3 \\ z &z &\ldots &z &z } $
and permutations of the columns.

%%%%%%%%%%%%%%%%%%%%%%%%%%%%%%%%%%%%%%%%%%%%%%%%%%%%%%%%%%%
\subsubsection{Eliminating shuffles}
\label{sec:5.8.2}
%%%%%%%%%%%%%%%%%%%%%%%%%%%%%%%%%%%%%%%%%%%%%%%%%%%%%%%%%%%

The next step in the quest for indecomposable eMGFs at fixed $|A|,|B|$ is to eliminate all shuffles
of $\beta^{\rm sv}$ encountered at smaller lattice weights from the lists in the
previous section. The depth-two example $\bsvtau{0 &0 \\ 2&2 \\ z &z }$ in 
table \ref{exsmallAB} at $|A|=|B|=2$ is just $\frac{1}{2}g(z|\tau)^2$.
Similarly, the sums of the depth-two entries $\bsvtau{0 &j \\ 2&3 \\ z &z }+\bsvtau{j &0 \\ 3&2 \\ z &z }
=\bsvtau{0 \\ 2 \\ z }\bsvtau{j  \\ 3 \\ z }$ at $|A|+|B|=5$ occur in the products
$g(z|\tau)\dplus{1 \\ 2}\!(z|\tau)$ and $g(z|\tau)\dplus{2 \\ 1}\!(z|\tau)$, see (\ref{simpins}).

It is straightforward to generate a list of all shuffle-products of simpler $\beta^{\rm sv}$,
but one has to pick a convention in enumerating shuffle-independent representatives:
One can for instance take either $\bsvtau{0 &j \\ 2&3 \\ z &z }$ or
$\bsvtau{j &0 \\ 3&2 \\ z &z }$ as independent under the shuffle
$\bsvtau{j  \\ 3 \\ z}\bsvtau{ 0 \\  2 \\ z}$ at depth two
and $k_1{+}k_2=|A|{+}|B|=5$. 

At depth $\ell$, there are $(\ell{-}1)!$ shuffle-independent permutations of the columns
$\begin{smallmatrix} \ldots &j_i &\ldots \\ \ldots &k_i&\ldots \\ \ldots &z_i &\ldots\end{smallmatrix}$ 
as long as they are pairwise distinct. For identical columns, the number of 
shuffle-independent permutations is reduced -- there are for instance two 
independent shuffle relations
\begin{align}
\bsvtau{0&1  \\ 2&3 \\ z&z}\bsvtau{ 0 \\  2 \\ z}
&=2 \bsvtau{0&0 &1 \\ 2&2&3 \\ z&z&z}+\bsvtau{0&1 &0 \\ 2&3&2 \\ z&z&z}
\notag \\
\bsvtau{1&0  \\ 3&2 \\ z&z}\bsvtau{ 0 \\  2 \\ z}
&= \bsvtau{0&1 &0 \\ 2&3&2 \\ z&z&z}+2\bsvtau{1&0 &0 \\ 3&2&2 \\ z&z&z}
\end{align}
among $ \bsvtau{0&0 &1 \\ 2&2&3 \\ z&z&z}, \bsvtau{0&1 &0 \\ 2&3&2 \\ z&z&z}$ and
$\bsvtau{1&0 &0 \\ 3&2&2 \\ z&z&z}$. In general, one can form a Lyndon-word
decomposition of the shuffle algebra after assigning an ordering convention for the columns
$\begin{smallmatrix} j\\k\\z \end{smallmatrix}$ and $\begin{smallmatrix} j\\k\\ \emptyslot\end{smallmatrix}$, see e.g.\ \cite{RADFORD1979432}.\footnote{One can for instance consider
an empty slot in the third line as ``smaller'' than $z$ and take the natural integer ordering for
$j$ and $k$. The triplet $\begin{smallmatrix} j_1\\k_1\\z_1 \end{smallmatrix}$
then counts as smaller than $\begin{smallmatrix} j_2\\k_2\\z_2 \end{smallmatrix}$
if either $j_1{<}j_2$ or $j_1{=}j_2, \, k_1{<}k_2$ or $(j_1,k_1){=}(j_2,k_2)$ with $z_1$
``smaller'' than $z_2$. This is the standard ordering of the cartesian product 
$\mathbb{Z} \times \mathbb{Z} \times \{0,z\}$.} 
The numbers of shuffle-independent $\beta^{\rm sv}$ of various depths $\ell$
relevant to $\cplus{A \\ B \\ Z} $ at fixed $|A|,|B| \leq 10$ in (\ref{againAB})
are gathered in table \ref{allshuffbsv}. The two shuffle-independent representatives
at $(|A|,|B|)=(2,3)$ can for instance be taken as $\bsvtau{1 \\ 5 \\ z }$
and $\bsvtau{0 &0 \\ 2&3 \\ z &z }$.

\begin{table}[h]
\begin{center}
\begin{tabular}{c||c|c|c|c|c|c|c|c|c}
\diagbox[]{$|A| \! \!$}{$\! \! |B|$} &1&2&3
&4&5&6 &7&8&9  \\\hline\hline
1 &1 &1 &2 &1 &2
&1 &2 &1 &2  \\\hline
2 &1 &2 &2 &4 &4 &6 &6 &9 &  \\\hline
3 &2 &2 &5 &7 &13 &15 &25 & &  \\\hline
4 &1 &4 &7 &15 &23 &42 & & &  \\\hline
5 &2 &4 &13 &23 &51 & & && \\\hline
6 &1 &6 &15 &42 & & &&& \\\hline
7 &2 &6 &25 & & &&& \\\hline
8 &1 &9 & & &&&& \\\hline
9 &2 & & &&&&&
\end{tabular}
\end{center}
\caption{\textit{The numbers of shuffle-independent $\beta^{\rm sv}$ relevant to
the counting of indecomposable ${\cal C}^+[\ldots]$ at given $|A|,|B|$.}}
\label{allshuffbsv}
\end{table}

%%%%%%%%%%%%%%%%%%%%%%%%%%%%%%%%%%%%%%%%%%%%%%%%%%%%%%%%%%%
\subsubsection{Modding out by the relations among $b_k,\epsilon_k$}
\label{sec:5.8.9}
%%%%%%%%%%%%%%%%%%%%%%%%%%%%%%%%%%%%%%%%%%%%%%%%%%%%%%%%%%%

The shuffle-independent $\beta^{\rm sv}$ in table \ref{allshuffbsv}
usually overcount the indecomposable eMGFs in one variable at fixed $|A|,|B|$: We still have
to take the relations among the $b_k,\epsilon_k$ in section~\ref{sec:5.2} into account
that lead to dropouts among the $\beta^{\rm sv}$ in the path-ordered exponentials
(\ref{gen.34c}) or (\ref{count.6}) that govern generating series of eMGFs.

One may wonder if these dropouts concern shuffles of $\beta^{\rm sv}$ or shuffle-independent
ones. This can be easily answered since all the known relations among the $b_k,\epsilon_k$
universal to $n$-point matrix representations are Lie-algebra valued, i.e.\ only involve nested commutators
of $b^{(j)}_k,\epsilon^{(j)}_k$. The same kinds of nested commutators arise when eliminating
all shuffles from generating series like (\ref{gen.34c}): Once the $\beta^{\rm sv}$ are modded out by
shuffles (e.g.\ setting $\bsvtau{0 &0 \\ 3&2 \\ z &z } = - \bsvtau{0 &0 \\ 2&3 \\ z &z } \ {\rm mod} \ \shuffle$), the accompanying operators $b^{(j)}_k,\epsilon^{(j)}_k$ will conspire to nested commutators by standard results in combinatorics \cite{Reutenauer}.

Hence, every relation among $b_k,\epsilon_k$ in table \ref{allrelcount} will lead to the
dropout of one shuffle-independent $\beta^{\rm sv}$ from the generating
series $Y(z|\tau)$ and obstruct one indecomposable eMGF. On these grounds,
the counting of indecomposable eMGFs in table \ref{allshuffir} is obtained from the
differences of the entries in table \ref{allshuffbsv} (for the shuffle-independent 
$\beta^{\rm sv}$) and those of table \ref{allrelcount} (for the relations among 
$b_k,\epsilon_k$ including derived ones). This counting includes both MGFs 
and one-variable eMGFs but excludes eMGFs in multiple variables
$z_1,z_2,\ldots,z_m$ at $m\geq 2$ and MZVs. We reiterate that the entries $j_i,k_i$
of the $\beta^{\rm sv}$ and those of the operators $\epsilon^{(j_i)}_{k_i}, b^{(j_i)}_{k_i}$
%$t^{(j_i)}_{k_i} \in \{ \epsilon^{(j_i)}_{k_i}, b^{(j_i)}_{k_i}\}$
are related to the exponents $|A|= \sum_i a_i$ and $|B|= \sum_i b_i$ of $\cplus{A \\ B \\ Z}$ via
\begin{align}
(|A|,|B|) \ &\longleftrightarrow \
\bsvtau{j_1 &j_2 &\ldots &j_\ell \\ k_1 &k_2 &\ldots &k_\ell \\ z_1&z_2 &\ldots &z_\ell} \ {\rm at} \
\left\{ \begin{array}{l}
 |A| =\sum_{i=1}^\ell (j_i{+}1)\\
 |B| =\sum_{i=1}^\ell (k_i{-}j_i{-}1)
\end{array} \right.
\notag \\
 (|A|,|B|) \ &\longleftrightarrow \
%\bsvtau{j_1 &j_2 &\ldots &j_\ell \\ k_1 &k_2 &\ldots &k_\ell \\ z_1&z_2 &\ldots &z_\ell}
%t^{(j_1)}_{k_1} t^{(j_2)}_{k_2} \ldots t^{(j_\ell)}_{k_\ell}
\ell \ {\rm letters} \ \epsilon^{(j_i)}_{k_i} \ {\rm or} \  b^{(j_i)}_{k_i}
 \ {\rm at} \
\left\{ \begin{array}{l}
 |A| = \sum_{i=1}^\ell (k_i{-}j_i{-}1)\\
 |B| = \sum_{i=1}^\ell (j_i{+}1)
\end{array} \right.
\label{dictio.2}
\end{align}
%

%%%%%%%%%%%%%%%%%%%%%%%%%%%%%%%%%%%%%%%%%%%%%%%%%%%%%%%%%%%
\subsubsection{Example eMGFs at $|A|=|B|=3$}
\label{sec:5.8.3}
%%%%%%%%%%%%%%%%%%%%%%%%%%%%%%%%%%%%%%%%%%%%%%%%%%%%%%%%%%%

As an example of the general counting strategy in the previous sections, we shall
now deduce the existence of four indecomposable one-variable eMGFs
$\cplus{A \\ B \\ Z} $ at $|A|=|B|=3$, i.e.\ at lattice weight $\sum_i k_i=6$
and modular weights $(0,0)$.
\begin{itemize}
\item According to section \ref{sec:5.8.1} and in particular (\ref{againAB}),
we enumerate all $\bsvtau{j_1 &j_2 &\ldots &j_\ell \\ k_1 &k_2 &\ldots &k_\ell
\\ z_1&z_2 &\ldots &z_\ell}$ with $\ell + \sum_{i=1}^\ell j_i = 3$ and
$\sum_{i=1}^\ell (k_i{-}j_i) = 3+\ell$. This has a total of nine solutions,
\begin{itemize}
\item[$\ast$] 2 solutions $\bsvtau{2 \\ 6 \\ z}$ and $\bsvtau{2 \\ 6 \\  \emptyslot}$ at depth $\ell=1$
\item[$\ast$] the following 6 solutions at depth $\ell= 2$
\[
\bsvtau{0 &1 \\ 2 &4 \\ z &z } , \ \bsvtau{1&0 \\ 4 &2 \\ z &z } , \
\bsvtau{0 &1 \\ 2 &4  \\ z &\emptyslot }, \ \bsvtau{1 &0 \\ 4 &2  \\ \emptyslot &z }, \
\bsvtau{1 &0 \\ 3 &3 \\ z &z } , \ \bsvtau{0 &1 \\ 3 &3 \\ z &z }
\]
\item[$\ast$] 1 solution $\bsvtau{0&0&0 \\ 2&2&2 \\ z&z&z}$ at depth $\ell=3$.
\end{itemize}
\item Next, we follow section \ref{sec:5.8.2} to eliminate the
four shuffles
\begin{align}
\bsvtau{0 \\ 2 \\ z} \bsvtau{1 \\ 4 \\ \emptyslot} &=  \bsvtau{0 &1 \\ 2 &4 \\ z &\emptyslot } + \bsvtau{1&0 \\ 4 &2 \\ \emptyslot &z }
\notag \\
\bsvtau{0 \\ 2 \\ z} \bsvtau{1 \\ 4 \\ z} &=  \bsvtau{0 &1 \\ 2 &4 \\ z &z } + \bsvtau{1&0 \\ 4 &2 \\ z &z }
\notag \\
\bsvtau{1   \\ 3  \\ z  }  \bsvtau{0  \\ 3  \\ z }
&= \bsvtau{1 &0 \\ 3 &3 \\ z &z } + \bsvtau{0 &1 \\ 3 &3 \\ z &z }
\label{comrel.23}\\
\bsvtau{0 \\ 2 \\ z}^3 &= 6\bsvtau{0&0&0 \\ 2&2&2 \\ z&z&z}
\notag
\end{align}
One possible choice of five shuffle-independent representatives is furnished by
\beq
\bsvtau{2 \\ 6 \\ z}, \ \bsvtau{2 \\ 6 \\ \emptyslot}, \
\bsvtau{0 &1 \\ 2 &4 \\ z &z } , \ \bsvtau{0 &1 \\ 2 &4  \\ z &\emptyslot }, \ \bsvtau{1 &0 \\ 3 &3 \\ z &z }
\label{comrel.22}
\eeq
\item Finally, by section \ref{sec:5.8.9}, it remains to mod out by the relation
$[b_2,b_4^{(1)}{+}\ep_4^{(1)}]=0$ which eliminates one linear
combination of $\bsvtau{0 &1 \\ 2 &4 \\ z &z }$ and $\bsvtau{0 &1 \\ 2 &4  \\ z &\emptyslot }$
from the generating series of eMGFs. More specifically, we can rewrite
the sector $\sim c_{0,2}c_{2,4}$ in the path-ordered exponential $Y(z|\tau)$ in (\ref{gen.34c}) via
\begin{align}
Y(z|\tau) \, \big|_{c_{0,2}c_{2,4}}  &= \bsvtau{0 &1 \\ 2 &4 \\ z &\emptyslot } R_{\vec{\eta}}(\ep^{(1)}_4 b_2)
+\bsvtau{0 &1 \\ 2 &4 \\ z &z } R_{\vec{\eta}}(b^{(1)}_4 b_2) \notag \\
&\ \
+ \bsvtau{1 &0 \\ 4&2 \\ \emptyslot &z} R_{\vec{\eta}}(b_2 \ep^{(1)}_4 )
+\bsvtau{1 &0 \\ 4 &2 \\ z &z } R_{\vec{\eta}}(b_2 b^{(1)}_4) \notag
\\
%%%%%%%%
&=R_{\vec{\eta}}(\ep^{(1)}_4 b_2) \bigg( \bsvtau{0 &1 \\ 2 &4 \\ z &\emptyslot }  +  \bsvtau{1 &0 \\ 4&2 \\ \emptyslot &z}  \bigg) \notag\\
&\quad + R_{\vec{\eta}}(b^{(1)}_4 b_2) \bigg( \bsvtau{0 &1 \\ 2 &4 \\ z &z }  +  \bsvtau{1 &0 \\ 4&2 \\ \emptyslot &z}  \bigg)
\notag \\
&\quad  + R_{\vec{\eta}}(  b_2 b^{(1)}_4 ) \bigg( \bsvtau{1 &0 \\ 4 &2 \\ z &z } -  \bsvtau{1 &0 \\ 4&2 \\ \emptyslot &z}  \bigg)
\label{comrel.21} \\
%
%%%%%%
& =R_{\vec{\eta}}(\ep^{(1)}_4 b_2) \bigg( \bsvtau{0  \\ 2  \\ z  }
\bsvtau{1 \\ 4 \\ \emptyslot } \bigg)
\notag
\\
&\quad  + R_{\vec{\eta}}(b^{(1)}_4 b_2) \bigg(  \bsvtau{1  \\ 4 \\ \emptyslot }  \bsvtau{0 \\ 2 \\ z}
 + \bsvtau{0 &1 \\ 2 &4 \\ z &z }  - \bsvtau{0 &1 \\ 2 &4 \\ z &\emptyslot }
  \bigg)
\notag  \\
&\quad+ R_{\vec{\eta}}(  b_2 b^{(1)}_4 ) \bigg(
\bsvtau{1  \\ 4 \\ z }  \bsvtau{0 \\ 2 \\ z} - \bsvtau{1  \\ 4 \\ \emptyslot }  \bsvtau{0 \\ 2 \\ z} \notag \\
&\quad \quad \quad \quad \quad
- \bsvtau{0 &1 \\ 2 &4 \\ z &z }  + \bsvtau{0 &1 \\ 2 &4 \\ z &\emptyslot }    \bigg)
\notag
\end{align}
where we used $R_{\vec{\eta}}(b_2 \ep^{(1)}_4) = R_{\vec{\eta}}(\ep^{(1)}_4 b_2 - b_2 b^{(1)}_4 
+ b^{(1)}_4 b_2)$ in passing to the third line. In the last step, we have eliminated
$ \bsvtau{1 &0 \\ 4 &2 \\  \ast &z } =
\bsvtau{1 \\ 4 \\ \ast  } \bsvtau{0 \\ 2 \\ z  } - \bsvtau{0 &1\\ 2 &4 \\ z & \ast   }$
with either $z$ or an empty slot in the place of $\ast$.
Hence, only the difference $ \bsvtau{0 &1 \\ 2 &4 \\ z &\emptyslot }
- \bsvtau{0 &1 \\ 2 &4 \\ z &z } $ is realized within eMGFs but not
$ \bsvtau{0 &1 \\ 2 &4 \\ z &\emptyslot } $ or $\bsvtau{0 &1 \\ 2 &4 \\ z &z } $
individually. Together with the remaining three shuffle-independent
$\bsvtau{2 \\ 6 \\ z},  \bsvtau{2 \\ 6 \\ \emptyslot}$ and $\bsvtau{1 &0 \\ 3 &3 \\ z &z }$
in (\ref{comrel.22}), we arrive at a total of four real-analytic iterated KE integrals
that can contribute to indecomposable one-variable eMGFs at $|A|=|B|=3$.
\end{itemize}
By (\ref{invbsv.7}) and (\ref{nwbsv.11}), the indecomposable eMGFs at depth one
are given by
\begin{align}
g_3(z|\tau) = -  30 \bsvtau{2 \\ 6 \\ z} \, , \ \ \ \ \ \
E_3(\tau) = -  30 \bsvtau{2 \\ 6 \\ \emptyslot} + \frac{3 \zeta_5}{4y^2}
\label{comrel.19}
\end{align}
where the contribution $ \frac{3 \zeta_5}{4y^2}$ to the MGF $E_3$ exemplifies
that we have consistently dropped admixtures of MZVs in the counting of $\beta^{\rm sv}$. 
The depth-two integral $\bsvtau{1 &0 \\ 3 &3 \\ z &z }$ signals the
eMGF $C_{1|1,1}(z|\tau)$ previewed with all its lower-depth contributions in (\ref{prevex.1}).
As will be detailed in section \ref{bassec.2.2}, the last indecomposable eMGF at $|A|=|B|=3$
in (\ref{comrel.21}) can be realized via
\begin{align}
\cplus{2 &0 &1 \\ 0 &2 &1 \\ z &0 &0} - \cplus{0 &2 &1 \\ 2 &0 &1 \\ z &0 &0} &=
12   \bsv{0& 1\\2& 4\\z& \emptyslot}   - 12 \bsv{0& 1\\2& 4\\z& z}
+2 \zeta_3 B_2(u)   \label{oddA.01}\\
&\quad
+ 6 \bsv{0 \\2 \\z } \bsv{ 1\\ 4\\ z}
- 6 \bsv{0 \\2 \\z } \bsv{ 1\\ 4\\ \emptyslot}
+\frac{  \zeta_3 }{y}  \bsv{0\\2\\z}
\notag
\end{align}
where the shuffles in the second line conspire to $g(z|\tau) \big( g_2(z|\tau) - E_2(\tau) \big)$.
A possible choice of basis for indecomposable eMGFs at more general $|A|{+}|B|\leq 10$ will
be discussed in section \ref{bassec}. Since the bases at these lattice weights can
be spanned via dihedral eMGFs, we conclude that any one-variable eMGF of trihedral or more
general topology at $|A|{+}|B|\leq 10$ can be rewritten in terms of dihedral eMGFs.

\newpage

%%%%%%%%%%%%%%%%%%%%%%%%%%%%%%%%%%%%%%%%%%%%%%%%%%%%%%%%%%%
%%%%%%%%%%%%%%%%%%%%%%%%%%%%%%%%%%%%%%%%%%%%%%%%%%%%%%%%%%%
\section{Converting eMGFs to iterated integrals}
\label{sec:9}
%%%%%%%%%%%%%%%%%%%%%%%%%%%%%%%%%%%%%%%%%%%%%%%%%%%%%%%%%%%
%%%%%%%%%%%%%%%%%%%%%%%%%%%%%%%%%%%%%%%%%%%%%%%%%%%%%%%%%%%

In this section, we discuss different strategies to convert one-variable eMGFs ${\cal C}^+[\ldots]$
into their canonical representations in terms of real-analytic iterated KE integrals $\beta^{\rm sv}$.
The first approach is to apply the expansion (\ref{gen.34c}) of the generating series of
eMGFs and to extract rich information from a simple sector of the
initial value $\widehat Y(i\infty)$ to be spelt out below. The second approach is to work with the
lattice-sum representation of individual eMGFs and to combine the differential equations
in $z,\tau$ with degenerations at $z\rightarrow0, \, \tau \rightarrow i\infty$, extending
the {\it sieve algorithm} known from MGFs \cite{DHoker:2016mwo, DHoker:2016quv}.
We will illustrate the advantages and drawbacks of both approaches and explain how
their interplay determines the ($\bar \tau,u,v$-dependent) integration constants
$\overline{ \alpha[\ldots;\tau]}$ in the expression (\ref{eq:ebsv2}) for 
$\beta^{\rm sv}$ of depth $\geq 2$.

%%%%%%%%%%%%%%%%%%%%%%%%%%%%%%%%%%%%%%%%%%%%%%%%%%%%%%%%%%%
\subsection{Partial knowledge of the initial value $\widehat Y(i\infty)$}
\label{sec:9.1}
%%%%%%%%%%%%%%%%%%%%%%%%%%%%%%%%%%%%%%%%%%%%%%%%%%%%%%%%%%%

The expansion (\ref{gen.34c}) of the generating series requires the knowledge
of the initial value $\widehat Y(i\infty)$ at $\tau \rightarrow i\infty$. In principle,
one can compute $\widehat Y(i\infty)$ from the Laurent polynomials of the eMGFs in the generating
series, i.e.\ their expansion around the cusp modulo terms ${\cal O}(q^u,\bar q^u)$
as in sections \ref{sec:2.4} and \ref{sec:3.3.3}: Similar to the initial value for generating
series of MGFs in section 3.4 of \cite{Gerken:2020yii}, the exponential of
$\frac{ R_{\vec{\eta}}(\epsilon_0) }{4y} $ in (\ref{gen.31}) eliminates all positive
powers of $y=\pi \Im \tau$ in the $q^0\bar q^0$ terms. In other words,
\beq
\widehat Y(\begin{smallmatrix} M \\ N \end{smallmatrix} |  \begin{smallmatrix} K \\ L \end{smallmatrix}  | \tau)\, \big|_{ q^0\bar q^0} = \sum_{P,Q}  \exp\bigg( \frac{ R_{\vec{\eta}}(\epsilon_0) }{4y} \bigg)_{\begin{smallmatrix} K \\ L \end{smallmatrix} \big|  \begin{smallmatrix} P \\ Q \end{smallmatrix} } Y(\begin{smallmatrix} M \\ N \end{smallmatrix} |  \begin{smallmatrix} P \\ Q \end{smallmatrix}  | \tau) \, \big|_{ q^0\bar q^0}
\label{inival.01}
\eeq
has a unique Taylor expansion in $y^{-1}$, and we define the initial value
$\widehat Y(i\infty)$ in (\ref{gen.34a}) and (\ref{gen.34c}) as the zeroth coefficient
$\widehat Y_{m=0}$ of
\beq
\widehat Y(\tau) \, \big|_{ q^0\bar q^0}  = \sum_{m=0}^{\infty}
\widehat Y_m y^{-m}
\, , \ \ \ \
\widehat Y(i\infty) = \widehat Y_0
\label{inival.02}
\eeq
We reiterate that the $\tau \rightarrow i\infty$ limit
(\ref{inival.01}) is taken at fixed $u,v$ in $z=u\tau{+}v$ such that $\widehat Y(i\infty)$
is a shorthand for $\widehat Y(u,v|i\infty)$. In fact, we find the expansion of the
initial value (\ref{inival.02}) to only depend on $u$ which renders all the ingredients
of (\ref{gen.34c}) invariant under the modular $T$ transformation.

As a major downside of computing the initial value via (\ref{inival.01}) and (\ref{inival.02}), it
already requires the Laurent polynomials of the eMGFs in the expansion of $Y(\tau)\,  |_{ q^0\bar q^0} $ w.r.t.\ $s_{ij},\eta_j,
\bar \eta_j$. For MGFs, this issue is mitigated by the closed formula for the two-point
version of their generating series at $ q^0\bar q^0$ in (4.2) of \cite{Gerken:2020yii}.
The analogous formula for the generating series $Y(\tau) \, |_{q^0\bar q^0}$ of eMGFs
is under investigation but currently unknown, even for the two-point instance
in (\ref{eq:Y2pt}). However, the subsector $\widehat Y(i\infty)$ without any MZVs
is easily accessible at all multiplicities as will be explained below.

The present difficulties to determine the MZVs in the expansion of $\widehat Y(i\infty)$
does not undermine the following key virtue of (\ref{gen.34c}): Irrespective of the detailed
form of the initial value, this structural result on the expansion sharply bounds the
$\tau$-dependent building blocks that can contribute to eMGFs. In the first place, it is the
counting of $\beta^{\rm sv}$ which determined the basis dimensions of indecomposable
one-variable eMGFs $\cplus{A \\ B \\ Z}$ at fixed $|A|,|B|$ in section \ref{sec:5}. Moreover,
each shuffle-independent $\beta^{\rm sv}$ at depth $\ell\geq 2$ (see e.g.\ (\ref{eq:ebsv2}))
contains an integration constant $\overline{ \alpha[\ldots;\tau]}$ depending on
$\bar \tau, u,v$ that we will infer from the reality properties of eMGFs.

%%%%%%%%%%%%%%%%%%%%%%%%%%%%%%%%%%%%%%%%%%%%%%%%%%%%%%%%%%%
\subsubsection{The non-MZV sector at two points}
\label{sec:9.1.1}
%%%%%%%%%%%%%%%%%%%%%%%%%%%%%%%%%%%%%%%%%%%%%%%%%%%%%%%%%%%

The expansion of the initial value (\ref{inival.02}) in $s_{ij}, \eta_j,\bar \eta_j$ again
yields a series in (conjecturally single-valued) MZVs. The subsector without
any MZVs turns out to be particularly simple and accessible to all multiplicity -- we
shall informally refer to it as the {\it non-MZV sector}. Among other things, our
results on the non-MZV sector at $n$ points underpin the correlation (\ref{count.1})
between lattice weights of ${\cal C}^+$ and entries of $\beta^{\rm sv}$.
We shall now start by explaining the reasoning to determine the non-MZV
sector of $\widehat Y(i\infty)$ at two points.

The two-point component integrals resulting from the $\eta,\bar \eta$-expansion of
(\ref{eq:Y2pt}) are characterized by integrands $f^{(k)}(z_{12}) \overline{ f^{(\ell)}(z_{12})}$
in the entry $Y_{11}$ and modified arguments $z_{12}\rightarrow z_{02}$ of
$f^{(k)}$ or $\overline{ f^{(\ell)} }$ in the remaining entries $Y_{ij}$ with $i,j=1,2$. 
As detailed in section 4.1.1 of \cite{Dhoker:2020gdz}, the expansion
of component integrals in $s_{02},s_{12}$ can be easily expressed in terms of eMGFs
in their lattice-sum representations ${\cal C}^+$ to all orders. In case of $(k,\ell) = (1,1)$,
one additionally
encounters kinematic poles $\frac{1}{s_{12}}$ and $\frac{1}{s_{02}}$ due to the poles
$|z_{j2}|^{-2}$ in the integrands $f^{(1)}(z_{12}) \overline{ f^{(1)}(z_{12})}$ and $f^{(1)}(z_{02}) \overline{ f^{(1)}(z_{02})}$, respectively, see appendix C of \cite{Dhoker:2020gdz} for the systematic
computation of their residues.

It is easy to pinpoint the orders of the two-point expansion of $Y_{ij}(\tau)$
in $s_{02},s_{12},\eta,\bar \eta$, where the lattice sums trivialize, i.e.\ where the
expansion coefficients are independent on $\tau$ and $z$: First, the
component integrals at $(k,\ell)=(0,0)$ with constant integrand $f^{(0)}  \overline{ f^{(0)}} = 1$
have an expansion of the form $1+{\cal O}(s_{ij}^2)$, see (4.16) of \cite{Dhoker:2020gdz},
leading to a trivial lattice sum
in the limit $s_{ij} \rightarrow0$ of all the $2\times 2$ matrix entries $Y_{ij}(\tau)$.
Second, the residues of the kinematic poles $f^{(1)}(z_{j2}) \overline{ f^{(1)}(z_{j2})} \leftrightarrow \frac{1}{s_{j2}}$ in the diagonal entries $Y_{11}(\tau)$ and $Y_{22}(\tau)$
contain trivial lattice sums at their leading order in $s_{ij}$. The remaining component
integrals with $(k,\ell) \notin \{(0,0),(1,1)\}$
only feature non-trivial lattice sums throughout their expansion in $s_{02},s_{12}$.

The trivial lattice sums are independent on $\tau$ and must therefore stem from the
contribution $ \exp( {-}\frac{ R_{\vec{\eta}}(\ep_0)}{4y}  ) \widehat Y(i\infty)$ to (\ref{gen.34c}).
In fact, already the expression
\begin{align}
\widehat Y_{ij}(i\infty)  &=
\ccb \frac{1}{ \eta \bar \eta} - \frac{ 2\pi i }{s_{12}} &\frac{1}{ \eta \bar \eta}
\\
\frac{1}{ \eta \bar \eta} &\frac{1}{ \eta \bar \eta} - \frac{ 2\pi i }{s_{02}}
\cce_{ij}  \ \modMZV
\label{inival.03}
\end{align}
for the non-MZV sector correctly reproduces the trivial lattice sums at $(k,\ell)=(0,0)$ and
$(k,\ell)=(1,1)$: One can easily check that (\ref{inival.03}) is annihilated by $R_{\vec{\eta}}(\ep_0)$
in (\ref{revsec.23}) such that the exponential reduces to $\exp( {-}\frac{ R_{\vec{\eta}}(\ep_0)}{4y}  ) \rightarrow 1$ upon matrix multiplication. This truncation of the exponential is consistent with the
fact that the desired trivial lattice sums at $(k,\ell)=(0,0)$ and $(k,\ell)=(1,1)$ occur
with a vanishing power of $y$ in the Laurent polynomial.

In summary, (\ref{inival.03}) is an exact expression for the non-MZV sector of the
two-point initial value $ \widehat Y(i\infty)\big|_{n=2}$ by the arguments in this section.

%%%%%%%%%%%%%%%%%%%%%%%%%%%%%%%%%%%%%%%%%%%%%%%%%%%%%%%%%%%
\subsubsection{Beyond the non-MZV sector at two points}
\label{sec:9.1.2}
%%%%%%%%%%%%%%%%%%%%%%%%%%%%%%%%%%%%%%%%%%%%%%%%%%%%%%%%%%%

In order to display more representative contributions to the initial values $\widehat Y(i\infty)$,
we shall now preview samples of the MZVs at two points which have been obtained from the
method in section \ref{sec:9.2}. In contrast to the non-MZV sector
(\ref{inival.03}), the coefficients of MZVs in the initial value depend on
$u$ via polynomials at fixed order in $s_{ij},\eta,\bar \eta$. At the leading order
in $\eta,\bar \eta$, for instance, we have
\begin{align}
\widehat Y_{ij}(i\infty) \, \Big|_{\eta^{-1} \bar \eta^{-1}} &= \ccb 1 & 1 \\ 1 &1 \cce_{ij}
\bigg\{1
+ \bigg( \frac{1}{6}(s_{12}^3{+}s_{02}^3) + s_{12}s_{02} (s_{12}{+}s_{02}) B_2(u) \bigg) \zeta_3 \notag \\
& \quad
+ \bigg( \frac{1}{360} (s_{12} {+} s_{02}) (43 s_{12}^4 {-} 47 s_{12}^3 s_{02}
{+} 56 s_{12}^2 s_{02}^2 {-} 47 s_{12} s_{02}^3 {+} 43 s_{02}^4) \notag \\
&\quad \quad
   + \frac{1}{4} s_{12} s_{02} (s_{12}{+} s_{02}) (3 s_{12}^2 {-} 2 s_{12} s_{02} {+} 3 s_{02}^2) B_2(u)
   \label{inival.04}  \\
&\quad \quad
   + \frac{1}{6} s_{12} s_{02} (s_{12}{+} s_{02}) (s_{12}^2 {+} s_{02}^2) B_4(u) \bigg) \zeta_5
   +\ldots \bigg\}
   \notag
\end{align}
with MZVs of weight $\geq 6$ in the ellipsis,
see appendix \ref{app:init.1} for subleading orders in $\eta,\bar\eta$. As one
can anticipate from the expansion (\ref{fkreps.1}) of the integration kernels $f^{(k)}$
around the cusp, Bernoulli polynomials $B_k(u)$ cast the $u$-dependence into
its most compact form.

On top of the leading order in $\eta,\bar \eta$, one can give a closed formula
at leading order $s_{ij}^{-1}$ and $s_{ij}^0$ which is exact in $\eta,\bar\eta$,
\begin{align}
\sum_{\ell=1}^2 \exp\bigg( {-}\frac{ R_{\vec{\eta}}(\ep_0)}{4y}  \bigg)_{j\ell}
&\widehat Y_{i\ell}(i\infty)=
\frac{1}{\eta \bar \eta} \ccb 1 &1 \\ 1 &1 \cce_{ij} - 2\pi i \ccb s_{12}^{-1} &0 \\ 0 &s_{02}^{-1} \cce_{ij} \label{inival.05} \\
&\quad
+ 4\pi i \sum_{k=1}^\infty  \zeta_{2k+1} \bigg( \eta + \frac{ i\pi  \bar \eta}{2y} \bigg)^{2k}
 \ccb1 &0 \\ 0 &1 \cce_{ij} +{\cal O}(s_{ij}) \notag
\end{align}
see appendix \ref{app:init.2} for subleading orders in $s_{ij}$ and
(C.7) of \cite{Gerken:2020yii} for the analogous formula for generating series of MGFs.
The expansion of the two-point initial value up to a more systematic cutoff
(``order 10'' in the lingo of section 3.4.2 of \cite{Gerken:2020yii}) can be
found in an ancillary file.

%%%%%%%%%%%%%%%%%%%%%%%%%%%%%%%%%%%%%%%%%%%%%%%%%%%%%%%%%%%
\subsubsection{The non-MZV sector at higher points}
\label{sec:9.1.3}
%%%%%%%%%%%%%%%%%%%%%%%%%%%%%%%%%%%%%%%%%%%%%%%%%%%%%%%%%%%

The identification of trivial lattice sums in the expansion of higher-point generating
series $Y(\begin{smallmatrix} M \\ N \end{smallmatrix} |  \begin{smallmatrix} K \\ L \end{smallmatrix})$ in (\ref{high.4}) follows the logic of section \ref{sec:9.1.1}: First, the case when
all KE integrands contribute via $\Omega(z_{ij}, (\tau{-}\bar\tau)\eta_A)
\rightarrow \frac{1}{ (\tau{-}\bar\tau)\eta_A }$ and
$\overline{\Omega(z_{kl},\eta_B)} \rightarrow \frac{1}{\bar \eta_B}$
leads to terms of homogeneity degree $\sim (\eta_A \bar \eta_B)^{1-n}$ at $n$ points.
Second, contributions of $\Omega(z_{ij}, (\tau{-}\bar\tau)\eta_A)
\overline{\Omega(z_{kl},\eta_B)} \rightarrow f^{(1)}_{ij} \overline{ f^{(1)}_{kl}  }$
again integrate to kinematic poles if the labels $i,j$ and $k,l$ match or overlap.
At $n= 3$ points, the prescription $f^{(1)}_{ij} \overline{ f^{(1)}_{ij}  } \rightarrow \frac{1}{s_{ij}}$
to identify poles in two-particle Mandelstam invariants is augmented by ``pinching
rules'' for three-particle Mandelstam invariants upon Koba-Nielsen integration,
\beq
f^{(1)}_{ij} f^{(1)}_{jk} \overline{ f^{(1)}_{ij}  f^{(1)}_{jk}  } \rightarrow \frac{1}{s_{ijk}} \bigg(
\frac{1}{s_{ij}} + \frac{1}{s_{jk}}
\bigg) \, , \ \ \ \
f^{(1)}_{ij} f^{(1)}_{jk} \overline{ f^{(1)}_{ik}  f^{(1)}_{jk}  } \rightarrow \frac{1}{ s_{jk} s_{ijk}}
\label{inival.06}
\eeq
see appendix D.2 of \cite{Gerken:2019cxz} for a derivation of similar rules from integration by parts.
As exemplified by the first row of the non-MZV sector of the initial value at $n=3$ points
\begin{align}
\widehat Y_{ \begin{smallmatrix}  23 \\ \emptyset \end{smallmatrix} |
\begin{smallmatrix}  K \\ L \end{smallmatrix}}(i\infty)   &=
\bigg(
 \frac{1}{ \eta_{23}\eta_{3} \bar \eta_{23} \bar \eta_3} - \frac{ 2\pi i }{s_{12} \eta_3 \bar \eta_3}
 - \frac{ 2\pi i }{s_{23} \eta_{23} \bar \eta_{23}} + \frac{(2\pi i)^2 }{s_{123}} \bigg[ \frac{1}{s_{12}} + \frac{1}{s_{23}} \bigg] , \notag \\
 &\quad  \frac{1}{ \eta_{23} \eta_2 \bar \eta_{23} \bar \eta_3} + \frac{ 2\pi i }{s_{23}\eta_{23} \bar \eta_{23}} - \frac{ (2\pi i)^2 }{s_{123} s_{23}} , \
  \frac{1}{ \eta_2 \eta_3 \bar \eta_{23} \bar \eta_3} - \frac{ 2\pi i }{s_{12} \eta_3 \bar \eta_3},
  \label{inival.07}  \\
&\quad  \frac{1}{ \eta_{2} \eta_3 \bar \eta_{23} \bar \eta_3}, \
\frac{1}{ \eta_{23} \eta_2 \bar \eta_{23} \bar \eta_3} + \frac{2\pi i }{s_{23} \eta_{23} \bar \eta_{23}}, \
\frac{1}{ \eta_{23} \eta_3 \bar \eta_{23} \bar \eta_3}  - \frac{ 2\pi i }{s_{23} \eta_{23} \bar \eta_{23}}  \bigg)_{\begin{smallmatrix}  K \\ L \end{smallmatrix}}  \ \modMZV
\notag
\end{align}
one can find up to $n{-}1$ simultaneous kinematic poles at $n$ points together with
all interpolating contributions $\pm (\frac{2\pi i }{s_C} )^p (\eta_A \bar \eta_B)^{p+1-n}$
with $p=0,1,2,\ldots,n{-}1$ to the non-MZV sector of
 $\widehat Y_{ \begin{smallmatrix}  M \\ N \end{smallmatrix} |
\begin{smallmatrix}  K \\ L \end{smallmatrix}}(i\infty)$.
The complete $6\times 6$ matrix for the non-MZV sector of the three-point initial value
can be found in an ancillary file. At any multiplicity,
the non-MZV sector of the initial value should be annihilated by $R_{\vec{\eta}}(\epsilon_0)$ in
the sense that
\beq
\sum_{P,Q} R_{\vec{\eta}}(\epsilon_0)_{ \begin{smallmatrix} K \\ L \end{smallmatrix} \big|  \begin{smallmatrix} P \\ Q \end{smallmatrix}}
\widehat Y_{ \begin{smallmatrix}  M \\ N \end{smallmatrix} |
\begin{smallmatrix}  P \\ Q \end{smallmatrix}}(i\infty) = 0 \ \modMZV
\label{inival.08}
\eeq
Otherwise, the exponential of $-\frac{R_{\vec{\eta}}(\epsilon_0)}{4y}$ in (\ref{gen.34c})
would introduce powers of $y^{-1}$ into the trivial lattice sums in the expansions of
the component integrals which do not arise in our conventions for factors of $\Im \tau$
in (\ref{high.4}). The vanishing $R_{\vec{\eta}}(\epsilon_0)$-action in (\ref{inival.08}) is a
useful crosscheck for the variety of relative signs of different kinematic poles at $n\geq 3$ points.

As detailed in section 3.4.2 of \cite{Gerken:2020yii}, the generating series of MGFs in the
reference have similarly approachable non-MZV sectors in their initial values. At four points,
examples of three simultaneous kinematic poles in the initial values can be found in
appendix D of \cite{Gerken:2020yii}.

In summary, the non-MZV sector of the initial value $\widehat Y(i\infty)$ is particularly
simple and accessible to all multiplicities. It can be used to obtain the $\beta^{\rm sv}$
without any accompanying MZVs for a huge number of eMGFs by matching the ${\cal C}^+$ in the
expansion of the generating series with (\ref{gen.34c}) as explained in the next section.
It would be interesting to systematically determine the MZV contributions
to initial values beyond two points, for instance from the $\alpha'$-expansion of
genus-zero integrals with unintegrated punctures \cite{Vanhove:2018elu, Britto:2021prf}.

%%%%%%%%%%%%%%%%%%%%%%%%%%%%%%%%%%%%%%%%%%%%%%%%%%%%%%%%%%%
\subsubsection{Exponents of lattice sums versus $\beta^{\rm sv}$ entries}
\label{sec:9.1.6}
%%%%%%%%%%%%%%%%%%%%%%%%%%%%%%%%%%%%%%%%%%%%%%%%%%%%%%%%%%%

With the above information on the non-MZV sector of $\widehat Y(z|i\infty)$, we can
now derive the correspondence (\ref{count.1}) between the sums $|A|,|B|$ of
(anti-)holomorphic exponents of $\cplus{A \\ B \\ Z}$ and the terms without MZV
coefficients in its $\beta^{\rm sv}$ representation. The entries of such
$\bsvtau{j_1 &j_2 &\ldots &j_\ell \\ k_1 &k_2 &\ldots &k_\ell \\ z_1&z_2 &\ldots &z_\ell}$
without accompanying MZVs were claimed to obey
\beq
|A|= \ell + \sum_{i=1}^\ell j_i
\, , \ \ \ \
|B|= {-}\ell+ \sum_{i=1}^\ell (k_i{-} j_i)
\label{proof41.01}
\eeq
which we shall now prove via simple observations on homogeneity degrees that
closely follow the lines of the analogous discussion for MGFs in section 6.2
of \cite{Gerken:2020yii}.

On the one hand, the expansion of the $n$-point generating series (\ref{high.4}) in $s_{ij},\eta_j,\bar\eta_j$ can be performed in terms of lattice sums. The constituents $g(z_{ij}|\tau)$,
$f^{(k)}(z_{ij}|\tau)$ and $\overline{f^{(k)}(z_{ij}|\tau)}$ of the integrand interlock the contributions
to $|A|,|B|$ of the resulting ${\cal C}^+$ with the homogeneity degrees in the expansion variables via
\begin{align}
 g(z_{ij}|\tau)&\leftrightarrow s_{ij} \ {\rm and} \ {\rm contributes} \ (1,1) \ {\rm to} \ \big(|A|,|B| \big)
 \notag \\
f^{(k)}(z_{ij}|\tau)&\leftrightarrow \eta^k_j \ {\rm and} \ {\rm contributes} \ (k,0) \ {\rm to} \ \big(|A|,|B| \big)
\label{proof41.02} \\
\overline{f^{(k)}(z_{ij}|\tau)}&\leftrightarrow \bar \eta^k_j \ {\rm and} \ {\rm contributes} \ (0,k) \ {\rm to} \ \big(|A|,|B| \big)
 \notag
\end{align}
This is evident for non-singular Koba-Nielsen integrals but also applies to the residues of
kinematic poles in $s_{i\ldots j}$ when following the subtraction scheme in appendix C of
\cite{Dhoker:2020gdz} or appendix D of \cite{Gerken:2019cxz}.

Once we rescale the expansion variables via
\beq
s_{ij} \rightarrow \alpha \beta s_{ij} \, , \ \ \ \
\eta_j \rightarrow \alpha \eta_j \, , \ \ \ \
\bar \eta_j \rightarrow \beta \bar \eta_j
\label{proof41.03}
\eeq
and multiply the $n$-point generating series by $(\alpha \beta)^{n-1}$ to account
for the poles of $\Omega(z,\eta|\tau) \sim \eta^{-1}$, then the homogeneity degrees 
in $\alpha$ and $\beta$ track the values of $|A|,|B|$ at each order of the expansion 
in $s_{ij},\eta_j,\bar\eta_j$.

The next step is to identify the coefficients of $\alpha^{|A|}$ and $\beta^{|B|}$
in the expansion (\ref{gen.34c}) of the $n$-point generating series in terms of
$\beta^{\rm sv}$. The above multiplication by $(\alpha \beta)^{n-1}$ is most
conveniently absorbed in to the non-MZV sector of the initial value -- as exemplified
by (\ref{inival.03}) and (\ref{inival.07}), all the contributions to $\widehat Y_{ \begin{smallmatrix}  M \\ N \end{smallmatrix} |
\begin{smallmatrix}  P \\ Q \end{smallmatrix}}(z|i\infty) \ \modMZV$
take the form of $ (s_C)^{-p} (\eta_A \bar \eta_B)^{p+1-n}$ with $p=0,1,\ldots,n{-}1$
and scale with a compensating power $(\alpha \beta)^{1-n}$ under (\ref{proof41.03}).
Hence, the $|A|,|B|$ of the lattice sums are in one-to-one correspondence with
the exponents of $\alpha,\beta$ upon applying (\ref{proof41.03}) to the path-ordered
exponential in (\ref{gen.34c}).

The matrix representations $R_{\vec{\eta}}(\epsilon_{k}) , R_{\vec{\eta}}(b_{k})$
in \cite{Broedel:2020tmd, Dhoker:2020gdz} have homogeneity degrees\footnote{We are counting
each derivative $\partial_{\eta_j}$ in $R_{\vec{\eta}}(\epsilon_{0})$ as contributing homogeneity
$-1$ in $\eta_j$.} $ \eta_j^{k-2}s_{ij}$
for $k\neq 0$ and a combination of $\frac{ s_{ij} }{\eta_j^2}$ and $\frac{ \bar\eta_j }{\eta_j}$
in case of $R_{\vec{\eta}}(\epsilon_{0})$. Accordingly, their transformation under
(\ref{proof41.03}) is
\beq
R_{\vec{\eta}}(\epsilon_{k}) \rightarrow \alpha^{k-1} \beta R_{\vec{\eta}}(\epsilon_{k})
\, , \ \ \ \
R_{\vec{\eta}}(b_{k}) \rightarrow \alpha^{k-1} \beta R_{\vec{\eta}}(b_{k})
\label{proof41.04}
\eeq
which, together with the special case $R_{\vec{\eta}}(\epsilon_{0}) \rightarrow \frac{\beta}{\alpha} R_{\vec{\eta}}(\epsilon_{0})$, implies that
\beq
\left( \begin{array}{c} R_{\vec{\eta}}({\rm ad}_{\epsilon_0}^{k-2-j}\epsilon_{k})
\\
R_{\vec{\eta}}({\rm ad}_{\epsilon_0}^{k-2-j}b_{k}) \end{array} \right) \rightarrow
\alpha^{j+1} \beta^{k-j-1}
\left( \begin{array}{c} R_{\vec{\eta}}({\rm ad}_{\epsilon_0}^{k-2-j}\epsilon_{k})
\\
R_{\vec{\eta}}({\rm ad}_{\epsilon_0}^{k-2-j}b_{k}) \end{array} \right)
\label{proof41.05}
\eeq
For the products of $\ell$ operators $R_{\vec{\eta}}({\rm ad}_{\epsilon_0}^{k_i-2-j_i}\epsilon_{k_i})$
or $R_{\vec{\eta}}({\rm ad}_{\epsilon_0}^{k_i-2-j_i}b_{k_i})$, we arrive
at a scaling by $\prod_i( \alpha^{j_i+1} \beta^{k_i-j_i-1}  ) =\alpha^{\ell + \sum_i j_i}\beta^{-\ell+\sum_i(k_i{-}j_i)}$.
By the discussion around (\ref{proof41.03}), the exponents of $\alpha$ and
$\beta$ can be identified with $|A|,|B|$, which finishes the proof of (\ref{proof41.01}).

Since we restricted the initial value to its non-MZV sector, the relations
(\ref{proof41.01}) do not apply to contributions with MZVs to the $\beta^{\rm sv}$ representations
of eMGFs. For instance, the depth-zero terms $\sim \zeta_{2k-1}$ in (\ref{nwbsv.11}) clearly
have different values of $\ell + \sum_i j_i$ and $-\ell+\sum_i(k_i{-}j_i)$ than the $\beta^{\rm sv}$
of depth one.

%%%%%%%%%%%%%%%%%%%%%%%%%%%%%%%%%%%%%%%%%%%%%%%%%%%%%%%%%%%
\subsection{eMGFs as iterated integrals from generating series}
\label{sec:9.7}
%%%%%%%%%%%%%%%%%%%%%%%%%%%%%%%%%%%%%%%%%%%%%%%%%%%%%%%%%%%

We shall now illustrate how the expansion (\ref{gen.34c}) of the generating series
together with the information on the initial value from the previous section
determine eMGFs in terms of $\beta^{\rm sv}$. Based on the reality properties of eMGFs,
these $\beta^{\rm sv}$ representations are used to compute the
integration constants $\overline{ \alpha[\ldots;\tau]}$ in the
expressions (\ref{eq:bsv1}) and (\ref{fkreps.18c}) for real-analytic iterated
KE integrals of depth $\geq 2$.

%%%%%%%%%%%%%%%%%%%%%%%%%%%%%%%%%%%%%%%%%%%%%%%%%%%%%%%%%%%
\subsubsection{Applications at depth one}
\label{sec:9.1.4}
%%%%%%%%%%%%%%%%%%%%%%%%%%%%%%%%%%%%%%%%%%%%%%%%%%%%%%%%%%%

As a first illustration of the generating-function method, we reproduce
the $\beta^{\rm sv}$ representation (\ref{invbsv.7}) of Zagier's
single-valued elliptic polylogarithms $\dplus{a \\ b}\!(z|\tau)$ as a
multiple of $\bsvtau{a-1\\ a+b \\ z }$. These single-valued elliptic polylogarithms occur in
the $s_{ij} \rightarrow 0$ limit of the off-diagonal component integrals $Y_{ij}^{(a|b)}$ at two points
\begin{align}
Y_{ij}(\eta,\bar \eta) &= \sum_{a,b = 0}^\infty \eta^{a-1} \bar \eta^{b-1} (2\pi i)^b Y_{ij}^{(a|b)}
\label{inival.09} \\
Y_{12}^{(a|b)} &=  (2i)^{a-b} \dplus{a \\ b}\!(z|\tau) + {\cal O}(s_{ij}) \, , \ \ \ \ a{+}b \geq 2
\notag
\end{align}
see section 4.1 of \cite{Dhoker:2020gdz} for details. We can alternatively
compute this $s_{ij} \rightarrow 0$ limit from the expansion (\ref{gen.34c})
of the generating series in terms of $\beta^{\rm sv}$:
With the explicit form of the two-point
$2\times 2$ matrix representations $R_{\eta}(\epsilon_k)$ and $R_{\eta}(b_k)$
in (\ref{revsec.23}), the depth-one
contributions to the path-ordered exponential are given by
\begin{align}
& \sum_{k=2}^\infty \sum_{j=0}^{k-2} c_{j,k}
\bigg(   \bsvtau{j \\ k \\  \emptyslot} R_{\eta}\big( {\rm ad}_{\ep_0}^{k-j-2}(\ep_{k}) \big)
+\bsvtau{j \\ k \\ z }  R_{\eta}\big( {\rm ad}_{\ep_0}^{k-j-2}(b_{k}) \big)\bigg)
\notag \\
&=  \ccb -s_{02} &s_{02} \\ s_{12} &-s_{12} \cce \bsvtau{0 \\ 2 \\ z} + \sum_{k=3}^\infty \sum_{j=0}^{k-2} \frac{ (k{-}1)! }{j!(k{-}2{-}j)!}
(2\pi i \bar \eta)^{k-j-2} \eta^j  \label{inival.10} \\
&\quad \times \bigg\{ \ccb s_{12} &0 \\ 0 &s_{02} \cce   \bsvtau{j \\ k \\  \emptyslot}
+ \ccb 0 &(-1)^k s_{02}  \\ s_{12} &0 \cce \bsvtau{j \\ k \\ z} \bigg\}
+ {\cal O}(s_{ij}^2,\partial_\eta) \notag
\end{align}
Since we are only tracking the linear order in $s_{02},s_{12}$, the adjoint action of
$R_{\eta}(\epsilon_0)$ reduces to the term $-2\pi i \bar \eta \partial_\eta$ in (\ref{revsec.23}).
When acting on the non-MZV sector of the initial value (\ref{inival.03}), only the poles
$\widehat Y_{ij}(i\infty) \rightarrow -2\pi i
\big(\begin{smallmatrix} s_{12}^{-1} &0 \\ 0 &s_{02}^{-1} \end{smallmatrix} \big)$ contribute
to the $s_{ij}\rightarrow 0$ limit of the component integrals (\ref{inival.09}).
For the diagonal entries $Y_{11}$ and $Y_{22}$, the $s_{ij}^0$ order additionally receives
contributions from the series in $(\eta + \frac{i \pi \bar \eta}{2y})^{2k} \zeta_{2k+1}$
in the second line of (\ref{inival.05}) multiplying the depth-zero part of the path-ordered
exponential. Since these zeta values do not
contribute to the off-diagonal entries $Y_{12}$ and $Y_{21}$ at $s_{ij} \rightarrow 0$, 
the contraction of (\ref{inival.10}) with the polar part of $\widehat Y_{ij}(i\infty)$ yields
\begin{align}
Y_{12}(\eta,\bar \eta) = \frac{1}{\eta \bar \eta}
- 2\pi i \sum_{k=2}^\infty \sum_{j=0}^{k-2} \frac{(k{-}1)!}{j! (k{-}2{-}j)!}
 (2\pi i \bar \eta)^{k-2-j} \eta^j \bsvtau{j \\ k \\ z}  + {\cal O}(s_{ij})
 \label{inival.11}
\end{align}
By matching the coefficients of $\bar \eta^{k-2-j} \eta^j$ in (\ref{inival.11})
with the component integrals in (\ref{inival.09}) at $a{+}b\geq 2$, we
reproduce the expression (\ref{invbsv.7}) for Zagier's
single-valued elliptic polylogarithms in terms of $\beta^{\rm sv}$.
Instead of the direct computation underlying (\ref{invbsv.7}), the method
of this section is based on the expansion (\ref{gen.34c}) of the generating
series of eMGFs.

Note that the same logic has been applied in appendix C.3 of \cite{Gerken:2020yii}
to derive the $\beta^{\rm sv}$ representation (\ref{nwbsv.11}) of non-holomorphic
Eisenstein series from the two-point generating series of MGFs. In a derivation
from the two-point generating series of eMGFs, the additional
odd zeta values in (\ref{nwbsv.11}) can be traced back to the second
line of (\ref{inival.05}) which does not contribute to the off-diagonal
$Y_{12}(\eta,\bar \eta)$ in (\ref{inival.11}).

Now that the $\beta^{\rm sv}$ representation (\ref{invbsv.7}) of Zagier's
single-valued elliptic polylogarithms has been derived from generating functions,
we shall justify the antiholomorphic contributions to ${\cal E}^{\rm sv}$ and hence
$\beta^{\rm sv}$ in the first line of (\ref{fkreps.18}): The lattice-sum representation
exposes the complex-conjugation properties
\beq
\overline{ \dplus{a \\ b}\!(z|\tau)  }   = (\pi \Im \tau)^{a-b} \dplus{b \\ a}\!({-}z|\tau)
= (-1)^{a+b} y^{a-b}\dplus{b \\ a}\!(z|\tau)
 \label{inival.12}
\eeq
whose compatibility with (\ref{invbsv.7}) requires
\beq
\overline{ \bsvtau{j \\ k \\ z}  }   = (-4y)^{2+2j-k} \bsvtau{k{-}j{-}2 \\ k \\ z}
 \label{inival.13}
\eeq
This uniquely fixes the depth-one contributions $\overline{  \EBR{j_1-r_1 \\ k_1 \\ z_1}{\tau} } $
to the first line of (\ref{fkreps.18}) as well as the integral over
$\overline{f^{(k_1)}(u_1 \tau_1{+}v_1|\tau_1)} $ in the second line of (\ref{eq:ebsv1}).
Moreover, (\ref{inival.13}) excludes any depth-one analogue of the antiholomorphic 
integration constants $\overline{ \alpha[\ldots;\tau]}$ entering ${\cal E}^{\rm sv}$ 
and $\beta^{\rm sv}$ at depth $\geq 2$ such as (\ref{fkreps.18}) and (\ref{eq:ebsv2}). Note 
that (\ref{inival.13}) lines up with the complex-conjugation properties
\beq
\overline{ \bsvtau{j \\ k \\ \emptyslot }  }   = (-4y)^{2+2j-k} \bsvtau{k{-}j{-}2 \\ k \\ \emptyslot }
 \label{inival.13alt}
\eeq
relevant to non-holomorphic Eisenstein series \cite{Gerken:2020yii}. Given that
$\bsvtau{\ldots &j &\ldots \\ \ldots &k &\ldots \\ \ldots &\emptyslot &\ldots}$ are in general
different from the $z\rightarrow 0$ limit of $\bsvtau{\ldots &j &\ldots \\ \ldots &k &\ldots \\ \ldots &z&\ldots}$,
see for instance (\ref{nwbsv.13}), 
the respective complex conjugates will be determined independently.

We emphasize that the non-MZV sector (\ref{inival.03}) of the initial value is sufficient
to fix the iterated KE integrals at depth one in the expression (\ref{eq:ebsv1}) for $\bsvtau{j \\ k \\ z}$.
In spelling out the second line of (\ref{inival.05}), we have already used results from the
sieve algorithm of section~\ref{sec:9.2} to fix the coefficients of MZVs in the initial value
$\widehat Y_{ij}(i\infty)$. Without this information, we would not exclude MZV corrections
to the $\beta^{\rm sv}$ representation (\ref{invbsv.7}) of $\dplus{a \\ b}\!(z|\tau) $
and possibly (\ref{inival.13}) as long as we solely rely on (\ref{inival.03}).
However, such MZV corrections do not interfere with the antiholomorphic terms in
(\ref{fkreps.18}) and (\ref{eq:ebsv1}) since they arise without any accompanying MZV.

Nevertheless, we shall keep on using the information on the MZV corrections to
$\widehat Y_{ij}(i\infty)$ in the remainder of this section in order to
illustrate the derivation of integration constants $\overline{ \alpha[\ldots;\tau]}$
at depth $\geq 2$.

%%%%%%%%%%%%%%%%%%%%%%%%%%%%%%%%%%%%%%%%%%%%%%%%%%%%%%%%%%%
\subsubsection{Applications at higher depth}
\label{sec:9.1.5}
%%%%%%%%%%%%%%%%%%%%%%%%%%%%%%%%%%%%%%%%%%%%%%%%%%%%%%%%%%%

The logic of the depth-one discussion of $\dplus{a \\ b}\!(z|\tau)$ extends to higher depth:
For each lattice-sum representation ${\cal C}^+$ of eMGFs encountered in the component
integrals of the $n$-point generating series, one can obtain the $\beta^{\rm sv}$ representation
from (\ref{gen.34c}) by isolating the appropriate orders in $s_{ij}$ and $\eta_j,\bar \eta_j$.
If we only supply the non-MZV sector of the initial value which is accessible to all multiplicity,
then the $\beta^{\rm sv}$ representations of eMGFs are only obtained modulo MZVs, and one
for instance misses the odd zeta values in (\ref{nwbsv.11}) and the preview examples
(\ref{prevex.1}).

As soon as shuffle-independent $\beta^{\rm sv}$ at depth two are involved, it remains to
fix the integration constants $\overline{ \alpha[\ldots;\tau]}$ in the kernel of $\nabla_\tau$.
Such antiholomorphic objects can be determined from the complex-conjugation
properties of eMGFs which are manifest in their lattice-sum representation
(\ref{gen.66}) \cite{Dhoker:2020gdz},
\beq
\overline{ \cplus{A \\ B \\ Z}\! (\tau) }
= y^{|A|-|B|} \cplus{B \\ A \\ -Z}\! (\tau)
= (-1)^{|A|+|B|} y^{|A|-|B|} \cplus{B \\ A \\ Z}\! (\tau)
 \label{inival.16}
\eeq
After inserting the $\beta^{\rm sv}$ representations on both sides, the complex conjugates
$\overline{ \bsvtau{j_1 &\ldots &j_\ell \\ k_1 &\ldots &k_\ell  \\ z_1 &\ldots &z_\ell  }  } $ are expressed in terms of the original $\beta^{\rm sv}$ with non-positive powers of $y$, MZVs
and polynomials in $u$ in their coefficients. At depth one, for instance, this procedure
leads to (\ref{inival.13}) and (\ref{inival.13alt}).

At depth two, one can for instance identify
\begin{align}
 \cplus{1 &1 &1 \\ 1 &1 &1 \\ z &0 &0}\! (\tau) &= 2 Y_{11}^{(0|0)} \big|_{s_{02}^2 s_{12}}
 \notag \\
  \cplus{1 &1 &1 \\ 0 &1 &1 \\ z &0 &0}\! (\tau) &= - i Y_{11}^{(1|0)} \big|_{s_{02}^2}
 \label{inival.17} \\
  \cplus{0 &1 &1 \\ 1 &1 &1 \\ z &0 &0}\! (\tau) &= - 4 i Y_{11}^{(0|1)} \big|_{s_{02}^2}
 \notag
\end{align}
in the expansion of component integrals (\ref{inival.09}) described in section 4.1 of
\cite{Dhoker:2020gdz}. By isolating the corresponding orders in $s_{02},s_{12},\eta,\bar \eta$
in the expansion (\ref{gen.34c}) of the two-point generating series, one infers the
$\beta^{\rm sv}$ representations
\begin{align}
 \cplus{1 &1 &1 \\ 1 &1 &1 \\ z &0 &0}\! (\tau) &= 8 \bsvtau{1& 0\\3& 3\\z& z}
- 10 \bsvtau{2\\6\\ \emptyslot} - 20 \bsvtau{2\\6\\z}
+ 2  \zeta_{3} B_{2}(u)+\frac{ \zeta_{5}}{4 y^2}
 \notag \\
  \cplus{1 &1 &1 \\ 0 &1 &1 \\ z &0 &0}\! (\tau) &=
   - 2 i \bsvtau{1& 0\\3& 2\\z& z}
+   4 i \bsvtau{2\\5\\z}
  - 2 i \zeta_3 B_1(u)
 \label{inival.18} \\
  \cplus{0 &1 &1 \\ 1 &1 &1 \\ z &0 &0}\! (\tau) &=
8 i \bsvtau{0& 0\\2& 3\\z& z}
- 16 i \bsvtau{1\\5\\z}
  -\frac{2 i}{y} \zeta_3 B_1(u)  \notag
\end{align}
and obtains the preview examples in the first line of (\ref{prevex.1}) and (\ref{prevex.2}). Note
that we have here used information on the initial values beyond the non-MZV sector,
e.g.\ the term $s_{02}^2 s_{12} \zeta_3 B_2(u)$ in the first line of (\ref{inival.04}).

By (\ref{inival.16}), the complex conjugation properties of these
lattice sums are
\begin{align}
\overline{ \cplus{1 &1 &1 \\ 1 &1 &1 \\ z &0 &0}\! (\tau) }=
\cplus{1 &1 &1 \\ 1 &1 &1 \\ z &0 &0}\! (\tau)  \, , \ \ \ \
\overline{  \cplus{1 &1 &1 \\ 0 &1 &1 \\ z &0 &0}\! (\tau) }&= -y \, \cplus{0 &1 &1 \\ 1 &1 &1 \\ z &0 &0}\! (\tau)
  \label{inival.19}
\end{align}
With the $\beta^{\rm sv}$ representations (\ref{inival.18}) and the depth-one result
in (\ref{inival.13}), one can solve for
\beq
 \overline{  \bsvtau{1& 0\\3& 3\\z& z} } =  \bsvtau{1& 0\\3& 3\\z& z}
 \, , \ \ \ \
\overline{  \bsvtau{1& 0\\3& 2\\z& z}  } = -4y  \bsvtau{0& 0\\2& 3\\z& z}
\label{inival.20}
\eeq
These complex-conjugation properties are only compatible with the expression (\ref{eq:ebsv2}) for
$\beta^{\rm sv}$ if the respective integration constants are set to zero. Hence, (\ref{inival.20})
together with the shuffle property (\ref{alpshffl}) of the $\overline{ \alpha[\ldots;\tau]}$ imply
\beq
 \overline{\alphaBR{j_1 &j_2 \\ 2 &3 \\ z &z}{\tau}}
 =  \overline{\alphaBR{j_1 &j_2 \\ 3 &2 \\ z &z}{\tau}}
 =  \overline{\alphaBR{j_1 &j_2 \\ 3 &3 \\ z &z}{\tau}}
 = 0
\label{inival.21}
\eeq
throughout the sectors with $(k_1,k_2) \in \{ (2,3),(3,2),(3,3)\}$.

%%%%%%%%%%%%%%%%%%%%%%%%%%%%%%%%%%%%%%%%%%%%%%%%%%%%%%%%%%%
\subsubsection{Extracting non-zero integration constants $\overline{ \alpha[\ldots;\tau]}$}
\label{sec:9.1.ex}
%%%%%%%%%%%%%%%%%%%%%%%%%%%%%%%%%%%%%%%%%%%%%%%%%%%%%%%%%%%

The non-vanishing integration constants $\overline{ \alpha[\ldots;\tau]}$
can be most compactly represented in terms of the ${\cal E}_0$ variant of the meromorphic
iterated KE integrals in section \ref{qexpsec.A} as well as the corresponding iterated
Eisenstein integrals in section \ref{sec:2.5.2}. In particular, by employing their instances
$\overline{\mathcal{E}_0(\begin{smallmatrix} k_1&\ldots \\
z_1&\ldots \end{smallmatrix};\tau)}$ with $k_1\neq 0$, one manifests the
$T$-invariance of the integration constants $\overline{ \alpha[\ldots;\tau]}$:
Both the complete ${\cal E}^{\rm sv}[\ldots]$ and the explicit
combinations of ${\cal E}[\ldots]$ and $\overline{{\cal E}[\ldots]}$ in (\ref{fkreps.18}) and
(\ref{fkreps.18c}) are $T$-invariant, so the same has
to be true for the $\overline{ \alpha[\ldots;\tau]}$.

Already the simplest non-vanishing $\overline{ \alpha[\ldots;\tau]}$ cannot be determined
individually since not all of the corresponding $\beta^{\rm sv}$ are realized as eMGFs.
For the depth-two sectors of the path-ordered exponential (\ref{gen.34c}) involving
$ b_2{\rm ad}_{\ep_0}\ep_4$ and $b_2 {\rm ad}_{\ep_0}b_4$, we have seen in
(\ref{comrel.21}) how the relation $[b_2,\epsilon_4{+}b_4]=0$ affects the
combinations of $\beta^{\rm sv}$ that enter eMGFs.
% the implications of the
%relation $[b_2,\epsilon_4{+}b_4]=0$  for the combinations of $\beta^{\rm sv}$ that enter eMGFs
%are spelt out in (\ref{comrel.21}).
Accordingly, the complex-conjugation properties (\ref{inival.16}) of eMGFs only
determine the complex conjugates of selected combinations of $\beta^{\rm sv}$, for instance
\begin{align}
\overline{ \bsvtau{0 &0 \\ 2 &4 \\ z&z }  }  - \overline{ \bsvtau{0 &0 \\ 2 &4 \\ z&\emptyslot }  }  &=
\frac{1}{16y^2} \Big\{ \bsvtau{2 &0 \\ 4 &2 \\ z&z }  - \bsvtau{2 &0 \\ 4 &2 \\ \emptyslot&z } \Big\}
+ \frac{ \zeta_3}{24y^2} \bsvtau{0 \\ 2 \\  z} + \frac{ \zeta_3 B_2(u)}{12y}
\notag\\
\overline{ \bsvtau{0 &1 \\ 2 &4 \\ z&z }  }  - \overline{ \bsvtau{0 &1 \\ 2 &4 \\ z&\emptyslot }  }  &=
\bsvtau{1 &0 \\ 4 &2 \\ z&z }  - \bsvtau{1 &0 \\ 4 &2 \\ \emptyslot&z }
+ \frac{ \zeta_3}{6y} \bsvtau{0 \\ 2 \\  z} + \frac{ \zeta_3 B_2(u)}{3}
\label{inival.32} \\
\overline{ \bsvtau{0 &2 \\ 2 &4 \\ z&z }  }  - \overline{ \bsvtau{0 &2 \\ 2 &4 \\ z&\emptyslot }  }  &=
16y^2 \Big\{ \bsvtau{0 &0 \\ 4 &2 \\ z&z }  - \bsvtau{0 &0 \\ 4 &2 \\ \emptyslot&z } \Big\}
+ \frac{2\zeta_3}{3}  \bsvtau{0 \\ 2 \\  z}  + \frac{4 y \zeta_3 B_2(u)}{3}
\notag
\end{align}
Compatibility with (\ref{eq:ebsv2}) and $\nabla_\tau \overline{ \alpha[\ldots;\tau]}=0$
then implies the expressions
\begin{align}
 \overline{\alphaBR{0 &0 \\ 2 &4 \\ z &z}{\tau}} - \overline{\alphaBR{0 &0 \\ 2 &4 \\ z& \emptyslot}{\tau}} &= 0
 \notag \\
 \overline{\alphaBR{0 &1 \\ 2 &4 \\ z &z}{\tau}} - \overline{\alphaBR{0 &1 \\ 2 &4 \\ z& \emptyslot}{\tau}} &= 0
 \label{inival.33}    \\
 \overline{\alphaBR{0 &2 \\ 2 &4 \\ z &z}{\tau}} - \overline{\alphaBR{0 &2 \\ 2 &4 \\ z& \emptyslot}{\tau}}
 &=    \frac{ 2\zeta_3}{3} \overline{\mathcal{E}_0\!\SM{2}{z }{\tau}} \notag
 \end{align}
By the second line of (\ref{inival.31}), each ${\cal E}_0(\begin{smallmatrix} k&\vec{0}^{p-1} \\
z&\vec{0}^{p-1} \end{smallmatrix};\tau) $ is composed of iterated KE integrals
${\cal E}[\ldots]$ of depth one and zero. Hence, it is a helpful crosscheck of $T$ invariance
in our setup that the complex-conjugation properties in (\ref{inival.32}) yield exactly the relative
factors between ${\cal E}[\ldots]$ of depth one and zero needed for recombination
to ${\cal E}_0(\begin{smallmatrix} k&\vec{0}^{p-1} \\z&\vec{0}^{p-1} \end{smallmatrix};\tau) $.

At lattice weight $\sum_i k_i\geq 7$, the combinations of $\overline{ \alpha[\ldots;\tau]}$ realized in
eMGFs also mix different depths such as for instance in
\beq
 \overline{\alphaBR{0 &0 &1 \\ 2 &2 &3 \\ z&z&z}{\tau}}
 +  5\overline{\alphaBR{0 &2 \\ 2 &5 \\ z &z}{\tau}}
 - \frac{15}{2}\overline{\alphaBR{0 &2 \\ 3 &4 \\ z&\emptyslot}{\tau}} =
    5 \zeta_3 \overline{ \mathcal{E}_0\!\SM{3}{z }{\tau}  }
  - \zeta_3 B_1(u) \overline{ \mathcal{E}_0\!\SM{2}{z }{\tau} }
  \label{inival.34}
  \eeq
This example also illustrates that the coefficients of $\overline{{\cal E}_0}$ may
depend polynomially on~$u$. 

Both of (\ref{inival.33}) and (\ref{inival.34}) exemplify that each contribution to
$\overline{ \alpha[\ldots;\tau]}$ involves MZVs of transcendental weight $\geq 3$, 
and we expect once more to only
encounter single-valued MZVs. A complete list of $\overline{ \alpha[\ldots;\tau]}$
relevant to lattice weight $\sum_i k_i=7$ can be found in appendix \ref{app:baral.A},
and the analogous results at lattice weights $\sum_i k_i\leq 10$ are included
into the ancillary files. At higher weight, we also encounter products of MZVs 
and $\overline{{\cal E}_0}$ with several non-zero entries, for instance
\begin{align}
&\overline{\alphaBR{0 &0 &2\\ 2 &3 &4 \\ z& z & z}{\tau}} - 
\overline{\alphaBR{0 &0 &2\\ 2 &3 &4 \\ z& z & \emptyslot}{\tau}}+ 
 \frac{110}{9} \bigg( \overline{\alphaBR{0 &3 \\ 3 &6 \\ z& z}{\tau}}
 - \overline{\alphaBR{0 &3 \\ 3 &6 \\ z& \emptyslot}{\tau}} \bigg) 
 + \frac{70}{3} \overline{\alphaBR{0 &3 \\ 4 &5 \\ \emptyslot& z}{\tau}} - 
\frac{ 31}{3} \overline{\alphaBR{0 &3 \\ 4 &5 \\ z & z}{\tau}} \notag \\
%%%%%
&\quad = -\frac{1}{9} \zeta_3 B_3(u)  
 \overline{ \mathcal{E}_0\!\SM{2 &0}{z &0}{\tau}  }
 +  \bigg( \frac{1}{6} + \frac{ B_2(u) }{3} \bigg)  \zeta_3  \overline{ \mathcal{E}_0\!\SM{3&0}{z &0}{\tau}  }
  +  \frac{  2}{3} \zeta_3
 \overline{ \mathcal{E}_0\!\SM{3 &2}{z &z }{\tau}  } 
 \label{exd2abar}
\end{align}
Up to corrections by MZVs, the complex conjugate $\beta^{\rm sv}$ at depth two can be
given in closed form
\beq
\overline{ \bsvtau{j_1 &j_2 \\ k_1 &k_2 \\ z_1 &z_2 }  }   = (-4y)^{4+2j_1+2j_2-k_1-k_2} \bsvtau{k_2{-}j_2{-}2 &k_1{-}j_1{-}2 \\ k_2 &k_1 \\ z_2 &z_1}  \ \modMZV
 \label{inival.35}
\eeq
where $z_1,z_2$ may be any combination of one variable $z$ and empty slots.
This is consistent with a variety of examples including (\ref{inival.20}), (\ref{inival.32}) and 
validates the terms excluding $\overline{ \alpha[\ldots;\tau]}$ in
the expressions (\ref{fkreps.18}) and (\ref{eq:ebsv2}) for ${\cal E}^{\rm sv}$ and $\beta^{\rm sv}$
at depth two. Still, the MZV corrections to (\ref{inival.35}) and equivalently 
the $\overline{ \alpha[\ldots;\tau]}$ need to be determined independently for 
any combination of $z$ and empty slots in the place of $(z_1,z_2)$.

Apart from exposing their relations, the $\beta^{\rm sv}$ representations of eMGFs
conveniently give access to their expansion around the cusp as detailed in section
\ref{qexpsec.B}. It is essential to determine
the complete set of contributing $\overline{ \alpha[\ldots;\tau]}$ as one would otherwise
miss infinite series in $\bar q^u,\bar q$ or $\bar q^{1\pm u}$. These expansions
around the cusp in turn lend themselves for efficient numerical evaluations
of eMGFs which would be considerably more challenging in their lattice-sum
representation.

%%%%%%%%%%%%%%%%%%%%%%%%%%%%%%%%%%%%%%%%%%%%%%%%%%%%%%%%%%%
\subsubsection{Leading terms of $\overline{ \alpha[\ldots;\tau]}$}
\label{sec:9.111}
%%%%%%%%%%%%%%%%%%%%%%%%%%%%%%%%%%%%%%%%%%%%%%%%%%%%%%%%%%%

The leading terms of iterated KE integrals discussed in section \ref{sec:3.3}
often provide a hint towards expressions for the {\it individual} $\overline{ \alpha[\ldots;\tau]}$,
even though they are not yet well-defined from the information on eMGFs.
By demanding that the leading terms of $\beta^{\rm sv}$ are expressible in terms
of single-valued polylogarithms at genus zero, we derived the predictions 
in (\ref{baralp.1}) for the leading terms of the $ \overline{\alphaBR{0 &j \\ 2 &4 \\ z &z}{\tau}}$
with $j=0,1,2$.
Indeed, setting $ \overline{\alphaBR{0 &2 \\ 2 &4 \\ z &z}{\tau}} \rightarrow  \frac{ 2\zeta_3}{3}  \overline{\mathcal{E}_0(\begin{smallmatrix} 2 \\
z \end{smallmatrix};\tau)}$ reproduces the predicted 
$\frac{2\zeta_3}{3}\overline{ G(1;e^{2\pi i z})}$, and (\ref{inival.33}) would
then imply $ \overline{\alphaBR{0 &2 \\ 2 &4 \\ z& \emptyslot}{\tau}} \rightarrow 0$.

However, $\overline{ \alpha[\ldots;\tau]}$ at higher weight
will generically comprise $z$-independent $\overline{{\cal E}_0}$ of section \ref{sec:2.5.2}
which do not contribute to the leading terms. Hence, a leading-term analysis 
cannot suffice to propose expressions for all the individual $\overline{ \alpha[\ldots;\tau]}$.
Instead, we expect the study of Poincar\'e series with $z$-dependent
seed functions involving $\overline{{\cal E}_0}$ to yield well-defined 
$ \overline{\alphaBR{j_1 &\ldots &j_\ell \\ k_1 &\ldots &k_\ell \\ z_1&\ldots&z_\ell}{\tau}} $ 
for all entries $0\leq j_i\leq k_i{-}2$. For the $z$-independent case of 
$\overline{\alphaBR{j_1 &j_2 \\ k_1 &k_2 }{\tau}}$, this extension beyond MGFs has been 
achieved in \cite{Dorigoni:2021jfr, Dorigoni:2021ngn}, and we leave the analogous extensions of 
eMGFs for the future.

Note that the properties of $\beta^{\rm sv}$ will predict
vanishing leading terms for integration constants at depth two
with only one entry $z$ in the last line,
\beq
\overline{\alphaBR{j_1 &j_2 \\ k_1 &k_2  \\ z &\emptyslot}{\tau}}
 = {\cal O}(\bar q^{1-u}) \, , \ \ \ \
 \overline{\alphaBR{j_1 &j_2 \\ k_1 &k_2  \\ \emptyslot &z}{\tau}}
 = {\cal O}(\bar q^{1-u})
\eeq
When expanding the corresponding 
$\bsvtau{j_1 &j_2 \\ k_1 &k_2  \\ z &\emptyslot }$ as in (\ref{d2bsv}),
their holomorphic building blocks (\ref{hybrid.5}) only feature polylogarithms
of depth one in their leading terms. The corresponding single-valued
polylogarithms at depth one do not involve any MZVs from the ${\cal Z}^{\rm sv}$ in
(\ref{svpolycl}) and (\ref{svpolyex}), so there is no need for 
$\overline{\alphaBR{j_1 &j_2 \\ k_1 &k_2  \\ z &\emptyslot}{\tau}}$ to
contribute leading terms. 

An overview of non-vanishing leading terms $\overline{\alphaBR{j_1 &j_2 \\ k_1 &k_2  \\ z &z}{\tau}}$
at depth two and lattice weights $k_1{+}k_2 =7,8$ can be found in appendix \ref{app:baral.B}.

%%%%%%%%%%%%%%%%%%%%%%%%%%%%%%%%%%%%%%%%%%%%%%%%%%%%%%%%%%%
\subsection{Sieve algorithm for eMGFs}
\label{sec:9.2}

%%%%%%%%%%%%%%%%%%%%%%%%%%%%%%%%%%%%%%%%%%%%%%%%%%%%%%%%%%%

To study relations between eMGFs and construct the initial values $\widehat{Y}(i\infty)$
beyond the non-MZV sector of section \ref{sec:9.1}, we seek a systematic 
method for rewriting eMGFs in terms of combinations of real-analytic iterated KE 
integrals $\beta^{\mathrm{sv}}$. We restrict ourselves here to the case of dihedral 
eMGFs in one variable $z$ though the main ideas apply to trihedral and 
more general graphs as well. We will develop a method based on studying the differential 
operators $\nabla_\tau, \nabla_z$ and
sometimes $\overline{\nabla}_z$, which in a certain sense have the effect of \lq simplifying\rq~eMGFs:
By their action (\ref{gen.66Ctau}), (\ref{revsec.9}) and (\ref{revsec.11}) on
a generic dihedral eMGF, $ \nabla_z$ and $\overline{\nabla}_z$ decrease 
the entries in the second or first row for the exponents
of the lattice momenta. 

When acting with $\nabla_\tau$, however, the sum of the entries or lattice weight $|A|{+}|B|$ is preserved.
Nonetheless, the $\nabla_\tau$-action on eMGFs has the same simplifying features that were already
used in the sieve algorithm \cite{DHoker:2016mwo, DHoker:2016quv} for MGFs: By
lowering entries in the second row, $\nabla_\tau$ paves the road for applying 
momentum conservation, holomorphic subgraph reduction, and factorization identities
of eMGFs. These simplifications will be elaborated on later in this section and closely
follow the sieve algorithm for MGFs \cite{DHoker:2016mwo, DHoker:2016quv} 
(see \cite{Gerken:2020aju} for a {\sc Mathematica} package). However, we point 
out that the approach taken here for eMGFs goes beyond
the techniques for MGFs in view of the information we shall extract from $\nabla_z$. 

Let us start by taking $F(z,\tau) = \cplus{A \\ B \\Z}\! (\tau)$ to be a generic dihedral eMGF, for which we are interested in obtaining a representation in terms of $\beta^{\mathrm{sv}}$-functions. Like before, we will assume that all non-zero entries in the third row $Z$ are equal to the same variable $z$. Furthermore, we assume that in the rows $A$ and $B$ at most one entry is equal to zero (one could otherwise simplify via holomorphic subgraph reduction), and that none of the entries is negative. Otherwise, we may first simplify such eMGFs using lattice-sum relations (to be elaborated on below). We also remind the reader that single-column eMGFs vanish, and that in (\ref{invbsv.7}) an explicit 
$\beta^{\rm sv}$ representation at depth one was already given for two-column eMGFs,
i.e.\ Zagier's single-valued elliptic polylogarithm. 
Hence, we will assume next that $F$ is an eMGF with three or more columns.

To start, our approach will be to act with the $\nabla_\tau$ operator and to integrate back up, in order to find an expression in terms of $\beta^{\mathrm{sv}}$-functions up to a $\tau$-independent integration constant. The $\nabla_\tau$ operator is suitable for this, as its action is known
on both the lattice-sum representation of eMGFs and iterated KE integrals
$\beta^{\rm sv}$ from (\ref{gen.66Ctau}) and (\ref{poesec.1}), (\ref{poesec.1a}),
respectively. In the lattice-sum representation, 
$\nabla_\tau$ has the effect of producing a combination of eMGFs with lowered entries in the second row. In the following two scenarios, this leads to straightforward simplifications:
\begin{itemize}
  \item There are eMGFs produced with an entry equal to $-1$ in the second row.
  \item There are eMGFs produced with two zeros in the second row.
\end{itemize}
In the first scenario, we can \lq trade\rq~the entry $-1$ in the $m^{\rm th}$ column for a zero, 
using the momentum-conservation identity in the first or second row
\begin{align}
    \label{eq:momconsreordered}
    \cplus{A \\B \\Z}\!(\tau) 
    = -\sum_{n =1 \atop{n \neq m}}^R \cplus{A \\B+S_{m}-S_{n} \\Z}\!(\tau)
    = -\sum_{n = 1 \atop{n \neq m}}^R \cplus{A+S_{m}-S_{n} \\B \\Z}\!(\tau)
\end{align}
at fixed $m$, where $R$ is the number of columns as in (\ref{basic.15}).
In the second scenario, we may use the holomorphic-subgraph-reduction formulae 
given in (3.19) and (3.21) of \cite{Dhoker:2020gdz}, to rewrite eMGFs with two zeros 
in the second row in terms of eMGFs with one fewer column. Besides the two cases 
discussed above, there may be opportunities for additional simplifications. For example, 
lattice sums with zeros along the same column $a_i=b_i=0$, may be simplified using the 
factorization property (\ref{compap.5}). Furthermore, we may minimize the number of 
nonzero $z$'s in the third row using the translation and reflection identities in (\ref{compap.2})
and (\ref{compap.8}). 

Now, suppose that no (further) straightforward simplifications can be performed. In this case, we may take additional derivatives in $\tau$ until we reach one of the two scenarios described above. Therefore it is clear that the $\nabla_\tau$ derivative always leads us towards simpler eMGFs than the one we started from. Therefore, let us proceed in a recursive fashion. We will assume that the $\beta^{\mathrm{sv}}$ representation is known for all lattice sums that appear on the right-hand side of $\nabla_{\tau} F$, and we will describe a method for finding the $\beta^{\mathrm{sv}}$ representation of $F$ itself.

\subsubsection{Integrating up the $\tau$-derivative}
\label{sec:sievedz.0}

The first step is to find a primitive $\tilde{F}$ of $\nabla_{\tau} F$ in terms of iterated KE
integrals $\beta^\mathrm{sv}$ which may depart from $F$ by an integration constant
$C(u,v,\bar \tau)$ in the kernel of $\nabla_\tau$: 
\begin{align}
    \label{eq:preprimitive}
    F(z,\tau) = C(u,v,\bar \tau) +  \tilde{F}(z,\tau) \, , \ \ \ \ \ \ 
   \tilde{F}(z,\tau)=  \sum_{w} \xi_w(u,y)\beta^{\mathrm{sv}}[w ; \tau]
\end{align}
The sum is over words $w$ made out of columns 
$\bsv{\ldots &j &\ldots  \\ \ldots &k &\ldots  \\ \ldots &z &\ldots }$
or $\bsv{\ldots &j &\ldots  \\ \ldots &k &\ldots  \\ \ldots & &\ldots }$
such that as usual $0 \leq j \leq k{-}2$. The combinations of $G_k,f^{(k)}$ and
simpler eMGFs in the lattice-sum representation of $\nabla_{\tau} F$
imply that the coefficients $\xi_w(u,y)$ are Laurent polynomials in $y=\pi \Im \tau$ 
and polynomials in $u = \frac{ \Im z }{\Im \tau}$.

In fact, the integration constants $C(u,v,\bar \tau)$ in (\ref{eq:preprimitive})
can only depend on $u$: First of all, antimeromorphic functions of $\bar \tau$ 
will automatically be accounted for by the quantities $\overline{\alpha[\ldots]}$ 
in $\beta^{\rm sv}$ at depth $\geq 2$, see e.g.\
(\ref{eq:ebsv2}) or (\ref{d3bsvLT}). Hence, we can assume the integration constant
$C(u,v,\bar \tau)$ to be independent on $\bar \tau$. Moreover, both $F$
and $ \tilde{F}$ are invariant under the modular $T$-transformation
$\tau \rightarrow \tau{+}1$: This is manifest for the lattice sum defining $F$,
and all the constituents $y=\pi \Im \tau, \ u = \frac{ \Im z }{\Im \tau}$ and
$\beta^{\rm sv}$ of $\tilde F$ are individually $T$-invariant. In order for the
integration constant $C(u,v,\bar \tau)=C(u,v)$ to preserve $T$-invariance, 
it cannot depend on $v$ by its $T$-transformation
$v \rightarrow v{-}u$, and we actually have $C(u,v,\bar \tau)=C(u)$. This argument 
is specific to the one-variable case since two elliptic variables $z_1=u_1\tau{+}v_1$ 
and $z_2=u_2\tau{+}v_2$ would admit a modular invariant $u_1 v_2-u_2 v_1$.

We may construct a suitable primitive $\tilde{F}$ in (\ref{eq:preprimitive}) by studying the 
$\nabla_{\tau}$-derivatives (\ref{poesec.1}) and (\ref{poesec.1a}) of $\beta^{\mathrm{sv}}$.
It is clear that the effect of the differential operator is to increase the entries $j_i$ of $\beta^{\rm sv}$, 
or to strip off a column when the rightmost $j_\ell$ has the maximal value $k_\ell {-} 2$. Thus, given 
an integrand composed of $\beta^{\mathrm{sv}}$ and non-positive powers of $y$, we can 
construct a suitable ansatz for the 
primitive consisting of similar combinations of $\beta^{\mathrm{sv}}$-functions
(with downshifted entries $j_i$ and/or shifted powers of $y$) 
while appending columns in case of $f^{(k)}$- and $G_k$-kernels. Upon writing down an
 ansatz of suitable $\beta^{\mathrm{sv}}$ with generic coefficients, we may fix their coefficients 
$\xi_w(u,y)$ in (\ref{eq:preprimitive}) by taking a derivative and equating the result 
to the integrand. 

In summary, recursive application of the above procedure yields representations
\begin{align}
    \label{eq:Ftildeprimitive}
    F(z,\tau) = C(u) + \sum_{w} \xi_w(u,y)\beta^{\mathrm{sv}}[w ; \tau]
\end{align}
for dihedral eMGFs with known Laurent polynomials $\xi_w(u,y)$
and integration constants $C(u)$ to be discussed below. Only finitely 
many words $w$ representing the entries $j_i,k_i$ and $z_i$ of $\beta^{\rm sv}$ are summed over for a given eMGF $F(z,\tau)$.

\subsubsection{Fixing the $u$-dependence of the integration constant $C(u)$}
\label{sec:sievedz}

Next, we describe a method for determining the integration constant $C(u)$
in (\ref{eq:Ftildeprimitive}), up to a final additive constant $C_0$ that is independent 
of $\tau,\bar \tau, u$ and $v$. 
To proceed with fixing $C(u)$, we study the result of applying the derivative operator $\nabla_z$ 
at the cusp. This is motivated as follows: We know the action of $\nabla_z$ on the lattice-sum representation of eMGFs. However, we do not have a closed formula for the action of $\nabla_z$ on a generic $\beta^\mathrm{sv}$. Luckily, at the cusp the $\beta^\mathrm{sv}$-functions degenerate into terms consisting of Bernoulli polynomials in $u$ 
and powers of $y$, see section \ref{sec:3.3.3}. The action
of $\nabla_z$ on these degenerations follows from
\begin{align}
    \label{eq:delzonBP}
    \nabla_z B_k(u) = \overline{\nabla}_z B_k(u) = k B_{k-1} (u)\,, \quad \text{for $k \geq 0$}
\end{align}
and will allow us to make a comparison. 
Thus, let us consider: 
\begin{align}
    \label{eq:cusplimitfinz1}
    \nabla_z  F(z,\tau) \, \big|_{q^0 \bar q^0} =   \nabla_z C(u) +  \sum_{w}  
     \nabla_z\big(  \xi_w(u,y) \beta^{\mathrm{sv}}[w ; \tau]  \, \big|_{q^0 \bar q^0}  \big)
\end{align}
The operation $ \big|_{q^0 \bar q^0} $ is understood as discarding any positive
power of $q$ and $\bar q$ in the expansion of eMGFs and $\beta^{\rm sv}$ around the
cusp. We are also discarding terms $q^u,\bar q^u$ and are left with Laurent polynomials
in $y,u$ as in sections \ref{sec:2.4} and \ref{sec:3.3.3}. In particular, (\ref{eq:cusplimitfinz1})
is insensitive to the polylogarithmic contributions in the leading terms of $\beta^{\rm sv}$
discussed in sections \ref{sec:3.3.2} and \ref{sec:3.3.2ex} since any $G^{\rm sv}(\ldots,1;e^{2\pi i z})$
is suppressed by $q^u$ or $\bar q^u$ at the cusp.

The key idea to determine the integration constant $C(u)$ is to
simplify the left-hand side of (\ref{eq:cusplimitfinz1})
from its lattice-sum representation and its $z$-derivatives (\ref{revsec.9}),
\begin{align}
 \nabla_z F(z,\tau) &=  \nabla_z\, \cplus{A \\ B \\Z}\! (\tau)  = 2 i   \sum_{r=1}^{R} \frac{\partial z_{r}}{\partial z} \cplus{A\\B-S_{r} \\Z}\! (\tau)
      \label{eq:cusplimitfinz2}
\end{align}
which is accessible without truncating to the leading terms $q^0 \bar q^0$
in an expansion around the cusp. 
By the recursive nature of our procedure, we may assume that all eMGFs on the right-hand 
side are already known in terms of $\beta^{\mathrm{sv}}$ since the lattice weight is 
reduced to $|A|{+}|B|{-}1$ by action of $\nabla_z$. In some cases, the lattice sums
resulting from $\nabla_z F$ may not be absolutely convergent or even diverge. These
situations will be considered in section \ref{sec:edgecase} below, and we will for the moment
assume that the lattice sums on the right-hand side of (\ref{eq:cusplimitfinz2}) are
absolutely convergent.

By identifying (\ref{eq:cusplimitfinz1}) with the $q^0 \bar q^0$-terms of (\ref{eq:cusplimitfinz2}), 
one can solve for $\nabla_z C(u)$. We may integrate the expression and
obtain $C(u) = \tilde{C}(u) + C_0$, where $ \tilde{C}(u)$ is a known primitive of
$\nabla_z C(u)$ subject to $\tilde C(0)$=0. Hence, the only missing information
resides in the additive constant $C_0 = C(u{=}0)$ independent of $\tau,\bar \tau, u,v$. 
Plugging the result for $C(u)$ into (\ref{eq:Ftildeprimitive}), 
straightforwardly gives:
\begin{align}
    F(z,\tau) =C_0 + \tilde{C}(u)+\sum_{w} \xi_{w}(u,y) \beta^{\mathrm{sv}}[w ; \tau]
    \label{sieveid}
\end{align}
where $C_0$ is the only remaining undetermined constant. 

Note that this procedure may be repeated for the $\overline{\nabla}_z$ derivative, 
in order to check for consistency: On the one hand, the action of $
\nabla_z=  \partial_u - \bar \tau \partial_v$ and $\overline{\nabla}_{ z}  = \partial _u - \tau \partial_v$
is identical on the $q^0\bar q^0$-expressions that only depend on $u$ but not on $v$.
On the other hand, the lattice sums encountered in intermediate steps are very different
when employing the antimeromorphic derivative $ \overline{\nabla}_z \, \cplus{A\\B\\Z}$ 
in (\ref{revsec.11}) instead of (\ref{eq:cusplimitfinz2}).

\subsubsection{Fixing the final integration constant $C_0$}
\label{sec:sievedz.2}

To fix the last unknown $C_0$ in (\ref{sieveid}), we will study the consecutive 
limit where we first set $z\rightarrow 0$ and then
expand around the cusp. In the lattice-sum representation this gives
\begin{align}
    \label{eq:emgfcuspvanishingzsieve}
  \Big( \lim_{u,v  \rightarrow 0}  F(z,\tau) \Big)  \, \Big|_{q^0 \bar q^0} 
  =   \Big( \lim_{u,v \rightarrow 0}  \cplus{A \\ B \\ Z}\! (\tau) \Big) \, \Big|_{q^0 \bar q^0} 
   = \cplus{A \\ B \\ 0}\! (\tau) \, \Big|_{q^0 \bar q^0}
\end{align}
where the subleading terms at $z=0$ are $q^1$ and $\bar q^1$ rather than $q^u$
and $\bar q^u$. In the limit $z\rightarrow 0$, eMGFs degenerate to MGFs whose
Laurent polynomials can be imported from the 
literature \cite{DHoker:2015gmr, DHoker:2016quv, DHoker:2017zhq},
see the {\sc Mathematica} package \cite{Gerken:2020aju} for all dihedral 
and trihedral cases up to and including $|A|{+}|B| = 12$. For the moment, the $z\rightarrow0$ limit
(\ref{eq:emgfcuspvanishingzsieve}) is assumed to be finite, and we will comment
on scenarios with logarithmic divergences in section \ref{sec:edgecase} below.

The counterpart of (\ref{eq:emgfcuspvanishingzsieve}) in the 
$\beta^\mathrm{sv}$ representation of eMGFs can be written as
\begin{align}
    \label{eq:emgfcuspvanishingzsieve2}
 \Big( \lim_{u,v  \rightarrow 0}  F(z,\tau) \Big)  \, \Big|_{q^0 \bar q^0} = C_0 + \bigg(
 \lim_{u,v \rightarrow 0} \sum_{w}    \xi_w(u,y) \beta^{\mathrm{sv}}[w ; \tau] \bigg) \, \Big|_{q^0 \bar q^0}
\end{align}
which follows from (\ref{sieveid}) with $\tilde C(0)=0$.
The bracketing on the right-hand side indicates that we now encounter the
{\it extended Laurent polynomials} of section \ref{sec:3.ext}: By taking
the limit $z\rightarrow 0$ {\it before} expanding around the cusp, the polylogarithms
in the leading terms of the $\beta^{\rm sv}$ (see sections \ref{sec:3.3.2} and
\ref{sec:3.3.2ex}) become of order $q^0  \bar q^0$ instead of series in $q^u$ and $\bar q^u$. 
In this way, the right-hand side of (\ref{eq:emgfcuspvanishingzsieve2}) generically
features the single-valued MZVs (\ref{svMZVlim}) from the contributions
$G^{\rm sv}(\ldots,1;e^{2\pi i z})$ to the leading terms at $z=0$. At depth one and two, for
instance, the explicit form of these leading terms can be assembled from (\ref{dpt1LT})
and by applying the substitution rule (\ref{svMZVlim}) to (\ref{d2bsv}), respectively.
Even though individual $\beta^{\mathrm{sv}}[w ; \tau]$ may introduce poles in $u$
into the coefficients of $G^{\rm sv}(\ldots,1;e^{2\pi i z})$, the sums over words $w$ 
in (\ref{eq:emgfcuspvanishingzsieve2}) realized by eMGFs will be manifestly free of negative
powers of $u$, see for instance the discussion around (\ref{comrel.36}).

By equating (\ref{eq:emgfcuspvanishingzsieve2}) with 
(\ref{eq:emgfcuspvanishingzsieve}) and importing the Laurent polynomials of
MGFs for $\cplus{A \\ B \\ 0}\! (\tau) \, \Big|_{q^0 \bar q^0}$, we may solve for $C_0$. 
As long as the Laurent polynomials of MGFs in (\ref{eq:emgfcuspvanishingzsieve}) exclusively feature 
single-valued MZVs as conjectured in \cite{Zerbini:2015rss, DHoker:2015wxz}, it 
follows from induction and (\ref{svMZVlim}) that also the desired
constants $C_0$ are single-valued MZVs.
This concludes the procedure of rewriting $F$ in terms of $\beta^{\mathrm{sv}}$-function. 
We will refer to this method as the sieve algorithm for eMGFs.

Unfortunately, there are some cases where the above algorithm does not work straightforwardly. For example, the $z\rightarrow 0$ limit of an eMGF may be plagued by logarithmic divergences, in which 
case we can not match onto the Laurent polynomial of an MGF. This happens when the eMGF contains pairs of (not necessarily adjacent) columns of the form:
\begin{align}
    \left[\begin{smallmatrix}
\ldots &1 & 0 &\ldots \\ \ldots &0 & 1 &\ldots \\ \ldots &z & 0 &\ldots
\end{smallmatrix}\right] \text{ or } \left[\begin{smallmatrix}
\ldots &1 & 0 &\ldots \\ \ldots &0 & 1 &\ldots \\ \ldots &0 & z &\ldots
\end{smallmatrix}\right]
\label{gfblock}
\end{align}
In this case, the $\nabla_z$ derivative is also divergent, so we are not able to fix the $u$-dependent part of $C(u)$ in the manner that was described above. 
We take a deeper dive into a variety such cases in section
\ref{sec:edgecase} below.

\subsubsection{Worked out example}
\label{sec:sievedz.3}

As an example of the sieve algorithm described in this section, we shall
determine the $\beta^{\rm sv}$ representation of the three-column eMGF
$C_{1|1,1}(z|\tau)$ defined by (\ref{cabc.2}). In order to illustrate the salient
points, we assume that the simpler objects
\begin{align}
\pi \nabla_\tau C_{1|1,1}(z|\tau)  &= \dplus{2  \\ 1 }\! (z|\tau)^2
+\frac{1}{3}\pi \nabla_\tau E_3(\tau) +\frac{2}{3}\pi \nabla_\tau g_3(z|\tau)
\notag \\
 &= - 2 \bsvtau{1& 1\\3& 3\\z& z} + 5 \bsvtau{3\\6\\ \emptyslot}
 + 10 \bsvtau{3\\6\\ z} - \frac{ \zeta_5}{2y}  \label{svexpl.1} \\
\nabla_z C_{1|1,1}(z|\tau) &= - 8 \bsvtau{2 \\ 5 \\ z} + 4 \bsvtau{1 &0 \\ 3 &2 \\ z &z} + 4 \zeta_3 B_1(u)
\notag
\end{align}
have already been determined by other means.\footnote{One can for instance 
apply the sieve algorithm to $\pi \nabla_\tau C_{1|1,1}(z|\tau)$ and deduce
the $\beta^{\rm sv}$ representation in (\ref{svexpl.1}) from holomorphic subgraph
reduction.} 

The first step of the sieve algorithm in section \ref{sec:sievedz.0} is
to find the primitive
\beq
C_{1|1,1}(z|\tau)  =
C(u) + 8 \bsvtau{1& 0\\3& 3\\z& z}
- 10 \bsvtau{2\\6\\ \emptyslot} 
- 20 \bsvtau{2\\6\\z}
+\frac{ \zeta_{5}}{4 y^2} 
 \label{svexpl.2}
\eeq
of $\pi \nabla_\tau C_{1|1,1}(z|\tau)$ in (\ref{svexpl.1}) w.r.t.\ $\tau$ which is easily
checked via (\ref{nabbsv.2}) and $\pi \nabla(y^m) = m y^{m+1}$.

As a second step, we determine the $u$-derivative of the integration constant
$C(u)$ in (\ref{svexpl.2}) by the method of section \ref{sec:sievedz}: The Laurent polynomial
of $\nabla_z C_{1|1,1}(z|\tau)$ in (\ref{svexpl.1}) is given by
\beq
\nabla_z C_{1|1,1}(z|\tau) =  - y^3\bigg( \frac{16}{5} B_5(u) + \frac{16}{9} B_3(u)\bigg)   + 4 \zeta_3 B_1(u) +{\cal O}(q^u,\bar q^u)
 \label{svexpl.3}
\eeq
where we have employed the techniques of section \ref{sec:3.3.3} to extract the Laurent polynomials
of the $\beta^{\rm sv}$. Given that $B_1(u)=u-\frac{1}{2}$,
this is consistent with the $u$-derivative of
\beq
C_{1 | 1,1}(z|\tau) =
C(u)- y^3 \bigg( \frac{8 }{15} B_{6}(u) +\frac{4 }{9}  B_{4}(u)\bigg) + \frac{ \zeta_{5}}{4 y^2}
+{\cal O}(q^u,\bar q^u)
 \label{svexpl.4}
\eeq
if the integration constant in (\ref{svexpl.2}) is chosen as $C(u)= 2\zeta_3(u^2 - u)+C_0$.

Finally, we follow section \ref{sec:sievedz.2} and determine $C_0$ by 
demanding that the well-known Laurent polynomial of the MGF
\beq
C_{1 ,1,1}(\tau)=C_{1 | 1,1}(0|\tau) =
\frac{ 2 y^3}{945} + \zeta_3  + \frac{3 \zeta_5}{4y^2} +{\cal O}(q,\bar q)
 \label{svexpl.6}
\eeq
is reproduced from the leading terms of (\ref{svexpl.2}),
\begin{align}
&C_{1|1,1}(z|\tau)  =
C(u)- y^3 \bigg( \frac{8 }{15} B_{6}(u) +\frac{4 }{9}  B_{4}(u)\bigg) + \frac{ \zeta_{5}}{4 y^2}\notag \\
&\quad + 8\, \bigg\{  {-}  \frac{y}{6} B_3(u)   G^{\rm sv}(0,1;e^{2\pi i z}) +
\frac{  G^{\rm sv}(0,1;e^{2\pi i z})^2 }{32 y} +  \bigg({-} \frac{ 1}{24}  +
 \frac{ u}{8}    - \frac{ u^2}{12} \bigg)  G^{\rm sv}(0,0,1;e^{2\pi i z}) \notag \\
 &\quad \quad \ \ \ \ +
\frac{ u}{4}  G^{\rm sv}(0,1,1;e^{2\pi i z}) \bigg\}   -20\, \bigg\{  \frac{ G^{\rm sv}(0,0,1,0,0;e^{2\pi i z}) }{480 y^2} \bigg\} +{\cal O}(q^{1-u},\bar q^{1-u})
 \label{svexpl.7}
\end{align}
The single-valued polylogarithms in the leading terms of $\bsvtau{1& 0\\3& 3\\z& z}$ and
$\bsvtau{2\\6\\z}$ can be found in (\ref{dpt1LT}) and (\ref{comrel.32}), respectively.
At $z=0$, the leading terms in (\ref{svexpl.7}) reduce to
\begin{align}
C_{1|1,1}(0|\tau)  &= C_0 + \frac{2 y^3}{945} + \frac{ \zeta_5 }{4y^2}
 - \frac{G^{\rm sv}(0,0,1;1)}{3} 
 + \frac{ G^{\rm sv}(0,1;1)^2 }{4y}
  - \frac{G^{\rm sv}(0,0,1,0,0;1)}{24y^2} 
 +{\cal O}(q,\bar q) \notag \\
 &=  \frac{2 y^3}{945} + C_0 + \frac{2}{3} \zeta_3 + \frac{3 \zeta_5 }{4y^2}
  +{\cal O}(q,\bar q) 
 \label{svexpl.8}
\end{align}
where we have used $ G^{\rm sv}(0,1;1)= 0,\ G^{\rm sv}(0,0,1;1)=-2\zeta_3$ and 
$G^{\rm sv}(0,0,1,0,0;1)=-12 \zeta_5$ in passing to the second line.
As a consistency check of our method, the orders $y^3$ and $y^{-2}$ 
are observed to match (\ref{svexpl.6}). By demanding the $y^0$ orders 
of (\ref{svexpl.6}) and (\ref{svexpl.8}) to agree, we infer
the final integration constant $ C_0=\frac{1}{3} \zeta_3$
such that $C(u)$ in (\ref{svexpl.4}) can be recombined to
the Bernoulli polynomial
\begin{align}
C(u) = 2\zeta_3\bigg(u^2 - u + \frac{1}{6} \bigg) = 2 \zeta_3 B_2(u)
 \label{svexpl.9}
\end{align}
In this way, we arrive at the $\beta^{\rm sv}$ representation of
$C_{1|1,1}(z|\tau)$ previewed in (\ref{prevex.1}).

\subsection{Edge cases for the sieve algorithm}
\label{sec:edgecase}

In this section, we discuss how to obtain the $\beta^{\mathrm{sv}}$ representation of eMGFs
where a naive application of the sieve algorithm introduces ill-defined lattice sums in intermediate
steps. These situations will be referred to as {\it edge cases}. In each case, the obstacle is due 
to lattice sums that do not converge absolutely on the entire torus and do
not necessarily obey the same identities as convergent ones.

While the regularization of iterated integrals is well-understood, see for instance
\cite{DeligneTBP, Brown:mmv, Panzer:2015ida} or section 4.3 of the review \cite{Abreu:2022mfk}, 
the procedure is more difficult for divergent lattice sums. This is 
discussed for example in \cite{Gerken:2020aju} and section 5.6 of
\cite{Gerken:review} for the case of MGFs. Our approach will be to avoid
introducing divergent lattice sums at any stage of the computation, which requires the need for 
small modifications to the sieve algorithm in certain cases.

\subsubsection{Prototypical example}
\label{sec:edgecase.1}

Let us illustrate the type of problems we may encounter using a simple example:
\begin{align}
    F(z,\tau) = \cplus{1 & 0 & 1 \\0 & 1 & 1 \\ z & 0 & 0}\! (\tau)
    \label{edgeeq.01}
\end{align}
Upon taking a derivative in $\nabla_z$ or $\overline{\nabla}_z$, we get a
pair of divergent lattice sums via
\begin{align}
 \cplus{1 & 0 & 1 \\-1 & 1 & 1 \\ z & 0 & 0}\! (\tau)
 = -  \cplus{1 & 0 & 1 \\0 & 0 & 1 \\ z & 0 & 0}\! (\tau) -  \cplus{1 & 0 & 1 \\0 & 1 & 0 \\ z & 0 & 0}\! (\tau)
    \label{edgeeq.00}
\end{align}
which prohibits us from applying the sieve algorithm as in section \ref{sec:9.2}. In 
fact, we even run into problems when looking at the $\nabla_\tau$ operator. For example, we have that:
\begin{align}
    \pi \nabla_\tau F(z,\tau) = \cplus{0 & 1 & 2 \\1 & 0 & 0 \\
0 & z & 0}\! (\tau)+\cplus{0 & 1 & 2 \\1 & 1 & -1 \\0 & 0 & z}\! (\tau)
\label{edgeeq.02}
\end{align}
The first eMGF on the right-hand side contains the block in (\ref{gfblock})
and shares the convergence properties of the closed-string Green function $g(z|\tau)
= \dplus{1 \\ 1} \! (z|\tau)$ with lattice-sum representation (\ref{elemlattice}) by the second 
Kronecker limit formula. Indeed, the eMGF (\ref{edgeeq.01}) under discussion will be found
in (\ref{eq:momentumconservationidentitytwoloops}) below to contain an 
additive term $\sim g(z|\tau)^2$. Since the first eMGF 
on the right-hand side of (\ref{edgeeq.02}) has two 
entries equal to zero in the second row, 
we may apply holomorphic subgraph reduction to obtain:
\begin{align}
    \cplus{0 & 1 & 2 \\1 & 0 & 0 \\ 0 & z & 0}\! (\tau) &= 2\dplus{3 \\ 1} \! (0|\tau) +\dplus{3 \\ 1} \! (z|\tau)+    (\Im \tau)^2 f^{(2)}(z|\tau) g(z|\tau) \notag \\
    &\quad -  \Im \tau f^{(1)}(z| \tau) \dplus{2 \\ 1} \! (z|\tau)
    \label{edgeeq.03}
\end{align}
The second eMGF on the right-hand side of (\ref{edgeeq.02}) is more problematic. It has an entry equal to $-1$ that we would like to rewrite using the momentum-conservation identity in (\ref{eq:momconsreordered}). However, this results in two divergent lattice
sums on the right-hand side of
\begin{align}
    \cplus{0 & 1 & 2 \\1 & 1 & -1 \\0 & 0 & z}\! (\tau) = -\cplus{0 & 1 & 2 \\0 & 1 & 0 \\0 & 0 & z}\! (\tau)
    -\cplus{0 & 1 & 2 \\1 & 0 & 0 \\0 & 0 & z}\! (\tau)
    \label{edgeeq.04}
\end{align}
We aim to avoid introducing divergent lattice sums, as they may obey unexpected relations 
(see \cite{Gerken:2020aju} and section 5.6 of
\cite{Gerken:review} for a discussion in the case of MGFs).

In the present case, the appearance of divergent lattice sums can be
sidestepped with a simple trick. Straightforward application of momentum conservation
to the first line yields
\begin{align}
    F(z,\tau) = \cplus{1 & 0 & 1 \\ 0 & 1 & 1 \\ z & 0 & 0}\! (\tau) 
    = -\cplus{0 & 1 & 1 \\0 & 1 & 1 \\z & 0 & 0}\! (\tau)
    -\cplus{0 & 1 & 1 \\1 & 0 & 1 \\0 & z & 0}\! (\tau)
    \label{edgeeq.05}
\end{align}
Since $|A|{+}|B|$ is even we may use that:
\begin{align}
    \cplus{0 & 1 & 1 \\1 & 0 & 1 \\0 & z & 0}\! (\tau) 
    = \overline{\cplus{0 & 1 & 1 \\1 & 0 & 1 \\0 & z & 0}\! (\tau)}
    = \cplus{0 & 1 & 1 \\1 & 0 & 1 \\ -z & 0 & 0}\! (\tau) 
    = \cplus{0 & 1 & 1 \\1 & 0 & 1 \\ z & 0 & 0}\! (\tau)
    \label{edgeeq.06}
\end{align}
Hence, upon rearranging the equations above we find that
\begin{align}
    \label{eq:momentumconservationidentitytwoloops}
    F(z,\tau) = -\frac{1}{2}\cplus{0 & 1 & 1 \\0 & 1 & 1 \\z & 0 & 0}\! (\tau) = \frac{1}{2}E_2(\tau) - \frac{1}{2}g(z|\tau)^2
\end{align}
where we have applied factorization (\ref{compap.5}) in the last step.
This example illustrates that a naive application of the sieve algorithm
may lead to difficulties even if the lattice sum under investigation
can be expressed in terms of depth-one eMGFs by means of algebraic identities.
And we certainly aim for extensions of the sieve algorithm to eMGFs that are
polynomials in the closed-string Green functions and eMGFs with a finite
$z \rightarrow 0$ limit as in (\ref{eq:momentumconservationidentitytwoloops}).

\subsubsection{Basic strategies for avoiding divergent lattice sums}
\label{sec:edgecase.2}

It is clear from the above that in certain cases the sieve algorithm fails, and here we discuss some first mitigation strategies. Let us denote the sieve algorithm by $S$, so that $S(F)$ gives the $\beta^\mathrm{sv}$ representation of an eMGF $F = \cplus{A \\ B \\ Z}$ when possible. If the sieve algorithm fails due to the appearance of divergent lattice sums, we let $S(F)$ return a failure message. 

\paragraph{Scenario: The $\nabla_\tau$ derivative gives divergent lattice sums (after simplifications):} Suppose that $S(F)$ initially fails, because the simplification of the $\nabla_\tau$-derivative gives divergent lattice sums. In this case, we look at the zeros in the entries of $A = (a_1,\ldots,a_R)$ and $B = (b_1,\ldots,b_R)$, and for each zero we generate a momentum-conservation identity of the form of (\ref{eq:momconsreordered}). In each generated identity, we solve for the original eMGF if it appears more than once. Thereafter, we try for each generated identity to apply the sieve algorithm on the new eMGFs until we succeed. To avoid running into a loop, we keep track of the eMGFs which we have already attempted to compute before.

\paragraph{Scenario: The $\nabla_z$-derivative gives divergent lattice sums:} Suppose that the sieve algorithm fails due to the appearance of divergent lattice sums coming from the $\nabla_z$ operator. The first thing to try is to fix $C(u)$ from the $\overline{\nabla}_z$ operator instead, see the
comment below (\ref{sieveid}). If both $\nabla_z$ and $\overline{\nabla}_z$ produce divergent lattice sums, we may attempt to take the derivative on a different representation of the lattice sum. In particular, we may apply translations $z_j \rightarrow z_j {-} z$ as in (\ref{compap.2}), and the reflection identity
(\ref{compap.8}), to obtain a form of the eMGF which has the $z$-arguments in different slots. Lastly, if neither of these tricks work, we may attempt to run the sieve algorithm on the complex conjugate eMGF $\overline{F}$. Since complex conjugation commutes with the
expansions around the cusp employed in section \ref{sec:9.2}, 
the integration constant $C(u)$ in $F$ is determined by $S(\overline{F})$.

\subsubsection{Regularization via extra punctures and combining $\nabla_z$, $\overline{\nabla}_z$}
\label{sec:edgecase.3}

For cases beyond the reach of the basic strategies outlined above, we choose to deform the eMGFs via additional punctures that regulate certain problematic lattice sums at intermediate stages. This approach will be illustrated via concrete examples in this section, and allows us to extend the Sieve algorithms to dihedral one-variable eMGFs with seven columns and $|A| + |B| = 10$.

\paragraph{Four-column eMGFs:} Our key example is the
following four-column eMGF in one variable at $|A|=|B|=3$:
\begin{align}
    F(z,\tau) = \cplus{0 & 1 & 1 & 1 \\ 1 & 0 & 1 & 1 \\ z & 0 & 0 & z}\! (\tau)
\end{align}
For this eMGF, we obtain a combination of divergent lattice sums after taking a derivative 
w.r.t.\ $\nabla_z$ or $\overline{\nabla}_z$. The same issue arises when trying to simplify 
$\nabla_\tau F(z,\tau)$ by applying momentum conservation to resolve an entry $-1$. 

Next, we show in detail how to simplify this eMGF using a combination of two tricks. First, let us resolve intermediate divergences in the $\nabla_\tau$-derivative. To do this, we add an additional puncture $z_2$, and we will take the limit $z_2\rightarrow z$ later on. In particular, consider the deformed eMGF
\begin{align}
    \tilde{F}(z,z_2,\tau) = \cplus{0 & 1 & 1 & 1 \\ 1 & 0 & 1 & 1 \\ z & 0 & 0 & z_2}\! (\tau)
\end{align}
which clearly reduces to $F(z,\tau) $ as $z_2\rightarrow z$.
We then have that: 
\begin{align}
    \label{eq:Ftildedeformedderivative}
   \pi &\nabla_\tau \tilde{F}(z,z_2,\tau) =  2 ( \Im \tau)^{3}  f^{(3)}(z_2| \tau) \dplus{1 \\2}\!(z | \tau)
   - (\Im \tau)^2 \cplus{0 & 1 & 1 \\1 & 0 & 1 \\z & 0 & 0}\! (\tau) f^{(2)}(z_2|\tau)
 \nonumber \\
    & \quad -  (\Im \tau)^2 \cplus{0 & 1 & 1 \\1 & 0 & 1 \\z & 0 & z_2}\! (\tau) f^{(2)}(z_2| \tau)
     - \cplus{0 & 1 & 3 \\ 1 & 1 & 0 \\ z & 0 & z_2}\! (\tau)
     +\cplus{0 & 1 & 3 \\ 1 & 1 & 0 \\ z & 0 & 0}\! (\tau)
     +\cplus{1 & 1 & 2 \\ 1 & 1 & 0 \\ 0 & z_2 & 0}\! (\tau) \nonumber \\ 
    & \quad +(\Im \tau)^2 g(z| \tau)  g(z{-}z_2| \tau) \big[f^{(2)}(z| \tau)-f^{(2)}(z_2|\tau) \big]
\end{align}
after applying a number of lattice-sum simplifications. Of particular relevance is the last term 
whose factor of $g(z{-}z_2|\tau)$ diverges logarithmically as $z_2 \rightarrow z$. 
However, the divergence is canceled out by 
the accompanying difference $f^{(2)}(z| \tau)-f^{(2)}(z_2|\tau)$, which vanishes
at a faster rate, i.e.\ linearly in $z{-}z_2$ and $\bar z {-} \bar z_2$.

Now, previously all the $\beta^{\mathrm{sv}}$ representations of two-loop lattice sums were determined, so we know the $\beta^{\mathrm{sv}}$ representation of (\ref{eq:Ftildedeformedderivative}) after taking the limit. We may integrate up using a suitable ansatz, and find that:
\begin{align}
&F(z,\tau)=C(u) -\frac{\zeta_3 }{y} \bsvtau{0 \\2 \\z}+\frac{1}{3} \bsvtau{0 \\2 \\z}^{3}\! +6 \bsvtau{1 \\4 \\
\emptyslot} \bsvtau{0 \\2 \\z}-6 \bsvtau{1 \\4 \\z} \bsvtau{0 \\2 \\z}\nonumber \\
&\quad  +12 \bsvtau{0 \\3 \\z} \bsvtau{1 \\3 \\z}+20 \bsvtau{2 \\6 \\z}-16 \bsvtau{0 & 1 \\3 & 3 \\z & z}-30 \bsvtau{2 \\6 \\ \emptyslot}+\frac{3 \zeta_5}{4 y^{2}}
\label{eq:edgecase3loopF}
\end{align}
It remains to determine the integration constant $C(u)$. Unfortunately, we are unable to fix it from a derivative in $z$, as this will yield divergent lattice sums. Instead, we use the following trick. First, we
identify a primitive of the target eMGF w.r.t.\ $\overline{\nabla}_z$:
\begin{align}
    \label{eq:nablaztrick}
   \overline{\nabla}_z \cplus{1 & 1 & 1 & 1 \\0 & 1 & 1 & 1 \\0 & 0 & z & z}\! (\tau) =
    4\pi i \Im \tau \, \cplus{0 & 1 & 1 & 1 \\ 1 & 0 & 1 & 1 \\ z & 0 & 0 & z}\! (\tau)=
    4\pi i \Im \tau\,  F(z,\tau)
\end{align}
Next, we note that the sieve algorithm can be applied to the primitive 
$\cplus{1 & 1 & 1 & 1 \\0 & 1 & 1 & 1 \\0 & 0 & z & z}$ without any difficulties. 
In particular, the $\nabla_z$-derivative gives an eMGF which can be simplified 
using holomorphic subgraph reduction. Furthermore, we do not encounter any 
intermediate divergences when taking the $\nabla_\tau$-derivative and simplifying. 
Unfortunately we do not have closed formulae for the $\overline{\nabla}_z$-derivative of the $\beta^{\mathrm{sv}}$ functions, so we can not use (\ref{eq:nablaztrick}) directly to obtain the $\beta^\mathrm{sv}$ representation of $\cplus{0 & 1 & 1 & 1 \\ 1 & 0 & 1 & 1 \\ z & 0 & 0 & z}$. However, we can first take the Laurent polynomial and then perform the $\overline{\nabla}_z$-derivative in order to fix $C(u)$. This way, we find that the missing piece is:
\begin{align}
    \label{eq:edgecase3loopFC}
    C(u) = 4 \zeta_3 B_{2}(u)+\frac{\zeta_3}{3}
\end{align}
The $\beta^{\rm sv}$ representation in (\ref{eq:edgecase3loopF}) together
with (\ref{eq:edgecase3loopFC}) implies the following simplification 
\begin{align}
\cplus{0 & 1 & 1 & 1 \\1 & 0 & 1 & 1 \\0 & z & 0 & z}\! (\tau) &=
2 \cplus{1 & 1 & 1 \\1 & 1 & 1 \\z & 0 & 0}\! (\tau)
+\frac{1}{3}\cplus{1 & 1 & 1 \\1 & 1 & 1 \\0 & 0 & 0}\! (\tau)
- \dplus{1 \\2}\!(z | \tau) \dplus{2 \\1}\!(z | \tau)\nonumber \\ &\quad 
+ E_{2}(\tau) g(z|\tau)
- \frac{1}{3} g(z| \tau)^{3}- g_{2}(z|\tau) g(z|\tau)-2 g_{3}(z|\tau)
\end{align}
to lattice sums with at most three columns on the right-hand side. This can be verified
by inserting the $\beta^{\rm sv}$ representations of the right-hand side and viewed
as a weight-three analogue of Basu's identity (\ref{teas.9}).

\paragraph{Beyond four columns:}
Using the central tricks in the previous section, 
\begin{itemize}
\item deforming $z \rightarrow z_2$ for some of the entries of $Z$ in $\cplus{A \\ B \\Z}$
to regularize individual terms in the $\nabla_\tau$-derivative
\item identifying a primitive w.r.t.\ either $\overline{\nabla}_z$ or 
$\overline{\nabla}_z$ in order to deduce the Laurent polynomial from a
better-behaved eMGF
\end{itemize}
we derived various special cases up to and including seven columns and $|A|{+}|B| = 10$. 
%The derivation of these cases is delegated to a {\sc Mathematica} notebook which can be found in the ancillary files.

\newpage

%%%%%%%%%%%%%%%%%%%%%%%%%%%%%%%%%
%%%%%%%%%%%%%%%%%%%%%%%%%%%%%%%%%
%%%%%%%%%%%%%%%%%%%%%%%%%%%%%%%%%
\section{Explicit bases of eMGFs at various lattice weights}
\label{bassec}
%%%%%%%%%%%%%%%%%%%%%%%%%%%%%%%%%
%%%%%%%%%%%%%%%%%%%%%%%%%%%%%%%%%
%%%%%%%%%%%%%%%%%%%%%%%%%%%%%%%%%

This section aims to present explicit bases of indecomposable
one-variable eMGFs $\cplus{A \\ B \\Z}$ with lattice weights $|A|{+}|B|\leq 10$ 
that cannot be reduced to combinations of MZVs and/or simpler eMGFs.
For all of these basis elements, we shall give representations in terms of both 
$\beta^{\rm sv}$ and dihedral lattice sums, and a machine-readable form of our
results can be found in the ancillary file. The existence of a dihedral basis choice
implies that all eMGFs of trihedral or more complicated topologies
at lattice weights $|A|{+}|B|\leq 10$ can be
expressed in terms of dihedral representatives.

The choice of bases is of course far from unique, and we will rely on the
following guiding principles, or the maximal set of mutually compatible ones:
\begin{align}
\textrm{(i)} &\ \textrm{prefer dihedral eMGFs (\ref{gen.66}) with small numbers $R$ of columns}
\notag \\
\textrm{(ii)} &\ \textrm{for a given number of columns, pick eMGFs (\ref{cabc.1}) with identical holomorphic} \notag \\ 
&\quad \textrm{and antiholomorphic exponents $a_j=b_j$ whenever possible}
\notag \\
\textrm{(iii)} &\ \textrm{for non-zero modular weight $|A| \neq |B|$, pick 
$\nabla_\tau,\overline{\nabla}_\tau$- and $\nabla_z, \overline{\nabla}_z$
derivatives}
\notag \\
&\quad \textrm{of modular invariant eMGFs whenever possible}
\label{guiding} \\
\textrm{(iv)}&\ \textrm{in modular invariant sectors with $|A| = |B|$, pick basis elements
that are even}
\notag \\
&\quad \textrm{or odd under the automorphism $(z,\tau) \rightarrow ({-}\bar z,{-}\bar \tau)$ 
explained in section \ref{bassec.1}}
\notag \\
\textrm{(v)}&\ \textrm{for modular invariant basis elements $|A| = |B|$, delay the appearance}
\notag \\
&\quad \textrm{of factors $f^{(k)}(z|\tau)$ or $G_k(\tau)$ to higher $\nabla_\tau$-derivatives}
\notag
\\
\textrm{(vi)}&\ \textrm{non-singular limit $z \rightarrow 0$}
\notag
\end{align}
The analogous bases of indecomposable {\it MGFs} with similar guiding principles 
can be found in \cite{Gerken:2020yii} for $|A|{+}|B|\leq 10$ and in \cite{Gerken:2020aju} 
for $|A|{+}|B|=12$.

The bases of indecomposable eMGFs at given $|A|=a$ and $|B|=b$ will be denoted by
${\cal V}_{a,b}$. In any sector with $a,b\geq 1$, the first guiding principle
(i) in (\ref{guiding}) requires the single-valued elliptic polylogarithms $\dplus{a \\ b}\!(z|\tau)$
in (\ref{basic.13}) to be basis elements, i.e.\ that $\dplus{a \\ b}\!(z|\tau) \in {\cal V}_{a,b}$.
Moreover, the MGFs obtained from the special values at $z \in \ZZ {+} \tau\ZZ$
are considered as independent eMGFs but vanish for odd $a{+}b$. Hence,
we have $\dplus{a \\ b}\!(0|\tau) \in {\cal V}_{a,b}$ if $a{+}b \in 2\mathbb N$
and $a{+}b \geq 4$ to avoid the logarithmic singularity of $g(z|\tau) = \dplus{1 \\ 1}\!(z|\tau)$
at the origin. 

The main efforts of this section are dedicated to completing the bases ${\cal V}_{a,b}$
at $a{+}b\leq 10$ with dihedral eMGFs of $\geq 3$ columns such that the basis
dimensions in table \ref{allshuffir} are attained. Generic eMGFs at fixed $(|A|,|B|)$ 
can be obtained from sums of products of basis elements in ${\cal V}_{a,b}$ at
$a\leq |A|$ and $b\leq |B|$ with $\mathbb Q$-linear combinations of (conjecturally
single-valued) MZVs as coefficients. The last statement incorporates the fact that
$\zeta_3$ and $\zeta_5$ are expressible in terms of MGFs \cite{DHoker:2015gmr},
and the same is expected for any single-valued MZV.

%%%%%%%%%%%%%%%%%%%%%%%%%%%%%%%%%
%%%%%%%%%%%%%%%%%%%%%%%%%%%%%%%%%
\subsection{Even versus odd eMGFs}
\label{bassec.1}
%%%%%%%%%%%%%%%%%%%%%%%%%%%%%%%%%
%%%%%%%%%%%%%%%%%%%%%%%%%%%%%%%%%

Modular invariant MGFs can be decomposed into an even and an odd
part w.r.t.\ the automorphism $\tau \rightarrow - \bar \tau$ of the upper half 
plane \cite{DHoker:2019txf}, where the odd parts vanish at lattice weight
$|A|{+}|B|<10$. We shall perform the analogous decompositions for modular 
invariant eMGFs under the extended automorphism $(z,\tau) \rightarrow ({-}\bar z,{-}\bar \tau)$ or
equivalently $(u,v,\tau) \rightarrow (u,{-}v,{-}\bar \tau)$. At fixed $|A|=|B|$,
one can perform an independent counting of indecomposable eMGFs that
are even and odd under this automorphism. As we will see, non-vanishing
odd eMGFs already occur at lattice weight $|A|{+}|B|=6$.

Odd eMGFs will in general be denoted by
\beq
\aplus{A \\ B \\ Z}\! (\tau) = \cplus{A \\ B \\ Z}\! (\tau)  - \cplus{B \\ A \\ Z}\! (\tau)\, , \ \ \ \ |A|=|B|
\label{dictio.6}
\eeq
and the superscript of ${\cal A}^+$ may as well be dropped since the normalization
factors $(\Im \tau)^{|A|}$ $\pi^{-|B|}$ of ${\cal C}^+$ in (\ref{gen.66}) cannot be distinguished
from those of ${\cal C}^-$ in (\ref{compap.6}) if $|A|=|B|$. The odd eMGFs in (\ref{dictio.6})
vanish for less than three columns since (\ref{revsec.6}) implies that 
modular invariant two-column eMGFs are proportional to $g_a(z|\tau)=\dplus{a \\ a}\!(z|\tau)$
which are even under $(z,\tau) \rightarrow ({-}\bar z,{-}\bar \tau)$. 

%%%%%%%%%%%%%%%%%%%%%%%%%%%%%%%%%
%%%%%%%%%%%%%%%%%%%%%%%%%%%%%%%%%
\subsubsection{A simple class of odd three-column eMGFs}
\label{bassec.1.1}

The simplest non-vanishing odd eMGFs have three columns, and we will preferably include
\begin{align}
A_{u,v|w}(z|\tau) &= \cplus{u &v &w \\ v &u &w \\ z &0 &0}\! (\tau) - \cplus{v &u &w \\ u &v &w \\ z &0 &0}\! (\tau)
= \aplus{u &v &w \\ v &u &w \\ z &0 &0}\! (\tau)
\label{adefs}
\\
B_{u|v,w}(z|\tau) &= \cplus{u &v &w \\ w &u &v \\ 0 &0 &z}\! (\tau) - \cplus{u &w &v \\ v &u &w \\ 0 &0 &z}\! (\tau) 
= \aplus{u &v &w \\ w &u &v \\ 0 &0 &z}\! (\tau)
\label{bdefs}
\end{align}
into the bases at $|A|=|B|=u{+}v{+}w$. While the first family
$A_{u,v|w} $ is engineered to have a vanishing $z \rightarrow 0$ limit,
the $B_{u|v,w}$ may reduce to odd MGFs as $z \rightarrow 0$ (with
$\lim_{z \rightarrow 0} B_{u|v,w} = 0$ if $u{+}v{+}w\leq 4$). The definitions
(\ref{adefs}) and (\ref{bdefs}) manifest the antisymmetries
\beq
A_{u,v|w}(z|\tau) 
= -A_{v,u|w}(z|\tau)\, , \ \ \ \ 
B_{u|v,w}(z|\tau)
= -B_{u|w,v}(z|\tau)
\label{aprops}
\eeq
Moreover, there are various choices of $u,v,w$ that we shall disregard
for the construction of bases for eMGFs:
\begin{itemize}
\item $A_{ u,v| w}$ with $w=0$ have a trivial third column and therefore boil down to $\dplus{a \\ b}$
\item in case of $(u,v)=(1,0)$ or $(0,1)$, the
$z\rightarrow 0$ limit of the individual ${\cal C}^+$ in the 
definition (\ref{adefs}) of $A_{ u,v| w}$ diverges
\item there should be no more than one of $u,v,w=0$ in $A_{ u,v| w}$ and $B_{ u | v,w}$
to prevent a trivial column with entries $\begin{smallmatrix} 0 \\ 0 \\ 0 \end{smallmatrix}$
or $\begin{smallmatrix} 0 \\ 0 \\ z \end{smallmatrix}$
\end{itemize}

%%%%%%%%%%%%%%%%%%%%%%%%%%%%%%%%%
%%%%%%%%%%%%%%%%%%%%%%%%%%%%%%%%%
\subsubsection{Expanding odd eMGFs around the cusp}
\label{bassec.1.2}

The expansion of ${\cal A}^+$ in (\ref{dictio.6}) around the cusp begins with terms
\beq
\aplus{A \\ B \\ Z} = {\cal O}(q^u,\bar q^u)
\label{dictio.91}
\eeq
since the coefficients of $q^0\bar q^0$ only depend on $y$ and $u$ which are even
under $(z,\tau) \rightarrow ({-}\bar z,{-}\bar \tau)$. However, the leading terms of
$\aplus{A \\ B \\ Z}$ in the terminology of early section~\ref{sec:3.3} may be non-zero
since single-valued polylogarithms $G^{\rm sv}(a_1,\ldots,a_w;e^{2\pi i z})$ with 
$a_j \in \{0,1\}$ generically have an odd part. As a simple example, the
leading terms of $A_{2,0|1}$ following from the $\beta^{\rm sv}$ representation 
previewed in (\ref{oddA.01}) are spelt out in appendix \ref{last:app.A} 
and demonstrated to flip sign under $(z,\tau) \rightarrow ({-}\bar z,{-}\bar \tau)$.
The example of (\ref{oddA.01}) also illustrates that the $\beta^{\rm sv}$ representations
of odd eMGFs may involve depth-zero terms that cancel the contributions of
$\beta^{\rm sv}$ to the Laurent polynomial along with $q^0\bar q^0$,
see section \ref{sec:3.3.3}.

%%%%%%%%%%%%%%%%%%%%%%%%%%%%%%%%%
%%%%%%%%%%%%%%%%%%%%%%%%%%%%%%%%%
\subsection{Warmup examples at $|A|{+}|B|\leq 6$}
\label{bassec.2}
%%%%%%%%%%%%%%%%%%%%%%%%%%%%%%%%%
%%%%%%%%%%%%%%%%%%%%%%%%%%%%%%%%%

In order to illustrate the guiding principles (\ref{guiding}) and
the appearance of odd eMGFs in our bases ${\cal V}_{|A|,|B|}$ of indecomposables, 
we shall now discuss the sectors with $|A|{+}|B|\leq 6$ in more detail.

%%%%%%%%%%%%%%%%%%%%%%%%%%%%%%%%%
%%%%%%%%%%%%%%%%%%%%%%%%%%%%%%%%%
\subsubsection{Lessons from $|A|{+}|B|\leq 4$}
\label{bassec.2.1}

\begin{figure}[h!]
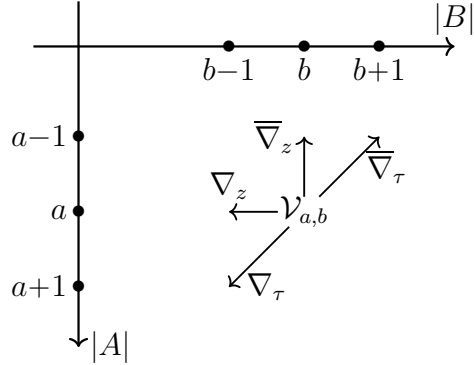

\begin{center}
\tikzpicture[line width=0.3mm]
\draw[->](0,0.6)--(0,-4)node[right]{$|A|$};
\draw[->](-0.6,0)--(5,0)node[above]{$|B|$};
\draw(0,-3.2)node{$\bullet$}node[left]{$a{+}1$};
\draw(0,-2.2)node{$\bullet$}node[left]{$a$};
\draw(0,-1.2)node{$\bullet$}node[left]{$a{-}1$};
\draw(2,0)node{$\bullet$}node[below]{$b{-}1$};
\draw(3,0)node{$\bullet$}node[below]{$b$};
\draw(4,0)node{$\bullet$}node[below]{$b{+}1$};
\draw(3,-2.2)node{${\cal V}_{a,b}$};
\draw[->,line width=0.25mm](3.2,-2.0) -- (4,-1.2);
\draw[->,line width=0.25mm](2.8,-2.4) -- (2,-3.2);
\draw(4.1,-1.6)node{$\overline{\nabla}_\tau$};
\draw(2.5,-3.2)node{$\nabla_\tau$};
\draw[->,line width=0.25mm](3,-2.0) -- (3,-1.2);
\draw[->,line width=0.25mm](2.65,-2.2) -- (2,-2.2);
\draw(2.6,-1.2)node{$\overline{\nabla}_z$};
\draw(2,-1.85)node{$\nabla_z$};
%\draw[->](0,-0.6)--(0,4)node[right]{$|A|$};
%\draw[->](-0.6,0)--(5,0)node[above]{$|B|$};
%%
%\draw(0,3.2)node{$\bullet$}node[left]{$a{+}1$};
%\draw(0,2.2)node{$\bullet$}node[left]{$a$};
%\draw(0,1.2)node{$\bullet$}node[left]{$a{-}1$};
%%
%\draw(2,0)node{$\bullet$}node[below]{$b{+}1$};
%\draw(3,0)node{$\bullet$}node[below]{$b$};
%\draw(4,0)node{$\bullet$}node[below]{$b{-}1$};
%%
%\draw(3,2.2)node{${\cal V}_{a,b}$};
%%
%\draw[->,line width=0.25mm](3.2,2.4) -- (4,3.2);
%\draw[->,line width=0.25mm](2.8,2) -- (2,1.2);
%\draw(4,2.75)node{$\nabla_\tau$};
%\draw(2.5,1.2)node{$\overline{\nabla}_\tau$};
%%
%\draw[->,line width=0.25mm](3,2.5) -- (3,3.2);
%\draw[->,line width=0.25mm](2.7,2.2) -- (2,2.2);
%\draw(2.6,3.2)node{$\nabla_z$};
%\draw(2,2.6)node{$\overline{\nabla}_z$};
\endtikzpicture
\end{center}
\caption{\textit{Action of the differential operators $\nabla_\tau,\overline{\nabla}_\tau,\nabla_z,\overline{\nabla}_z$ on the contributions $|A|,|B|$ to the lattice weight of eMGFs, where 
$\overline{\nabla}_\tau $ and $\overline{\nabla}_z$ are understood to be followed by
$y^{-2}$ and $y^{-1}$ multiplication.}}
\label{diffops}
\end{figure}

By the discussion around table \ref{exsmallAB} and
$\bsvtau{0 &0 \\ 2&2 \\ z &z } = \frac{1}{2} g(z|\tau)^2$, the bases of indecomposable 
eMGFs at $|A|{+}|B|\leq 4$ are entirely determined by two-column ${\cal C}^+$, namely 
\begin{align}
{\cal V}_{1,1} &= \big\{  \dplus{1 \\ 1}\!(z|\tau)\big\}=  \big\{  g(z|\tau)\big\}
\notag \\
{\cal V}_{2,1} &= \big\{  \dplus{2 \\ 1}\!(z|\tau)\big\}=  \big\{ \tfrac{1}{2i} \nabla_z g_2(z|\tau)\big\}
\label{tobasis.01} \\
{\cal V}_{1,2} &= \big\{  \dplus{1 \\ 2}\!(z|\tau)\big\}=  \big\{ \tfrac{1}{2iy} \overline{\nabla}_z g_2(z|\tau)\big\}
\notag
\end{align}
and
\begin{align}
{\cal V}_{3,1} &= \big\{  \dplus{3 \\ 1}\!(z|\tau),\  \dplus{3 \\ 1}\!(0 |\tau)\big\}
= \big\{ \tfrac{1}{2} \pi \nabla_\tau g_2(z|\tau) , \ \tfrac{1}{2} \pi \nabla_\tau E_2(\tau) \big\} 
\notag \\
{\cal V}_{2,2} &= \big\{  \dplus{2 \\ 2}\!(z|\tau),\  \dplus{2 \\ 2}\!(0 |\tau)\big\}
= \big\{   g_2(z|\tau) , \   E_2(\tau) \big\} 
\label{tobasis.02} \\
{\cal V}_{1,3} &= \big\{  \dplus{1 \\ 3}\!(z|\tau),\  \dplus{1 \\ 3}\!(0 |\tau)\big\}
= \big\{ \tfrac{1}{2y^2}  \pi \overline{  \nabla}_\tau g_2(z|\tau) , \ \tfrac{1}{2y^2}  \pi\overline{  \nabla}_\tau E_2(\tau) \big\} 
\notag 
\end{align}
The rewriting of these bases in terms of $g_k(z|\tau)$, $E_k(\tau)$ in (\ref{basic.13a}),
(\ref{revsec.5}) and their derivatives illustrates guiding principle (iii) in (\ref{guiding}):
The modular invariant objects $g_2(z|\tau) ,  E_2(\tau) \ni {\cal V}_{2,2}$ carry the
complete information on all the bases ${\cal V}_{a,b}$ with $a{+}b=3,4$. In principle,
one could also write $\dplus{3 \\ 1}\!(z|\tau) = -\frac{1}{4} \nabla_z^2 g_3(z|\tau)$, but
this is considered as less economic than $\dplus{3 \\ 1}\!(z|\tau) = \tfrac{ \pi }{2}\nabla_\tau 
g_2(z|\tau)$. More generally, we refine guiding principle (iii) in (\ref{guiding})~to
\begin{align}
\textrm{(iiia)} &\ \textrm{for non-zero modular weight $|A| \neq |B|$, pick $\nabla_\tau,
\overline{\nabla}_\tau$- and $\nabla_z,\overline{\nabla}_z$ derivatives of}
\notag\\
&\quad \textrm{modular invariant eMGFs whenever possible while minimizing the order}
\label{tobasis.03}\\
&\quad \textrm{of the differential operator and the lattice weight of the modular invariant}
\notag
\end{align}
The bases (\ref{tobasis.01}) and (\ref{tobasis.02}) clearly reproduce the entries $1$ and
$2$ in table \ref{allshuffir} with $|A|{+}|B|\leq 4$. The differential operators $\nabla_\tau,
\nabla_z$ and their complex conjugates employed in the rewritings of the basis elements
generically map entries of the table to those in the adjacent cells as visualized 
in figure \ref{diffops}. While $\nabla_\tau,\nabla_z$ preserve the 
vanishing holomorphic modular weight of eMGFs in the ${\cal C}^+$ convention
(\ref{gen.66}), the action of $\overline{\nabla}_\tau$ and $\overline{\nabla}_z$ 
on modular invariants introduces holomorphic modular weights
$-2$ and $-1$, respectively. This needs to be compensated by factors of
$y^{-2}$ and $y^{-1}$ (cf.\ the rightmost representations of ${\cal V}_{1,2}$ and ${\cal V}_{1,3}$
in (\ref{tobasis.01}) and (\ref{tobasis.02})) to attain vanishing holomorphic modular weights.

%%%%%%%%%%%%%%%%%%%%%%%%%%%%%%%%%
%%%%%%%%%%%%%%%%%%%%%%%%%%%%%%%%%
\subsubsection{Modular invariant eMGFs at $|A|{+}|B| = 6$}
\label{bassec.2.2}

In the same way as the bases ${\cal V}_{2,1},{\cal V}_{1,2}$ can be conveniently
spanned by the $\nabla_z,\overline{\nabla}_z$-derivatives of $g_2(z|\tau) \in {\cal V}_{2,2}$,
we shall first spell out bases at lattice weight six before filling in the bases at $|A|{+}|B| = 5$.
Similarly, the basis of modular invariants at $|A|=|B| = 3$ will determine
the remaining cases with $|A|{+}|B| = 6$ and $|A|\neq |B|$ via $\nabla_\tau$
and $\overline{\nabla}_\tau$ action, cf.\ (\ref{tobasis.02}).
By the discussion in section \ref{sec:5.8.3}, the counting of admissible $\beta^{\rm sv}$ and 
the relation $[b_2,b_4{+}\epsilon_4]=0$ lead to a four-dimensional basis ${\cal V}_{3,3}$. Apart
from the obvious two-column basis elements $g_3(z|\tau)= \dplus{3 \\ 3}\!(z|\tau)$
and $E_3(\tau)=  \dplus{3 \\ 3}\!(0 |\tau)$, one needs two representatives with three
or more columns, corresponding to $\beta^{\rm sv}$ of depth two. 

By applying the complex-conjugation properties (\ref{inival.35}) of $\beta^{\rm sv}$ 
to the shuffle-independent representatives of depth two in (\ref{comrel.22}),
one can foresee the $(z,\tau) \rightarrow({-}\bar z,{-}\bar \tau)$ parity of the 
missing elements of ${\cal V}_{3,3}$: While $\bsvtau{1& 0\\3& 3\\z& z}$ is mapped
to itself under (\ref{inival.35}), both of $\bsvtau{0& 1\\2& 4\\z& z},\bsvtau{0& 1\\2& 4\\z& \emptyslot}$
are mapped to minus themselves modulo shuffles and MZVs, e.g.\ 
\begin{align}
\overline{ \bsvtau{0& 1\\2& 4\\z& \emptyslot}} &= \bsvtau{1& 0\\4& 2\\ \emptyslot &z} \ \modMZV
\label{tobasis.04}
\\
&=  \bsvtau{1\\4\\ \emptyslot}  \bsvtau{0\\2\\ z} -  \bsvtau{0& 1\\2& 4\\z& \emptyslot} \ \modMZV
\notag
\end{align}
This odd parity modulo shuffle and MZVs applies to the coefficients of both
$b_2b^{(j)}_4$ and $b_2\epsilon^{(j)}_4$ with $j=0,1,2$ such that the 
relation $[b_2,b^{(1)}_4{+}\epsilon^{(1)}_4]=0$
effectively removes one odd eMGF. As a consequence, the indecomposable 
eMGFs in ${\cal V}_{3,3}$ with $\geq 3$ columns can be chosen as one even and
one odd representative under $(z,\tau) \rightarrow({-}\bar z,{-}\bar \tau)$ (see
guiding principle (iv) in (\ref{guiding}), though it
is in principle possible to take linear combinations). 
Since the simple classes of three-column eMGFs in (\ref{cabc.2}) and
(\ref{adefs}) suffice to span the basis, we found it most beneficial to pick
\begin{align}
C_{1|1,1}(z|\tau)&= 8 \bsvtau{1& 0\\3& 3\\z& z}
- 10 \bsvtau{2\\6\\ \emptyslot} -
 20 \bsvtau{2\\6\\z} 
+ 2  \zeta_{3} B_{2}(u)+\frac{ \zeta_{5}}{4 y^2} 
\notag
\\
A_{2,0|1}(z|\tau)&=
12   \bsvtau{0& 1\\2& 4\\z& \emptyslot}   - 12 \bsvtau{0& 1\\2& 4\\z& z}  
+2 \zeta_3 B_2(u)  \label{tobasis.05} \\
&\quad
+ 6 \bsvtau{0 \\2 \\z } \bsvtau{ 1\\ 4\\ z} 
- 6 \bsvtau{0 \\2 \\z } \bsvtau{ 1\\ 4\\ \emptyslot} 
+\frac{  \zeta_3 }{y}  \bsvtau{0\\2\\z} 
\notag
\end{align}
These $\beta^{\rm sv}$ representations were previewed in (\ref{prevex.1}) and (\ref{oddA.01}), respectively, and follow from the methods of section \ref{sec:9}, see in particular
section \ref{sec:sievedz.3} for $C_{1|1,1}(z|\tau)$. The instances of
$B_{ u | v,w}$ in (\ref{bdefs}) with $u{+}v{+}w=3$ are related to the 
basis element $A_{ 2, 0| 1}$ via 
\beq
B_{ 2| 0, 1}(z|\tau) = A_{ 2, 0| 1}(z|\tau)
\, , \ \ \ \
B_{ 1| 0, 2}(z|\tau) = 0
\, , \ \ \ \
B_{ 0| 1, 2}(z|\tau) = -A_{ 2, 0| 1}(z|\tau)
\label{ABrels}
\eeq
While the lattice-sum 
definition (\ref{adefs}) of $A_{2,0|1}$ manifests its vanishing as $z\rightarrow 0$, 
the appearance of $\zeta_3 \bsv{0\\2\\z}  = - \zeta_3g(z|\tau)$ in the last line of (\ref{tobasis.05})
gives rise to worry about a logarithmic singularity as $z\rightarrow 0$. 
By adding a term $g(z|\tau)   \big(E_2(\tau)-g_2(z|\tau) \big) $ that preserves
the vanishing as $z\rightarrow 0$, one can shorten the $\beta^{\rm sv}$ representation
and remove the term $\sim \zeta_3g(z|\tau)$
\begin{align}
  \tilde A_{2,0|1}(z|\tau) &= A_{2,0|1}(z|\tau) + g(z|\tau)   \big(E_2(\tau)-g_2(z|\tau) \big) 
  \label{oddA.3} \\
&=   12 \bsvtau{0& 1\\ 2&4\\ z& \emptyslot} -  12 \bsvtau{0&1\\ 2&4\\ z& z} + 2  B_2(u) \zeta_3
\notag
\end{align}
The extended Laurent polynomial (\ref{lpmero.85}) then implies
that the $z\rightarrow 0$ behaviour of the $\beta^{\rm sv}$ in the second line of (\ref{oddA.3})
compensates that of $ 2  B_2(u) \zeta_3 |_{z=0} = \frac{   \zeta_3 }{3}$. However,
since the eMGF (\ref{oddA.3}) no longer has definite parity under $(z,\tau) 
\rightarrow({-}\bar z,{-}\bar \tau)$, the following basis choice appears more natural to us
by guiding principle (iv) in (\ref{guiding}):
\beq
{\cal V}_{3,3} =  \big\{ 
g_3(z|\tau), \
E_3(\tau), \
C_{1|1,1}(z|\tau), \
A_{2,0|1}(z|\tau)
\big\}
\label{tobasis.06}
\eeq

%%%%%%%%%%%%%%%%%%%%%%%%%%%%%%%%%
%%%%%%%%%%%%%%%%%%%%%%%%%%%%%%%%%
\subsubsection{Remaining eMGFs at $|A|{+}|B| = 6$}
\label{bassec.2.3}

The remaining bases ${\cal V}_{4,2},{\cal V}_{2,4}$ and ${\cal V}_{5,1},{\cal V}_{1,5}$
at lattice weight six have dimensions 3 and 2, respectively, according to table \ref{allshuffir}.
Two inevitable elements of ${\cal V}_{4,2}$ and ${\cal V}_{5,1}$ by guiding principle (i)
in (\ref{guiding}) are furnished by
$\pi \nabla_\tau g_3(z|\tau),\pi \nabla_\tau E_3(\tau)$ and $(\pi \nabla_\tau)^2 g_3(z|\tau),
(\pi \nabla_\tau )^2 E_3(\tau)$, respectively. Hence, it remains to identify one indecomposable
eMGF with more than two columns in ${\cal V}_{4,2}$, and both of $\pi \nabla_\tau C_{1|1,1}(z|\tau)$
and $\pi \nabla_\tau A_{2,0|1}(z|\tau)$ are obvious candidates. 

The first option was already spelt out in (\ref{svexpl.1}) and can be expressed
entirely in terms of simpler indecomposable eMGFs,
\beq
\pi \nabla_\tau C_{1|1,1}(z|\tau)  = - \frac{1}{4} \big( \nabla_z g_2(z|\tau)\big) ^2
+\frac{1}{3}\pi \nabla_\tau E_3(\tau) +\frac{2}{3}\pi \nabla_\tau g_3(z|\tau)
\label{tobasis.07}
\eeq
Hence, the indecomposable three-column eMGF for ${\cal V}_{4,2}$ needs
to be derived from
\begin{align}
\pi \nabla_\tau  A_{2,0|1}(z|\tau) &= 3 \bsvtau{0&2 \\2&4\\z&z}-3 \bsvtau{0&2 \\2&4 \\z&\emptyslot} + \dplus{2 \\ 0}\!(z |\tau)
\bigg( 6 \bsvtau{1 \\4\\ \emptyslot}  {-} 6  \bsvtau{1 \\4\\ z} {-} \frac{ \zeta_3}{y} \bigg) \notag \\
&\quad
+ \bsvtau{0 \\2\\ z} \bigg( \frac{3}{2} \bsvtau{2 \\4\\ \emptyslot}  {-} \frac{3}{2}  \bsvtau{2 \\4\\ z} {-}  \zeta_3 \bigg) 
\label{tobasis.08}
\end{align}
However, one may prefer a choice of basis elements without any purely holomorphic factors
$\dplus{k \\ 0}\!(z|\tau)  = - (\Im \tau)^k f^{(k)}(z|\tau)$ or
$\dplus{k \\ 0}\!(0|\tau)  = (\Im \tau)^k G_k(\tau)$ as functions of $\tau$ at fixed $u,v$.
With this criterion in mind, the shifted version (\ref{oddA.3}) of $\tilde A_{2,0|1}$
with indefinite parity under $(z,\tau) \rightarrow ({-}\bar z,{-}\bar \tau)$
is a preferable primitive of the leftover basis element
\beq
\pi \nabla_\tau \tilde A_{2,0|1}(z|\tau) =
3 \bsvtau{0&2 \\2&4\\z&z}-3 \bsvtau{0&2 \\2&4 \\z&\emptyslot}
\label{tobasis.09}
\eeq
which illustrates a conflict between guiding principles (iv) and (v) in (\ref{guiding}).
The outcome of the next $\nabla_\tau$ derivative
\begin{align}
(\pi \nabla_\tau)^2 \tilde A_{2,0|1}(z|\tau) &=
12 \big( \dplus{4 \\ 0}\!(0 |\tau) - \dplus{4 \\ 0}\!(z |\tau) \big) g(z|\tau)
\label{tobasis.10}
\end{align}
is not indecomposable and as expected does not contribute a basis element for
${\cal V}_{5,1}$. 

In summary, the discussion of this section suggests to employ the bases
\begin{align}
{\cal V}_{4,2} &=  \big\{ 
\pi \nabla_\tau g_3(z|\tau), \
\pi \nabla_\tau E_3(\tau), \
\pi \nabla_\tau \tilde A_{2,0|1}(z|\tau)
\big\} \notag \\
{\cal V}_{5,1} &=  \big\{ 
(\pi \nabla_\tau)^2 g_3(z|\tau), \
(\pi \nabla_\tau)^2 E_3(\tau)
\big\}
\label{tobasis.11}
\end{align}
at lattice weight 6 which complement the modular invariant ${\cal V}_{3,3}$ in (\ref{tobasis.06}).
The analogous bases ${\cal V}_{2,4}$ and ${\cal V}_{1,5}$ are determined by complex 
conjugation and multiplication by powers of $y^{-2}$ as in (\ref{tobasis.02}).

%%%%%%%%%%%%%%%%%%%%%%%%%%%%%%%%%
%%%%%%%%%%%%%%%%%%%%%%%%%%%%%%%%%
\subsubsection{Bases with $|A|{+}|B| = 5$}
\label{bassec.2.4}

At lattice weight $|A|{+}|B| = 5$, table \ref{allshuffir} lists two-dimensional bases ${\cal V}_{3,2},{\cal V}_{2,3}$ as well as one-dimensional bases ${\cal V}_{4,1},{\cal V}_{1,4}$. Apart from
the inevitable $\dplus{a \\ b}\!(z|\tau) \ni {\cal V}_{a,b}$ it remains to identify one
indecomposable eMGF beyond two columns to span ${\cal V}_{3,2}$ and
${\cal V}_{2,3}$. By analogy with ${\cal V}_{2,1}$ and
${\cal V}_{1,2}$ in (\ref{tobasis.01}), natural candidates
for elements of ${\cal V}_{n,n-1}$ and ${\cal V}_{n-1,n}$  arise from $\nabla_z$ and
$\overline{\nabla}_z$-derivatives of ${\cal V}_{n,n}$-elements. 

In the present case, this reasoning is extended to mixed derivatives 
$\dplus{4 \\ 1}\!(z|\tau) = \frac{1}{ 6i} \nabla_z \nabla_\tau g_3(z|\tau) $ which is
considered as preferable to $\dplus{4 \\ 1}\!(z|\tau) = \frac{1}{ (2i)^3} \nabla_z^3 g_4(z|\tau)$, 
see the discussion around (\ref{tobasis.03}). Moreover, we need to
investigate the three-column elements $C_{1|1,1}(z|\tau),
A_{2,0|1}(z|\tau)$ of ${\cal V}_{3,3}$ in (\ref{tobasis.06}) and indeed find an indecomposable
\begin{align}
\nabla_z C_{1|1,1}(z|\tau) 
= 2i \cplus{1 & 1 & 1 \\0 & 1 & 1 \\ z & 0 & 0}\! (\tau) 
= 4 \bsvtau{1 &0 \\ 3 &2 \\ z &z} - 8 \bsvtau{2 \\ 5 \\ z}  + 4 \zeta_3 B_1(u)
\label{tobasis.12}
\end{align}
whose $\beta^{\rm sv}$ representation was previewed in (\ref{prevex.2}).
One could have alternatively obtained this basis element from
\begin{align}
\nabla_z A_{2,0|1}(z|\tau) &= 
2i\Big\{  \dplus{3 \\ 2}\!(z|\tau) - g(z|\tau)  \dplus{2 \\ 1}\!(z|\tau)  
+ \dplus{1 \\ 0}\!(z|\tau) \big( g_2(z|\tau) {-} E_2(\tau) \big) \Big\} \notag \\
&\quad + \nabla_z C_{1|1,1}(z|\tau)
\label{tobasis.13}
\end{align}
but the expression in (\ref{tobasis.12}) is preferable since it does not contain
any factor of $\dplus{1 \\ 0}\!(z|\tau)= - (\Im \tau)f^{(1)}(z|\tau)$. This choice
leads to the bases
\begin{align}
{\cal V}_{3,2} &=  \big\{ 
\nabla_z g_3(z|\tau), \
\nabla_z C_{1|1,1}(z|\tau)
\big\} \notag \\
{\cal V}_{4,1} &=  \big\{ 
\pi \nabla_\tau \nabla_z  g_3(z|\tau)\big\}
\label{tobasis.14}
\end{align}
with similar bases ${\cal V}_{2,3}, {\cal V}_{1,4}$ obtained from complex conjugation.

%%%%%%%%%%%%%%%%%%%%%%%%%%%%%%%%%
%%%%%%%%%%%%%%%%%%%%%%%%%%%%%%%%%
\subsection{Extensions to $|A|{+}|B|=7$ and $8$}
\label{bassec.3}
%%%%%%%%%%%%%%%%%%%%%%%%%%%%%%%%%
%%%%%%%%%%%%%%%%%%%%%%%%%%%%%%%%%

In this section, we present a possible choice of basis ${\cal V}_{a,b}$ for 
indecomposable eMGFs in one variable at $a{+}b=7,8$. We will no longer
make explicit reference to the ubiquitous argument $\tau$ and for instance
write $g_k(z)$ in the place of $g_k(z|\tau)$ to save space.

%%%%%%%%%%%%%%%%%%%%%%%%%%%%%%%%%
%%%%%%%%%%%%%%%%%%%%%%%%%%%%%%%%%
\subsubsection{Modular invariant eMGFs at $|A|{+}|B| = 8$}
\label{bassec.3.1}

Following the strategy of the previous section, the basis elements of
${\cal V}_{4,4}$ will later suggest elements of ${\cal V}_{a,b}$
with $a\neq b$ and $a{+}b=7,8$. Hence, we start by stating our choice
\begin{align}
{\cal V}_{4,4} &= \Big\{
g_4(z),\, E_4,\, C_{2|1,1}(z),\,C_{1|2,1}(z), \, C_{2,1,1},\,
C_{1|1,1,1}(z),\, 
\cplus{0&1&1&2 \\
2&1&1&0 \\
0&z&z&0} \notag \\
&\quad \quad 
A_{2,0|2}(z),\, A_{3,0|1}(z), \,
\aplus{0 &1 &1 &2 \\ 1 &1 &1 &1 \\ z &0 &0 &0}
\Big\}
\label{tobasis.21}
\end{align}
and elaborating on its elements that were once more selected with the guiding principles
(\ref{guiding}) in mind. The ten basis elements expected from table \ref{allshuffir} are
assembled from seven even eMGFs in the first line of (\ref{tobasis.21}) 
and three odd ones in the second line.

Most obviously, the basis (\ref{tobasis.21}) contains two-column eMGFs $g_4(z),E_4$
as usual for even lattice weights. One then maximizes the number of three-column
representatives, focussing on the even and odd prototypes in (\ref{cabc.2}) and
(\ref{adefs}). The even three-column eMGFs in ${\cal V}_{4,4}$ can be spanned
by one MGF $C_{2,1,1}$ and the two $z$-dependent $C_{2|1,1}(z),C_{1|2,1}(z)$,
\begin{align}
C_{2 | 1,1}(z ) &= {-} 18 \bsv{2& 0\\4& 4\\ \emptyslot& z} -
 18 \bsv{2& 0\\4& 4\\z& \emptyslot} +
 18 \bsv{2& 0\\4& 4\\z& z}
 \notag \\
 & \quad + 14 \bsv{3\\8\\ \emptyslot} -
 140 \bsv{3\\8\\z} +
 12 \zeta_{3} \bsv{0\\4\\z}
 - \frac{ \zeta_{5}}{2 y} B_{2}(u)  - \frac{ \zeta_{7}}{16 y^3}  \notag \\
C_{1 | 2,1}(z ) &=
{-} 18 \bsv{2& 0\\4& 4\\z& z} +
 24 \bsv{1& 1\\3& 5\\z& z} +
 24 \bsv{2& 0\\5& 3\\z& z}  
\label{tobasis.22} \\
 & \quad - 70 \bsv{3\\8\\ \emptyslot} - 56 \bsv{3\\8\\z} 
 +\frac{3  \zeta_{5}}{2 y} B_{2}(u) + \frac{5 \zeta_{7}}{16 y^3}  \notag \\
%%%
 C_{2,1,1}  &= -18  \bsv{2&0\\4&4\\ \emptyslot}  - 126 \bsv{3\\8\\ \emptyslot} 
 + 12 \zeta_3  \bsv{0\\4\\ \emptyslot} 
  +  \frac{ 5 \zeta_5 }{12y} - \frac{ \zeta_3^2}{4y^2}
  + \frac{ 9 \zeta_7 }{16y^3}
 \notag 
\end{align}
see (\ref{prevex.1}) and (\ref{prevex.2}) for previews of the first two $\beta^{\rm sv}$ representations
and \cite{Gerken:2020yii} for that of the closely related MGF $E_{2,2} = C_{2,1,1} - \frac{9}{10} E_4$ \cite{Broedel:2018izr} or $F^{+(2)}_{2,2} = -C_{2,1,1} + \frac{9}{10} E_4$ \cite{Dorigoni:2021jfr}.

The odd three-column eMGFs in (\ref{tobasis.21}) boil down to the following $\beta^{\rm sv}$
\begin{align}
A_{ 2, 0|2}(z) &=
-24 \zeta_3 \bsv{0\\4\\z} - \frac{
 12 \zeta_3 }{y}  \bsv{1\\4\\z}+
 48 \bsv{1& 1\\3& 5\\z& z}-
 48 \bsv{2& 0\\5& 3\\z& z}  \notag \\
 &\quad +
 72 \bsv{1& 1\\4& 4\\ \emptyslot& z} -
 72 \bsv{1& 1\\4& 4\\z& \emptyslot} +
 36 \bsv{2& 0\\4& 4\\ \emptyslot& z} -
 36 \bsv{2& 0\\4& 4\\z& \emptyslot}
\notag  \\
 %%%%%%%
 A_{ 3, 0| 1}(z) &=
 -\frac{ 3 B_2 \zeta_5}{2 y} - \frac{ 3 \zeta_5 }{ 4 y^2} \bsv{0\\2\\z}
  - 6 \zeta_3 \bsv{0\\4\\z} - \frac{
 6 \zeta_3 }{y} \bsv{1\\4\\z} - \frac{
 3 \zeta_3 }{8 y^2}  \bsv{2\\4\\z} \label{tobasis.23} \\
 &\quad -
 30 \bsv{0& 2\\2& 6\\z& \emptyslot} +
 30 \bsv{0& 2\\2& 6\\z& z}+
 30 \bsv{2& 0\\6& 2\\ \emptyslot& z} -
 30 \bsv{2& 0\\6& 2\\z& z} \notag \\
 &\quad +
 9 \bsv{0& 2\\4& 4\\ \emptyslot& z} -
 9 \bsv{0& 2\\4& 4\\z& \emptyslot} +
 36 \bsv{1& 1\\4& 4\\ \emptyslot& z} -
 36 \bsv{1& 1\\4& 4\\z& \emptyslot} +
 9 \bsv{2& 0\\4& 4\\ \emptyslot& z} -
 9 \bsv{2& 0\\4& 4\\z& \emptyslot}
 \notag
 \end{align}
and other choices of $A_{u,v|w}(z)$ and $B_{u|v,w}(z)$ in (\ref{adefs}) 
with $u{+}v{+}w=4$ are related to $A_{2,0|2},A_{ 3, 0| 1}$
as well as one-column eMGFs via (\ref{tobasis.appA}) and (\ref{tobasis.appB}).

Finally, the basis of eMGFs at $|A|=|B|=4$ necessitates
three elements beyond three columns as in the second line of (\ref{tobasis.21}).
This can be anticipated from the combinations of $\beta^{\rm sv}$ at depth three 
\beq
\bsv{1&0&0\\3&2&3\\z&z&z} , \ \bsv{0&1&0\\2&3&3\\z&z&z}, \ 
\bsv{0&0&1\\2&2&4\\z&z&z} {-} \bsv{0&0&1\\2&2&4\\z&z&\emptyslot}
 \label{tobasis.24}
\eeq
which still enter the generating series (\ref{gen.34c})\footnote{More specifically,
the relations (\ref{count.16}) and (\ref{count.17}) enforce $\bsv{0&0&1\\2&2&4\\z&z&z}$
and $\bsv{0&0&1\\2&2&4\\z&z&\emptyslot}$ to arise with relative factor $-1$ and
dictate admixtures of depth-two terms to all the three combinations in (\ref{tobasis.24}).}
and cannot arise from eMGFs with three or less columns. The depth-three $\beta^{\rm sv}$
in (\ref{tobasis.24}) can be obtained from two even eMGFs and one odd eMGF. For
the even ones, the preferred class of $C_{a_1,\ldots,a_s|a_{s+1},\ldots,a_R}(z)$ in
the guiding principles (\ref{guiding}) only contributes one linearly independent representative
\begin{align}
C_{1|1,1,1}(z) &=
-24 \bsv{1&0&0\\3&2&3\\z&z&z}
-54 \bsv{2&0\\4&4\\\emptyslot&\emptyslot}+48 \bsv{1&1\\3&5\\z&z}+108 \bsv{1&1\\4&4\\\emptyslot&z}+108 \bsv{1&1\\4&4\\z&\emptyslot} \notag \\
&\quad -162 \bsv{2&0\\4&4\\\emptyslot&z}-162 \bsv{2&0\\4&4\\z&\emptyslot}-54 \bsv{2&0\\4&4\\z&z}+48 \bsv{2&0\\5&3\\z&z}
 \label{dictio.11}\\
&\quad
-252 \bsv{3\\8\\\emptyslot}-252 \bsv{3\\8\\z}
 -\frac{18 \zeta_3 }{y}  \bsv{1\\4\\z}
 +108 \zeta_3 \bsv{0\\4\\z}+36 \zeta_3 \bsv{0\\4\\\emptyslot}
\notag \\
  &\quad
    -24 \zeta_3 B_1(u) \bsv{0\\3\\z}  + \frac{\zeta_5}{4 y}+\frac{3 \zeta_5 B_2(u)}{2 y}
    - \frac{3 \zeta_3^2}{4 y^2}+\frac{9 \zeta_7}{8 y^3}
\notag
\end{align}
so one has to pick another real eMGF. Also the choice of odd eMGF for ${\cal V}_{4,4}$ 
beyond three columns is not straightforward from the guiding principles in earlier
sections. We picked $\cplus{0&1&1&2 \\
2&1&1&0 \\
0&z&z&0}$ and $\aplus{0 &1 &1 &2 \\ 1 &1 &1 &1 \\ z &0 &0 &0}$ but cannot exclude
that more convenient choices are possible. The respective $\beta^{\rm sv}$ representations
can be found in appendix \ref{last:app.C}.

%%%%%%%%%%%%%%%%%%%%%%%%%%%%%%%%%
%%%%%%%%%%%%%%%%%%%%%%%%%%%%%%%%%
\subsubsection{Application to an eMGF relation}
\label{bassec.3.2}

Given the $\beta^{\rm sv}$ representations of all the elements in ${\cal V}_{4,4}$,
the relation (\ref{teas.9}) among modular invariant eMGFs \cite{Basu:2020pey} is manifest:
The only eMGF in the relation which we did not include into any ${\cal V}_{a,b}$ is
\begin{align}
C_{1,1|1,1}(z) &= 
 -  32 \bsv{1 &0 &0 \\ 3 &2 &3 \\  z &z&z}
 +  72 \bsv{1 &1 \\ 4 &4 \\ \emptyslot & \emptyslot} + 
 144 \bsv{1 &1 \\ 4 &4 \\ z &z} - 
 72 \bsv{2 &0 \\ 4 &4 \\ \emptyslot & \emptyslot} - 
 72 \bsv{2 &0 \\ 4 &4 \\ \emptyslot & z} \notag \\
 &\quad - 
 72 \bsv{2 &0 \\ 4 &4 \\ z&\emptyslot} - 
 216 \bsv{2 &0 \\ 4 &4\\ z &z} + 
 64 \bsv{2 &0 \\ 5 &3 \\ z &z}  
 + 64 \bsv{1 &1 \\ 3 &5 \\ z &z}
 \label{tobasis.25} \\
&\quad
- 168 \bsv{3\\ 8\\ \emptyslot}  -  336 \bsv{3\\ 8\\ z}
 + 48 \zeta_3 \bsv{0 \\ 4\\ \emptyslot} + 48 \zeta_3 \bsv{0\\ 4\\ z} 
 - \frac{  12 \zeta_3}{y} \bsv{1\\ 4\\ \emptyslot} \notag \\
 &\quad -  32 B_1(u) \zeta_3 \bsv{0 \\ 3 \\ z}
+\frac{\zeta_5}{3 y} + \frac{6 B_2(u) \zeta_5}{y} + \frac{3 \zeta_7}{4 y^3}
\notag
 \end{align}
 where the $\beta^{\rm sv}$ representation is once more found from the methods
 of section \ref{sec:9.2}. Upon comparison with the $\beta^{\rm sv}$ representations
 (\ref{tobasis.22}) and (\ref{dictio.11}) of $C_{2|1,1}(z)$ and $C_{1|1,1,1}(z)$ (also
 see (\ref{invbsv.7}) and (\ref{nwbsv.11}) for depth-one results),
the relation (\ref{teas.9}) is found to hold at the level of the coefficients of all the
$\beta^{\rm sv}$. This illustrates the virtue of iterated KE integrals to expose the
relations among eMGFs.
 
%%%%%%%%%%%%%%%%%%%%%%%%%%%%%%%%%
%%%%%%%%%%%%%%%%%%%%%%%%%%%%%%%%%
\subsubsection{Remaining eMGFs at $|A|{+}|B| = 8$}
\label{bassec.3.3}

For the remaining ${\cal V}_{a,b}$ at $a{+}b=8$, table
\ref{allshuffir} lists basis dimensions
\beq
\dim \, {\cal V}_{5,3} = 9 \, , \ \ \ \ 
\dim \, {\cal V}_{6,2} = 4 \, , \ \ \ \
\dim \, {\cal V}_{7,1} = 2
 \label{tobasis.27}
\eeq
and we make the following basis choice based on $\nabla_\tau$-derivatives
of (\ref{tobasis.21}),
\begin{align}
{\cal V}_{5,3} &= \Big\{  \pi \nabla_\tau g_4(z), \, \pi \nabla_\tau E_4 , \, 
\pi \nabla_\tau C_{2|1,1}(z) , \, 
\pi \nabla_\tau C_{1|2,1}(z) , \,
\pi \nabla_\tau C_{2,1,1}, \notag \\
&\quad \quad
\pi \nabla_\tau A_{2,0|2}(z), \,
\pi \nabla_\tau \widehat A_{3,0|1}(z ), \, 
\pi \nabla_\tau C_{1|1,1,1}(z),\, 
\pi \nabla_\tau  \chatplus{0&1&1&2 \\
2&1&1&0 \\
0&z&z&0} \Big\}
\notag \\
{\cal V}_{6,2} &= \Big\{ (\pi \nabla_\tau)^2 g_4(z), \,  (\pi \nabla_\tau)^2 E_4, \, \pi \nabla_\tau \widehat {\cal C}_{(5,3)}(z)
, \, \pi \nabla_\tau  \widehat {\cal C}'_{(5,3)}(z)
 \Big\}
 \label{tobasis.28}\\
{\cal V}_{7,1} &= \Big\{ (\pi \nabla_\tau)^3 g_4(z), \,  (\pi \nabla_\tau)^3 E_4 \Big\}
\notag 
\end{align}
In order to remove factors of $\dplus{a \\ 0}$ from their $\nabla_\tau$-derivatives,
the objects $\widehat A_{3,0|1}(z ),$ $\chatplus{0&1&1&2 \\
2&1&1&0 \\
0&z&z&0} ,\, \widehat {\cal C}_{(5,3)}(z)
, \,\widehat {\cal C}'_{(5,3)}(z)$ are obtained by adding products of simpler terms
to the eMGFs encoded by unhatted symbols. Their definitions will be spelt out
in the following bulletpoints:
\begin{itemize}
\item Similar to (\ref{oddA.3}), we obtain the element $\pi \nabla_\tau \widehat A_{3,0|1}(z )$
of ${\cal V}_{5,3}$ by shifting
\begin{align}
\widehat A_{3,0|1}(z ) &= A_{3,0|1}(z ) + g(z )   \big( g_3(z ) - E_3  \big)
+ \frac{  (\pi \nabla_\tau E_2) \pi \overline \nabla_\tau g_2(z)
{-} ( \pi \overline \nabla_\tau E_2) \pi \nabla_\tau g_2(z)
}{4 y^2}
\notag \\
&=
{-} 60 \bsv{0& 2\\2& 6\\z& \emptyslot} +
 60 \bsv{0& 2\\2& 6\\z& z}  +
 36 \bsv{1& 1\\4& 4\\\emptyslot& z} -
 36 \bsv{1& 1\\4& 4\\z& \emptyslot}   \label{oddA.54} \\
 &\ \ \ \ +
 18 \bsv{2& 0\\4& 4\\\emptyslot& z} -
 18 \bsv{2& 0\\4& 4\\z& \emptyslot}
  - 12 \zeta_3 \bsv{0 \\ 4 \\ z}
 - \frac{6 \zeta_3}{y} \bsv{1 \\ 4 \\z} 
  - \frac{ 3 B_2(u) \zeta_5}{2 y}
\notag
 \end{align}
which ensures that no factors of $\dplus{a \\ 0}$ arise from the first $\tau$-derivative.
Just like in (\ref{oddA.3}), the addition of the real term $g(z ) ( g_3(z ) - E_3)$
in the first line of (\ref{oddA.54}) spoils the odd parity of $A_{3,0|1}(z ) $
under $(z,\tau) \rightarrow ({-}\bar z,{-}\bar \tau)$. A similar tension between clean 
differential equations in $\nabla_\tau$ and definite parity was observed for MGFs, 
e.g.\ when constructing bases of lattice sums $\cplus{A\\B\\0}$ at $|A|=|B|=5$ 
in section 5.1 of \cite{Gerken:2020yii}.
\item In order to realize the desired $\beta^{\rm sv}$ of depth three in ${\cal V}_{5,3}$
without any factors of $\dplus{a \\ 0}$, we shift the basis eMGF 
$ \cplus{0&1&1&2 \\ 2&1&1&0 \\0&z&z&0} \ni {\cal V}_{4,4}$ with $\beta^{\rm sv}$ representation 
in (\ref{dictio.12}) by the following products of simpler objects:
\begin{align}
 \chatplus{0&1&1&2 \\
2&1&1&0 \\
0&z&z&0}  &=  \cplus{0&1&1&2 \\
2&1&1&0 \\
0&z&z&0}
+ 5 g(z) \big(g_3(z)- E_3 \big)
 + 2  g(z)^2 \big(g_2(z)- E_2 \big) \notag \\
&\quad
+ \frac{ \big(\nabla_z C_{1|1,1}(z)\big) \overline{\nabla}_z g_2(z)}{4y}
+ \frac{ \big(\overline{\nabla}_z C_{1|1,1}(z)\big)  \nabla_z g_2(z)}{4y} 
\label{tobasis.31}\\
&\quad 
- \frac{ \big(\pi \nabla_\tau g_2(z) \big)\pi \overline \nabla_\tau g_2(z)}{2y^2}
- g(z) A_{2,0|1}(z) \notag
\end{align}
We employ its $\nabla_\tau$-derivative as a four-column basis element
of ${\cal V}_{5,3}$ in (\ref{tobasis.28}).
\item The basis of ${\cal V}_{6,2}$ in (\ref{tobasis.28}) no longer necessitates any four-column
eMGFs or $\beta^{\rm sv}$ of depth three. The desired depth-two $\beta^{\rm sv}$ can
be separated from the admixtures of $\dplus{a \\ 0}$ by taking the $\nabla_\tau$-derivative
of the shifted combinations
\begin{align}
 \widehat {\cal C}_{(5,3)}(z) &= \pi \nabla_\tau A_{2,0|2}(z) - \frac{1}{2} \big( \nabla_z g_2(z)\big) \nabla_z g_3(z) - 2 E_2 \pi \nabla_\tau g_2(z) + 2 g_2(z) \pi \nabla_\tau E_2
\notag \\
\widehat {\cal C}'_{(5,3)}(z)  &= \pi \nabla_\tau \widehat A_{3,0|1}(z)
+g_2(z) \pi \nabla_\tau E_2 - E_2 \pi \nabla_\tau g_2(z)
\label{tobasis.32}
\end{align}
with $\widehat A_{3,0|1}(z)$ defined in (\ref{oddA.54}). 
\end{itemize}
The $\beta^{\rm sv}$ representations of $\pi \nabla_\tau C_{2|1,1}(z),\pi \nabla_\tau C_{1|2,1}(z),
\pi \nabla_\tau A_{2,0|2}(z)$ and $\pi \nabla_\tau \widehat A_{3,0|1}(z)$ can be 
found in appendix \ref{last:app.F}, and those of the remaining basis elements
of ${\cal V}_{5,3},{\cal V}_{6,2},{\cal V}_{7,1}$ are given in an ancillary file.

It would have been tempting to include second $\nabla_\tau$-derivatives of the
real eMGFs $C_{2,1,1},$ $C_{2|1,1}(z),\, C_{1|2,1}(z)$ into the basis ${\cal V}_{6,2}$
in (\ref{tobasis.28}). However, these second derivatives are entirely expressible
in terms of depth-one objects
\begin{align}
(\pi \nabla_\tau)^2 C_{2|1,1}(z) &= \frac{1}{2} \big(\pi \nabla_\tau g_2(z) \big)^2 
- (\pi \nabla_\tau E_2) \pi \nabla_\tau g_2(z)
- \frac{1}{10} (\pi \nabla_\tau)^2 E_4+  (\pi \nabla_\tau)^2 g_4(z) \notag \\
(\pi \nabla_\tau)^2 C_{1|2,1}(z) &= 
-\frac{1}{4} \big( \nabla_z g_3(z) \big) \pi \nabla_\tau \nabla_z g_2(z) 
-\frac{1}{2} \big( \nabla_z g_2(z) \big) \pi \nabla_\tau \nabla_z g_3(z)  \notag \\
 &\quad   -  \frac{1}{2} \big(\pi \nabla_\tau g_2(z) \big)^2
 +\frac{1}{2} (\pi \nabla_\tau)^2 E_4+ \frac{2}{5} (\pi \nabla_\tau)^2 g_4(z) 
 \label{tobasis.33} \\
(\pi \nabla_\tau)^2 C_{2,1,1}&=  {-}\frac{1}{2} \big(\pi \nabla_\tau E_2 \big)^2
+ \frac{9}{10} (\pi \nabla_\tau)^2 E_4  \notag 
\end{align}
see (\ref{tobasis.07}) for the analogous identity at lattice weight six 
that excluded $\pi \nabla_\tau C_{1|1,1}$
from the indecomposables in~${\cal V}_{4,2}$.

%%%%%%%%%%%%%%%%%%%%%%%%%%%%%%%%%
%%%%%%%%%%%%%%%%%%%%%%%%%%%%%%%%%
\subsubsection{Bases with $|A|{+} |B| = 7$}
\label{bassec.3.4}

For the bases ${\cal V}_{a,b}$ of indecomposable one-variable
eMGFs at lattice weight $a{+}b=7$, table \ref{allshuffir} 
lists dimensions
\beq
\dim \, {\cal V}_{4,3} = 5 \, , \ \ \ \ 
\dim \, {\cal V}_{5,2} = 3 \, , \ \ \ \
\dim \, {\cal V}_{6,1} = 1
 \label{tobasis.41} 
\eeq
We shall construct these bases from suitably chosen 
derivatives of modular invariant eMGFs in ${\cal V}_{4,4}$
\begin{align}
{\cal V}_{4,3} &= \big\{ \nabla_z g_4(z), \, \nabla_z C_{2|1,1}(z), \,
\nabla_z C_{1|2,1}(z) ,\, \nabla_z A_{2,0|2}(z) , \, \nabla_z C_{1|1,1,1}(z) \big\} \notag \\
{\cal V}_{5,2} &= \big\{ \pi \nabla_\tau \nabla_z g_4(z),\,
\pi \nabla_\tau \nabla_z C_{2|1,1}(z), \,
\pi \nabla_\tau \widehat {\cal C}_{(4,3)}(z)
\big\}
\label{tobasis.42} \\
{\cal V}_{6,1}  &= \big\{
(\pi \nabla_\tau)^2 \nabla_z g_4(z)
\big\}
\notag
\end{align}
where the absence of $\dplus{a\\0}$ from $\pi \nabla_\tau \widehat {\cal C}_{(4,3)}(z)$
is again tied to the shift in
\beq
 \widehat {\cal C}_{(4,3)}(z) = \nabla_z A_{2,0|2}(z) - 2 E_2 \nabla_z g_2(z) + 2 g(z) \nabla_z g_3(z)
 \label{tobasis.43}
\eeq
The $\beta^{\rm sv}$ representations of the elements of ${\cal V}_{4,3}$ beyond depth one
are given by 
\begin{align}
\nabla_z C_{2|1,1}(z) &=
12 \bsv{2 &0 \\ 4 &3 \\ z &z} 
 -12 \bsv{2	&0 \\ 4 &3 \\ \emptyslot &z}
  -60 \bsv{3 \\ 7 \\ z}
  + 8 \zeta_3 \bsv{ 0 \\3 \\ z} - \frac{ \zeta_5}{y} B_1(u)
\notag \\
\nabla_z C_{1|2,1}(z) &=   12 \bsv{1 &1 \\ 3 &4 \\ z &z}
-6  \bsv{2 &0 \\ 4 &3 \\ z &z} +12  \bsv{2 &0 \\ 5 &2 \\ z &z}
- 24 \bsv{3 \\ 7 \\ z} + \frac{3  \zeta_5}{y} B_1(u) \notag\\
\nabla_z A_{2,0|2}(z)&= -24 \bsv{1&1 \\ 3 &4 \\ z &\emptyslot}
+24 \bsv{1 &1 \\ 3 &4 \\ z &z}+24 \bsv{1 &1 \\ 4 &3 \\ \emptyslot &z}+24 \bsv{2	&0 \\ 4 &3 \\ \emptyslot &z}
-12 \bsv{2 &0 \\ 4 &3 \\ z &z} \notag \\
& \ \ -24 \bsv{2 &0 \\ 5 &2 \\ z &z} + 168 \bsv{3 \\ 7 \\ z}
-16 \zeta_3 \bsv{0 \\ 3 \\ z}- \frac{ 4 \zeta_3}{z} \bsv{1 \\ 3 \\ z}
 \label{tobasis.44} \\
\nabla_z C_{1|1,1,1}(z) &= -12 \bsv{1 &0 &0 \\
3 &2 &2 \\
z &z &z}
+36  \bsv{1 &1 \\ 4 &3 \\ \emptyslot &z}
-72 \bsv{2 &0 \\ 4 &3 \\ \emptyslot &z}
+24 \bsv{2 &0 \\ 5 &2 \\ z &z}
-108 \bsv{3 \\ 7 \\z} \notag \\
&\ \ -12 \zeta_3 B_1(u)  \bsv{0 \\ 2 \\ z}
- \frac{6 \zeta_3}{y}  \bsv{1 \\ 3 \\ z}
+48 \zeta_3 \bsv{ 0 \\ 3 \\ z}
+\frac{ 3 \zeta_5}{y} B_1(u)
\notag
\end{align}
Similarly, the $\beta^{\rm sv}$ representations for ${\cal V}_{5,2}$ beyond depth one
are given by
\begin{align}
\pi \nabla_\tau \nabla_z C_{2|1,1}(z)&= 30 \bsv{4 \\ 7 \\ z}+3 \bsv{2 &1 \\ 4 &3 \\ \emptyslot&z}-3 \bsv{2 &1 \\
4 &3 \\ z &z}-2 \zeta_3 \bsv{1 \\ 3 \\ z}+ \zeta_5 B_1(u)
\notag \\
\pi \nabla_\tau 
 \widehat {\cal C}_{(4,3)}(z)  &=
12 \bsv{1 &2 \\ 3 &4 \\ z &\emptyslot}-6 \bsv{1 &2 \\ 3 &4 \\ z&z}
-6 \bsv{2 &1 \\ 4 &3 \\ \emptyslot &z}+3 \bsv{2 &1 \\ 4 &3 \\ z &z} 
 \label{tobs.45} \\
&\quad -6 \bsv{0 &3 \\ 2 &5 \\ z &z}
-84 \bsv{4 \\ 7 \\ z}+4 \zeta_3  \bsv{1 \\ 3 \\z}
\notag
\end{align}
%

%%%%%%%%%%%%%%%%%%%%%%%%%%%%%%%%%
%%%%%%%%%%%%%%%%%%%%%%%%%%%%%%%%%
\subsection{Extensions to $|A|{+}|B|=9$ and $10$}
\label{bassec.4}
%%%%%%%%%%%%%%%%%%%%%%%%%%%%%%%%%
%%%%%%%%%%%%%%%%%%%%%%%%%%%%%%%%%

We have also applied the strategy of the previous sections to construct
bases ${\cal V}_{a,b}$ of eMGFs at lattice weight $a{+}b=9$ and $10$. 
While most of the expressions at these lattice weights are rather lengthy and 
relegated to ancillary files, this section aims to give an overview and to
highlight salient points.

%%%%%%%%%%%%%%%%%%%%%%%%%%%%%%%%%
%%%%%%%%%%%%%%%%%%%%%%%%%%%%%%%%%
\subsubsection{Modular invariant eMGFs at $|A|{+}|B| = 10$}
\label{bassec.4.1}

As before, we start by spanning the 29-dimensional basis ${\cal V}_{5,5}$ of 
indecomposable modular invariant eMGFs in order to later on assemble
the remaining ${\cal V}_{a,b}$ with $a{+}b=9$ and $10$ from their $\nabla_\tau$- and $\nabla_z$
derivatives. One can decompose ${\cal V}_{5,5}= {\cal V}_{5,5}^{\rm even} \cup {\cal V}_{5,5}^{\rm odd}$ into 16 even and 13 odd representatives
under $(z,\tau) \rightarrow ({-}\bar z,{-}\bar \tau)$, and our counting will include the
four indecomposable MGFs $\cplus{A \\ B}$ at $|A|=|B|=5$ 
(two even ones $E_5,C_{3,1,1}$ and two odd ones 
$\aplus{0 &2 &3 \\ 3 &0 &2 \\ 0 &0 &0},\aplus{0 &1 &2 &2 \\ 1 &1 &0 &3 \\ 0&0&0&0}$
as in \cite{Gerken:2020yii}).

Let us first state and briefly describe our choice of even basis
\begin{align}
{\cal V}_{5,5}^{\rm even} &= \Big\{g_5(z),\,E_5,\,C_{3|1,1}(z),\, C_{1|3,1}(z),\,
C_{2|2,1}(z),\, C_{1|2,2}(z), \, C_{3,1,1},\,  \notag \\
&\quad \quad 
C_{2|1,1,1}(z),\,
C_{1|2,1,1}(z), \, C_{2,1|1,1}(z),\, \cplus{1 &1 &1 &2 \\ 1 &1 &2 &1 \\ 0 &0 &z &z},\,
\cplus{0 &1 &1 &3 \\ 3 &1 &1 &0 \\ 0 &z &z &0}, 
\label{tobasis.45} \\
&\quad \quad \cplus{1 &1 &1 &2 \\ 1 &1 &2 &1 \\ 0 &z &0 &0},\,
\cplus{0 &1 &2 &2 \\ 2 &1 &0 &2 \\ 0&z&0 &0},\, 
C_{1|1,1,1,1}(z),\, 
\Re \cplus{0 &1 &1 &1 &2 \\ 1 &1 &1 &1 &1 \\ z &0 &0&0&0} \Big\}
\notag
\end{align}
%%%
\begin{itemize}
\item The usual two-column eMGFs $g_5(z),E_5$.
\item A total of five even three-column eMGFs $C_{3|1,1}(z),C_{1|3,1}(z),C_{2|2,1}(z),C_{1|2,2}(z)$ and $C_{3,1,1}$, the latter is often represented in the combinations $E_{2,3}=C_{3,1,1}-\frac{43}{35} E_5$ \cite{Broedel:2018izr} and ${\rm F}^{+(3)}_{2,3}= - \frac{1}{4}C_{3,1,1} + \frac{43}{140} E_5 
- \frac{\zeta_5}{240}$ \cite{Dorigoni:2021jfr} in the MGF literature.
\item A total of seven even four-column eMGFs, three of them
 $C_{2|1,1,1}(z), C_{1|2,1,1}(z),$ $C_{2,1|1,1}(z)$ in the more special class
 (\ref{cabc.2}) and four of them
 in a more general form $ \cplus{1 &1 &1 &2 \\ 1 &1 &2 &1 \\ 0 &0 &z &z},
\cplus{0 &1 &1 &3 \\ 3 &1 &1 &0 \\ 0 &z &z &0}, \cplus{1 &1 &1 &2 \\ 1 &1 &2 &1 \\ 0 &z &0 &0},
\cplus{0 &1 &2 &2 \\ 2 &1 &0 &2 \\ 0&z&0 &0}$.
\item Two even five-column eMGFs $C_{1|1,1,1,1}(z)$ and 
$\Re \cplus{0 &1 &1 &1 &2 \\ 1 &1 &1 &1 &1 \\ z &0 &0&0&0}$.
\end{itemize}
Their odd counterparts admit the following choice of basis
\begin{align}
{\cal V}_{5,5}^{\rm odd} &= \Big\{
B_{0|2,3}(0),\, \aplus{0 &1 &2 &2 \\ 1 &1 &0 &3 \\ 0&0&0&0},\,
A_{2,0|3}(z),\, A_{3,0|2}(z), \, A_{4,0|1}(z), \, B_{0|2,3}(z),
\notag \\
&\quad \quad 
\aplus{0 &1 &2 &2 \\ 1 &1 &0 &3 \\ 0&0&0&z} ,\,
\aplus{0 &1 &2 &2 \\ 1 &1 &0 &3 \\ 0&z&0&0},\,
\aplus{0 &1 &2 &2 \\ 1 &1 &0 &3 \\ 0&0&z&0},\,
\aplus{0 &1 &2 &2 \\ 1 &1 &0 &3 \\ 0&z&0&z} ,\,
\aplus{0 &1 &1 &3 \\ 1 &1 &1 &2 \\ z&0&0&0},
\label{tobasis.46} \\
&\quad \quad
\aplus{0 &1 &1 &1 &2 \\ 1 &1 &1 &1 &1 \\ z &0 &0&0&0},\,
\aplus{0 &1 &1 &1 &2 \\ 1 &1 &1 &1 &1 \\ z &0 &0&0&z}
\Big\}
\notag
\end{align}
\begin{itemize}
\item The two MGFs  $\aplus{0 &2 &3 \\ 3 &0 &2 \\ 0 &0 &0}=B_{0|2,3}(0)$
and $ \aplus{0 &1 &2 &2 \\ 1 &1 &0 &3 \\ 0&0&0&0}$.
\item Four odd three-column eMGFs $A_{2,0|3}(z),A_{3,0|2}(z), A_{4,0|1}(z), B_{0|2,3}(z)$ 
from the two classes in (\ref{adefs}) and (\ref{bdefs}).
\item Five odd eMGFs $\aplus{0 &1 &2 &2 \\ 1 &1 &0 &3 \\ 0&0&0&z} ,
\aplus{0 &1 &2 &2 \\ 1 &1 &0 &3 \\ 0&z&0&0},
\aplus{0 &1 &2 &2 \\ 1 &1 &0 &3 \\ 0&0&z&0},
\aplus{0 &1 &2 &2 \\ 1 &1 &0 &3 \\ 0&z&0&z} ,$
$\aplus{0 &1 &1 &3 \\ 1 &1 &1 &2 \\ z&0&0&0}$ with four columns as in 
(\ref{dictio.6}); the first four of them reduce to
the basis MGF $ \aplus{0 &1 &2 &2 \\ 1 &1 &0 &3 \\ 0&0&0&0}$ as $z\rightarrow 0$.
\item Two odd five-column eMGFs $\aplus{0 &1 &1 &1 &2 \\ 1 &1 &1 &1 &1 \\ z &0 &0&0&0},
\aplus{0 &1 &1 &1 &2 \\ 1 &1 &1 &1 &1 \\ z &0 &0&0&z}$.
\end{itemize}
The eMGFs in (\ref{tobasis.45}) and (\ref{tobasis.46}) are dihedral and span all
the combinations of $\beta^{\rm sv}$ encountered in the path-ordered exponential (\ref{gen.34c}) of
the generating series $Y(\begin{smallmatrix} M \\ N \end{smallmatrix} |  \begin{smallmatrix} K \\ L \end{smallmatrix} )$. Hence, {\it any} eMGF at $|A| =|B|=5$ encountered in the expansion of
(\ref{high.4}) in $s_{ij},\eta_j,\bar \eta_j$ including arbitrary non-dihedral topologies
is expressible in a dihedral basis (generically involving products of MZVs and 
eMGFs in ${\cal V}_{a,b}$ at $a{+}b\leq 8$). As will be detailed below, we arrive 
at the same conclusion for the remaining ${\cal V}_{a,b}$ at $a{+}b= 9$ or $10$, so 
any $\cplus{A \\ B\\Z}$ at $|A|{+}|B|\leq 10$ will therefore have a dihedral lattice-sum 
representation with MZV coefficients.

The $\beta^{\rm sv}$ representations of $C_{2 | 2,1}(z), C_{1 | 2,2}(z),
C_{3 | 1,1}(z),C_{1 | 3,1}(z)$ and $C_{1|1,1,1,1}(z)$ can be found in
appendix \ref{last:app.D}, and those of the remaining elements of ${\cal V}_{5,5}$
are relegated to an ancillary file.

%%%%%%%%%%%%%%%%%%%%%%%%%%%%%%%%%
%%%%%%%%%%%%%%%%%%%%%%%%%%%%%%%%%
\subsubsection{Remaining eMGFs at $|A|{+}|B| = 10$}
\label{bassec.4.2}

According to table \ref{allshuffir}, the remaining bases 
${\cal V}_{a,b}$ at $a{+}b=10$ have dimensions
\begin{align}
\dim \, {\cal V}_{6,4} &=24\,  ,&
\dim \, {\cal V}_{8,2} &=6 \notag \\
\dim \, {\cal V}_{7,3} &=15\,  ,&
\dim \, {\cal V}_{9,1} &=2
\label{tobasis.51}
\end{align}
Similar to the strategy of section \ref{bassec.3.3}, we assemble
bases ${\cal V}_{5+k,5-k}$ from the $k^{\rm th}$ $\nabla_{\tau}$-
derivatives of the modular invariants in ${\cal V}_{5,5}$ while
adding products of simpler terms to eliminate the $\dplus{a \\ 0}$:
\begin{itemize}
\item For ${\cal V}_{6,4}$, there is no need to remove any $\dplus{a \\ 0}$ from the
first 14 entries of
\begin{align}
{\cal V}_{6,4} &= \Big\{ \pi \nabla_\tau g_5(z), \, \pi \nabla_\tau E_5, \,
\pi \nabla_\tau C_{3|1,1}(z), \, \pi \nabla_\tau C_{1|3,1},\,
 \pi \nabla_\tau C_{2|2,1}(z),\,  \pi \nabla_\tau C_{3,1,1},\,  \notag \\
&\quad \quad 
 \pi \nabla_\tau C_{2|1,1,1}(z),\,
 \pi \nabla_\tau C_{1|2,1,1}(z), \, 
  \pi \nabla_\tau C_{2,1|1,1}(z),\, 
   \pi \nabla_\tau A_{2,0|3}(z) , \,
    \pi \nabla_\tau A_{3,0|2}(z), \notag\\
 &\quad \quad 
 \pi \nabla_\tau B_{0|2,3}(z), \,  
  \pi \nabla_\tau \aplus{0 &2 &3 \\ 3 &0 &2} , \,
  \pi \nabla_\tau \cplus{0 &1 &1 &1 &2 \\ 1 &1 &1 &1 &1 \\ z &0 &0&0&0}, \,
  10 \ {\rm others} \Big\}
   \label{tobasis.61} 
\end{align}
%%%%%%
\item For ${\cal V}_{7,3}$, there is no need to remove any $\dplus{a \\ 0}$ from the
first 5 entries of
\begin{align}
{\cal V}_{7,3} &= \big\{ (\pi \nabla_\tau )^2g_5(z), \, (\pi \nabla_\tau)^2 E_5, \,
(\pi \nabla_\tau)^2 C_{3|1,1}(z),\notag\\
 &\quad \quad 
( \pi \nabla_\tau)^2 C_{2|2,1}(z),\, ( \pi \nabla_\tau)^2 C_{3,1,1},\, 
  10 \ {\rm others} \big\}
   \label{tobasis.62} 
\end{align}
%%%%%%
\item For ${\cal V}_{8,2}$ and ${\cal V}_{9,1}$, there is no need to remove any $\dplus{a \\ 0}$ 
from the first 2 entries of
\begin{align}
{\cal V}_{8,2} &= \big\{ (\pi \nabla_\tau )^3g_5(z), \, (\pi \nabla_\tau)^3 E_5, \,
  4 \ {\rm others} \big\} \notag \\
{\cal V}_{9,1} &= \big\{ (\pi \nabla_\tau )^4g_5(z), \, (\pi \nabla_\tau)^4 E_5\big\}   
 \label{tobasis.63} 
\end{align}
\end{itemize}
The ``other'' $10+10+4$ basis elements suppressed in (\ref{tobasis.61})
to (\ref{tobasis.63}) resemble the hatted eMGFs in (\ref{oddA.54})
to (\ref{tobasis.32}) and can be found in an ancillary file.

%%%%%%%%%%%%%%%%%%%%%%%%%%%%%%%%%
%%%%%%%%%%%%%%%%%%%%%%%%%%%%%%%%%
\subsubsection{Bases with $|A|{+}|B| = 9$}
\label{bassec.4.3}

At lattice weight $a{+}b=9$, the basis dimensions $\dim \, {\cal V}_{a,b}$ of
table \ref{allshuffir} are
\begin{align}
\dim \, {\cal V}_{5,4} &=14\,  ,&
\dim \, {\cal V}_{7,2} &=4 \notag \\
\dim \, {\cal V}_{6,3} &=9\,  ,&
\dim \, {\cal V}_{8,1} &=1
\label{tobasis.52}
\end{align}
In close analogy with section \ref{bassec.3.4}, we construct elements of ${\cal V}_{5+k,4-k}$ from the
$k^{\rm th}$ $\nabla_\tau$-derivative of $\nabla_z {\cal V}_{5,5}$ and form suitable linear combinations
with products to eliminate tentative factors of $\dplus{a\\0}$ in the derivatives. 
\begin{itemize}
\item For ${\cal V}_{5,4}$, there is no need to remove any $\dplus{a \\ 0}$ from the
first 13 entries of
\begin{align}
{\cal V}_{5,4} &= \Big\{ \nabla_z g_5(z),\, \nabla_z C_{3|1,1}(z),\,   \nabla_zC_{1|3,1}(z),\,
 \nabla_zC_{2|2,1}(z),\, \notag \\
&\quad \quad 
 \nabla_z C_{2|1,1,1}(z),\,
 \nabla_z C_{1|2,1,1}(z), \,  \nabla_z C_{2,1|1,1}(z),\,
  \nabla_z \cplus{1 &1 &1 &2 \\ 1 &1 &2 &1 \\ 0 &0 &z &z},\,
   \nabla_z C_{1|1,1,1,1}(z),
\notag \\
&\quad \quad 
 \nabla_z A_{2,0|3}(z),\,
 \nabla_z  A_{3,0|2}(z), \, 
 \nabla_z B_{0|2,3}(z),\,
 \nabla_z \aplus{0 &1 &2 &2 \\ 1 &1 &0 &3 \\ 0&0&0&z} ,\,
 1 \ {\rm other}
  \Big\}
\label{tobasis.65}
\end{align}
%%%%
\item For ${\cal V}_{6,3}$, there is no need to remove any $\dplus{a \\ 0}$ from the
first 3 entries of
\begin{align}
{\cal V}_{6,3} &= \big\{  \pi \nabla_\tau \nabla_z g_5(z),\,
 \pi \nabla_\tau  \nabla_z C_{3|1,1}(z),\,  
  \pi \nabla_\tau \nabla_z C_{2|1,1,1}(z),\,
   6 \ {\rm others}  \big\}
\label{tobasis.66}
\end{align}
%%%%
\item For ${\cal V}_{7,2}$, there is no need to remove any $\dplus{a \\ 0}$ from the
first 2 entries of
\begin{align}
{\cal V}_{7,2} &= \big\{ ( \pi \nabla_\tau)^2 \nabla_z g_5(z),\,
( \pi \nabla_\tau)^2  \nabla_z C_{3|1,1}(z),\,  
   2 \ {\rm others}  \big\}
\label{tobasis.67}
\end{align}
%%%%
\item Finally, the one-dimensional basis ${\cal V}_{8,1}$
is given by
\begin{align}
{\cal V}_{8,1} &= \big\{ ( \pi \nabla_\tau)^3 \nabla_z g_5(z) \big\}
\label{tobasis.68}
\end{align}
\end{itemize}
The ``other'' $1+6+2$ basis elements suppressed in (\ref{tobasis.65})
to (\ref{tobasis.67}) resemble $ \widehat {\cal C}_{(4,3)}(z) $ in (\ref{tobasis.43})
and can be found in an ancillary file.

\newpage

%%%%%%%%%%%%%%%%%%%%%%%%%%%%%%%%%%%%%%%%%%%%%%%%%%%%%%%%%%%
%%%%%%%%%%%%%%%%%%%%%%%%%%%%%%%%%%%%%%%%%%%%%%%%%%%%%%%%%%%
\section{Conclusion and outlook}
\label{sec:concl}
%%%%%%%%%%%%%%%%%%%%%%%%%%%%%%%%%%%%%%%%%%%%%%%%%%%%%%%%%%%
%%%%%%%%%%%%%%%%%%%%%%%%%%%%%%%%%%%%%%%%%%%%%%%%%%%%%%%%%%%

In this work, we have developed a description of elliptic modular graph forms (eMGFs) in
terms of iterated integrals over modular parameters $\tau$. The original definition of eMGFs 
in terms of lattice sums
over discrete momenta on a torus \cite{DHoker:2018mys, Dhoker:2020gdz} manifests
their transformations as non-holomorphic modular forms under $SL(2,\mathbb Z)$. 
However, the wealth of algebraic relations among eMGFs and their expansion around the 
cusp is obscured in this picture. The iterated-integral representations of eMGFs constructed 
in this work no longer expose their modular properties but are shown in detail to pinpoint 
both the systematics of algebraic \& differential relations and the explicit form of their 
$q$-expansions.

The central building blocks for eMGFs in this work are real-analytic variants $\beta^{\rm sv}$
of iterated integrals over Kronecker-Eisenstein kernels $f^{(k)}(u\tau{+}v|\tau)$ at 
fixed comoving coordinates $u,v \in [0,1]$ of a marked point $z=u\tau{+}v$ on a torus.
These $\beta^{\rm sv}$ generalize the real-analytic iterated Eisenstein integrals
which elucidated the systematics of modular graph forms
\cite{Gerken:2020yii, Gerken:2020xfv} and furnish canonical representations
of eMGFs. In fact, the $\beta^{\rm sv}$ representations
provided in this work mix integration kernels $f^{(k)}(u\tau{+}v|\tau)$ at $k\geq 2$ with holomorphic
Eisenstein series $G_k(\tau) = - f^{(k)}(0|\tau)$ at $k\geq 4$. The counting of
such iterated integrals then sets upper bounds to the numbers of independent eMGFs at given
transcendental and modular weights.

In fact, only a subset of the iterated (Kronecker-)Eisenstein integrals are realized as eMGFs, i.e.\ the
upper bound from the counting of $\beta^{\rm sv}$ is in general not saturated.
We have described a method to anticipate the dropouts of $\beta^{\rm sv}$ from eMGFs by using
(i) the absence of poles in $u$ in eMGFs and (ii) generating series of eMGFs via Koba-Nielsen-type
integrals known from closed-string genus-one amplitudes \cite{Dhoker:2020gdz}. 
The dropouts of $\beta^{\rm sv}$ from eMGFs are governed by a generalization of 
Tsunogai's algebra of derivations $\epsilon_k$ dual to holomorphic Eisenstein 
series \cite{Tsunogai, Pollack}. We have combined the information from (i) and (ii) 
to infer large classes of commutation relations among the algebra generators. 

These commutators in turn encode the basis dimensions of indecomposable
eMGFs, see table \ref{allshuffir} for the counting for a variety of transcendental and
modular weights. Moreover, we gave a detailed construction of the respective bases
in terms of both lattice sums and iterated integrals, see section \ref{bassec} and
the ancillary file. An upcoming {\sc Mathematica} package \cite{Hidding:2022zzz}
will make an efficient implementation of the methods and results in this work 
as well as numerical evaluations of eMGFs publicly available.

This work falls into the research agenda of exploring the periods of
configuration spaces encountered in open- and closed-string amplitudes at various
loop and leg orders. Closed-string amplitudes at tree level naturally introduced single-valued
polylogarithms \cite{svpolylog, Broedel:2016kls, DelDuca:2016lad} as 
subsets of the marked points on a Riemann sphere are integrated out
order by order in the inverse string tension $\alpha'$ \cite{Stieberger:2013wea, Schlotterer:2018abc, Vanhove:2018elu}. Accordingly, eMGFs which originate from the configuration-space
integrals in closed-string one-loop amplitudes \cite{DHoker:2018mys, 
Dhoker:2020gdz} provide an organizing principle for single-valued elliptic polylogarithms,
extending those of Zagier \cite{Ramakrish} beyond depth one.

While this work's perspective on single-valued elliptic polylogarithms is governed by
iterated integrals over modular parameters $\tau$, one can alternatively find canonical
representations via iterated integrals over marked points $z$. As will be explored in future work, 
multiple elliptic polylogarithms of Brown and Levin \cite{BrownLev} together with their 
complex conjugates and (elliptic) multiple zeta values yield an equivalent description of eMGFs.
It would be interesting to extend the depth-one formulae of this type in \cite{Broedel:2019tlz}
to higher depth, possibly employing single-valued integration as outlined in \cite{Panzertalk} or generating-function methods akin to the
construction at genus zero \cite{svpolylog, DelDuca:2016lad}.

Another promising line of follow-up research concerns Poincar\'e-series representations
of eMGFs as a variant of the manifestly modular lattice sums. Poincar\'e series
naturally extended modular graph forms at depth two by integrals of holomorphic
cusp forms \cite{Dorigoni:2021jfr, Dorigoni:2021ngn} that should yield 
Brown's modular iterated integrals at depth two \cite{Brown:mmv, Brown:2017qwo},
see \cite{Diamantis:2020} for closely related work.
Similarly, Poincar\'e series involving the $T$-invariant building blocks ${\cal E}_0$ of
eMGFs from section \ref{sec:qeMGFs} in their seed functions are expected to extend eMGFs
in a controlled way: By analogy with the extension of modular graph forms in
\cite{Dorigoni:2021jfr, Dorigoni:2021ngn}, this may associate non-holomorphic modular 
forms to {\it all} of the iterated integrals $\beta^{\rm sv}$ described in this work.

Both the $\beta^{\rm sv}$ representations of eMGFs in this work and their tentative
extensions by Poincar\'e series call for a study of the modular properties of their meromorphic
building blocks. In particular, it could be rewarding to investigate if (suitably regularized) 
iterated integrals of $G_k(\tau)$ 
and $f^{(k)}(u\tau{+}v|\tau)$ from $0$ to $i\infty$ qualify as $z$-dependent analogues 
of multiple modular values \cite{Brown:mmv}. This line of thought is hoped to generalize the fruitful
connection between multiple modular values and holomorphic cusp 
forms \cite{Pollack, Brown:mmv, Brown2019, Dorigoni:2021ngn}.

Finally, this work aims to set the stage for systematic studies of configuration-space integrals with
more unintegrated marked points and at higher genus: For eMGFs at depth two depending on two
marked points $z_1,z_2$, the counting of independent representatives will
be governed by multivariable generalizations of Tsunogai's derivation algebra which
has been pioneered in an open-string context in \cite{Kaderli:2022qeu}. 
At higher genus, pilot studies of modular graph forms 
\cite{DHoker:2017pvk, DHoker:2018mys} and tensors \cite{DHoker:2020uid}
as well as their differential structure \cite{DHoker:2014oxd, Pioline:2015qha, Basu:2018bde, 
Basu:2020goe, Basu:2021xdt} motivate detailed investigations of their systematics and
iterated-integral representations.
Moreover, generalizations of higher-genus modular graph tensors that depend on
marked points raise even more ambitious long-term questions concerning
their degenerations and differential relations as well as decompositions into
meromorphic building blocks.

\appendix

\newpage

%%%%%%%%%%%%%%%%%%%%%%%%%%%%%%%%%%%%%%%%%%%%%%%%%%%%%%%%%%%
%%%%%%%%%%%%%%%%%%%%%%%%%%%%%%%%%%%%%%%%%%%%%%%%%%%%%%%%%%%
\section{Elementary definitions}
\label{app:def}
%%%%%%%%%%%%%%%%%%%%%%%%%%%%%%%%%%%%%%%%%%%%%%%%%%%%%%%%%%%
%%%%%%%%%%%%%%%%%%%%%%%%%%%%%%%%%%%%%%%%%%%%%%%%%%%%%%%%%%%

This appendix gathers elementary definitions related to theta functions, 
multiple polylogarithms and iterated Eisenstein integrals that support the discussions
in the main text.

%%%%%%%%%%%%%%%%%%%%%%%%%%%%%%%%%%%%%%%%%%%%%%%%%%%%%%%%%%%
%%%%%%%%%%%%%%%%%%%%%%%%%%%%%%%%%%%%%%%%%%%%%%%%%%%%%%%%%%%
\subsection{Kronecker-Eisenstein sums from theta functions}
\label{app:theta}
%%%%%%%%%%%%%%%%%%%%%%%%%%%%%%%%%%%%%%%%%%%%%%%%%%%%%%%%%%%
%%%%%%%%%%%%%%%%%%%%%%%%%%%%%%%%%%%%%%%%%%%%%%%%%%%%%%%%%%%

The closed-string Green function $g(z|\tau)$ with lattice-sum representation 
in (\ref{elemlattice}) is defined in terms of the
Dedekind eta function $\eta (\tau) = q^{{1 \over 24}} \prod _{n=1}^\infty (1-q^n)$
with $q=e^{2\pi i \tau}$ and the odd Jacobi theta function in normalization conventions,
\beq
\tet_1(z|\tau) = q^{1/8} (e^{i\pi z} - e^{-i\pi z} ) \prod_{n=1}^\infty (1-q^n)(1-e^{2\pi i z} q^n)(1-e^{-2\pi i z} q^n)
\label{app:theta.1}
\eeq
namely
\bea
g(z|\tau) = - \log \left| \frac{ \tet_1(z|\tau) }{\eta(\tau)} \right|^2 + \frac{ 2\pi (\Im z)^2 }{\Im \tau}
\label{app:theta.2}
\eea
The doubly-periodic Kronecker-Eisenstein series is given by
\cite{Kronecker, BrownLev}
\bea
\label{Omega}
\Omega(z,\eta|\tau) = \exp \Big( 2\pi i \eta\, \frac{ \Im z }{\Im \tau} \Big)
\frac{ \tet_1'(0|\tau) \,  \tet_1(z{+}\eta|\tau) }{\tet_1(z|\tau) \, \tet_1(\eta|\tau)}
= \sum_{k=0}^{\infty} \eta^{k-1} f^{(k)}(z|\tau)
\label{app:theta.3}
\eea
where the last step defines the Kronecker-Eisenstein coefficients such as
\begin{align}
f^{(0)}(z|\tau)&=1 \, , \ \ \ \ \ \
f^{(1)}(z|\tau) = \partial_z \log \tet_1(z|\tau) + 2\pi i \frac{\Im z}{\Im \tau}
\notag\\
f^{(2)}(z|\tau) &= \frac{1}{2} \bigg\{ f^{(1)} (z|\tau)^2  + \partial_z^2  \log \tet_1(z|\tau)
- \frac{   \tet'''_1(0|\tau) }{3 \tet'_1(0|\tau)} \bigg\}
\label{app:theta.4}
\end{align}
with $\tet'_1(0|\tau) = \partial_z \tet_1(z|\tau)|_{z=0}$.
The first two non-trivial instances $f^{(1)},f^{(2)}$ are related to the
closed-string Green function (\ref{app:theta.1}) via
\beq
f^{(1)}(z|\tau) = - \partial_z g(z|\tau) \, , \ \ \ \ \ \
f^{(2)}(u\tau{+}v|\tau) = - 2\pi i \partial_\tau g(u\tau{+}v|\tau)
\label{gfderiv}
\eeq
where the $\tau$-derivative is taken at constant comoving coordinates $u,v$,
and the lattice-sum representation of all $f^{(k\geq 1)}$ can be found
in (\ref{elemlattice}).

%%%%%%%%%%%%%%%%%%%%%%%%%%%%%%%%%%%%%%%%%%%%%%%%%
%%%%%%%%%%%%%%%%%%%%%%%%%%%%%%%%%%%%%%%%%%%%%%%%%
\subsection{Bernoulli numbers and polynomials}
\label{app:Bern}
%%%%%%%%%%%%%%%%%%%%%%%%%%%%%%%%%%%%%%%%%%%%%%%%%
%%%%%%%%%%%%%%%%%%%%%%%%%%%%%%%%%%%%%%%%%%%%%%%%%

Bernoulli polynomials $B_k(x)$ are defined by the generating function,
\begin{equation}
\frac{t e^{x t}}{e^{t}-1}=\sum_{k=0}^{\infty} B_{k}(x) \frac{t^{k}}{k !}
 \label{fkreps.2}
\end{equation}
such that the simplest instances at $0\leq k \leq 4$ are given by
\begin{align}
B_0(x) &=1 \, , 
&B_3(x) &=   x^3   - \frac{ 3}{2} x^2 + \frac{ x}{2}
 \notag \\
B_1(x) &= x - \frac{1}{2}\, , 
&B_4(x) &=  x^4 - 2 x^3 + x^2 - \frac{1}{30}  \label{fkreps.3}  \\
B_2(x) &= x^2 - x + \frac{ 1}{6} \, ,  
&B_5(x) &= x^5 - \frac{5}{2} x^4
+\frac{5}{3}x^3 -\frac{x}{6}
  \notag
\end{align}
They obey the differential equation
\beq
\partial_x B_k(x) = k B_{k-1}(x)
\eeq
and reduce to Bernoulli numbers $B_k= B_k(0)$
at vanishing argument, e.g.
\begin{align}
B_0 &= 1 \, , &B_4 &= - \frac{1}{30} \notag \\
B_1 &= - \frac{1}{2} \, , &B_6 &= \frac{1}{42}
\label{fkreps.3a} \\
B_2 &= \frac{1}{6} \, , &B_{2m+1} &= 0 \, , \ \ \ \ m \geq 1
\notag
\end{align}
%

%%%%%%%%%%%%%%%%%%%%%%%%%%%%%%%%%%%%%%%%%%%%%%%%%
%%%%%%%%%%%%%%%%%%%%%%%%%%%%%%%%%%%%%%%%%%%%%%%%%
\subsection{Multiple polylogarithms and MZVs}
\label{app:poly}
%%%%%%%%%%%%%%%%%%%%%%%%%%%%%%%%%%%%%%%%%%%%%%%%%
%%%%%%%%%%%%%%%%%%%%%%%%%%%%%%%%%%%%%%%%%%%%%%%%%

Multiple polylogarithms in one variable $z$ can be recursively defined via
\beq
G(a_1,a_2,\ldots,a_w;z) = \int^z_0 \frac{ \dd t}{t-a_1} \, G(a_2,\ldots,a_w;t)
\label{fkreps.10}
\eeq
with $G(\emptyset;z)=1$, and the $a_i$ are constants commonly referred to as letters. The total number $w$ of $a_i$ is known as the (transcendental) weight of the polylogarithm, and for polylogarithms where $a_i \in \{ 0,1 \}$ the number of $a_i = 1$ is referred to as the \emph{depth}. Multiple polylogarithms obey
the usual shuffle relations of iterated integrals
\beq
G(A;z) G(B;z) = \sum_{C \in A\shuffle B} G(C;z)
\label{Gshuffle}
\eeq
where the shuffle $A\shuffle B$ comprises all permutations of the ordered
sets $A=(a_1,a_2,\ldots,a_n)$ and $B=(b_1,b_2,\ldots,b_m)$ that preserve
the order within $A$ and $B$. The simplest examples of polylogarithms are given by
\begin{align}
G(\vec{0}^{p};z) &= \frac{1}{p!} \big[ \log (z)\big]^p
\notag \\
G(\vec{1}^{p};z) &= \frac{1}{p!} \big[ \log (1- z) \big]^p
\label{fkreps.11}\\
G(\vec{0}^{p-1},1;z) &= -  {\rm Li}_p(z)
\notag
\end{align}
where the expression for $G(\vec{0}^{p};z)$ is understood as a regularized value.
More generally, multiple polylogarithms $G(\ldots,0;z)$ with an endpoint divergence can be
shuffle-regularized on the basis of (\ref{fkreps.11}) by
imposing (\ref{Gshuffle}) to hold for their regularized values,
e.g.\ $G(1,0;z)= G(0;z) G(1;z)-G(0,1;z)$, see for instance
\cite{Panzer:2015ida, Abreu:2022mfk} for further details.

At unit argument $z=1$, multiple polylogarithms are well-known to reduce to MZVs
\begin{align}
\zeta_p &= {\rm Li}_p(1) = - G(\vec{0}^{p-1},1;1)
\label{fkreps.12} \\
\zeta_{p_1,p_2,\ldots,p_r} &= (-1)^r G(\vec{0}^{p_r-1},1, \ldots,\vec{0}^{p_2-1},1,\vec{0}^{p_1-1},1;1)
\notag
\end{align}
and one can again define shuffle-regularized MZVs by imposing (\ref{Gshuffle})
at $z=1$ together with $G(1;1)=G(0;1)=0$. By geometric-series expansion of
the kernels $\frac{\dd t}{t-1}$, these integral representations are equivalent to
the nested sums
\beq
\zeta_{p_1,p_2,\ldots,p_r} = \sum_{0<k_1<k_2<\ldots<k_r}^{\infty}
k_1^{-p_1} k_2^{-p_2} \ldots k_r^{-p_r}\, , \ \ \ \ \ \ p_r\geq 2
\eeq

%%%%%%%%%%%%%%%%%%%%%%%%%%%%%%%%%%%%%%%%%%%%%%%%%
%%%%%%%%%%%%%%%%%%%%%%%%%%%%%%%%%%%%%%%%%%%%%%%%%
\subsection{Single-valued multiple polylogarithms}
\label{app:svpoly}
%%%%%%%%%%%%%%%%%%%%%%%%%%%%%%%%%%%%%%%%%%%%%%%%%
%%%%%%%%%%%%%%%%%%%%%%%%%%%%%%%%%%%%%%%%%%%%%%%%%

The monodromies of multiple polylogarithms (\ref{fkreps.10}) around
the singular points $z=a_i$ of the integration kernels can be compensated
by means of complex conjugate polylogarithms. More precisely, Brown constructed
single-valued polylogarithms in one variable \cite{svpolylog} that obey the same holomorphic
differential equations as their meromorphic counterparts in (\ref{fkreps.10})
\beq
\partial_z G^{\rm sv}(a_1,a_2,\ldots,a_w;z)  =  \frac{G^{\rm sv}(a_2,\ldots,a_w;z) }{z-a_1}
\eeq
but their antiholomorphic derivatives $\partial_{\bar z} G^{\rm sv}(a_1,\ldots,a_w;z)$ are much more involved.
What is more, they are engineered to preserve the shuffle relations (\ref{Gshuffle})
\beq
G^{\rm sv}(A;z) G^{\rm sv}(B;z) = \sum_{C \in A\shuffle B} G^{\rm sv}(C;z)
\label{Gsvshuffle}
\eeq
At general weight and depth, $G^{\rm sv}(a_1,\ldots,a_w;z)$ are built from products of
$G(\ldots;z)$, complex conjugates $\overline{G(\ldots;z)}$ and MZVs such
that the monodromies cancel and the weight of each term adds up to $w$.
In general, they can be constructed by making use of the Hopf-algebra structure of polylogarithms~\cite{goncharov2001multiple}, the coproduct $\Delta$ of which is given by
\beq
\Delta G(\vec{a};z) = \sum_{\vec{b}\subseteq \vec{a}} G(\vec{b};z) \otimes G_{\vec{b}}(\vec{a};z)
\eeq
The sum runs over ordered subsets $\vec{b}$ of $\vec{a}$, and $G_{\vec{b}}(\vec{a};z)$ 
denotes an iterated integral with the same integrand as $G(\vec{a};z)$, but integrated over 
a contour encircling the poles in $\vec{b}$ in order.
Having defined a coproduct, the antipode $S$ can be inferred from the constraint $\mu (S \otimes {\rm id})\Delta = 0$ where the operator $\mu$ denotes multiplication $\mu(a \otimes b) = a \cdot b$.
Using this structure, one may define a single-valued map~\cite{Brown:2013gia,brown2015notes}
\beq
{\rm sv} = \mu (\tilde{S} \otimes {\rm id})\Delta
\eeq
where $\tilde{S}$ denotes a modified antipode such that
\beq
\tilde{S}(G(\vec{a};z)) = (-1)^{|\vec{a}|}\overline{S(G(\vec{a};z))}
\eeq
This single-valued map sends every polylogarithm to its single-valued analogue
\beq
{\rm sv}(G(\vec{a};z)) = G^{\rm sv}(\vec{a};z)
\eeq
Using this construction, it is straightforward to obtain single-valued polylogarithms at generic weight, even for a higher number of independent variables, see e.g.~\cite{DelDuca:2016lad}.

Let us present some examples of single-valued polylogarithms relevant to the discussions in the main text of this paper for the letters $a_i \in \{0,1\}$. The contributions without MZVs can be given in closed form through the first line of
\begin{align}
G^{\rm sv}(a_1,a_2,\ldots,a_w;z) &= \sum_{j=0}^w G(a_1,a_2,\ldots,a_j;z) \overline{  G(a_w,a_{w-1}\ldots,a_{j+1};z)}  \notag \\
&\quad
+\mathcal{Z}^{\rm sv}(a_1,a_2,\ldots,a_w;z)
\label{svpolycl}
\end{align}
while the leftover contribution $\mathcal{Z}^{\rm sv}(a_1,\ldots,a_w;z)$ in the second line
carries at least three units of weight via MZVs. At depth zero and one, these extra terms
vanish,
\beq
\mathcal{Z}^{\rm sv}(\vec{0}^a;z) =0 = \mathcal{Z}^{\rm sv}(\vec{0}^a,1,\vec{0}^b;z)
\label{svpolyex}
\eeq
and we are led to simple expressions such as
\begin{align}
G^{\rm sv}(\vec{0}^{p};z) &= \frac{1}{p!} (\log |z|^2)^p
\notag \\
G^{\rm sv}(\vec{1}^{p};z) &= \frac{1}{p!} (\log |1{-} z|^2)^p
\label{comrel.27e} \\
G^{\rm sv}(\vec{0}^{b-1},1,\vec{0}^{a-1};z)&=(-1)^{a}\sum_{k=b}^{a+b-1} {k-1 \choose b-1} \frac{(-\log|z|^2)^{a+b-1-k}}{(a+b-1-k)!}\text{Li}_k(z)\nonumber\\
&\quad +(-1)^{b}\sum_{k=a}^{a+b-1} {k-1 \choose a-1} \frac{(-\log|z|^2)^{a+b-1-k}}{(a+b-1-k)!} \overline{\text{Li}_k(z) }
\notag
\end{align}
At depth two, the $\mathcal{Z}^{\rm sv}$ in (\ref{svpolycl}) can be neatly given in closed form
\begin{align}
\mathcal{Z}^{\rm sv}(\vec{0}^a,1,\vec{0}^b,1;z) &=2 (-1)^{b} \sum_{k=\lfloor\frac{b}{2}\rfloor+1}^{\lfloor\frac{a+b}{2}\rfloor}  {2k \choose b}\zeta_{2k+1} \big[ G^{\rm sv}(\vec{0}^{a+b-2k},1;z) - G(\vec{0}^{a+b-2k},1;z) \big]
\label{comrel.26}
\end{align}
which we have not yet encountered in the literature. 
The simplest non-vanishing instances of (\ref{comrel.26}) are
 \small
\begin{align}
\mathcal{Z}^{\rm sv}(0,1,0,1;z) &= -4 \zeta_3 \overline{ G(1;z)}   \, , \ 
&\mathcal{Z}^{\rm sv}(0,0,1,0,1;z) &=  -4 \zeta_3 \big[ \overline{ G(1,0;z)}  + \overline{ G(1;z)}  G(0;z)\big]
\notag\\
\mathcal{Z}^{\rm sv}(0,0,1,1;z) &=  2 \zeta_3\overline{ G(1;z)}   \, ,
&\mathcal{Z}^{\rm sv}(0,0,0,1,1;z) &=  2 \zeta_3 \big[\overline{ G(1,0;z)} + \overline{ G(1;z)}  G(0;z)\big]
\label{comrel.27}
\end{align} \normalsize
and higher-weight examples may involve several odd zeta values, e.g.
\begin{align}
&\mathcal{Z}^{\rm sv}(0,0,0,1,0,1;z)  = - 8 \zeta_5 \overline{ G(1;z)}   \label{comrel.27.a}
\\
&\ \ \ \
 -4  \zeta_3\big[\overline{ G(1,0,0;z)} + \overline{ G(1,0;z)}  G(0;z) +
    \overline{ G(1;z)}  G(0,0;z)\big]  \notag
\end{align}
Depth-two instances of $\mathcal{Z}^{\rm sv}(\ldots,0;z)$ with a zero in the last entry
are determined from (\ref{comrel.26}) through the shuffle relations (\ref{Gsvshuffle}).

%%%%%
\subsection{Alternative bases ${\cal E}^{\rm sv}$ of $\beta^{\rm sv}$}
\label{sec:appesv}

The original construction of $\beta^{\rm sv}[\ldots]$ in \cite{Gerken:2020yii}
employed another type of real-analytic combinations ${\cal E}^{\rm sv}[\ldots]$ of Brown's
meromorphic iterated Eisenstein integrals (\ref{gen.36a}). Their depth $\ell\leq 2$
instances are given by
\begin{align}
\EsvBR{j_1 \\ k_1}{\tau} &= \sum_{r_1=0}^{j_1} (-2\pi i \bar\tau)^{r_1} \binom{j_1}{r_1}
\Big\{ \EBR{j_1-r_1 \\ k_1}{\tau}
+(-1)^{j_1-r_1} \overline{  \EBR{j_1-r_1 \\ k_1}{\tau} } \Big\}
\notag \\
\EsvBR{j_1 &j_2 \\ k_1 &k_2}{\tau} &= \overline{\alphaBR{j_1 &j_2 \\ k_1 &k_2}{\tau}}+ \sum_{r_1=0}^{j_1}\sum_{r_2=0}^{j_2} (-2\pi i \bar\tau)^{r_1+r_2}
\binom{j_1}{r_1}\binom{j_2}{r_2}
\Big\{ \EBR{j_1 -r_1&j_2-r_2 \\ k_1 &k_2}{\tau}
\label{fkreps.17}  \\
&\hspace{10mm} +(-1)^{j_1-r_1} \overline{ \EBR{j_1-r_1 \\ k_1}{\tau} }  \EBR{j_2-r_2 \\ k_2}{\tau}
+(-1)^{j_1+j_2-r_1-r_2} \overline{ \EBR{j_2-r_2 &j_1-r_1 \\ k_2 &k_1}{\tau} }
\Big\}   \notag
\end{align}
with antiholomorphic $T$-invariants $\overline{\alphaBR{j_1&j_2 \\ k_1 &k_2}{\tau}}$ as 
in (\ref{fkreps.19}), and their holomorphic differential equation at general depth $\ell$ reads
\begin{align}
2\pi i \partial_\tau
\esvtau{j_1 &j_2 &\ldots &j_\ell \\ k_1 &k_2 &\ldots &k_\ell }
&=
-(2\pi i)^{2-k_\ell + j_\ell} (\tau {-} \bar \tau)^{j_\ell} G_{k_\ell}( \tau)
\esvtau{j_1 &j_2 &\ldots &j_{\ell-1} \\ k_1 &k_2 &\ldots &k_{\ell-1} }
\label{gen.35}
\end{align}
They are related to the $\beta^{\rm sv}[\ldots]$ with identical entries $k_1,k_2,\ldots,k_\ell$ 
through the simple change of basis
\begin{align}
\bsvtau{j_1 &j_2 &\ldots &j_\ell \\ k_1 &k_2 &\ldots &k_\ell }
&= \sum_{p_1=0}^{k_1-j_1-2}  \sum_{p_2=0}^{k_2-j_2-2} \ldots  \sum_{p_\ell=0}^{k_\ell-j_\ell-2}
{ k_1{-}j_1{-}2 \choose p_1}  { k_2{-}j_2{-}2 \choose p_2} \ldots { k_\ell{-}j_\ell{-}2 \choose p_\ell} \notag \\
&\ \ \ \ \times \Big( \frac{1}{4y} \Big)^{p_1+p_2+\ldots +p_\ell}
\esvtau{j_1+p_1 &j_2+p_2 &\ldots &j_\ell + p_\ell \\ k_1 &k_2 &\ldots &k_\ell}
\label{gen.38a}
\end{align}
as one can verify at $\ell \leq 2$ by comparing (\ref{eq:bsv1}) with (\ref{fkreps.17}).
This change of basis restricts the appearance of holomorphic Eisenstein series
in the $\tau$-derivatives to cases with $j_\ell=k_\ell{-}2$, and it furthermore simplifies
the leading-depth terms in the modular $S$-transformations.

\newpage

%%%%%%%%%%%%%%%%%%%%%%%%%%%%%%%%%%%%%%%%%%%%%%%%%%%%%%%%%%%
%%%%%%%%%%%%%%%%%%%%%%%%%%%%%%%%%%%%%%%%%%%%%%%%%%%%%%%%%%%
\section{Brief review of eMGF identities}
\label{app:A.1}
%%%%%%%%%%%%%%%%%%%%%%%%%%%%%%%%%%%%%%%%%%%%%%%%%%%%%%%%%%%
%%%%%%%%%%%%%%%%%%%%%%%%%%%%%%%%%%%%%%%%%%%%%%%%%%%%%%%%%%%

In this section, we review various types of identities among eMGFs
following the discussion and conventions in \cite{Dhoker:2020gdz}.

%%%%%%%%%%%%%%%%%%%%%%%%%%%%%%%%%%%%%%%%%%%%%%%%%
%%%%%%%%%%%%%%%%%%%%%%%%%%%%%%%%%%%%%%%%%%%%%%%%%
\subsection{Momentum conservation and shift identities}
\label{app:A.1.1}
%%%%%%%%%%%%%%%%%%%%%%%%%%%%%%%%%%%%%%%%%%%%%%%%%
%%%%%%%%%%%%%%%%%%%%%%%%%%%%%%%%%%%%%%%%%%%%%%%%%

By the momentum-conserving delta function in (\ref{gen.66}), dihedral eMGFs
with entries shifted by unit vectors $S_r=[\vec{0}^{r-1},1, \vec{0}^{R-r}]$ are related by
\beq
 \sum_{r=1}^R \cplus{A-S_r \\ B \\ Z} = 0 \, , \ \ \ \ \ \
  \sum_{r=1}^R \cplus{A \\ B-S_r \\ Z} = 0
  \label{compap.1}
\eeq
The first and second identity follow from insertion of
$\sum_{r=1}^R p_r=0$ and $\sum_{r=1}^R \bar p_r=0$ into the summands,
respectively. Similarly, inserting the characters $\chi_{  \sum_{r=1}^R p_r }(- z|\tau)=1$
into the summands of (\ref{gen.66}) yields the shift invariance
\beq
\cplus{a_1 &a_2& \ldots &a_R \\ b_1 &b_2& \ldots &b_R \\ z_1 &z_2& \ldots &z_R}
= \cplus{a_1 &a_2& \ldots &a_R \\ b_1 &b_2& \ldots &b_R \\ z_1-z &z_2-z& \ldots &z_R-z}
 \label{compap.2}
\eeq

%%%%%%%%%%%%%%%%%%%%%%%%%%%%%%%%%%%%%%%%%%%%%%%%%
%%%%%%%%%%%%%%%%%%%%%%%%%%%%%%%%%%%%%%%%%%%%%%%%%
\subsection{Integral representations and factorization}
\label{app:A.1.2}
%%%%%%%%%%%%%%%%%%%%%%%%%%%%%%%%%%%%%%%%%%%%%%%%%
%%%%%%%%%%%%%%%%%%%%%%%%%%%%%%%%%%%%%%%%%%%%%%%%%

Given that products of characters integrate to a momentum-conserving delta
function, it is easy to see that dihedral eMGFs (\ref{gen.66}) arise from the integral
over the torus $\Sigma$
\beq
\cplus{A \\ B \\ Z}\!(\tau) = \int_{\Sigma} \frac{ \dd^2 z}{\Im \tau} \prod_{r=1}^R
\dplus{a_r \\ b_r} \! (z_r{-}z|\tau)
 \label{compap.3}
\eeq
Together with the special case of (\ref{basic.13}) at $a=b=0$,
\beq
\dplus{0 \\ 0}\!(z|\tau) = \sum_{p \in \Lambda'} \chi_p(z|\tau) = (\Im \tau) \delta^2(z,\bar z) - 1
 \label{compap.4}
\eeq
we arrive at the following factorization identity for eMGFs with a vanishing column
(its lowest entry can always be shifted to zero via (\ref{compap.2}))
\beq
\cplus{a_1 &a_2 &\ldots &a_R &0 \\ b_1 &b_2 &\ldots &b_R &0 \\ z_1 &z_2 &\ldots &z_R &0}
= \bigg(
 \prod_{r=1}^R\dplus{a_r \\ b_r} \! (z_r|\tau) \bigg) - \cplus{a_1 &a_2 &\ldots &a_R \\ b_1 &b_2 &\ldots &b_R \\ z_1 &z_2 &\ldots &z_R}
 \label{compap.5}
 \eeq

%%%%%%%%%%%%%%%%%%%%%%%%%%%%%%%%%%%%%%%%%%%%%%%%%
%%%%%%%%%%%%%%%%%%%%%%%%%%%%%%%%%%%%%%%%%%%%%%%%%
\subsection{Complex conjugation and reflection}
\label{app:A.1.3}
%%%%%%%%%%%%%%%%%%%%%%%%%%%%%%%%%%%%%%%%%%%%%%%%%
%%%%%%%%%%%%%%%%%%%%%%%%%%%%%%%%%%%%%%%%%%%%%%%%%

The normalization factors of $(\Im \tau)^{|A|} \pi^{-|B|}$ in the definition (\ref{gen.66})
of dihedral eMGFs ${\cal C}^{+}$ are tailored to attain vanishing holomorphic modular
weight and therefore a simple action of the Cauchy-Riemann derivative $\nabla_\tau$
in (\ref{crdrv}). However, these properties are not preserved by complex conjugation,
and one can introduce an alternative variant
\beq
\cminus{A \\ B \\ Z} = (\pi \Im \tau)^{|B|-|A|} \cplus{A \\ B \\ Z}
 \label{compap.6}
\eeq
of eMGFs with modular weight $(|A|{-}|B|,0)$. In this way, we can compactly write
complex conjugate eMGFs as
\beq
\overline{ \cplus{A \\ B \\ Z}  } = \cminus{B \\ A \\ -Z}
\label{compap.7}
\eeq
where the arguments $z_r$ can be reflected via
\beq
 \cplus{A \\ B \\ -Z}  = (-1)^{|A|+|B|}  \cplus{A \\ B \\ Z}  \, , \ \ \ \ \ \
  \cminus{A \\ B \\ -Z}  = (-1)^{|A|+|B|}  \cminus{A \\ B \\ Z}
\label{compap.8}
\eeq

\newpage

\section{Iterated KE integrals at depth three}
\label{app:exps}

This appendix gathers various identities
related to iterated KE integrals of depth three.

\subsection{Path-ordered exponential at depth three}
\label{POEd3}

The explicit form of the depth-three contributions to the path-ordered exponential in (\ref{gen.34c})
 is (see (\ref{gen.34b}) for the $c_{j_i,k_i}$) \small
\begin{align}
Y&(\begin{smallmatrix} M \\ N \end{smallmatrix} |  \begin{smallmatrix} K \\ L \end{smallmatrix} | \tau) \, \big|_{{\rm depth} \ 3} = \sum_{P,Q,R,S}  \sum_{k_1,k_2,k_3=2}^\infty \sum_{j_1=0}^{k_1-2} \sum_{j_2=0}^{k_2-2}
 \sum_{j_3=0}^{k_3-2} c_{j_1,k_1}c_{j_2,k_2}c_{j_3,k_3} \notag \\
 &\times \bigg\{  \bsvtau{j_1 &j_2 &j_3 \\ k_1 &k_2 &k_3 \\ \emptyslot &\emptyslot &\emptyslot } R_{\vec{\eta}}\big(
 {\rm ad}_{\ep_0}^{k_3-j_3-2}(\ep_{k_3})  {\rm ad}_{\ep_0}^{k_2-j_2-2}(\ep_{k_2}) {\rm ad}_{\ep_0}^{k_1-j_1-2}(\ep_{k_1}) \big) \notag \\
&\ \
+ \bsvtau{j_1 &j_2 &j_3 \\ k_1 &k_2 &k_3 \\ z &\emptyslot &\emptyslot } R_{\vec{\eta}}\big( {\rm ad}_{\ep_0}^{k_3-j_3-2}(\ep_{k_3})  {\rm ad}_{\ep_0}^{k_2-j_2-2}(\ep_{k_2}) {\rm ad}_{\ep_0}^{k_1-j_1-2}(b_{k_1}) \big) \notag \\
&\ \
+ \bsvtau{j_1 &j_2 &j_3 \\ k_1 &k_2 &k_3 \\ \emptyslot &z &\emptyslot } R_{\vec{\eta}}\big( {\rm ad}_{\ep_0}^{k_3-j_3-2}(\ep_{k_3})
 {\rm ad}_{\ep_0}^{k_2-j_2-2}(b_{k_2}) {\rm ad}_{\ep_0}^{k_1-j_1-2}(\ep_{k_1}) \big) \notag \\
 &\ \
+ \bsvtau{j_1 &j_2 &j_3 \\ k_1 &k_2 &k_3 \\ z &z &\emptyslot } R_{\vec{\eta}}\big( {\rm ad}_{\ep_0}^{k_3-j_3-2}(\ep_{k_3})
  {\rm ad}_{\ep_0}^{k_2-j_2-2}(b_{k_2}) {\rm ad}_{\ep_0}^{k_1-j_1-2}(b_{k_1}) \big)
\label{POEd3.1} \\
%%%%
%%%%
%%%%
& \ \
+ \bsvtau{j_1 &j_2 &j_3 \\ k_1 &k_2 &k_3 \\ \emptyslot &\emptyslot &z } R_{\vec{\eta}}\big( {\rm ad}_{\ep_0}^{k_3-j_3-2}(b_{k_3})
 {\rm ad}_{\ep_0}^{k_2-j_2-2}(\ep_{k_2}) {\rm ad}_{\ep_0}^{k_1-j_1-2}(\ep_{k_1}) \big) \notag \\
&\ \
+ \bsvtau{j_1 &j_2 &j_3 \\ k_1 &k_2 &k_3 \\ z &\emptyslot  &z} R_{\vec{\eta}}\big( {\rm ad}_{\ep_0}^{k_3-j_3-2}(b_{k_3})
 {\rm ad}_{\ep_0}^{k_2-j_2-2}(\ep_{k_2}) {\rm ad}_{\ep_0}^{k_1-j_1-2}(b_{k_1}) \big) \notag \\
&\ \
+ \bsvtau{j_1 &j_2 &j_3 \\ k_1 &k_2 &k_3 \\ \emptyslot &z &z } R_{\vec{\eta}}\big(   {\rm ad}_{\ep_0}^{k_3-j_3-2}(b_{k_3})
 {\rm ad}_{\ep_0}^{k_2-j_2-2}(b_{k_2}) {\rm ad}_{\ep_0}^{k_1-j_1-2}(\ep_{k_1}) \big) \notag \\
 &\ \
+ \bsvtau{j_1 &j_2 &j_3 \\ k_1 &k_2 &k_3 \\ z &z &z } R_{\vec{\eta}}\big(  {\rm ad}_{\ep_0}^{k_3-j_3-2}(b_{k_3})  {\rm ad}_{\ep_0}^{k_2-j_2-2}(b_{k_2}) {\rm ad}_{\ep_0}^{k_1-j_1-2}(b_{k_1}) \big)
\bigg\}_{\begin{smallmatrix} K \\ L \end{smallmatrix} \big|  \begin{smallmatrix} P \\ Q \end{smallmatrix}}   \notag \\
&\times \exp\bigg( {-}\frac{ R_{\vec{\eta}}(\ep_0)}{4y} \bigg)_{\begin{smallmatrix} P \\ Q \end{smallmatrix} \big|  \begin{smallmatrix} R \\ S \end{smallmatrix}}
 \widehat Y(
\begin{smallmatrix} M \\ N \end{smallmatrix} |  \begin{smallmatrix} R \\ S \end{smallmatrix} | i\infty) \notag
\end{align}  \normalsize
where we have again suppressed the dependence of both
$Y(\begin{smallmatrix} M \\ N \end{smallmatrix} |  \begin{smallmatrix} K \\ L \end{smallmatrix} | \tau)$
and $ \widehat Y(
\begin{smallmatrix} M \\ N \end{smallmatrix} |  \begin{smallmatrix} R \\ S \end{smallmatrix} | i\infty)$ on $u,v$ to avoid cluttering.

\subsection{Converting different types of iterated KE integrals}
\label{app:exps1}

The two types of iterated KE integrals ${\cal E}[\ldots]$ into ${\cal E}(\ldots)$
at depth $\leq 3$ are related by the following conversion formulae
which streamline and extend (\ref{revsec.41a}) and (\ref{fkreps.8}). 
\small
\begin{align}
\eeetau{j_1  \\ k_1  \\ z_1 } &= j_1 ! \sum_{n_1=0}^{j_1}  \frac{ (-1)^{n_1}  \log(q)^{j_1-n_1}}{(j_1{-}n_1)!}
\mathcal{E}\!\SM{  k_1 &\vec{0}^{n_1}}{  z_1 &\vec{0}^{n_1}  }{\tau}
\label{toezer} 
\end{align}
\begin{align}
\eeetau{j_1 &j_2 \\ k_1 &k_2 \\ z_1 &z_2} &= \sum_{n_1=0}^{j_1} \sum_{n_2=0}^{j_1+j_2-n_1}
\frac{ (-1)^{n_1+n_2} j_1! (j_1{+}j_2{-}n_1)! \log(q)^{j_1+j_2-n_1-n_2} }{(j_1{-}n_1)! (j_1{+}j_2{-}n_1{-}n_2)!}
\mathcal{E}\!\SM{  k_1 &\vec{0}^{n_1} &k_2 &\vec{0}^{n_2}}{  z_1 &\vec{0}^{n_1} &z_2 &\vec{0}^{n_2} }{\tau}
\notag \\
\eeetau{j_1 &j_2 &j_3 \\ k_1 &k_2 &k_3 \\ z_1 &z_2 &z_3} &=
\sum_{n_1=0}^{j_1} \sum_{n_2=0}^{j_1+j_2-n_1} \sum_{n_3= 0}^{j_1+j_2+j_3-n_1-n_2}
\frac{ (-1)^{n_1+n_2+n_3} j_1! (j_1{+}j_2{-}n_1)! (j_1{+}j_2{+}j_3{-}n_1{-}n_2)!  }{(j_1{-}n_1)!(j_1{+}j_2{-}n_1{-}n_2)!(j_1{+}j_2{+}j_3{-}n_1{-}n_2{-}n_3)! }
\notag  \\
&\ \ \ \ \times \log(q)^{j_1+j_2+j_3-n_1-n_2-n_3}\mathcal{E}\!\SM{  k_1 &\vec{0}^{n_1} &k_2 &\vec{0}^{n_2} &k_3 &\vec{0}^{n_3}}{  z_1 &\vec{0}^{n_1} &z_2 &\vec{0}^{n_2} &z_3 &\vec{0}^{n_3}}{\tau}
\end{align} \normalsize

\subsection{Leading orders in the $q$-expansion at depth three}
\label{app:exps2}

For iterated KE integrals at depth one and two, the leading terms in the expansion
around the cusp have been spelt out in (\ref{fkreps.9}). The depth-three analogue 
of these expressions is given by
\begin{align}
&\mathcal{E}\!\SM{k_1 & \vec{0}^{p_1-1} & k_2 &\vec{0}^{p_2-1} & k_3 &\vec{0}^{p_3-1}}{z &\vec{0}^{p_1-1} & z &\vec{0}^{p_2-1}& z &\vec{0}^{p_3-1}}{\tau} = \frac{1}{\left(k_{1}{-}1\right) !\left(k_{2}{-}1\right) !\left(k_{3}{-}1\right) !}  \notag \\
&\ \ \times \bigg\{
\frac{ B_{k_1}(u) B_{k_2}(u) B_{k_3}(u) }{k_1k_2k_3} \frac{ \log(q)^{p_1+p_2+p_3} }{(p_1{+}p_2{+}p_3)!} \notag \\
& \ \ \ \
 - \frac{ u^{ k_3 -1 - p_3} B_{k_1}(u) B_{k_2}( u) }{k_1 k_2 (p_3{-}1)!}
 \sum_{n_3=0}^{p_1+p_2} (-u)^{-n_3}  \frac{ (n_3 {+} p_3{-}1)! }{n_3! (p_1{+}p_2{-}n_3)!}
  \log(q)^{p_1 + p_2-n_3} {\rm Li}_{ n_3 + p_3}(e^{2\pi i z}) \notag \\
& \ \ \ \
-  \frac{ u^{ k_2 - 1 - p_2 - p_3} B_{k_1}(u) B_{k_3}(u) }{k_1 k_3 (p_2{+}p_3{-}1)!}
  \sum_{n_2=0}^{p_1} (-u)^{-n_2} \frac{ (n_2 {+} p_2{+}p_3{-}1)!  }{n_2! (p_1{-}n_2)!}   \log(q)^{p_1 - n_2} {\rm Li}_{n_2+p_2+p_3}(e^{2\pi i z})
\notag \\
& \ \ \ \
-\frac{ u^{ k_1-1 - p_1 - p_2 - p_3}B_{k_2}( u) B_{k_3}(u) }{k_2k_3}
{\rm Li}_{ p_1+ p_2 + p_3}(e^{2\pi i z}) \label{appd3} \\
&\ \ \ \ + \frac{ u^{k_2+k_3-2-p_2-p_3} B_{k_1}(u)}{k_1(p_2{-}1)! (p_3{-}1)!}
\sum_{n_2=0}^{p_1}\sum_{n_3=0}^{p_1-n_2} (-u)^{-n_2-n_3}  \frac{ (n_2{+}p_2{-}1)! (n_3{+}p_3{-}1)! }{n_2! n_3! (p_1{-}n_2{-}n_3)!} \notag \\
&\ \ \ \ \ \ \ \ \ \ \ \ \ \ \ \ \ \ \ \ \ \ \ \  \times \log(q)^{p_1-n_2-n_3}
 G(\vec{0}^{n_3+p_3-1},1,\vec{0}^{n_2+p_2-1},1;e^{2\pi i z})
\notag \\
&\ \ \ \ + \frac{ u^{k_1+k_3-2-p_1-p_2-p_3} B_{k_2}(u) }{k_2}
 G(\vec{0}^{p_3-1},1,\vec{0}^{p_1+p_2-1},1;e^{2\pi i z})
\notag \\
&\ \ \ \ + \frac{ u^{k_1+k_2-2-p_1-p_2-p_3} B_{k_3}(u) }{k_3}
 G(\vec{0}^{p_2+p_3-1},1,\vec{0}^{p_1-1},1;e^{2\pi i z})
\notag \\
& \ \ \ \ + u^{k_1+k_2+k_3-3-p_1-p_2-p_3}  G(\vec{0}^{p_3-1},1,\vec{0}^{p_2-1},1,\vec{0}^{p_1-1},1;e^{2\pi i z})
  + {\cal O}(q^{1-u})  \bigg\} \notag
\end{align}

\subsection{Real-analytic iterated KE integrals ${\cal E}^{\rm sv}$}
\label{app:exps3}

The ${\cal E}^{\rm sv}$-versions of real-analytic iterated KE integrals at 
depth one and two in (\ref{fkreps.18}) generalize as follows to depth three
\begin{align}
&\EsvBR{j_1 &j_2 &j_3 \\ k_1 &k_2 &k_3 \\ z_1 &z_2 &z_3}{\tau} =
\overline{\alphaBR{j_1 &j_2 &j_3 \\ k_1 &k_2 &k_3 \\ z_1 &z_2 &z_3}{\tau}}
\notag \\
&\ \ \ \
+\overline{\alphaBR{j_1 &j_2 \\ k_1 &k_2 \\ z_1 &z_2}{\tau}} \sum_{r_3=0}^{j_3}  (-2\pi i \bar\tau)^{r_3}
\binom{j_3}{r_3}  \bigg\{ \EBR{j_3-r_3 \\ k_3 \\ z_3}{\tau} +
(-1)^{j_3-r_3} \overline{ \EBR{j_3-r_3 \\ k_3 \\ z_3}{\tau} }
\bigg\} \notag \\
& \ \ \ \
+ \sum_{r_1=0}^{j_1}\sum_{r_2=0}^{j_2}\sum_{r_3=0}^{j_3} (-2\pi i \bar\tau)^{r_1+r_2+r_3}
\binom{j_1}{r_1}\binom{j_2}{r_2}\binom{j_3}{r_3}
\bigg\{ \EBR{j_1 -r_1&j_2-r_2 &j_3-r_3 \\ k_1 &k_2 &k_3 \\ z_1 &z_2 &z_3}{\tau}
\label{fkreps.18c}  \\
&\hspace{10mm} +(-1)^{j_1-r_1} \overline{ \EBR{j_1-r_1 \\ k_1 \\ z_1}{\tau} }  \EBR{j_2-r_2 &j_3-r_3 \\ k_2 &k_3 \\ z_2 &z_3}{\tau}
+(-1)^{j_1+j_2-r_1-r_2} \overline{ \EBR{j_2-r_2 &j_1-r_1 \\ k_2 &k_1 \\ z_2 &z_1}{\tau} }   \EBR{j_3-r_3 \\ k_3 \\ z_3}{\tau}
\notag \\
&\hspace{10mm}+(-1)^{j_1+j_2+j_3-r_1-r_2-r_3} \overline{ \EBR{j_3-r_3 &j_2-r_2 &j_1-r_1 \\ k_3 &k_2 &k_1 \\ z_3 &z_2 &z_1}{\tau} }
\bigg\}   \notag
\end{align}
The product of the integration constants $\overline{\alphaBR{j_1 &j_2 \\ k_1 &k_2 \\ z_1 &z_2}{\tau}}$ of the depth-two cases in (\ref{fkreps.18}) and ${\cal E}[\ldots]$
at depth one is required by the differential equation (\ref{gen.35z}). The combination of  ${\cal E}[\ldots]$ and
their complex conjugates along with the $\overline{\alpha[\ldots]}$ at depth two is engineered to preserve the $\tau \rightarrow \tau {+} 1$ invariance of ${\cal E}^{\rm sv}$ at depth one. Finally, the new integration constants
$\overline{\alphaBR{j_1 &j_2 &j_3 \\ k_1 &k_2 &k_3 \\ z_1 &z_2 &z_3}{\tau}}$ are again purely antiholomorphic and vanish in shuffles, see (\ref{alpshffl}).

\subsection{Decomposition into meromorphic objects}
\label{app:exps7}

We shall finally spell out the depth-three analogue of the
decompositions (\ref{discont.6}) and (\ref{d2bsv}) of $\beta^{\rm sv}$ 
into meromorphic and antimeromorphic constituents:
\begin{align}
\bsvtau{j_1 &j_2 &j_3\\ k_1 &k_2 &k_3 \\ z_1 &z_2 &z_3}&=\sum_{p_1=0}^{k_1{-}2{-}j_1} \sum_{p_2=0}^{k_2{-}2{-}j_2} \sum_{p_3=0}^{k_3{-}2{-}j_3} \frac{\binom{k_1{-}2{-}j_1}{p_1}\binom{k_2{-}2{-}j_2}{p_2}\binom{k_3{-}2{-}j_3}{p_3}}{(4y)^{p_1+p_2+p_3}} \overline{\alphaBR{j_1 +p_1&j_2+p_2 &j_3+p_3\\k_1 &k_2 &k_3 \\ z_1 &z_2 &z_3}{\tau}}
\notag \\
&\hspace{-1.3cm}+\sum_{p_1=0}^{k_1{-}2{-}j_1} \sum_{p_2=0}^{k_2{-}2{-}j_2} \frac{\binom{k_1{-}2{-}j_1}{p_1}\binom{k_2{-}2{-}j_2}{p_2}}{(4y)^{p_1+p_2}} \overline{\alphaBR{j_1 +p_1&j_2+p_2\\k_1 &k_2 \\ z_1 &z_2}{\tau}} \bsvtau{ j_3\\  k_3 \\ z_3} \notag \\
%%%%
%%%%
&\hspace{-1.3cm}+ (-1)^{k_1+k_2+k_3} \sum_{c_1=0}^{k_1-j_1-2} {k_1{-}j_1{-}2 \choose c_1}
\sum_{d_1=0}^{j_1} {j_1\choose d_1}
\sum_{c_2=0}^{k_2-j_2-2+c_1} {k_2{-}j_2{-}2{+}c_1 \choose c_2} \notag \\
&\hspace{-1.3cm} \ \ \times
\sum_{d_2=0}^{j_2+d_1} {j_2{+}d_1 \choose d_2}
\sum_{a=0}^{j_3+d_2} {j_3{+}d_2 \choose a}
(k_3{-}j_3{-}2{+}a{+}c_2)!(k_2{-}2{+}c_1{-}c_2{+}d_1{-}d_2)! \notag \\
&\hspace{-1.3cm} \ \ \times
(k_1{-}2{-}c_1{-}d_1)! (4y)^{6+j_1+j_2+2j_3+d_2-a-k_1-k_2-k_3} \notag \\
&\hspace{-1.3cm} \ \ \times \mathcal{E}\!\SM{k_1 & \vec{0}^{k_1{-}2{-}c_1{-}d_1}
&k_2 & \vec{0}^{k_2{-}2{-}c_2{-}d_2{+}c_1{+}d_1}
&k_3 &\vec{0}^{k_3-2-j_3+c_2+a} }{ z_1 & \vec{0}^{k_1{-}2{-}c_1{-}d_1}
&z_2 & \vec{0}^{k_2{-}2{-}c_2{-}d_2{+}c_1{+}d_1}
&z_3 &\vec{0}^{k_3-2-j_3+c_2+a}  }{\tau}
\notag \\
%%%%
%%%%
&\hspace{-1.3cm}+ (-1)^{k_2+k_3} \sum_{b=0}^{k_1-2-j_1} {k_1{-}2{-}j_1\choose b} \sum_{c_2=0}^{k_2-j_2-2}
{k_2{-}2{-}j_2 \choose c_2} \sum_{d_2=0}^{j_2} {j_2 \choose d_2}  \sum_{a=0}^{j_3+d_2} {j_3{+}d_2 \choose a}\notag \\
&\hspace{-1.3cm} \ \ \times  (j_1{+}b)! (k_3{-}2{-}j_3{+}c_2{+}a)!
(k_2{-}2{-}c_2{-}d_2)! (4y)^{4-a-b+d_2-k_2-k_3+j_2+2j_3}  \label{d3bsvLT} \\
&\hspace{-1.3cm} \ \ \times  \mathcal{E}\!\SM{k_2 & \vec{0}^{k_2{-}2{-}c_2{-}d_2}
&k_3 &\vec{0}^{k_3-2-j_3+c_2+a} }{z_2 & \vec{0}^{k_2{-}2{-}c_2{-}d_2}
&z_3 &\vec{0}^{k_3-2-j_3+c_2+a}  }{\tau}
\overline{\mathcal{E}\!\SM{k_1 & \vec{0}^{j_1+b} }{z_1 & \vec{0}^{j_1+b}}{\tau} }
\notag \\
%%%%
%%%%
&\hspace{-1.3cm}+ (-1)^{k_3}  \sum_{c_2=0}^{k_2-2-j_2} {k_2{-}2{-}j_2 \choose c_2}
\sum_{d_2=0}^{j_2} {j_2 \choose d_2} \sum_{b=0}^{k_1-2-j_1+c_2} { k_1{-}2{-}j_1{+}c_2 \choose b}
\sum_{a=0}^{j_3} {j_3 \choose a} \notag \\
&\hspace{-1.3cm} \ \ \times(k_3{-}2{-}j_3{+}a)! (k_2{-}2{-}c_2{-}d_2)!
(j_1{+}d_2{+}b)! (4y)^{4-a-b+c_2+j_2+2j_3-k_2-k_3}  \notag \\
&\hspace{-1.3cm} \ \ \times  \mathcal{E}\!\SM{k_3 & \vec{0}^{k_3{-}2{-}j_3{+}a} }{z_3 & \vec{0}^{k_3{-}2{-}j_3{+}a} }{\tau}
\overline{ \mathcal{E}\!\SM{ k_2 &\vec{0}^{k_2{-}2{-}c_2{-}d_2} &k_1 &\vec{0}^{j_1{+}d_2{+}b}}{z_2 &\vec{0}^{k_2{-}2{-}c_2{-}d_2} &z_1 &\vec{0}^{j_1{+}d_2{+}b}}{\tau} }
\notag \\
%%%%
%%%%
&\hspace{-1.3cm}+ \sum_{c_3=0}^{k_3-j_3-2} { k_3{-}j_3{-}2 \choose c_3} \sum_{d_3=0}^{j_3} {j_3 \choose d_3}
\sum_{c_2=0}^{k_2-j_2-2+c_3} { k_2{-}j_2{-}2{+}c_3 \choose c_2} \sum_{d_2=0}^{d_3+j_2} {d_3{+}j_2 \choose d_2} \notag \\
&\hspace{-1.3cm} \ \ \times \sum_{b=0}^{k_1-j_1-2+c_2} {k_1{-}j_1{-}2{+}c_2 \choose b} (j_1{+}b{+}d_2)! (k_2{-}2{+}c_3{-}c_2{+}d_3{-}d_2)! (k_3{-}2{-}c_3{-}d_3)! \notag \\
&\hspace{-1.3cm} \ \ \times (4y)^{4-b+c_2+j_2+j_3-k_2-k_3} \overline{ \mathcal{E}\!\SM{k_3 & \vec{0}^{k_3{-}2{-}c_3{-}d_3} & k_2 &\vec{0}^{k_2{-}2{+}c_3{-}c_2{+}d_3{-}d_2} &k_1 &\vec{0}^{j_1+b+d_2}}{z_3 & \vec{0}^{k_3{-}2{-}c_3{-}d_3} &z_2 &\vec{0}^{k_2{-}2{+}c_3{-}c_2{+}d_3{-}d_2} &z_1 &\vec{0}^{j_1+b+d_2}}{\tau} }
\notag
\end{align}
In order to obtain its leading terms in terms of single-valued polylogarithms $G^{\rm sv}$, one can
apply the effective rules $\log(q) \rightarrow \log|q|^2 = -4y$ and $G(\ldots;e^{2\pi i z}) \rightarrow 
G^{\rm sv}(\ldots;e^{2\pi i z})$ to the leading terms of the ${\cal E}$ on the right-hand 
side (see (\ref{appd3}) and its lower-depth analogues in section \ref{sec:3.3.1}). 
As in (\ref{effrule.2}), the effective rules also instruct to discard
antimeromorphic iterated KE integrals $\overline{ {\cal E} }$ as well
as the integration constants $\overline{ \alpha[\ldots] }$ in the first
two lines of (\ref{d3bsvLT}). These effective rules implement the single-valued
map at genus zero which should be compatible with its tentative extension to
genus one at the level of ${\cal E}$.

\subsection{Conversion of ${\cal E}$ to ${\cal E}_0$}
\label{app:etoe0}

This appendix is dedicated to the conversion between the two variants
${\cal E}$ and ${\cal E}_0$ of meromorphic iterated KE integrals, extending
the discussion of section \ref{qexpsec.A}. The conversion formulae (\ref{inival.31}) and
(\ref{etoeod2}) at depth one and two generalize as follows to depth three:
\begin{align}
&\mathcal{E}\!\SM{k_1 &\vec{0}^{p_1-1} &k_2 &\vec{0}^{p_2-1}
&k_3 &\vec{0}^{p_3-1}}{z_1 &\vec{0}^{p_1-1} &z_2 &\vec{0}^{p_2-1}
&z_3 &\vec{0}^{p_3-1}}{\tau}
= \mathcal{E}_0\!\SM{k_1 &\vec{0}^{p_1-1} &k_2 &\vec{0}^{p_2-1}
&k_3 &\vec{0}^{p_3-1}}{z_1 &\vec{0}^{p_1-1} &z_2 &\vec{0}^{p_2-1}
&z_3 &\vec{0}^{p_3-1}}{\tau} \notag \\
&\ + \frac{ B_{k_3}(u_3)}{k_3!}  \mathcal{E}_0\!\SM{k_1 &\vec{0}^{p_1-1} &k_2 &\vec{0}^{p_2+p_3-1}  }{z_1 &\vec{0}^{p_1-1} &z_2 &\vec{0}^{p_2+p_3-1} }{\tau}
+ \frac{ B_{k_2}(u_2)}{k_2!}  \mathcal{E}_0\!\SM{k_1 &\vec{0}^{p_1+p_2-1} &k_3 &\vec{0}^{p_3-1}  }{z_1 &\vec{0}^{p_1+p_2-1} &z_3 &\vec{0}^{p_3-1}   }{\tau}
\notag \\
&\   
+ \frac{ B_{k_1}(u_1)}{k_1! (p_2{-}1)! (p_3{-}1)!}   
  \sum_{n_2=0}^{p_1} \sum_{n_3=0}^{p_1-n_2}  (-1)^{n_2+n_3} \,\frac{(n_2{+}p_2{-}1)!
  (n_3{+}p_3{-}1)!}{n_2! n_3! (p_1{-}n_2{-}n_3)!} \notag \\
  &\ \quad \quad \quad \quad \times
 \log(q)^{p_1-n_2-n_3}
\mathcal{E}_0\!\SM{k_2 &\vec{0}^{p_2+n_2-1} &k_3 &\vec{0}^{p_3+n_3-1}  }{z_2 &\vec{0}^{p_2+n_2-1} &z_3 &\vec{0}^{p_3+n_3-1}  }{\tau} 
\label{etoeod3}  \\
&\ + \frac{ B_{k_1}(u_1) B_{k_3}(u_3) }{k_1! k_3!(p_2{+}p_3{-}1)!} 
\sum_{n_2=0}^{p_1} (-1)^{n_2} \frac{ (p_2{+}p_3{+}n_2{-}1)! }{ n_2!(p_1{-}n_2)!}
\log(q)^{p_1-n_2}
\mathcal{E}_0\!\SM{k_2 &\vec{0}^{p_2+p_3+n_2-1} }{
z_2 &\vec{0}^{p_2+p_3+n_2-1} }{\tau}
\notag \\
&\ + \frac{ B_{k_1}(u_1) B_{k_2}(u_2) }{k_1! k_2!(p_3{-}1)!} 
\sum_{n_3=0}^{p_1+p_2} (-1)^{n_3} \frac{ (p_3{+}n_3{-}1)! }{ n_3!(p_1{+}p_2{-}n_3)!}
\log(q)^{p_1+p_2-n_3}
\mathcal{E}_0\!\SM{k_3 &\vec{0}^{p_3+n_3-1} }{
z_3 &\vec{0}^{p_3+n_3-1} }{\tau}
\notag \\
%%%%%%%
&\ 
+ \frac{ B_{k_2}(u_2) B_{k_3}(u_3) }{k_2! k_3!}
\mathcal{E}_0\!\SM{k_1 &\vec{0}^{p_1+p_2+p_3-1} }{
z_1 &\vec{0}^{p_1+p_2+p_3-1} }{\tau}
%%%
+ \frac{ B_{k_1}(u_1)B_{k_2}(u_2) B_{k_3}(u_3)}{k_1! k_2! k_3!} \frac{ \log(q)^{p_1+p_2+p_3} }{(p_1{+}p_2{+}p_3)!}
 \notag
\end{align} \normalsize
One can check that the one-variable case $z_1=z_2=z_3=z$ is consistent with the
leading terms of ${\cal E}$ and ${\cal E}_0$ in (\ref{appd3}) and (\ref{closedE0}),
respectively.

\newpage

%%%%%%%%%%%%%%%%
\section{Leading terms beyond depth one}
\label{app.lead}

The purpose of this appendix is to provide further details on the leading
terms of meromorphic and real-analytic iterated KE integrals, supplementing
the discussion of section~\ref{sec:3.3}.

%%%%%%%%%%%%%%%%%%%%%%%%%%%%%%%%%%%%%%%%%%%%%%%%%%%%%%%%%%%
\subsection{Meromorphic leading terms at arbitrary depth}
\label{sec:laurentpolynomialsarbitrarydepth}

Meromorphic iterated KE integrals are recursively defined in (\ref{fkreps.5}).
Their leading terms are obtained by iterated integration of the leading terms
\begin{align}
f^{(k)}(u\tau{+} v|\tau) &=  (2 \pi i)^{k} \big[  f_A^{(k)}(u) + f_B^{(k)}(u,v,\tau) \big] + {\cal O}(q^{1-u})   \label{D1app.01} \\
  f_A^{(k)}(u) &= \frac{B_k(u)}{k!} \, , \ \ \ \ 
f_B^{(k)}(u,v,\tau) = 
  \frac{u^{k-1}}{(k{-}1)!}  \, \frac{e^{2\pi i (u\tau+v)}}{e^{2\pi i (u\tau + v)}{-}1}
\notag
\end{align}
at fixed $u,v$. The recursion for the leading terms with $k_r\neq 0$ then simplifies to
\begin{align}
\mathcal{E}\!\SM{k_1 & k_2 & \ldots & k_r}{z_1 & z_2 & \ldots & z_r }{\tau} &=  
f_A^{(k_r)}(u) \int_{0}^{q} \frac{  \mathrm{d} q_{r} }{q_r}\, 
 \mathcal{E}\!\SM{k_1 & k_2 & \ldots & k_{r-1}}{z_1 & z_2 & \ldots & z_{r-1} }{\tau_{r}}    
 \label{D1app.02} \\
 &\quad
 + \int_{0}^{q} \frac{  \mathrm{d} q_{r} }{q_r}\, 
f_B^{(k_r)}(u,v,\tau_r)  \,
 \mathcal{E}\!\SM{k_1 & k_2 & \ldots & k_{r-1}}{z_1 & z_2 & \ldots & z_{r-1} }{\tau_{r}} 
+ {\cal O}(q^{1-u}) 
\notag
\end{align}
In the remainder of this section, we will describe a procedure to perform these $q_r$-integrals 
in terms of $\log(q)$ and multiple polylogarithms $G(a_1,\ldots,a_w;e^{2\pi i z})$ with 
$a_j \in \{0,1\}$ and $a_w=1$ reviewed in appendix \ref{app:poly}. It will be 
convenient to transform the integration variable to
\beq
\sigma_r = e^{2\pi i (u\tau_r+v)} = (q_r)^u e^{2\pi i v} \ \ \Rightarrow \ \ 
\frac{ \dd q_r}{q_r} = \frac{1}{u} \frac{ \dd \sigma_r}{\sigma_r}
\, , \ \ \ \
f_B^{(k_r)}(u,v,\tau_r) =   \frac{u^{k-1}}{(k{-}1)!}  \frac{ \sigma_r}{\sigma_r{-}1}
 \label{D1app.03}
\eeq
and to introduce the following shorthand for repeated integration of some
function $\varphi(q)$
\begin{align}
\bigg( \int_0^q \frac{\dd q'}{q'} \bigg)^n \varphi(q) = \int_0^q \frac{\dd q_1}{q_1} \int_0^{q_1} \frac{\dd q_2}{q_2} \ldots \int_0^{q_{n-1}} \frac{\dd q_{n}}{q_{n}} \, \varphi(q_n)
 \label{D1app.04}
\end{align}
In this setting, the leading terms at depth one in (\ref{fkreps.9.0}) are readily obtained via
\begin{align}
 \mathcal{E}\!\SM{k_1 &\vec{0}^{p_1-1}}{z &\vec{0}^{p_1-1}}{\tau}
&= \bigg( \int_0^q \frac{\dd q'}{q'} \bigg)^{p_1}  \big[  f_A^{(k)}(u) + f_B^{(k)}(u,v,\tau) \big] + {\cal O}(q^{1-u}) 
 \label{D1app.05} \\
&= \frac{ B_{k_1}(u)}{k_1!} \frac{ \log(q)^{p_1} }{p_1!}
+ \frac{ u^{k_1-p_1-1} }{(k_1{-}1)!} G(\vec{0}^{p_1-1},1;e^{2\pi i z}) + {\cal O}(q^{1-u}) 
\notag
\end{align}
by applying the following elementary integration rules (see (\ref{D1app.03}) for the $\sigma$ variable
and the transformation of the measures)
\begin{align}
 \bigg( \int_0^q \frac{\dd q'}{q'} \bigg)^{p_1} c &= c \, \frac{ \log(q)^{p_1} }{p_1!} \, , \ \ \ \ % 
 \bigg( \int_0^q \frac{\dd q'}{q'} \bigg)^{p_1} \frac{ \sigma }{\sigma{-}1} = \frac{1}{u^{p_1}} \, G(\vec{0}^{p_1-1},1;\sigma)
  \label{D1app.06} 
\end{align}
Throughout this appendix, we assume $u>0$, and integration rules such as
(\ref{D1app.06}) are understood to be valid within the scheme of tangential-basepoint 
regularization \cite{DeligneTBP, Brown:mmv}. In practice, this means that we set to zero any 
logarithms that appear at the basepoint of integration which is equivalent
to shuffle-regularizing the integrals on the left before performing the integration.

%%%%%%%%%
\subsubsection{Towards depth two}

The next step is to derive the depth-two result (\ref{fkreps.9}) from
\begin{align}
\mathcal{E}\!\SM{k_1 & \vec{0}^{p_1-1} & k_2 &\vec{0}^{p_2-1}}{z &\vec{0}^{p_1-1} & z &\vec{0}^{p_2-1}}{\tau}  &= \bigg( \int_0^q \frac{\dd q'}{q'} \bigg)^{p_2}  \big[  f_A^{(k_2)}(u) + f_B^{(k_2)}(u,v,\tau) \big] \label{D1app.07} \\
&\! \! \! \! \! \! \! \! \! \! \! \!
\! \! \! \! \! \! \! \! \! \! \! \! \times \bigg\{  \frac{ B_{k_1}(u)}{k_1!} \frac{ \log(q)^{p_1} }{p_1!}
+ \frac{ u^{k_1-p_1-1} }{(k_1{-}1)!} G(\vec{0}^{p_1-1},1;e^{2\pi i z}) \bigg\} + {\cal O}(q^{1-u}) 
\notag
\end{align}
and suitable extensions of the integration rules (\ref{D1app.06}). The
contributions involving $f_A^{(k_2)}(u)$ and the integral of
$f_B^{(k_2)}(u,v,\tau)G(\vec{0}^{p_1-1},1;e^{2\pi i z})$
are obtained from
\begin{align}
 \bigg( \int_0^q \frac{\dd q'}{q'} \bigg)^{p} \log(q)^r &= \frac{ r!}{(r{+}p)!} \log(q)^{r+p}
\notag \\
 \bigg( \int_0^q \frac{\dd q'}{q'} \bigg)^{p} G(\vec{0}^{r},1;\sigma)
 &= 
  \frac{1}{u^p} \, G(\vec{0}^{p+r},1;\sigma)
 \label{D1app.08} \\
 \bigg( \int_0^q \frac{\dd q'}{q'} \bigg)^{p} \frac{ \sigma }{\sigma{-}1} \, G(\vec{0}^{r},1;\sigma)
 &=  
   \frac{1}{u^p} \, G(\vec{0}^{p-1},1,\vec{0}^{r},1;\sigma)
\notag
\end{align}
which are simple consequences of (\ref{D1app.03}) and the recursive
definition (\ref{fkreps.10}) of polylogarithms. The integral over
$f_B^{(k_2)}(u,v,\tau)\log(q)^{p_1}$ is more challenging due to the
translation between $u \log (q) = \log(\sigma)-2\pi i v$. 
For integrands $F(q)$ composed of $G(\vec{a};\sigma)$ and possibly a factor
of $\frac{\sigma}{\sigma{-}1}$, repeated application of integration by parts along with
tangential-base-point regularization yields
\begin{align}
 \bigg( \int_0^q \frac{\dd q'}{q'} \bigg)^{p} \log(q)^r \, F(q)&= \frac{ r! }{ (p{-}1)!}
 \sum_{n=0}^{r} \frac{ (-1)^n(n{+}p{-}1)! }{n! (r{-}n)!} \log(q)^{r-n}  
\,  \bigg( \int_0^q \frac{\dd q'}{q'} \bigg)^{p+n} F(q)
 \label{D1app.master}
\end{align}
Specializing to $F(q) \rightarrow \frac{ \sigma}{\sigma{-}1}$ addresses the most
challenging integration in (\ref{D1app.07}),
\begin{align}
 \bigg( \int_0^q \frac{\dd q'}{q'} \bigg)^{p} \log(q)^r \, \frac{ \sigma }{\sigma{-}1}&= \frac{ r! }{u^p (p{-}1)!}
 \sum_{n=0}^{r} \frac{(n{+}p{-}1)! }{n! (r{-}n)!} \frac{ \log(q)^{r-n} }{(-u)^n} 
\, G(\vec{0}^{n+p-1},1;\sigma)
 \label{D1app.09}
\end{align}
where the conversion of $p{+}n$ integrations w.r.t.\ $\frac{\dd q'}{q'}$ in (\ref{D1app.master}) to
$\frac{1}{u}\frac{\dd \sigma'}{\sigma'}$ gives rise to the factors of $u^{-p-n}$ on the right-hand side.

Throughout this appendix, the guiding principle is to remove zeros
in the right-most entries of  $G(\ldots,1,\vec{0}^{n};\sigma), \ n\geq 1$
via shuffles with powers of $G(0;\sigma) = u \log(q)+2\pi i v$. The above
integration rules then lead to the following equivalent of (\ref{fkreps.9}): \small
\begin{align}
    \mathcal{E}\!\SM{k_1 & \vec{0}^{p_1-1} & k_2 &\vec{0}^{p_2-1}}{z &\vec{0}^{p_1-1} & z &\vec{0}^{p_2-1}}{\tau} &= 
    \frac{B_{k_{1}}(u) B_{k_{2}}(u)}{k_{1}! k_{2}!} \frac{\log (q)^{p_{1}+p_2}}{(p_1{+}p_2)!}    +  \frac{ B_{k_{2}}(u) u^{k_{1}-1-p_{1}-p_2} }{  k_{2}! (k_1{-}1)!} \, G(\vec{0}^{p_{1}+p_2-1},1;e^{2 \pi i z})   \notag \\
&\quad +   \frac{ B_{k_{1}}(u)u^{k_{2}-p_2-1}}{ k_{1}! (k_2{-}1)! \left(p_2{-}1\right)!} \sum_{n=0}^{p_1}
\frac{\left(n{+}p_2{-}1\right)!}{n! \left(p_1{-}n\right)!} \frac{ \log(q)^{p_1-n} }{(-u)^{n} } \,G(\vec{0}^{n+p_2},1;e^{2 i \pi  z}) \notag \\
&\quad +\frac{u^{k_{1}+k_{2}-p_{1}-p_{2}-2}}{ (k_1{-}1)! (k_2{-}1)! } \, G(\vec{0}^{p_{2}-1}, 1, \vec{0}^{p_{1}-1}, 1 ; e^{2 \pi i z})
  + {\cal O}(q^{1-u})  \label{D1app.41}
  \end{align} \normalsize
 
%%%%%%%%%
\subsubsection{Towards higher depth}

In order to extend the algorithmic integrations to higher depth $\geq 3$, it remains to
supply an integration rule for the terms $\log(q)^r G(\vec{a};\sigma)$ in the second
line of (\ref{D1app.41}) which do not arise in the leading terms (\ref{D1app.05}) at
depth one. By specializing (\ref{D1app.master}) to $F(q) \rightarrow G(\vec{a};\sigma)$ and 
$\frac{\sigma}{\sigma{-}1}G(\vec{a};\sigma)$ for arbitrary entries $\vec{a}=a_1,\ldots ,a_w$ 
with $a_j \in \{0,1\}$, we find
\begin{align}
 \bigg( \int_0^q \frac{\dd q'}{q'} \bigg)^{p} \log(q)^r G(\vec{a};\sigma)&= \frac{ r! }{u^p (p{-}1)!}
 \sum_{n=0}^{r} \frac{(n{+}p{-}1)! }{n! (r{-}n)!} \frac{ \log(q)^{r-n} }{(-u)^n} 
\, G(\vec{0}^{n+p},\vec{a};\sigma)
 \label{D1app.10} \\
  \bigg( \int_0^q \frac{\dd q'}{q'} \bigg)^{p} \log(q)^r G(\vec{a};\sigma) \,  \frac{\sigma}{\sigma{-}1} &= \frac{ r! }{u^p (p{-}1)!}
 \sum_{n=0}^{r} \frac{(n{+}p{-}1)! }{n! (r{-}n)!} \frac{ \log(q)^{r-n} }{(-u)^n} 
\, G(\vec{0}^{n+p-1},1,\vec{a};\sigma)
\notag
\end{align}
where the factors of $u^{-p-n}$ arise in the same way as explained below (\ref{D1app.09}). 
We conclude that expressions $\log(q)^r G(\vec{a};\sigma)$ seen on the right-hand side of (\ref{D1app.10}) are closed under the above integration rules
 -- also in presence of $\frac{ \sigma }{\sigma{-}1}$-insertions from $f_B^{(k)}$. 
Hence, (\ref{D1app.06}), (\ref{D1app.08}), (\ref{D1app.09}) and
(\ref{D1app.10}) are sufficient to systematically compute leading terms of arbitrary 
depth including (\ref{appd3}) at depth three.

%%%%%%%%%%%%%%%%
\subsection{Depth-two examples with poles in $u$}
\label{app.lead.1}

As discussed in section \ref{sec:3.3.2ex}, the leading
terms of $\beta^{\rm sv}$ at depth two may require
contributions from the integration constants $\overline{\alpha[\ldots]}$
in order to form single-valued genus-zero polylogarithms.
From the example (\ref{comrel.34}) together with
\begin{align}
\bsvtau{0 &1 \\ 2 &4  \\ z &z } &=
-\frac{y^3}{9}   B_2(u) B_4(u) - \frac{uy}{3}  B_2(u) G^{\rm sv}(0, 1;e^{2\pi i z}) -
  \frac{ B_2(u)}{3} G^{\rm sv}(0, 0, 1;e^{2\pi i z})  \notag \\
  &\ \ +
\frac{ B_4(u) G^{\rm sv}(0, 0, 1;e^{2\pi i z})}{24 u^2} +
 \frac{u}{6}  G^{\rm sv}(0, 1, 1;e^{2\pi i z}) -
\frac{ B_2(u) G^{\rm sv}(0, 0, 0, 1;e^{2\pi i z})}{8 u y}  \notag \\
& \ \ +
\frac{ B_4(u) G^{\rm sv}(0, 0, 0, 1;e^{2\pi i z})}{48 u^3 y} +
\frac{ G^{\rm sv}(0, 0, 1, 1;e^{2\pi i z})}{12 y}+ {\cal O}(q^{1-u},\bar q^{1-u}) 
\label{comrel.37}
 \end{align}
and
 \begin{align}
\bsvtau{0 &2 \\ 2 &4  \\ z &z } &=
 \frac{4y^4}{3}  B_2(u) B_4(u) - \frac{ 16}{3} u^2 y^3 B_2(u) G^{\rm sv}(1;e^{2\pi i z}) +
\frac{ 4}{3} u^2 y^2 G^{\rm sv}(1;e^{2\pi i z})^2 \notag \\
&\ \ -
 4 u y^2 B_2(u) G^{\rm sv}(0, 1;e^{2\pi i z}) +
\frac{ 2 y^2 B_4(u) G^{\rm sv}(0, 1;e^{2\pi i z})}{3 u} \notag \\
&\ \ -
 2 y B_2(u) G^{\rm sv}(0, 0, 1;e^{2\pi i z}) +
\frac{ y B_4(u) G^{\rm sv}(0, 0, 1;e^{2\pi i z})}{3 u^2}  \label{comrel.40} \\
&\ \ +
\frac{ 4}{3} u y G^{\rm sv}(0, 1, 1;e^{2\pi i z}) -
\frac{ B_2(u) G^{\rm sv}(0, 0, 0, 1;e^{2\pi i z})}{2 u}  \notag \\
&\ \ +
\frac{ B_4(u) G^{\rm sv}(0, 0, 0, 1;e^{2\pi i z})}{12 u^3} +
 \frac{1}{3} G^{\rm sv}(0, 0, 1, 1;e^{2\pi i z}) + {\cal O}(q^{1-u},\bar q^{1-u}) 
\notag
 \end{align}
 one can infer the leading terms of $\overline{\alphaBR{0 &j \\ 2 &4 \\ z &z}{\tau}}$ given in (\ref{baralp.1}).

%%%%%%%%%%%%%%%%
\subsection{Examples of $u$-pole cancellation}
\label{app.lead.2}

While the leading terms of individual $\beta^{\rm sv}$ may have poles in $u$,
their combinations entering the eMGFs in the expansion (\ref{gen.34c}) of the
generating series $Y$ are non-singular. This is exemplified by (\ref{comrel.36}) as well as
\begin{align}
&\bsvtau{0 &1 \\ 2 &4  \\ z &z } -  \bsvtau{0 &1 \\ 2 &4  \\ z &\emptyslot } =
\frac{y^3}{9}  \big(B_4(u)- B_4 \big) B_2(u)  +
 \frac{uy }{3}  B_2(u) G^{\rm sv}(0, 1;e^{2\pi i z})  \notag \\
 &\ \ \ \ \ \ + \bigg( \frac{7 u^2}{24} - \frac{ u}{4} + \frac{1}{72}   \bigg)  G^{\rm sv}(0, 0, 1;e^{2\pi i z})
  -   \frac{u}{6}  G^{\rm sv}(0, 1, 1;e^{2\pi i z})    \label{comrel.39} \\
  &\ \ \ \ \ \
  + \bigg(   \frac{ 5 u}{48 y}- \frac{1}{12 y}  \bigg) G^{\rm sv}(0, 0, 0, 1;e^{2\pi i z})
   -  \frac{ G^{\rm sv}(0, 0, 1, 1;e^{2\pi i z})}{12 y}+ {\cal O}(q^{1-u},\bar q^{1-u}) 
\notag
 \end{align}
and
\begin{align}
&\bsvtau{0 &2 \\ 2 &4  \\ z &z } -  \bsvtau{0 &2 \\ 2 &4  \\ z &\emptyslot } =
\frac{4}{3} y^4 \big( B_4-B_4(u) \big) B_2(u)
+ \frac{ 16}{3} u^2 y^3 B_2(u) G^{\rm sv}(1;e^{2\pi i z})  \notag \\
&\ \ \ \ \ \
-  \frac{4}{3} u^2 y^2 G^{\rm sv}(1;e^{2\pi i z})^2
 + \bigg( \frac{10 u }{3} -\frac{8}{3}    \bigg) u^2 y^2 G^{\rm sv}(0, 1;e^{2\pi i z}) \notag \\
 &\ \ \ \ \ \ +
 \bigg( \frac{5 u }{3} - \frac{4  }{3}  \bigg) u y G^{\rm sv}(0, 0, 1;e^{2\pi i z})  -
 \frac{4}{3} u y G^{\rm sv}(0, 1, 1;e^{2\pi i z})    \label{comrel.42}\\
 &\ \ \ \ \ \ +  \bigg( \frac{5 u}{12} -\frac{1}{3} \bigg) G^{\rm sv}(0, 0, 0, 1;e^{2\pi i z})
  -  \frac{1}{3} G^{\rm sv}(0, 0, 1, 1;e^{2\pi i z}) + {\cal O}(q^{1-u},\bar q^{1-u}) 
\notag
 \end{align}

\newpage

%%%%%%%%%%%%%%%%%%%%%%%%%%%%
%%%%%%%%%%%%%%%%%%%%%%%%%%%%
\section{Initial value $\widehat Y_{ij}(i\infty)$ at two points}
\label{app:init}

The purpose of this appendix is to display additional samples of
the MZVs in the two-point initial value $\widehat Y_{ij}(i\infty)$ 
besides those in section \ref{sec:9.1.2}. A machine-readable form
of the expressions in this appendix and higher orders in $s_{ij},\eta,\bar \eta$
can be found in an ancillary file.

%%%%%%%%%%%%%%%%%%%%%%%%%%%%
%%%%%%%%%%%%%%%%%%%%%%%%%%%%
\subsection{Subleading orders in $\eta$}
\label{app:init.1}

For the leading orders $\sim (\eta \bar \eta)^{-1}$ of $\widehat Y_{ij}(i\infty)$, 
the expansion in $s_{02},s_{12}$ up to the fifth order can be found in (\ref{inival.04}).
We shall now give the analogous $s_{ij}$-expansions for the subleading orders of
$\widehat Y_{ij}(i\infty)$ in $\eta$ and $\bar \eta$, i.e.\
\begin{align}
\widehat Y_{11}(i\infty) \, \Big|_{\eta^{-1} \bar \eta^{0}} &= \widehat Y_{12}(i\infty) \, \Big|_{\eta^{-1} \bar \eta^{0}}
= \frac{ i \pi }{3} s_{02} \big( (s_{12}{+}s_{02})^2 B_1(u) + 8 s_{12}s_{02}B_3(u) \big) \zeta_3 +
\ldots 
\\
\widehat Y_{21}(i\infty) \, \Big|_{\eta^{-1} \bar \eta^{0} } &=\widehat Y_{22}(i\infty) \, \Big|_{\eta^{-1} \bar \eta^{0} }
= - \frac{ i \pi }{3} s_{12} \big( (s_{12}{+}s_{02})^2 B_1(u) + 8 s_{12}s_{02}B_3(u) \big) \zeta_3 +\ldots
\notag
\end{align}
with MZVs of weight $\geq 5$ in the ellipsis and
\begin{align}
\widehat Y_{11}(i\infty) \, \Big|_{\eta^{0} \bar \eta^{0}} &= - \frac{2\pi i }{s_{12}}
+ \frac{ i \pi }{3}   (2 s_{02} {-} s_{12}) (s_{02} {+} s_{12}) \zeta_3
+2 \pi i   s_{02} (3 s_{02} {-} s_{12}) B_2(u) \zeta_3 \notag \\
&\quad + \frac{ i \pi}{180}   (s_{02} {+} s_{12}) (48 s_{02}^3 {-} 55 s_{02}^2 s_{12}
{+} 90 s_{02} s_{12}^2 {-}  43 s_{12}^3) \zeta_5 \notag \\
&\quad +\frac{i \pi}{6}   s_{02} (23 s_{02}^3 {-} 3 s_{02}^2 s_{12}
{+} 29 s_{02} s_{12}^2 {-} 9 s_{12}^3) B_2(u) \zeta_5 \notag \\
&\quad + \frac{ i \pi}{3}  s_{02} (13 s_{02}^3 {-} 13 s_{02}^2 s_{12}
{+} 13 s_{02} s_{12}^2 {-} s_{12}^3) B_4(u) \zeta_5+\ldots
\\
%%%
%%%
\widehat Y_{12}(i\infty) \, \Big|_{\eta^{0} \bar \eta^{0}} &=
 - \frac{ i \pi}{3} (s_{02} {+} s_{12})^2 \zeta_3 - 8 \pi i  s_{02} s_{12} B_2(u) \zeta_3 \notag \\
 &\quad - \frac{i \pi}{180}  (s_{02} {+} s_{12})^2 (43 s_{02}^2 {-} 42 s_{02} s_{12} {+} 43 s_{12}^2) \zeta_5 \notag \\
 &\quad - \frac{16}{3} i \pi s_{02} s_{12} (s_{02}^2 {+} s_{12}^2) B_2(u) \zeta_5 \notag \\
 &\quad - \frac{2}{3} i \pi s_{02} s_{12} (7 s_{02}^2 {-} 6 s_{02} s_{12} {+} 7 s_{12}^2) B_4(u) \zeta_5
+\ldots\notag
\end{align}
with MZVs of weight $\geq 6$ in the ellipsis.

%%%%%%%%%%%%%%%%%%%%%%%%%%%%
%%%%%%%%%%%%%%%%%%%%%%%%%%%%
\subsection{Subleading orders in $s_{ij}$}
\label{app:init.2}

The orders $s_{ij}^{\leq 0}$ of $\widehat Y_{ij}(i\infty)$ can be found in
(\ref{inival.05}), and we shall here display the linear order in $s_{ij}$ up to
certain cutoffs in transcendentality and $\bar \eta$. The matrix entries $\widehat Y_{21}(i\infty)$
and $\widehat Y_{22}(i\infty)$ can be reconstructed by suitable relabellings of
\begin{align}
\widehat Y_{11}(i\infty)&= \frac{1}{\eta \bar \eta} - \frac{2\pi i }{s_{12}}
+ 4\pi i (\eta^2 \zeta_3+\eta^4 \zeta_5+\eta^6 \zeta_7+\eta^8 \zeta_9+\ldots) \notag \\
&\quad + s_{12} \bigg\{ {-} \frac{ 2 \eta \zeta_3}{\bar \eta} - \frac{ 2 \eta^3 \zeta_5}{\bar \eta}
- 4 \pi i \eta^4 \zeta_3^2  - \frac{ 2 \eta^5 \zeta_7}{\bar \eta}
 - 8 \pi i \eta^6 \zeta_3 \zeta_5  - \frac{ 2 \eta^7 \zeta_9}{\bar \eta} +\ldots
 \bigg\} \notag \\
%%%%%
&\quad - 8\pi i s_{02} \bigg\{ \eta B_1(u) \zeta_3
+2\eta B_1(u) \zeta_5
+3\eta B_1(u) \zeta_7 +\ldots \bigg\}
+ {\cal O}(s_{ij}^2)
\end{align}
and \small
\begin{align}
\widehat Y_{12}(i\infty)&= \frac{1}{\eta \bar \eta} - \frac{2\pi i }{s_{12}}
+ 4\pi i (\eta^2 \zeta_3+\eta^4 \zeta_5+\eta^6 \zeta_7+\eta^8 \zeta_9+\ldots) \notag \\
&\hspace{-1cm}+ 4\pi i s_{12} \bigg\{
\eta \big( B_1(u) + \tfrac{1}{2} 2 i\pi \bar \eta B_2(u)
 + \tfrac{1}{3!} (2 i\pi \bar \eta)^2 B_3(u)
  + \tfrac{1}{4!} (2 i\pi \bar \eta)^3 B_4(u) \notag \\
  &\quad  
   + \tfrac{1}{5!} (2 i\pi \bar \eta)^4 B_5(u)
    + \tfrac{1}{6!} (2 i\pi \bar \eta)^5 B_6(u)+\ldots \big) \zeta_3 \notag \\
& + \eta^3  \big( B_1(u) + \tfrac{1}{2} 2 i\pi \bar \eta B_2(u)
 + \tfrac{1}{3!} (2 i\pi \bar \eta)^2 B_3(u)
  + \tfrac{1}{4!} (2 i\pi \bar \eta)^3 B_4(u) +\ldots \big) \zeta_5 \notag \\
& + \eta^5  \big( B_1(u) + \tfrac{1}{2} 2 i\pi \bar \eta B_2(u) + \ldots \big) \zeta_7 + \ldots
 \bigg\} \notag \\
 %%%
&\hspace{-1cm}+ s_{02} \bigg\{ {-} \frac{2 \eta \zeta_3}{\bar \eta}
\big(1+ 2 i\pi \bar \eta B_1(u) + \tfrac{1}{2!} (2 i\pi \bar \eta)^2 B_2(u)
+ \tfrac{1}{3!} (2 i\pi \bar \eta)^3 B_3(u) + \tfrac{1}{4!} (2 i\pi \bar \eta)^4 B_4(u) \notag \\
&\quad  + \tfrac{1}{5!} (2 i\pi \bar \eta)^5 B_5(u)+ \tfrac{1}{6!} (2 i\pi \bar \eta)^6 B_6(u) +\ldots\big) \notag \\
&  - \frac{2 \eta^3 \zeta_5}{\bar \eta}
 \big(1+ 2 i\pi \bar \eta B_1(u) + \tfrac{1}{2!} (2 i\pi \bar \eta)^2 B_2(u)
+ \tfrac{1}{3!} (2 i\pi \bar \eta)^3 B_3(u) + \tfrac{1}{4!} (2 i\pi \bar \eta)^4 B_4(u)+\ldots\big) \notag \\
&  - \frac{2 \eta^5 \zeta_7}{\bar \eta}
 \big(1+ 2 i\pi \bar \eta B_1(u) + \tfrac{1}{2!} (2 i\pi \bar \eta)^2 B_2(u)+\ldots \big)
 - \frac{2 \eta^7 \zeta_9}{\bar \eta}
 +\ldots \bigg\}
+ {\cal O}(s_{ij}^2)
\end{align} \normalsize

\newpage

%%%%%%%%%%%%%%%%%%%%%%%%%%%%%%%%%%%%%%%%%%%%%%%%%%%%%%%%%%%
%%%%%%%%%%%%%%%%%%%%%%%%%%%%%%%%%%%%%%%%%%%%%%%%%%%%%%%%%%%
\section{Further examples of integration constants}
\label{app:baral}
%%%%%%%%%%%%%%%%%%%%%%%%%%%%%%%%%%%%%%%%%%%%%%%%%%%%%%%%%%%
%%%%%%%%%%%%%%%%%%%%%%%%%%%%%%%%%%%%%%%%%%%%%%%%%%%%%%%%%%%

In this appendix, we supplement the discussion of integration constants
$\overline{\alpha[\ldots;\tau]}$ in section \ref{sec:9.1.ex} by further examples.

\subsection{All $\overline{ \alpha[\ldots;\tau]}$ at lattice weight 7}
\label{app:baral.A}

At lattice weight $\sum_i k_i=7$, a particularly rich example of the $\overline{ \alpha[\ldots;\tau]}$
was given in (\ref{inival.34}), and the remaining combinations realized in eMGFs
are determined by
\begin{align}
 \overline{\alphaBR{0 &0 \\ 3 &4 \\ z& z}{\tau}} - 
  \frac{2}{3}\overline{\alphaBR{0 &0\\ 2 &5 \\ z & z}{\tau}} &= 0\, ,
  & \overline{\alphaBR{1 &0 \\ 3 &4 \\ z &z}{\tau}} 
 - \overline{\alphaBR{1 &0 \\ 3 &4 \\ z &\emptyslot}{\tau}} &= 0 
    \notag \\
 \overline{\alphaBR{0 & 0 \\ 3 & 4 \\ z&z}{\tau}} 
 -  \overline{\alphaBR{0 &0 \\ 3 &4 \\ z &\emptyslot}{\tau}}
 &= 0 \, ,
 &2\overline{\alphaBR{0 &1 \\ 2 &5 \\ z & z}{\tau}} 
- 2\overline{\alphaBR{0 &1 \\ 3 &4 \\ z &\emptyslot}{\tau}}
- \overline{\alphaBR{1 &0 \\ 3 &4 \\ z &\emptyslot}{\tau}} &=0 \label{appal.01}\\ 
 %%%
 \overline{\alphaBR{0 &1 \\ 3 &4 \\ z &z}{\tau}} 
 -  \overline{\alphaBR{0 &1 \\ 3 &4 \\ z &\emptyslot}{\tau}} &= 0 \, ,
 & \overline{\alphaBR{0 &0 &0 \\ 2 &2 &3 \\ z& z &z}{\tau}} 
 +10\overline{\alphaBR{0 &1 \\ 2 &5 \\ z &z}{\tau}} 
 -  15\overline{\alphaBR{0 &1 \\ 3 &4 \\ z &\emptyslot }{\tau}} &= 0
 \notag 
\end{align}
and
\begin{align}
 \overline{\alphaBR{0 &2 \\ 3 &4 \\  z &z}{\tau}}  - \overline{\alphaBR{0 &2 \\ 3 &4 \\ z &\emptyslot}{\tau}} &=    \frac{2 \zeta_3}{3}  \overline{\mathcal{E}_0\!\SM{3}{z }{\tau}} \notag \\
  \overline{\alphaBR{1 &1 \\ 3 &4 \\ z &z}{\tau}} -  \overline{\alphaBR{1 &1 \\ 3 &4 \\ z &\emptyslot}{\tau}} &=0 \notag \\
 2 \overline{\alphaBR{0 &2 \\ 2 &5 \\ z &z}{\tau}}
- 2 \overline{\alphaBR{1 &1 \\ 3 &4 \\ z &\emptyslot}{\tau}} 
- \overline{\alphaBR{0 &2 \\ 3 & 4 \\ z & \emptyslot}{\tau}} &= \frac{ 2 \zeta_3}{3} \overline{ \mathcal{E}_0\!\SM{3}{z }{\tau} }
\label{appal.02} \\
 \overline{\alphaBR{0 &0 &1 \\ 2 &2 &3 \\ z&z&z}{\tau}} 
 +  5\overline{\alphaBR{0 &2 \\ 2 &5 \\ z &z}{\tau}}
 - \frac{15}{2}\overline{\alphaBR{0 &2 \\ 3 &4 \\ z&\emptyslot}{\tau}} &=
    5 \zeta_3 \overline{ \mathcal{E}_0\!\SM{3}{z }{\tau}  }
  - \zeta_3 B_1(u) \overline{ \mathcal{E}_0\!\SM{2}{z }{\tau} }
\notag \\
%%%%%%
 \overline{\alphaBR{1 &2 \\ 3 &4 \\ z &z}{\tau}} - \frac{2}{3}\overline{\alphaBR{0 &3 \\ 2 &5 \\ z&z}{\tau}} &=0
 \notag\\
 \overline{\alphaBR{1 &2 \\ 3 &4 \\ z &z}{\tau}}  - \overline{\alphaBR{1 &2 \\ 3 &4 \\ z&\emptyslot}{\tau}} &= 
    \frac{2\zeta_3}{3}  \overline{ \mathcal{E}_0\!\SM{3 &0}{z &0}{\tau}  }
    \notag
\end{align}
This collection of $\overline{ \alpha[\ldots;\tau]}$ completely fixes the $\beta^{\rm sv}$
relevant to all eMGFs at lattice weight $|A|{+}|B|=7$: After subtracting the $\dplus{a \\ b}\!(z|\tau)$ 
of depth one from the counting in table~\ref{allshuffir}, the above
$\overline{ \alpha[\ldots;\tau]}$ at depth $\geq 2$ contribute to
$2,4,4,2$ indecomposable eMGFs at $(|A|,|B|)=(2,5),(3,4),(4,3),(5,2)$. Indeed, (\ref{appal.01}) and
(\ref{appal.02}) comprise $2,4,4,2$ independent equations involving 
$ \overline{\alphaBR{j_1 &j_2 \\ k_1 &k_2 \\ z_1& z_2}{\tau}}$ with $j_1{+}j_2 = 0,1,2,3$
which match the expected counting.

\subsection{Leading terms at lattice weight 7 and 8}
\label{app:baral.B}

As detailed below (\ref{inival.33}), the leading terms of $\beta^{\rm sv}$
in section \ref{sec:3.3} necessitate contributions from the $\overline{ \alpha[\ldots;\tau]}$
to be expressible in terms of single-valued polylogarithms. The resulting predictions for
the leading terms at lattice weight $\sum_i k_i=7$
\begin{align}
 \overline{\alphaBR{j_1 &j_2 \\ 3 &4 \\ z &z}{\tau}} &= \frac{ \zeta_3}{3} \delta_{j_2,2} \times
 \left\{ \begin{array}{cl}  u \overline{ G(1; e^{2\pi i z}) } + {\cal O}(\bar q^{1-u}) &: \ j_1 = 0 \\
   \overline{ G(0,1; e^{2\pi i z}) } + {\cal O}(\bar q^{1-u}) &: \ j_1 = 1
 \end{array} \right.  \notag \\
 \overline{\alphaBR{j_1 &j_2 \\ 3 &4 \\ z &\emptyslot}{\tau}} &=  {\cal O}(\bar q^{1-u}) 
  \label{baralp.3} \\
%%%%
 \overline{\alphaBR{0 &j_2 \\ 2 &5 \\ z &z}{\tau}} &=   \zeta_3  \times
 \left\{ \begin{array}{cl}
 {\cal O}(\bar q^{1-u})  &: \ j_2 = 0,1 \\
 \frac{1}{6} u \overline{ G(1; e^{2\pi i z}) } + {\cal O}(\bar q^{1-u})  &: \ j_2 = 2 \\
 \frac{1}{2}  \overline{ G(0,1; e^{2\pi i z}) } + {\cal O}(\bar q^{1-u})  &: \ j_2 = 3
 \end{array} \right.
 \notag
\end{align}
are consistent with the expressions in (\ref{appal.01}) and (\ref{appal.02}).

At lattice weight $\sum_i k_i=8$, the leading terms of the $\beta^{\rm sv}$ at
depth two yield
\begin{align}
 \overline{\alphaBR{j_1 &j_2 \\ 4 &4 \\ z &\emptyslot }{\tau}} &=   {\cal O}(\bar q^{1-u})  \, , \ \ \ \ \ \
  &\overline{\alphaBR{0 &2 \\ 4 &4 \\ z &z }{\tau}} &= \frac{ \zeta_3}{9} u^2  \overline{ G(1; e^{2\pi i z}) }  + {\cal O}(\bar q^{1-u})   \label{baralp.4} \\
  \overline{\alphaBR{0 &1 \\ 4 &4 \\ z &z }{\tau}}&= {\cal O}(\bar q^{1-u}) \, , \ \ \ \ \ \
  &\overline{\alphaBR{1&2 \\ 4 &4 \\ z &z }{\tau}}&= \frac{ \zeta_3}{9} u  \overline{ G(0,1; e^{2\pi i z}) } + {\cal O}(\bar q^{1-u}) \notag
\end{align}
as well as
\begin{align}
 \overline{\alphaBR{0 &j_2 \\ 3 &5 \\ z &z}{\tau}} &=   \zeta_3  \times
 \left\{ \begin{array}{cl}
{\cal O}(\bar q^{1-u}) &: \ j_2 = 0,1 \\
 \frac{1}{12} u^2 \overline{ G(1; e^{2\pi i z}) } + {\cal O}(\bar q^{1-u})&: \ j_2 = 2 \\
 \frac{1}{4} u  \overline{ G(0,1; e^{2\pi i z}) } + {\cal O}(\bar q^{1-u})&: \ j_2 = 3
 \end{array} \right.
  \label{baralp.5} \\
   \overline{\alphaBR{1 &j_2 \\ 3 &5 \\ z &z}{\tau}} &=   \zeta_3  \times
 \left\{ \begin{array}{cl}
{\cal O}(\bar q^{1-u})&: \ j_2 = 0,1 \\
 \frac{1}{12} u \overline{ G(0,1; e^{2\pi i z}) } + {\cal O}(\bar q^{1-u}) &: \ j_2 = 2 \\
 \frac{1}{2}  \overline{ G(0,0,1; e^{2\pi i z}) } + {\cal O}(\bar q^{1-u}) &: \ j_2 = 3
 \end{array} \right.
 \notag
\end{align}
and
\begin{align}
 \overline{\alphaBR{0 &j_2 \\ 2 &6 \\ z &\emptyslot }{\tau}} &= {\cal O}(\bar q^{1-u})  \, , \ \ \ \ \ \
  &\overline{\alphaBR{0 &2 \\ 2 &6 \\ z &z }{\tau}} &= \frac{ \zeta_3}{30} u^2  \overline{ G(1; e^{2\pi i z}) }  +{\cal O}(\bar q^{1-u})  \label{baralp.6} \\
  \overline{\alphaBR{0 &0 \\ 2 &6 \\ z &z }{\tau}}&= {\cal O}(\bar q^{1-u}) \, , \ \ \ \ \ \
  &\overline{\alphaBR{0 &3 \\ 2 &6 \\ z &z }{\tau}}&=  \frac{ \zeta_3}{10} u  \overline{ G(0,1; e^{2\pi i z}) }  +{\cal O}(\bar q^{1-u})
  \notag \\
  \overline{\alphaBR{0 &1 \\ 2 &6 \\ z &z }{\tau}}&=  {\cal O}(\bar q^{1-u}) \, , \ \ \ \ \ \
  &\overline{\alphaBR{0&4 \\ 2 &6 \\ z &z }{\tau}}&= \frac{2 \zeta_3}{5}  \overline{ G(0,0,1; e^{2\pi i z}) }
  + \frac{2 \zeta_5}{5}  \overline{ G(1; e^{2\pi i z}) } + {\cal O}(\bar q^{1-u})  \notag
\end{align}

\newpage

%%%%%%%%%%%%%%%%%%%%%%%%%%%%%%%%%%%%%%%%%%%%%%%%%%%%%%%%%%%
%%%%%%%%%%%%%%%%%%%%%%%%%%%%%%%%%%%%%%%%%%%%%%%%%%%%%%%%%%%
\section{More on the bases of indecomposable eMGFs}
\label{last:app}
%%%%%%%%%%%%%%%%%%%%%%%%%%%%%%%%%%%%%%%%%%%%%%%%%%%%%%%%%%%
%%%%%%%%%%%%%%%%%%%%%%%%%%%%%%%%%%%%%%%%%%%%%%%%%%%%%%%%%%%

This appendix gathers further details on the bases of indecomposable
eMGFs at lattice weights $|A|{+}|B|\leq 10$ constructed in section \ref{bassec}.

%%%%%%%%%%%%%%%%%%%%%%%%%%%%%%%%%%%%%%%%%%%%%%%%%%%%%%%%%%%
%%%%%%%%%%%%%%%%%%%%%%%%%%%%%%%%%%%%%%%%%%%%%%%%%%%%%%%%%%%
\subsection{Non-vanishing leading terms of an odd eMGFs}
\label{last:app.A}
%%%%%%%%%%%%%%%%%%%%%%%%%%%%%%%%%%%%%%%%%%%%%%%%%%%%%%%%%%%
%%%%%%%%%%%%%%%%%%%%%%%%%%%%%%%%%%%%%%%%%%%%%%%%%%%%%%%%%%%

We shall here exemplify that eMGFs (\ref{dictio.6}) with odd parity under
$(z,\tau) \rightarrow ({-}\bar z,{-}\bar \tau)$ may have non-vanishing leading terms.
This will be done through the object $A_{2,0|1}$ defined by (\ref{adefs}),
one of the simplest non-trivial odd eMGFs.

As a consequence of (\ref{svpolycl}),
the automorphism $(z,\tau) \rightarrow ({-}\bar z,{-}\bar \tau)$ reverses the entries
$G^{\rm sv}(a_1,a_2,\ldots,a_w;e^{2\pi i z}) \rightarrow G^{\rm sv}(a_w,\ldots,a_2,a_1;e^{2\pi i z})$
up to corrections ${\cal Z}^{\rm sv}$ by MZVs. The $\beta^{\rm sv}$ representation
of $A_{2,0|1}$ in (\ref{tobasis.05}) together with $G^{\rm sv}(0;e^{2\pi i z})=-4uy$ and the
shuffle property of $G^{\rm sv}$ identify leading terms
\begin{align}
A_{2,0|1}(z|\tau) &= \bigg(\frac{ u}{16} - \frac{1}{8} \bigg) \big[  G^{\rm sv}(0,0,0,1;e^{2\pi i z})
- G^{\rm sv}(1,0,0,0;e^{2\pi i z}) \big] \notag \\
&\quad + \bigg({-}\frac{ 3u}{16} + \frac{1}{8} \bigg) \big[  G^{\rm sv}(0,0,1,0;e^{2\pi i z})
- G^{\rm sv}(0,1,0,0;e^{2\pi i z}) \big]
\label{ccleading.01} \\
&\quad
+ \frac{1}{4}  \big[  G^{\rm sv}(0,1,0,1;e^{2\pi i z})
- G^{\rm sv}(1,0,1,0;e^{2\pi i z}) + 4 \zeta_3 G^{\rm sv}(1;e^{2\pi i z}) \big]
\notag \\
&\quad - \frac{y}{12} \big[  G^{\rm sv}(0,0,1;e^{2\pi i z})
- G^{\rm sv}(1,0,0;e^{2\pi i z}) \big]  + {\cal O}(q^{1-u},\bar q^{1-u}) 
\notag
\end{align}
where each line is individually odd under $(z,\tau) \rightarrow ({-}\bar z,{-}\bar \tau)$. 
The third line is the only place where complex
conjugation introduces MZVs in intermediate steps: We have 
\beq
\overline{ G^{\rm sv}(0,1,0,1;e^{2\pi i z}) } = G^{\rm sv}(1,0,1,0;e^{2\pi i z}) - 4 \zeta_3 G^{\rm sv}(1;e^{2\pi i z})
\label{ccleading.02} 
\eeq
by (\ref{comrel.27}), and the $\zeta_3$ correction ensures that the third line of
(\ref{ccleading.01}) indeed flips sign under $(z,\tau) \rightarrow ({-}\bar z,{-}\bar \tau)$.

%%%%%%%%%%%%%%%%%%%%%%%%%%%%%%%%%%%%%%%%%%%%%%%%%%%%%%%%%%%
%%%%%%%%%%%%%%%%%%%%%%%%%%%%%%%%%%%%%%%%%%%%%%%%%%%%%%%%%%%
\subsection{Odd three-column eMGFs at $|A|=|B|=4$}
\label{last:app.B}
%%%%%%%%%%%%%%%%%%%%%%%%%%%%%%%%%%%%%%%%%%%%%%%%%%%%%%%%%%%
%%%%%%%%%%%%%%%%%%%%%%%%%%%%%%%%%%%%%%%%%%%%%%%%%%%%%%%%%%%

This appendix gathers relations among $A_{u,v|w}(z)$ and $B_{u|v,w}(z)$ at
with $u{+}v{+}w=4$ that generalize the weight-three relations in
(\ref{ABrels}). All the weight-four $A_{u,v|w}(z)$ and $B_{u|v,w}(z)$
that satisfy the criteria below (\ref{aprops}) reduce to the elements
$A_{ 2, 0|2}(z)$ and $A_{ 3, 0|1}(z)$ of ${\cal V}_{4,4}$ in (\ref{tobasis.21})
and products of single-valued elliptic polylogarithms,
\beq
A_{ 2, 1| 1} = \frac{ 1}{2} A_{ 2, 0|2} \, , \ \ \ \ 
B_{ 0| 1, 3} = - A_{ 3, 0| 1} +   \frac{ 1}{2} A_{ 2, 0| 2}\, , \ \ \ \
B_{ 1| 1, 2} = \frac{ 1}{2} A_{ 2, 0| 2} \, , \ \ \ \
B_{ 2| 0, 2}  =  A_{ 2, 0| 2}
\label{tobasis.appA} 
\eeq
and
\begin{align}
B_{ 1| 0, 3}  &=
 \frac{ (\pi \nabla_\tau E_2) (\pi \overline \nabla_\tau g_2) - (\pi \overline \nabla_\tau E_2) (\pi \nabla_\tau g_2)}{4y^2}
 \notag \\
 B_{ 3| 0, 1} &=
 A_{ 3, 0| 1} - \frac{ 1}{2} A_{ 2, 0| 2} +
  \frac{ (\pi \nabla_\tau E_2) (\pi \overline \nabla_\tau g_2) -   (\pi \overline \nabla_\tau E_2) (\pi \nabla_\tau g_2)}{4y^2} \label{tobasis.appB}  \\
  &\quad  - (\dplus{1\\ 2} \dplus{3\\ 2} -
    \dplus{2\\ 1} \dplus{2\\ 3})
    \notag 
\end{align}

%%%%%%%%%%%%%%%%%%%%%%%%%%%%%%%%%%%%%%%%%%%%%%%%%%%%%%%%%%%
%%%%%%%%%%%%%%%%%%%%%%%%%%%%%%%%%%%%%%%%%%%%%%%%%%%%%%%%%%%
\subsection{Completing the $\beta^{\rm sv}$ representations for ${\cal V}_{4,4}$ elements}
\label{last:app.C}
%%%%%%%%%%%%%%%%%%%%%%%%%%%%%%%%%%%%%%%%%%%%%%%%%%%%%%%%%%%
%%%%%%%%%%%%%%%%%%%%%%%%%%%%%%%%%%%%%%%%%%%%%%%%%%%%%%%%%%%

This appendix displays the $\beta^{\rm sv}$ representation
of the four-column elements
$\cplus{0&1&1&2 \\
2&1&1&0 \\
0&z&z&0} $ and $\aplus{0 &1 &1 &2 \\ 1 &1 &1 &1 \\ z &0 &0 &0}$
 in the basis ${\cal V}_{4,4}$ in (\ref{tobasis.21}):
\begin{align}
\cplus{0&1&1&2 \\
2&1&1&0 \\
0&z&z&0} &=
-12 \bsv{0&0&1\\2&2&4\\z&z&\emptyslot}+12 \bsv{0&0&1\\2&2&4\\z&z&z}-8 \bsv{0&0&1\\2&3&3\\z&z&z}-8 \bsv{0&1&0\\2&3&3\\z&z&z}-24 \bsv{0&1&0\\2&4&2\\z&\emptyslot&z} \notag \\
&+24 \bsv{0&1&0\\2&4&2\\z&z&z}-8 \bsv{0&1&0\\3&3&2\\z&z&z}-8 \bsv{1&0&0\\3&3&2\\z&z&z}-12 \bsv{1&0&0\\4&2&2\\\emptyslot&z&z}+12 \bsv{1&0&0\\4&2&2\\z&z&z} \notag \\
&+ 150 \bsv{0&2\\2&6\\z&\emptyslot}-150 \bsv{0&2\\2&6\\z&z}+16 \bsv{0&2\\3&5\\z&z}+18 \bsv{0&2\\4&4\\z&z}+176 \bsv{1&1\\3&5\\z&z} \notag \\
&+36 \bsv{1&1\\4&4\\\emptyslot&z}+36 \bsv{1&1\\4&4\\z&\emptyslot}-144 \bsv{1&1\\4&4\\z&z}+16 \bsv{1&1\\5&3\\z&z}+18 \bsv{2&0\\4&4\\\emptyslot&z}\label{dictio.12}  \\
&+18 \bsv{2&0\\4&4\\z&\emptyslot}-90 \bsv{2&0\\4&4\\z&z}+176 \bsv{2&0\\5&3\\z&z}+150 \bsv{2&0\\6&2\\\emptyslot&z}-150 \bsv{2&0\\6&2\\z&z} \notag \\
&+840 \bsv{3\\8\\z} -1344 \bsv{3\\8\\\emptyslot}
+\frac{2 \zeta_3 }{y}\bsv{0&0\\2&2\\z&z}
 -2 \bigg( \zeta_3 B_2(u)+ \frac{15 \zeta_5 }{4 y^2}  \bigg) \bsv{0\\2\\z} \notag\\
&+8 \zeta_3 B_1(u) \bsv{0\\3\\z}
+\frac{2 \zeta_3 B_1(u) }{y} \bsv{1\\3\\z}
-\frac{6 \zeta_3}{y}  \bsv{1\\4\\z} -12 \zeta_3 \bsv{0\\4\\z}  \notag \\
&+\frac{7 \zeta_5}{6 y}
+\frac{27 \zeta_5 B_2(u)}{2 y}+\frac{6 \zeta_7}{y^3}
\notag
\end{align}
as well as
\begin{align}
\aplus{0 &1 &1 &2 \\ 1 &1 &1 &1 \\ z &0 &0 &0} &=
-12 \bsv{0&0&1\\2&2&4\\z&z&\emptyslot}+12 \bsv{0&0&1\\2&2&4\\z&z&z}+8 \bsv{0&1&0\\2&3&3\\z&z&z}-8 \bsv{1&0&0\\3&3&2\\z&z&z}+12 \bsv{1&0&0\\4&2&2\\\emptyslot&z&z} \notag \\
&-12 \bsv{1&0&0\\4&2&2\\z&z&z}
+50 \bsv{0&2\\2&6\\z&\emptyslot}-50 \bsv{0&2\\2&6\\z&z}-16 \bsv{1&1\\3&5\\z&z}+36 \bsv{1&1\\4&4\\\emptyslot&z} \notag \\
&-36 \bsv{1&1\\4&4\\z&\emptyslot}-162 \bsv{2&0\\4&4\\\emptyslot&z}+162 \bsv{2&0\\4&4\\z&\emptyslot}+16 \bsv{2&0\\5&3\\z&z}-50 \bsv{2&0\\6&2\\\emptyslot&z}\! \! \label{dictio.13}\\
&+50 \bsv{2&0\\6&2\\z&z}
-\frac{2 \zeta_3 }{y}\bsv{0&0\\2&2\\z&z}
-4 \zeta_3 B_2(u) \bsv{0\\2\\z}
+\zeta_3 \bsv{0\\2\\z}
+\frac{5 \zeta_5}{4 y^2}  \bsv{0\\2\\z} \notag \\
&-8 \zeta_3 B_1(u) \bsv{0\\3\\z}
-\frac{6 \zeta_3}{y} \bsv{1\\4\\z}
+108 \zeta_3 \bsv{0\\4\\z}
+\frac{5 \zeta_5 B_2(u)}{2 y}
\notag
\end{align}

%%%%%%%%%%%%%%%%%%%%%%%%%%%%%%%%%%%%%%%%%%%%%%%%%%%%%%%%%%%
%%%%%%%%%%%%%%%%%%%%%%%%%%%%%%%%%%%%%%%%%%%%%%%%%%%%%%%%%%%
\subsection{Selected $\beta^{\rm sv}$ representations for ${\cal V}_{5,3}$ elements}
\label{last:app.F}
%%%%%%%%%%%%%%%%%%%%%%%%%%%%%%%%%%%%%%%%%%%%%%%%%%%%%%%%%%%
%%%%%%%%%%%%%%%%%%%%%%%%%%%%%%%%%%%%%%%%%%%%%%%%%%%%%%%%%%%

In this appendix, we display $\beta^{\rm sv}$ representations for some
of the ${\cal V}_{5,3}$ elements in (\ref{tobasis.28})
\begin{align}
\pi \nabla_\tau  C_{2|1,1}(z) &= 
9 \bsv{2 &1 \\ 4 &4 \\ \emptyslot &z}+9 \bsv{2 &1 \\ 4 &4 \\ z &\emptyslot}-9 \bsv{2	 &1 \\ 4 &4 \\ z &z} - \frac{21}{2} \bsv{4 \\8 \\ \emptyslot }+105 \bsv{4 \\ 8 \\z}
\notag \\
&\quad - 6 \zeta_3   \bsv{1 \\ 4 \\ z}+\frac{\zeta_5}{2}  B_2(u)+ \frac{ 3 \zeta_7}{16 y^2}
\notag \\
\pi \nabla_\tau  C_{1|2,1}(z)  &=
9 \bsv{2 &1 \\ 4 &4 \\ z &z}-12 \bsv{1 &2 \\ 3 &5 \\ z &z} -6 \bsv{2 &1 \\5 &3 \\ z &z}-6 \bsv{3 &0 \\ 5 &3 \\ z &z} \notag \\
&\quad 
+\frac{105}{2} \bsv{4 \\ 8 \\ \emptyslot } +42 \bsv{4 \\ 8 \\ z}
-\frac{3}{2} \zeta_5 B_2(u) - \frac{ 15 \zeta_7}{  16 y^2}
\label{somebsv53}
\\
%%%%%
%%%%%
\pi \nabla_\tau A_{2,0|2}(z) &=
-24 \bsv{1	&2 \\ 3&5 \\ z&z}-18 \bsv{1 &2 \\ 4& 4 \\ \emptyslot &z}+18 \bsv{1 &2 \\ 4 &4 \\ z &\emptyslot}
-36 \bsv{2 &1 \\ 4 &4 \\ \emptyslot &z}+36 \bsv{2 &1 \\ 4 &4 \\ z &\emptyslot} \notag \\
&\quad +12 \bsv{2 &1 \\ 5 &3 \\ z &z}
+12 \bsv{3 &0 \\ 5 &3 \\ z &z} +24 \zeta_3 \bsv{1 \\ 4 \\z}+\frac{ 3 \zeta_3}{y} \bsv{2 \\ 4 \\ z}
\notag \\
\pi \nabla_\tau \widehat A_{3,0|1}(z) &=30 \bsv{0 &3 \\2 &6 \\ z &\emptyslot}-30 \bsv{0 &3 \\ 2 &6 \\ z &z}
-9 \bsv{1 &2 \\ 4 &4 \\ \emptyslot &z}+9 \bsv{1 &2 \\ 4 &4 \\ z &\emptyslot}-18 \bsv{2 &1 \\ 4 &4 \\ \emptyslot &z}
\notag \\
&\quad +18 \bsv{2 &1 \\ 4 &4 \\ z &\emptyslot}+12 \zeta_3 \bsv{1 \\ 4 \\ z}
+\frac{ 3 \zeta_3}{2 y} \bsv{2 \\ 4 \\ z}+\frac{3}{2} \zeta_5 B_2(u)
\notag
\end{align}
%

%%%%%%%%%%%%%%%%%%%%%%%%%%%%%%%%%%%%%%%%%%%%%%%%%%%%%%%%%%%
%%%%%%%%%%%%%%%%%%%%%%%%%%%%%%%%%%%%%%%%%%%%%%%%%%%%%%%%%%%
\subsection{Selected $\beta^{\rm sv}$ representations for ${\cal V}_{5,5}$ elements}
\label{last:app.D}
%%%%%%%%%%%%%%%%%%%%%%%%%%%%%%%%%%%%%%%%%%%%%%%%%%%%%%%%%%%
%%%%%%%%%%%%%%%%%%%%%%%%%%%%%%%%%%%%%%%%%%%%%%%%%%%%%%%%%%%

This appendix lists the $\beta^{\rm sv}$ representation of selected even eMGFs
in the basis (\ref{tobasis.45}) of indecomposables. The four instances of
three-column eMGFs $C_{a|b,c}$ in (\ref{cabc.2}) are given~by
\begin{align}
C_{2 | 2,1}(z)&= 60 \bsv{2& 1\\4& 6\\z& z} - 60 \bsv{2& 1\\4& 6\\z& \emptyslot}  -
 60 \bsv{3& 0\\6& 4\\\emptyslot& z} +
 60 \bsv{3& 0\\6& 4\\z& z} -
 64 \bsv{3& 0\\5& 5\\z& z} 
  \notag \\
 &\quad +126 \bsv{4\\10\\ \emptyslot} - 378 \bsv{4\\10\\z} 
 + \frac{ 6 \zeta_{5} }{y} \bsv{0\\4\\z} 
-   \frac{  \zeta_{5} }{6} B_{4}(u) - \frac{ 5 \zeta_{7}}{8 y^2}  B_{2}(u)
 - \frac{7 \zeta_{9}}{ 64 y^4}  \notag \\
C_{1 | 2,2}(z)&= 120 \bsv{2& 1\\4& 6\\z& \emptyslot} -
 120 \bsv{2& 1\\4& 6\\z& z} +
 120 \bsv{3& 0\\6& 4\\ \emptyslot& z} -
 120 \bsv{3& 0\\6& 4\\z& z}
  \notag \\
 &\quad  +  288 \bsv{2& 1\\5& 5\\z& z} + 32 \bsv{3& 0\\5& 5\\z& z} - 504 \bsv{4\\10\\ \emptyslot} + 252 \bsv{4\\10\\z}
 \notag \\
 &\quad   - \frac{12 \zeta_{5} }{y}  \bsv{0\\4\\z}
 -   \frac{2  \zeta_{5}}{3} B_{4}(u) + \frac{5  \zeta_{7}}{4 y^2}  B_{2}(u)+ \frac{7 \zeta_{9}}{ 16 y^4}
  \label{w5cabc} \\
C_{3 | 1,1}(z)&=  32 \bsv{3& 0\\5& 5\\z& z}
- 120 \bsv{2& 1\\4& 6\\ \emptyslot& z}  - 120 \bsv{3& 0\\6& 4\\z& \emptyslot} 
- 18 \bsv{4\\10\\ \emptyslot} -
 756 \bsv{4\\10\\z}   \notag \\
 &\quad  + 80 \zeta_{3} \bsv{1\\6\\z}
 +\frac{\zeta_{5}}{6}   B_{4}(u) + \frac{ \zeta_{7}}{8 y^2}  B_{2}(u)+ \frac{ \zeta_{9}}{64 y^4}
 \notag \\
C_{1 | 3,1}(z)&= 32 \bsv{3& 0\\5& 5\\z& z}  +
 120 \bsv{1& 2\\3& 7\\z& z} +
 120 \bsv{3& 0\\7& 3\\z& z}-
 60 \bsv{2& 1\\4& 6\\z& \emptyslot} -
 60 \bsv{2& 1\\4& 6\\z& z}
 \notag \\
 &\quad   -
 60 \bsv{3& 0\\6& 4\\ \emptyslot& z}  -
 60 \bsv{3& 0\\6& 4\\z& z}  -
 378 \bsv{4\\10\\ \emptyslot} - 396 \bsv{4\\10\\z}\notag \\
 &\quad 
 + \frac{6 \zeta_{5} }{y} \bsv{0\\4\\z}
+ \frac{\zeta_{5}}{3}   B_{4}(u) + \frac{5 \zeta_{7}}{4 y^2} B_{2}(u)
+ \frac{21 \zeta_{9}}{ 64 y^4} 
 \notag
 \end{align}
and as an example of $\beta^{\rm sv}$ at depth four, we shall also display the 
iterated-KE-integral representation of the five-column eMGF
\begin{align}
C_{1|1,1,1,1}(z) &=
96 \bsv{1&0&0&0\\3&2&2&3\\z&z&z&z}
-192 \bsv{1&0&1\\3&2&5\\z&z&z}
-288 \bsv{1&0&1\\3&3&4\\z&z&\emptyslot}
+576 \bsv{1&1&0\\3&3&4\\z&z&\emptyslot}
 \notag \\
&\hspace{-1.5cm}
-288 \bsv{1&1&0\\4&3&3\\\emptyslot&z&z}+216 \bsv{2&0&0\\4&2&4\\\emptyslot&z&\emptyslot} 
-216 \bsv{2&0&0\\4&2&4\\\emptyslot&z&z}+576 \bsv{2&0&0\\4&3&3\\\emptyslot&z&z} 
-216 \bsv{2&0&0\\4&2&4\\z&z&\emptyslot}
 \notag \\
&\hspace{-1.5cm}
+216 \bsv{2&0&0\\4&2&4\\z&z&z}
-192 \bsv{2&0&0\\5&2&3\\z&z&z}
+360 \bsv{1&2\\4&6\\\emptyslot&\emptyslot}
-2520 \bsv{2&1\\4&6\\\emptyslot&\emptyslot}
+360 \bsv{2&1\\6&4\\\emptyslot&\emptyslot} 
\notag \\
&\hspace{-1.5cm}
-2520 \bsv{3&0\\6&4\\\emptyslot&\emptyslot}
+864 \bsv{1&2\\3&7\\z&z}
+720 \bsv{1&2\\4&6\\\emptyslot&z}
+720 \bsv{1&2\\4&6\\z&\emptyslot}
-1800 \bsv{2&1\\4&6\\\emptyslot&z}
 \notag \\
&\hspace{-1.5cm}
-1800 \bsv{2&1\\4&6\\z&\emptyslot}
-1080 \bsv{2&1\\4&6\\z&z}
+384 \bsv{2&1\\5&5\\z&z}
+720 \bsv{2&1\\6&4\\\emptyslot&z} 
+720 \bsv{2&1\\6&4\\z&\emptyslot}
\notag \\
&\hspace{-1.5cm}
+768 \bsv{3&0\\5&5\\z&z}
-1800 \bsv{3&0\\6&4\\\emptyslot&z}
-1800 \bsv{3&0\\6&4\\z&\emptyslot}
-1080 \bsv{3&0\\6&4\\z&z} 
+864 \bsv{3&0\\7&3\\z&z}
\label{dpth4.1} \\
&\hspace{-1.5cm} 
 -9720 \bsv{4\\10\\\emptyslot}-6480 \bsv{4\\10\\z}
 +96 \zeta_3 B_1(u) \bsv{0&0\\2&3\\z&z}
-144 \zeta_3 \bsv{0&0\\2&4\\z&\emptyslot}
 +144 \zeta_3 \bsv{0&0\\2&4\\z&z}
 \notag \\
&\hspace{-1.5cm}
 +\frac{48 \zeta_3 }{y} \bsv{1&0\\3&3\\z&z}
-384 \zeta_3 \bsv{0&0\\3&3\\z&z}
 - \frac{ 24 \zeta_5}{y}  B_1(u)  \bsv{0\\3\\z} 
 -192 \zeta_3 B_1(u) \bsv{1\\5\\z}
 \notag \\
 &\hspace{-1.5cm}
 -72 \zeta_3 B_2(u) \bsv{1\\4\\\emptyslot} 
  -24 \zeta_3 \bsv{1\\4\\z}
   -\frac{18 \zeta_5}{y^2}  \bsv{1\\4\\z} 
   -\frac{9 \zeta_5 }{y^2} \bsv{1\\4\\\emptyslot}
 +\frac{180 \zeta_5}{y}  \bsv{0\\4\\z} 
\notag \\
 &\hspace{-1.5cm}
 +\frac{252 \zeta_5 }{y} \bsv{0\\4\\\emptyslot} 
  -\frac{120 \zeta_3 }{y} \bsv{2\\6\\z}  
 -\frac{60 \zeta_3}{y}  \bsv{2\\6\\\emptyslot}
 +1680 \zeta_3 \bsv{1\\6\\\emptyslot} 
 +1200 \zeta_3 \bsv{1\\6\\z}
 \notag \\
&\hspace{-1.5cm}
-\frac{4 \zeta_5}{15}+18 \zeta_5 B_2(u)+4 \zeta_5 B_4(u)
- \frac{\zeta_3^2}{y}-\frac{6 \zeta_3^2 }{y} B_2(u) 
+\frac{6 \zeta_7}{y^2}+\frac{9 \zeta_7 }{2 y^2} B_2(u) 
-\frac{9 \zeta_3 \zeta_5}{y^3} +\frac{135 \zeta_9}{16 y^4}
\notag
\end{align}

\newpage

%\bibliography{citesEMGFintegrals}

\begin{thebibliography}{100}

\bibitem{Green:1999pv}
M.~B. Green and P.~Vanhove, ``{The Low-energy expansion of the one loop type II
  superstring amplitude},''
  \href{http://dx.doi.org/10.1103/PhysRevD.61.104011}{{\em Phys.Rev.} {\bf D61}
  (2000)  104011},
\href{http://arxiv.org/abs/hep-th/9910056}{{\tt arXiv:hep-th/9910056
  [hep-th]}}.
%%CITATION = HEP-TH/9910056;%%.

\bibitem{Green:2008uj}
M.~B. Green, J.~G. Russo, and P.~Vanhove, ``{Low energy expansion of the
  four-particle genus-one amplitude in type II superstring theory},''
  \href{http://dx.doi.org/10.1088/1126-6708/2008/02/020}{{\em JHEP} {\bf 02}
  (2008)  020},
\href{http://arxiv.org/abs/0801.0322}{{\tt arXiv:0801.0322 [hep-th]}}.
%%CITATION = ARXIV:0801.0322;%%.

\bibitem{DHoker:2015gmr}
E.~D'Hoker, M.~B. Green, and P.~Vanhove, ``{On the modular structure of the
  genus-one Type II superstring low energy expansion},''
  \href{http://dx.doi.org/10.1007/JHEP08(2015)041}{{\em JHEP} {\bf 08} (2015)
  041},
\href{http://arxiv.org/abs/1502.06698}{{\tt arXiv:1502.06698 [hep-th]}}.
%%CITATION = ARXIV:1502.06698;%%.

\bibitem{DHoker:2015wxz}
E.~D'Hoker, M.~B. Green, {\"O}.~G{\"u}rdogan, and P.~Vanhove, ``Modular graph
  functions,'' \href{http://dx.doi.org/10.4310/CNTP.2017.v11.n1.a4}{{\em
  Commun. Num. Theor. Phys.} {\bf 11} (2017)  165--218},
\href{http://arxiv.org/abs/1512.06779}{{\tt arXiv:1512.06779 [hep-th]}}.
%%CITATION = ARXIV:1512.06779;%%.

\bibitem{DHoker:2016mwo}
E.~D'Hoker and M.~B. Green, ``Identities between modular graph forms,''
  \href{http://dx.doi.org/10.1016/j.jnt.2017.11.015}{{\em J. Number Theory}
  {\bf 189} (2018)  25--80},
\href{http://arxiv.org/abs/1603.00839}{{\tt arXiv:1603.00839 [hep-th]}}.
%%CITATION = ARXIV:1603.00839;%%.

\bibitem{Green:2013bza}
M.~B. Green, C.~R. Mafra, and O.~Schlotterer, ``{Multiparticle one-loop
  amplitudes and S-duality in closed superstring theory},''
  \href{http://dx.doi.org/10.1007/JHEP10(2013)188}{{\em JHEP} {\bf 10} (2013)
  188},
\href{http://arxiv.org/abs/1307.3534}{{\tt arXiv:1307.3534 [hep-th]}}.
%%CITATION = ARXIV:1307.3534;%%.

\bibitem{DHoker:2015sve}
E.~D'Hoker, M.~B. Green, and P.~Vanhove, ``{Proof of a modular relation between
  1-, 2- and 3-loop Feynman diagrams on a torus},''
  \href{http://dx.doi.org/10.1016/j.jnt.2017.07.022}{{\em J.\ Number Theory}
  (2018)  381},
\href{http://arxiv.org/abs/1509.00363}{{\tt arXiv:1509.00363 [hep-th]}}.
%%CITATION = ARXIV:1509.00363;%%.

\bibitem{Basu:2015ayg}
A.~Basu, ``{Poisson equation for the Mercedes diagram in string theory at genus
  one},'' \href{http://dx.doi.org/10.1088/0264-9381/33/5/055005}{{\em Class.
  Quant. Grav.} {\bf 33} (2016) no.~5, 055005},
\href{http://arxiv.org/abs/1511.07455}{{\tt arXiv:1511.07455 [hep-th]}}.
%%CITATION = ARXIV:1511.07455;%%.

\bibitem{Basu:2016xrt}
A.~Basu, ``{Poisson equation for the three loop ladder diagram in string theory
  at genus one},'' \href{http://dx.doi.org/10.1142/S0217751X16501694}{{\em Int.
  J. Mod. Phys.} {\bf A31} (2016) no.~32, 1650169},
\href{http://arxiv.org/abs/1606.02203}{{\tt arXiv:1606.02203 [hep-th]}}.
%%CITATION = ARXIV:1606.02203;%%.

\bibitem{Basu:2016kli}
A.~Basu, ``{Proving relations between modular graph functions},''
  \href{http://dx.doi.org/10.1088/0264-9381/33/23/235011}{{\em Class. Quant.
  Grav.} {\bf 33} (2016) no.~23, 235011},
\href{http://arxiv.org/abs/1606.07084}{{\tt arXiv:1606.07084 [hep-th]}}.
%%CITATION = ARXIV:1606.07084;%%.

\bibitem{Basu:2016mmk}
A.~Basu, ``{Simplifying the one loop five graviton amplitude in type IIB string
  theory},'' \href{http://dx.doi.org/10.1142/S0217751X17500749}{{\em Int. J.
  Mod. Phys.} {\bf A32} (2017) no.~14, 1750074},
\href{http://arxiv.org/abs/1608.02056}{{\tt arXiv:1608.02056 [hep-th]}}.
%%CITATION = ARXIV:1608.02056;%%.

\bibitem{DHoker:2016quv}
E.~D'Hoker and J.~Kaidi, ``{Hierarchy of Modular Graph Identities},''
  \href{http://dx.doi.org/10.1007/JHEP11(2016)051}{{\em JHEP} {\bf 11} (2016)
  051},
\href{http://arxiv.org/abs/1608.04393}{{\tt arXiv:1608.04393 [hep-th]}}.
%%CITATION = ARXIV:1608.04393;%%.

\bibitem{Kleinschmidt:2017ege}
A.~Kleinschmidt and V.~Verschinin, ``{Tetrahedral modular graph functions},''
  \href{http://dx.doi.org/10.1007/JHEP09(2017)155}{{\em JHEP} {\bf 09} (2017)
  155},
\href{http://arxiv.org/abs/1706.01889}{{\tt arXiv:1706.01889 [hep-th]}}.
%%CITATION = ARXIV:1706.01889;%%.

\bibitem{Basu:2017nhs}
A.~Basu, ``{Low momentum expansion of one loop amplitudes in heterotic string
  theory},'' \href{http://dx.doi.org/10.1007/JHEP11(2017)139}{{\em JHEP} {\bf
  11} (2017)  139},
\href{http://arxiv.org/abs/1708.08409}{{\tt arXiv:1708.08409 [hep-th]}}.
%%CITATION = ARXIV:1708.08409;%%.

\bibitem{Broedel:2018izr}
J.~Broedel, O.~Schlotterer, and F.~Zerbini, ``{From elliptic multiple zeta
  values to modular graph functions: open and closed strings at one loop},''
  \href{http://dx.doi.org/10.1007/JHEP01(2019)155}{{\em JHEP} {\bf 01} (2019)
  155},
\href{http://arxiv.org/abs/1803.00527}{{\tt arXiv:1803.00527 [hep-th]}}.
%%CITATION = ARXIV:1803.00527;%%.

\bibitem{Ahlen:2018wng}
O.~Ahl{\'e}n and A.~Kleinschmidt, ``{$D^{6}R^{4}$ curvature corrections,
  modular graph functions and Poincar{\'e} series},''
  \href{http://dx.doi.org/10.1007/JHEP05(2018)194}{{\em JHEP} {\bf 05} (2018)
  194},
\href{http://arxiv.org/abs/1803.10250}{{\tt arXiv:1803.10250 [hep-th]}}.
%%CITATION = ARXIV:1803.10250;%%.

\bibitem{Gerken:2018zcy}
J.~E. Gerken and J.~Kaidi, ``{Holomorphic subgraph reduction of higher-point
  modular graph forms},'' \href{http://dx.doi.org/10.1007/JHEP01(2019)131}{{\em
  JHEP} {\bf 01} (2019)  131},
\href{http://arxiv.org/abs/1809.05122}{{\tt arXiv:1809.05122 [hep-th]}}.
%%CITATION = ARXIV:1809.05122;%%.

\bibitem{Gerken:2018jrq}
J.~E. Gerken, A.~Kleinschmidt, and O.~Schlotterer, ``{Heterotic-string
  amplitudes at one loop: modular graph forms and relations to open strings},''
  \href{http://dx.doi.org/10.1007/JHEP01(2019)052}{{\em JHEP} {\bf 01} (2019)
  052},
\href{http://arxiv.org/abs/1811.02548}{{\tt arXiv:1811.02548 [hep-th]}}.
%%CITATION = ARXIV:1811.02548;%%.

\bibitem{DHoker:2019txf}
E.~D'Hoker and J.~Kaidi, ``{Modular graph functions and odd cuspidal functions.
  Fourier and Poincar{\'e} series},''
  \href{http://dx.doi.org/10.1007/JHEP04(2019)136}{{\em JHEP} {\bf 04} (2019)
  136},
\href{http://arxiv.org/abs/1902.04180}{{\tt arXiv:1902.04180 [hep-th]}}.
%%CITATION = ARXIV:1902.04180;%%.

\bibitem{Dorigoni:2019yoq}
D.~Dorigoni and A.~Kleinschmidt, ``{Modular graph functions and asymptotic
  expansions of Poincar\'e series},'' {\em Commun. Num. Theor. Phys.} {\bf 13}
  (2019)  569--617,
\href{http://arxiv.org/abs/1903.09250}{{\tt arXiv:1903.09250 [hep-th]}}.
%%CITATION = ARXIV:1903.09250;%%.

\bibitem{DHoker:2019xef}
E.~D'Hoker and M.~B. Green, ``{Absence of irreducible multiple zeta-values in
  melon modular graph functions},''
  \href{http://dx.doi.org/10.4310/CNTP.2020.v14.n2.a2}{{\em Commun. Num. Theor.
  Phys.} {\bf 14} (2020) no.~2, 315--324},
\href{http://arxiv.org/abs/1904.06603}{{\tt arXiv:1904.06603 [hep-th]}}.
%%CITATION = ARXIV:1904.06603;%%.

\bibitem{DHoker:2019mib}
E.~D'Hoker, ``{Integral of two-loop modular graph functions},''
  \href{http://dx.doi.org/10.1007/JHEP06(2019)092}{{\em JHEP} {\bf 06} (2019)
  092},
\href{http://arxiv.org/abs/1905.06217}{{\tt arXiv:1905.06217 [hep-th]}}.
%%CITATION = ARXIV:1905.06217;%%.

\bibitem{DHoker:2019blr}
E.~D'Hoker and M.~B. Green, ``{Exploring transcendentality in superstring
  amplitudes},'' \href{http://dx.doi.org/10.1007/JHEP07(2019)149}{{\em JHEP}
  {\bf 07} (2019)  149},
\href{http://arxiv.org/abs/1906.01652}{{\tt arXiv:1906.01652 [hep-th]}}.
%%CITATION = ARXIV:1906.01652;%%.

\bibitem{Basu:2019idd}
A.~Basu, ``{Eigenvalue equation for the modular graph $C_{a,b,c,d}$},''
  \href{http://dx.doi.org/10.1007/JHEP07(2019)126}{{\em JHEP} {\bf 07} (2019)
  126},
\href{http://arxiv.org/abs/1906.02674}{{\tt arXiv:1906.02674 [hep-th]}}.
%%CITATION = ARXIV:1906.02674;%%.

\bibitem{Gerken:2019cxz}
J.~E. Gerken, A.~Kleinschmidt, and O.~Schlotterer, ``{All-order differential
  equations for one-loop closed-string integrals and modular graph forms},''
  \href{http://dx.doi.org/10.1007/JHEP01(2020)064}{{\em JHEP} {\bf 01} (2020)
  064},
\href{http://arxiv.org/abs/1911.03476}{{\tt arXiv:1911.03476 [hep-th]}}.
%%CITATION = ARXIV:1911.03476;%%.

\bibitem{Hohenegger:2019tii}
S.~Hohenegger, ``{From Little String Free Energies Towards Modular Graph
  Functions},'' \href{http://dx.doi.org/10.1007/JHEP03(2020)077}{{\em JHEP}
  {\bf 03} (2020)  077},
\href{http://arxiv.org/abs/1911.08172}{{\tt arXiv:1911.08172 [hep-th]}}.
%%CITATION = ARXIV:1911.08172;%%.

\bibitem{Gerken:2020yii}
J.~E. Gerken, A.~Kleinschmidt, and O.~Schlotterer, ``{Generating series of all
  modular graph forms from iterated Eisenstein integrals},''
  \href{http://dx.doi.org/10.1007/JHEP07(2020)190}{{\em JHEP} {\bf 07} (2020)
  no.~07, 190}, \href{http://arxiv.org/abs/2004.05156}{{\tt arXiv:2004.05156
  [hep-th]}}.

\bibitem{Basu:2020kka}
A.~Basu, ``{Zero mode of the Fourier series of some modular graphs from
  Poincare series},''
  \href{http://dx.doi.org/10.1016/j.physletb.2020.135715}{{\em Phys. Lett. B}
  {\bf 809} (2020)  135715}, \href{http://arxiv.org/abs/2005.07793}{{\tt
  arXiv:2005.07793 [hep-th]}}.

\bibitem{Vanhove:2020qtt}
P.~Vanhove and F.~Zerbini, ``{Building blocks of closed and open string
  amplitudes},'' in {\em {MathemAmplitudes 2019: Intersection Theory and
  Feynman Integrals}}.
\newblock 7, 2020.
\newblock \href{http://arxiv.org/abs/2007.08981}{{\tt arXiv:2007.08981
  [hep-th]}}.

\bibitem{Basu:2020pey}
A.~Basu, ``{Poisson equations for elliptic modular graph functions},''
  \href{http://dx.doi.org/10.1016/j.physletb.2021.136086}{{\em Phys. Lett. B}
  {\bf 814} (2021)  136086}, \href{http://arxiv.org/abs/2009.02221}{{\tt
  arXiv:2009.02221 [hep-th]}}.

\bibitem{Basu:2020iok}
A.~Basu, ``{Relations between elliptic modular graphs},''
  \href{http://dx.doi.org/10.1007/JHEP12(2020)195}{{\em JHEP} {\bf 12} (2020)
  195}, \href{http://arxiv.org/abs/2010.08331}{{\tt arXiv:2010.08331
  [hep-th]}}. [Erratum: JHEP 03, 061 (2021)].

\bibitem{Gerken:2020xfv}
J.~E. Gerken, A.~Kleinschmidt, C.~R. Mafra, O.~Schlotterer, and B.~Verbeek,
  ``{Towards closed strings as single-valued open strings at genus one},''
  \href{http://dx.doi.org/10.1088/1751-8121/abe58b}{{\em J. Phys. A} {\bf 55}
  (2022) no.~2, 025401}, \href{http://arxiv.org/abs/2010.10558}{{\tt
  arXiv:2010.10558 [hep-th]}}.

\bibitem{Hohenegger:2020slq}
S.~Hohenegger, ``{Diagrammatic Expansion of Non-Perturbative Little String Free
  Energies},'' \href{http://dx.doi.org/10.1007/JHEP04(2021)275}{{\em JHEP} {\bf
  04} (2021)  275}, \href{http://arxiv.org/abs/2011.06323}{{\tt
  arXiv:2011.06323 [hep-th]}}.

\bibitem{Dorigoni:2021jfr}
D.~Dorigoni, A.~Kleinschmidt, and O.~Schlotterer, ``{Poincar\'e series for
  modular graph forms at depth two. Part I. Seeds and Laplace systems},''
  \href{http://dx.doi.org/10.1007/JHEP01(2022)133}{{\em JHEP} {\bf 01} (2022)
  133}, \href{http://arxiv.org/abs/2109.05017}{{\tt arXiv:2109.05017
  [hep-th]}}.

\bibitem{Dorigoni:2021ngn}
D.~Dorigoni, A.~Kleinschmidt, and O.~Schlotterer, ``{Poincar\'e series for
  modular graph forms at depth two. Part II. Iterated integrals of cusp
  forms},'' \href{http://dx.doi.org/10.1007/JHEP01(2022)134}{{\em JHEP} {\bf
  01} (2022)  134}, \href{http://arxiv.org/abs/2109.05018}{{\tt
  arXiv:2109.05018 [hep-th]}}.

\bibitem{DHoker:2021ous}
E.~D'Hoker and N.~Geiser, ``{Integrating three-loop modular graph functions and
  transcendentality of string amplitudes},''
  \href{http://dx.doi.org/10.1007/JHEP02(2022)019}{{\em JHEP} {\bf 02} (2022)
  019}, \href{http://arxiv.org/abs/2110.06237}{{\tt arXiv:2110.06237
  [hep-th]}}.

\bibitem{Brown:mmv}
F.~Brown, ``{Multiple modular values and the relative completion of the
  fundamental group of ${\cal M}_{1,1}$},''
  \href{http://arxiv.org/abs/1407.5167}{{\tt arXiv:1407.5167 [math.NT]}}.

\bibitem{Zerbini:2015rss}
F.~Zerbini, ``{Single-valued multiple zeta values in genus 1 superstring
  amplitudes},'' \href{http://dx.doi.org/10.4310/CNTP.2016.v10.n4.a2}{{\em
  Commun. Num. Theor. Phys.} {\bf 10} (2016)  703--737},
\href{http://arxiv.org/abs/1512.05689}{{\tt arXiv:1512.05689 [hep-th]}}.
%%CITATION = ARXIV:1512.05689;%%.

\bibitem{Brown:2017qwo}
F.~Brown, ``{A class of non-holomorphic modular forms I},''
  \href{http://dx.doi.org/10.1007/s40687-018-0130-8}{{\em Res. Math. Sci.} {\bf
  5} (2018)  5:7}, \href{http://arxiv.org/abs/1707.01230}{{\tt arXiv:1707.01230
  [math.NT]}}.

\bibitem{Brown:2017qwo2}
F.~Brown, ``{A class of non-holomorphic modular forms II : equivariant iterated
  Eisenstein integrals},'' \href{http://dx.doi.org/10.1017/fms.2020.24}{{\em
  Forum~of~Mathematics,~Sigma} {\bf 8} (2020)  1},
\href{http://arxiv.org/abs/1708.03354}{{\tt arXiv:1708.03354 [math.NT]}}.
%%CITATION = ARXIV:1707.01230;%%.

\bibitem{DHoker:2017zhq}
E.~D'Hoker and W.~Duke, ``Fourier series of modular graph functions,''
  \href{http://dx.doi.org/10.1016/j.jnt.2018.04.012}{{\em J. Number Theory}
  {\bf 192} (2018)  1--36}, \href{http://arxiv.org/abs/1708.07998}{{\tt
  arXiv:1708.07998 [math.NT]}}.

\bibitem{Zerbini:2018sox}
F.~Zerbini, {\em {Elliptic multiple zeta values, modular graph functions and
  genus 1 superstring scattering amplitudes}}.
\newblock PhD thesis, Bonn U., 2017.
\newblock
\href{http://arxiv.org/abs/1804.07989}{{\tt arXiv:1804.07989 [math-ph]}}.
\newblock
%%CITATION = ARXIV:1804.07989;%%.

\bibitem{Zerbini:2018hgs}
F.~Zerbini, ``{Modular and holomorphic graph function from superstring
  amplitudes},'' in {\em {KMPB Conference: Elliptic Integrals, Elliptic
  Functions and Modular Forms in Quantum Field Theory Zeuthen, Germany, October
  23-26, 2017}}.
\newblock 2018.
\newblock
\href{http://arxiv.org/abs/1807.04506}{{\tt arXiv:1807.04506 [math-ph]}}.
\newblock
%%CITATION = ARXIV:1807.04506;%%.

\bibitem{Zagier:2019eus}
D.~Zagier and F.~Zerbini, ``{Genus-zero and genus-one string amplitudes and
  special multiple zeta values},''
  \href{http://dx.doi.org/10.4310/CNTP.2020.v14.n2.a4}{{\em Commun. Num. Theor.
  Phys.} {\bf 14} (2020) no.~2, 413--452},
\href{http://arxiv.org/abs/1906.12339}{{\tt arXiv:1906.12339 [math.NT]}}.
%%CITATION = ARXIV:1906.12339;%%.

\bibitem{Berg:2019jhh}
M.~Berg, K.~Bringmann, and T.~Gannon, ``{Massive deformations of Maass forms
  and Jacobi forms},''
  \href{http://dx.doi.org/10.4310/CNTP.2021.v15.n3.a4}{{\em Commun. Num. Theor.
  Phys.} {\bf 15} (2021) no.~3, 575--603},
  \href{http://arxiv.org/abs/1910.02745}{{\tt arXiv:1910.02745 [math.NT]}}.

\bibitem{Drewitt:2021}
J.~Drewitt, ``Laplace-eigenvalue equations for length three modular iterated
  integrals,''
  \href{http://dx.doi.org/https://doi.org/10.1016/j.jnt.2021.11.005}{{\em
  Journal of Number Theory} {\bf 239} (2022)  78--112},
  \href{http://arxiv.org/abs/2104.09916}{{\tt arXiv:2104.09916 [math.NT]}}.

\bibitem{Gerken:review}
J.~E. Gerken, ``{Modular Graph Forms and Scattering Amplitudes in String
  Theory},'' \href{http://arxiv.org/abs/2011.08647}{{\tt arXiv:2011.08647
  [hep-th]}}.

\bibitem{Berkovits:2022ivl}
N.~Berkovits, E.~D'Hoker, M.~B. Green, H.~Johansson, and O.~Schlotterer,
  ``{Snowmass White Paper: String Perturbation Theory},'' in {\em {2022
  Snowmass Summer Study}}.
\newblock 3, 2022.
\newblock \href{http://arxiv.org/abs/2203.09099}{{\tt arXiv:2203.09099
  [hep-th]}}.

\bibitem{Dorigoni:2022iem}
D.~Dorigoni, M.~B. Green, and C.~Wen, ``{The SAGEX Review on Scattering
  Amplitudes, Chapter 10: Modular covariance of type IIB string amplitudes and
  their $\mathcal{N}=4$ supersymmetric Yang-Mills duals},''
  \href{http://arxiv.org/abs/2203.13021}{{\tt arXiv:2203.13021 [hep-th]}}.

\bibitem{DHoker:2022dxx}
E.~D'Hoker and J.~Kaidi, ``{Lectures on modular forms and strings},''
  \href{http://arxiv.org/abs/2208.07242}{{\tt arXiv:2208.07242 [hep-th]}}.

\bibitem{Gerken:2020aju}
J.~E. Gerken, ``{Basis Decompositions and a Mathematica Package for Modular
  Graph Forms},'' \href{http://dx.doi.org/10.1088/1751-8121/abbdf2}{{\em J.
  Phys. A} {\bf 54} (2021) no.~19, 195401},
  \href{http://arxiv.org/abs/2007.05476}{{\tt arXiv:2007.05476 [hep-th]}}.

\bibitem{DHoker:2013fcx}
E.~D'Hoker and M.~B. Green, ``{Zhang-Kawazumi Invariants and Superstring
  Amplitudes},'' \href{http://dx.doi.org/10.1016/j.jnt.2014.03.021}{{\em J.
  Number Theor.} {\bf 144} (2014)  111},
\href{http://arxiv.org/abs/1308.4597}{{\tt arXiv:1308.4597 [hep-th]}}.
%%CITATION = ARXIV:1308.4597;%%.

\bibitem{DHoker:2014oxd}
E.~D'Hoker, M.~B. Green, B.~Pioline, and R.~Russo, ``{Matching the $D^{6}R^{4}$
  interaction at two-loops},''
  \href{http://dx.doi.org/10.1007/JHEP01(2015)031}{{\em JHEP} {\bf 01} (2015)
  031},
\href{http://arxiv.org/abs/1405.6226}{{\tt arXiv:1405.6226 [hep-th]}}.
%%CITATION = ARXIV:1405.6226;%%.

\bibitem{Pioline:2015qha}
B.~Pioline, ``{A Theta lift representation for the Kawazumi-Zhang and Faltings
  invariants of genus-two Riemann surfaces},''
  \href{http://dx.doi.org/10.1016/j.jnt.2015.12.021}{{\em J. Number Theor.}
  {\bf 163} (2016)  520--541},
\href{http://arxiv.org/abs/1504.04182}{{\tt arXiv:1504.04182 [hep-th]}}.
%%CITATION = ARXIV:1504.04182;%%.

\bibitem{DHoker:2017pvk}
E.~D'Hoker, M.~B. Green, and B.~Pioline, ``{Higher genus modular graph
  functions, string invariants, and their exact asymptotics},''
  \href{http://dx.doi.org/10.1007/s00220-018-3244-3}{{\em Commun. Math. Phys.}
  {\bf 366} (2019) no.~3, 927--979},
\href{http://arxiv.org/abs/1712.06135}{{\tt arXiv:1712.06135 [hep-th]}}.
%%CITATION = ARXIV:1712.06135;%%.

\bibitem{DHoker:2018mys}
E.~D'Hoker, M.~B. Green, and B.~Pioline, ``Asymptotics of the {$D^8{\cal R}^4$}
  genus-two string invariant,''
  \href{http://dx.doi.org/10.4310/CNTP.2019.v13.n2.a3}{{\em Commun. Num. Theor.
  Phys.} {\bf 13} (2019) no.~2, 351--462},
\href{http://arxiv.org/abs/1806.02691}{{\tt arXiv:1806.02691 [hep-th]}}.
%%CITATION = ARXIV:1806.02691;%%.

\bibitem{Basu:2018bde}
A.~Basu, ``{Eigenvalue equation for genus two modular graphs},''
  \href{http://dx.doi.org/10.1007/JHEP02(2019)046}{{\em JHEP} {\bf 02} (2019)
  046},
\href{http://arxiv.org/abs/1812.00389}{{\tt arXiv:1812.00389 [hep-th]}}.
%%CITATION = ARXIV:1812.00389;%%.

\bibitem{DHoker:2020tcq}
E.~D'Hoker, C.~R. Mafra, B.~Pioline, and O.~Schlotterer, ``{Two-loop
  superstring five-point amplitudes. Part II. Low energy expansion and
  S-duality},'' \href{http://dx.doi.org/10.1007/JHEP02(2021)139}{{\em JHEP}
  {\bf 02} (2021)  139}, \href{http://arxiv.org/abs/2008.08687}{{\tt
  arXiv:2008.08687 [hep-th]}}.

\bibitem{DHoker:2020uid}
E.~D'Hoker and O.~Schlotterer, ``{Identities among higher genus modular graph
  tensors},'' \href{http://dx.doi.org/10.4310/CNTP.2022.v16.n1.a2}{{\em Commun.
  Num. Theor. Phys.} {\bf 16} (2022) no.~1, 35--74},
  \href{http://arxiv.org/abs/2010.00924}{{\tt arXiv:2010.00924 [hep-th]}}.

\bibitem{Basu:2020goe}
A.~Basu, ``{Integrating simple genus two string invariants over moduli
  space},'' \href{http://dx.doi.org/10.1007/JHEP03(2021)158}{{\em JHEP} {\bf
  03} (2021)  158}, \href{http://arxiv.org/abs/2012.14006}{{\tt
  arXiv:2012.14006 [hep-th]}}.

\bibitem{Basu:2021xdt}
A.~Basu, ``{Poisson equation for genus two string invariants: a conjecture},''
  \href{http://dx.doi.org/10.1007/JHEP04(2021)050}{{\em JHEP} {\bf 04} (2021)
  050}, \href{http://arxiv.org/abs/2101.04597}{{\tt arXiv:2101.04597
  [hep-th]}}.

\bibitem{Nilsnewarticle}
N.~Matthes, ``On the algebraic structure of iterated integrals of quasimodular
  forms,'' \href{http://dx.doi.org/10.2140/ant.2017.11.2113}{{\em Algebra \&
  Number Theory} {\bf 11-9} (2017)  2113--2130},
  \href{http://arxiv.org/abs/1708.04561}{{\tt arXiv:1708.04561}}.

\bibitem{Dhoker:2020gdz}
E.~D'Hoker, A.~Kleinschmidt, and O.~Schlotterer, ``{Elliptic modular graph
  forms. Part I. Identities and generating series},''
  \href{http://dx.doi.org/10.1007/JHEP03(2021)151}{{\em JHEP} {\bf 03} (2021)
  151}, \href{http://arxiv.org/abs/2012.09198}{{\tt arXiv:2012.09198
  [hep-th]}}.

\bibitem{brown2015notes}
F.~Brown, ``Notes on motivic periods,''
  \href{http://arxiv.org/abs/1512.06410}{{\tt arXiv:1512.06410 [math]}}.

\bibitem{Brown:2018omk}
F.~Brown and C.~Dupont, ``{Single-valued integration and double copy},''
  \href{http://dx.doi.org/10.1515/crelle-2020-0042}{{\em J. Reine Angew. Math.}
  {\bf 2021} (2021) no.~775, 145--196},
  \href{http://arxiv.org/abs/1810.07682}{{\tt arXiv:1810.07682 [math.NT]}}.

\bibitem{Ramakrish}
D.~Zagier, ``The {B}loch-{W}igner-{R}amakrishnan polylogarithm function,'' {\em
  Math. Ann.} {\bf 286} (1990)  613.

\bibitem{svpolylog}
F.~Brown, ``{Polylogarithmes multiples uniformes en une variable},'' {\em C. R.
  Acad. Sci. Paris} {\bf Ser. I 338} (2004)  527--532.

\bibitem{Broedel:2016kls}
J.~Broedel, M.~Sprenger, and A.~Torres~Orjuela, ``{Towards single-valued
  polylogarithms in two variables for the seven-point remainder function in
  multi-Regge-kinematics},''
  \href{http://dx.doi.org/10.1016/j.nuclphysb.2016.12.016}{{\em Nucl. Phys. B}
  {\bf 915} (2017)  394--413}, \href{http://arxiv.org/abs/1606.08411}{{\tt
  arXiv:1606.08411 [hep-th]}}.

\bibitem{DelDuca:2016lad}
V.~Del~Duca, S.~Druc, J.~Drummond, C.~Duhr, F.~Dulat, R.~Marzucca,
  G.~Papathanasiou, and B.~Verbeek, ``{Multi-Regge kinematics and the moduli
  space of Riemann spheres with marked points},''
  \href{http://dx.doi.org/10.1007/JHEP08(2016)152}{{\em JHEP} {\bf 08} (2016)
  152}, \href{http://arxiv.org/abs/1606.08807}{{\tt arXiv:1606.08807
  [hep-th]}}.

\bibitem{Kawai:1985xq}
H.~Kawai, D.~C. Lewellen, and S.~H.~H. Tye, ``{A Relation Between Tree
  Amplitudes of Closed and Open Strings},''
\href{http://dx.doi.org/10.1016/0550-3213(86)90362-7}{{\em Nucl. Phys.} {\bf
  B269} (1986)  1--23}.
%%CITATION = NUPHA,B269,1;%%.

\bibitem{Schlotterer:2012ny}
O.~Schlotterer and S.~Stieberger, ``{Motivic Multiple Zeta Values and
  Superstring Amplitudes},''
  \href{http://dx.doi.org/10.1088/1751-8113/46/47/475401}{{\em J. Phys.} {\bf
  A46} (2013)  475401},
\href{http://arxiv.org/abs/1205.1516}{{\tt arXiv:1205.1516 [hep-th]}}.
%%CITATION = ARXIV:1205.1516;%%.

\bibitem{Stieberger:2013wea}
S.~Stieberger, ``{Closed superstring amplitudes, single-valued multiple zeta
  values and the Deligne associator},''
  \href{http://dx.doi.org/10.1088/1751-8113/47/15/155401}{{\em J. Phys.} {\bf
  A47} (2014)  155401},
\href{http://arxiv.org/abs/1310.3259}{{\tt arXiv:1310.3259 [hep-th]}}.
%%CITATION = ARXIV:1310.3259;%%.

\bibitem{Stieberger:2014hba}
S.~Stieberger and T.~R. Taylor, ``{Closed String Amplitudes as Single-Valued
  Open String Amplitudes},''
  \href{http://dx.doi.org/10.1016/j.nuclphysb.2014.02.005}{{\em Nucl. Phys.}
  {\bf B881} (2014)  269--287},
\href{http://arxiv.org/abs/1401.1218}{{\tt arXiv:1401.1218 [hep-th]}}.
%%CITATION = ARXIV:1401.1218;%%.

\bibitem{Schlotterer:2018abc}
O.~Schlotterer and O.~Schnetz, ``{Closed strings as single-valued open strings:
  A genus-zero derivation},''
  \href{http://dx.doi.org/10.1088/1751-8121/aaea14}{{\em J. Phys.} {\bf A52}
  (2019) no.~4, 045401},
\href{http://arxiv.org/abs/1808.00713}{{\tt arXiv:1808.00713 [hep-th]}}.
%%CITATION = ARXIV:1808.00713;%%.

\bibitem{Vanhove:2018elu}
P.~Vanhove and F.~Zerbini, ``{Single-valued hyperlogarithms, correlation
  functions and closed string amplitudes},'' {\em Adv. Theor. Math. Phys.} {\bf
  26} (2022) no.~2, , \href{http://arxiv.org/abs/1812.03018}{{\tt
  arXiv:1812.03018 [hep-th]}}.

\bibitem{Brown:2019wna}
F.~Brown and C.~Dupont, ``{Single-valued integration and superstring amplitudes
  in genus zero},'' \href{http://dx.doi.org/10.1007/s00220-021-03969-4}{{\em
  Commun. Math. Phys.} {\bf 382} (2021) no.~2, 815--874},
  \href{http://arxiv.org/abs/1910.01107}{{\tt arXiv:1910.01107 [math.NT]}}.

\bibitem{Schnetz:2013hqa}
O.~Schnetz, ``{Graphical functions and single-valued multiple
  polylogarithms},'' \href{http://dx.doi.org/10.4310/CNTP.2014.v8.n4.a1}{{\em
  Commun. Num. Theor. Phys.} {\bf 08} (2014)  589--675},
\href{http://arxiv.org/abs/1302.6445}{{\tt arXiv:1302.6445 [math.NT]}}.
%%CITATION = ARXIV:1302.6445;%%.

\bibitem{Brown:2013gia}
F.~Brown, ``{Single-valued Motivic Periods and Multiple Zeta Values},''
  \href{http://dx.doi.org/10.1017/fms.2014.18}{{\em SIGMA} {\bf 2} (2014)
  e25},
\href{http://arxiv.org/abs/1309.5309}{{\tt arXiv:1309.5309 [math.NT]}}.
%%CITATION = ARXIV:1309.5309;%%.

\bibitem{Panzertalk}
E.~Panzer, ``{Talk ``Modular graph functions as iterated Eisenstein integrals''
  given at the workshop ``Elliptic Integrals in Mathematics and Physics''
  (Ascona, Switzerland)}.''
  \url{https://indico.cern.ch/event/700233/contributions/3112451/attachments/1712442/2761239/elliptic.pdf},
  2018.

\bibitem{Enriquez:Emzv}
B.~Enriquez, ``Analogues elliptiques des nombres multiz\'etas,''
  \href{http://dx.doi.org/10.24033/bsmf.2718}{{\em Bull. Soc. Math. France}
  {\bf 144} (2016) no.~3, 395--427}, \href{http://arxiv.org/abs/1301.3042}{{\tt
  1301.3042}}.

\bibitem{Broedel:2015hia}
J.~Broedel, N.~Matthes, and O.~Schlotterer, ``{Relations between elliptic
  multiple zeta values and a special derivation algebra},''
  \href{http://dx.doi.org/10.1088/1751-8113/49/15/155203}{{\em J. Phys.} {\bf
  A49} (2016) no.~15, 155203},
\href{http://arxiv.org/abs/1507.02254}{{\tt arXiv:1507.02254 [hep-th]}}.
%%CITATION = ARXIV:1507.02254;%%.

\bibitem{MGFinprogress}
D.~Dorigoni, M.~Doroudiani, J.~Drewitt, M.~Hidding, A.~Kleinschmidt,
  N.~Matthes, O.~Schlotterer, and B.~Verbeek, ``Modular graph forms from
  equivariant iterated Eisenstein integrals,'' \href{http://arxiv.org/abs/(to
  appear)}{{\tt (to appear)}}.

\bibitem{Broedel:2014vla}
J.~Broedel, C.~R. Mafra, N.~Matthes, and O.~Schlotterer, ``{Elliptic multiple
  zeta values and one-loop superstring amplitudes},''
  \href{http://dx.doi.org/10.1007/JHEP07(2015)112}{{\em JHEP} {\bf 07} (2015)
  112},
\href{http://arxiv.org/abs/1412.5535}{{\tt arXiv:1412.5535 [hep-th]}}.
%%CITATION = ARXIV:1412.5535;%%.

\bibitem{Broedel:2017jdo}
J.~Broedel, N.~Matthes, G.~Richter, and O.~Schlotterer, ``{Twisted elliptic
  multiple zeta values and non-planar one-loop open-string amplitudes},''
  \href{http://dx.doi.org/10.1088/1751-8121/aac601}{{\em J. Phys.} {\bf A51}
  (2018) no.~28, 285401},
\href{http://arxiv.org/abs/1704.03449}{{\tt arXiv:1704.03449 [hep-th]}}.
%%CITATION = ARXIV:1704.03449;%%.

\bibitem{BrownLev}
F.~Brown and A.~Levin, ``{Multiple elliptic polylogarithms},''
  \href{http://arxiv.org/abs/1110.6917}{{\tt arXiv:1110.6917 [math]}}.

\bibitem{Broedel:2019tlz}
J.~Broedel and A.~Kaderli, ``{Functional relations for elliptic
  polylogarithms},'' \href{http://dx.doi.org/10.1088/1751-8121/ab81d7}{{\em J.
  Phys. A} {\bf 53} (2020) no.~24, 245201},
  \href{http://arxiv.org/abs/1906.11857}{{\tt arXiv:1906.11857 [hep-th]}}.

\bibitem{Broedel:2018iwv}
J.~Broedel, C.~Duhr, F.~Dulat, B.~Penante, and L.~Tancredi, ``{Elliptic symbol
  calculus: from elliptic polylogarithms to iterated integrals of Eisenstein
  series},'' \href{http://dx.doi.org/10.1007/JHEP08(2018)014}{{\em JHEP} {\bf
  08} (2018)  014},
\href{http://arxiv.org/abs/1803.10256}{{\tt arXiv:1803.10256 [hep-th]}}.
%%CITATION = ARXIV:1803.10256;%%.

\bibitem{Tsunogai}
H.~Tsunogai, ``On some derivations of {L}ie algebras related to {G}alois
  representations,'' \href{http://dx.doi.org/10.2977/prims/1195164794}{{\em
  Publ. Res. Inst. Math. Sci.} {\bf 31} (1995) no.~1, 113--134}.

\bibitem{Pollack}
A.~Pollack, ``{Relations between derivations arising from modular forms}.''
  \url{https://dukespace.lib.duke.edu/dspace/handle/10161/1281}, 2009.
\newblock Undergraduate thesis, Duke University.

\bibitem{Broedel:2020tmd}
J.~Broedel, A.~Kaderli, and O.~Schlotterer, ``{Two dialects for KZB equations:
  generating one-loop open-string integrals},''
  \href{http://dx.doi.org/10.1007/JHEP12(2020)036}{{\em JHEP} {\bf 12} (2020)
  036}, \href{http://arxiv.org/abs/2007.03712}{{\tt arXiv:2007.03712
  [hep-th]}}.

\bibitem{Kaderli:2022qeu}
A.~Kaderli and C.~Rodriguez, ``{Open-string integrals with multiple
  unintegrated punctures at genus one},''
  \href{http://arxiv.org/abs/2203.09649}{{\tt arXiv:2203.09649 [hep-th]}}.

\bibitem{KZB}
D.~Calaque, B.~Enriquez, and P.~Etingof, ``Universal {KZB} equations: the
  elliptic case,'' in {\em Algebra, arithmetic, and geometry: in honor of {Y}u.
  {I}. {M}anin. {V}ol. {I}}, vol.~269 of {\em Progr. Math.}, pp.~165--266.
\newblock Birkh\"auser Boston, Inc., Boston, MA, 2009.

\bibitem{EnriquezEllAss}
B.~Enriquez, ``Elliptic associators,''
  \href{http://dx.doi.org/10.1007/s00029-013-0137-3}{{\em Selecta Math. (N.S.)}
  {\bf 20} (2014) no.~2, 491--584}.

\bibitem{Hain}
R.~Hain, ``Notes on the universal elliptic {KZB} connection,''
  \href{http://dx.doi.org/10.4310/PAMQ.2020.v16.n2.a2}{{\em Pure Appl. Math.
  Q.} {\bf 16} (2020) no.~2, 229--312},
  \href{http://arxiv.org/abs/1309.0580}{{\tt arXiv:1309.0580 [math.AG]}}.

\bibitem{Hidding:2022zzz}
M.~Hidding, ``Algorithms for the study of elliptic modular graph forms,''
  \href{http://arxiv.org/abs/(to appear)}{{\tt (to appear)}}.

\bibitem{Mafra:2019ddf}
C.~R. Mafra and O.~Schlotterer, ``{All-order alpha'-expansion of one-loop
  open-string integrals},''
  \href{http://dx.doi.org/10.1103/PhysRevLett.124.101603}{{\em Phys. Rev.
  Lett.} {\bf 124} (2020) no.~10, 101603},
\href{http://arxiv.org/abs/1908.09848}{{\tt arXiv:1908.09848 [hep-th]}}.
%%CITATION = ARXIV:1908.09848;%%.

\bibitem{Mafra:2019xms}
C.~R. Mafra and O.~Schlotterer, ``{One-loop open-string integrals from
  differential equations: all-order $\alpha$'-expansions at $n$ points},''
  \href{http://dx.doi.org/10.1007/JHEP03(2020)007}{{\em JHEP} {\bf 03} (2020)
  007},
\href{http://arxiv.org/abs/1908.10830}{{\tt arXiv:1908.10830 [hep-th]}}.
%%CITATION = ARXIV:1908.10830;%%.

\bibitem{Broedel:2019gba}
J.~Broedel and A.~Kaderli, ``{Amplitude recursions with an extra marked
  point},'' \href{http://dx.doi.org/10.4310/CNTP.2022.v16.n1.a3}{{\em Commun.
  Num. Theor. Phys.} {\bf 16} (2022) no.~1, 75--158},
  \href{http://arxiv.org/abs/1912.09927}{{\tt arXiv:1912.09927 [hep-th]}}.

\bibitem{DeligneTBP}
P.~P. Deligne, ``Le groupe fondamental de la droite projective moins trois
  points,'' in {\em Galois Groups over $\mathbb Q$}, Y.~Ihara, K.~Ribet, and
  J.-P. Serre, eds., pp.~79--297.
\newblock Springer US, New York, NY, 1989.

\bibitem{Wilhelm:2022wow}
M.~Wilhelm and C.~Zhang, ``{Symbology for elliptic multiple polylogarithms and
  the symbol prime},'' \href{http://arxiv.org/abs/2206.08378}{{\tt
  arXiv:2206.08378 [hep-th]}}.

\bibitem{Bauer:2000cp}
C.~W. Bauer, A.~Frink, and R.~Kreckel, ``{Introduction to the GiNaC framework
  for symbolic computation within the C++ programming language},''
  \href{http://dx.doi.org/10.1006/jsco.2001.0494}{{\em J. Symb. Comput.} {\bf
  33} (2002)  1--12}, \href{http://arxiv.org/abs/cs/0004015}{{\tt
  arXiv:cs/0004015}}.

\bibitem{Vollinga:2004sn}
J.~Vollinga and S.~Weinzierl, ``{Numerical evaluation of multiple
  polylogarithms},'' \href{http://dx.doi.org/10.1016/j.cpc.2004.12.009}{{\em
  Comput. Phys. Commun.} {\bf 167} (2005)  177},
  \href{http://arxiv.org/abs/hep-ph/0410259}{{\tt arXiv:hep-ph/0410259}}.

\bibitem{LNT}
J.-G. Luque, J.-C. Novelli, and J.-Y. Thibon, ``{Period polynomials and Ihara
  brackets},'' {\em J. Lie Theory} {\bf 17} (2007)  229--239,
  \href{http://arxiv.org/abs/math/0606301}{{\tt arXiv:math/0606301
  [math.CO,math.NT]}}.

\bibitem{RADFORD1979432}
D.~E. Radford, ``A natural ring basis for the shuffle algebra and an
  application to group schemes,''
  \href{http://dx.doi.org/https://doi.org/10.1016/0021-8693(79)90171-6}{{\em
  Journal of Algebra} {\bf 58} (1979) no.~2, 432--454}.
  \url{https://www.sciencedirect.com/science/article/pii/0021869379901716}.

\bibitem{Reutenauer}
C.~Reutenauer, \href{http://dx.doi.org/10.1016/S1570-7954(03)80075-X}{``Free
  {L}ie algebras,''} in {\em Handbook of algebra, {V}ol. 3}, vol.~3 of {\em
  Handb. Algebr.}, pp.~887--903.
\newblock Elsevier/North-Holland, Amsterdam, 2003.
\newblock \url{https://doi.org/10.1016/S1570-7954(03)80075-X}.

\bibitem{Britto:2021prf}
R.~Britto, S.~Mizera, C.~Rodriguez, and O.~Schlotterer, ``{Coaction and
  double-copy properties of configuration-space integrals at genus zero},''
  \href{http://dx.doi.org/10.1007/JHEP05(2021)053}{{\em JHEP} {\bf 05} (2021)
  053}, \href{http://arxiv.org/abs/2102.06206}{{\tt arXiv:2102.06206
  [hep-th]}}.

\bibitem{Panzer:2015ida}
E.~Panzer, \href{http://dx.doi.org/10.18452/17157}{{\em {Feynman integrals and
  hyperlogarithms}}}.
\newblock PhD thesis, Humboldt U., 2015.
\newblock \href{http://arxiv.org/abs/1506.07243}{{\tt arXiv:1506.07243
  [math-ph]}}.

\bibitem{Abreu:2022mfk}
S.~Abreu, R.~Britto, and C.~Duhr, ``{The SAGEX Review on Scattering Amplitudes,
  Chapter 3: Mathematical structures in Feynman integrals},''
  \href{http://arxiv.org/abs/2203.13014}{{\tt arXiv:2203.13014 [hep-th]}}.

\bibitem{Diamantis:2020}
N.~Diamantis, ``Modular iterated integrals associated with cusp forms,''
  \href{http://dx.doi.org/10.1515/forum-2021-0224}{{\em Forum Mathematicum}
  {\bf 34} (2022)  157--174}, \href{http://arxiv.org/abs/2009.07128}{{\tt
  arXiv:2009.07128 [math.NT]}}.

\bibitem{Brown2019}
F.~Brown, ``{From the Deligne-Ihara conjecture to multiple modular values},''
  \href{http://arxiv.org/abs/1904.00179}{{\tt arXiv:1904.00179 [math.AG]}}.

\bibitem{Kronecker}
L.~Kronecker, ``{Zur Theorie der elliptischen Funktionen},'' {\em Mathematische
  Werke} {\bf IV} (1881)  313--318.

\bibitem{goncharov2001multiple}
A.~B. Goncharov, ``Multiple polylogarithms and mixed tate motives,''
  \href{http://arxiv.org/abs/0103059}{{\tt arXiv:0103059 [math]}}.

\end{thebibliography}
%\bibliographystyle{utphys}

\providecommand{\href}[2]{#2}\begingroup\raggedright\endgroup

\end{document}